\documentclass[nofootinbib,floats,aps,superscriptaddress,preprint]{revtex4}
\usepackage{amssymb,amsmath,amsbsy}
\usepackage{dsfont}
\usepackage{mathrsfs}
\usepackage[title]{appendix}
\usepackage{color}
\usepackage{threeparttable}
\usepackage[hypertexnames=false,hyperfootnotes=false,pdfencoding =auto]{hyperref}
\usepackage[all]{hypcap}       
\numberwithin{equation}{section}
\renewcommand\theequation{\arabic{section}.\arabic{equation}}
\makeatletter
\g@addto@macro\bfseries{\boldmath}
\makeatother	

\newenvironment{Eqnarray}%
     {\arraycolsep 0.14em\begin{eqnarray}}{\end{eqnarray}}

\DeclareTextCommand{\directprod}{PU}{\9052\002}
\DeclareTextCommand{\blackboardZ}{PU}{\9041\044}
\DeclareTextCommand{\uppercasePi}{PU}{\83\240}
\DeclareTextCommand{\uppercasePhi}{PU}{\83\246}
\DeclareTextCommand{\textxi}{PU}{\83\276}
\DeclareTextCommand{\textbeta}{PU}{\83\262}
\DeclareTextCommand{\textlambda}{PU}{\83\273}

\let\Re\relax
\let\Im\relax
\DeclareMathOperator{\Re}{Re}
\DeclareMathOperator{\Im}{Im}
\DeclareMathOperator{\Tr}{Tr}

\newcommand{\Lower}[1]{\smash{\lower 2ex \hbox{#1}}}
\def\bed{\begin{description}}
\def\eed{\end{description}}

\def\newcdot{\kern.06em{\cdot}\kern.06em}

\def\to{\rightarrow}

\def\nn{\nonumber}

\def\bea{\begin{Eqnarray}}
\def\eea{\end{Eqnarray}}
\def\lsup#1{^{\lower 6pt\hbox{$\scriptstyle#1$}}}
\def\l2sup#1{^{\lower 4pt\hbox{$\scriptstyle#1$}}}
\def\llsup#1{^{\lower 2pt\hbox{$\scriptstyle#1$}}}

\newcommand{\ug}{\mathrm{U}(1)}
\def\T{{\mathsf T}}
\def\lt{\left}
\def\rt{\right}

\def\tm1{{\mathcal{T}}^{-1}}

\def\ttm1{\widetilde{{\mathcal{T}}}^{-1}}
\def\abar{{\bar a}}
\def\bbar{{\bar b}}
\def\cbar{{\bar c}}
\def\dbar{{\bar d}}
\def\ebar{{\bar e}}

\def\gbar{{\bar g}}

\def\lam{\lambda}

\def\ben{\begin{enumerate}}
\def\een{\end{enumerate}}
\def\beq{\begin{equation}}
\def\eeq{\end{equation}}
\def\beqa{\begin{eqnarray}}
\def\eeqa{\end{eqnarray}}

\def\ifmath#1{\relax\ifmmode #1\else $#1$\fi}
\def\lsim{\mathrel{\raise.3ex\hbox{$<$\kern-.75em\lower1ex\hbox{$\sim$}}}}
\def\gsim{\mathrel{\raise.3ex\hbox{$>$\kern-.75em\lower1ex\hbox{$\sim$}}}}

\def\eq#1{eq.~(\ref{#1})}

\def\eqs#1#2{eqs.~(\ref{#1}) and~(\ref{#2})}

\def\eqss#1#2#3{eqs.~(\ref{#1}), (\ref{#2}) and (\ref{#3})}
\def\eqst#1#2{eqs.~(\ref{#1})--(\ref{#2})}
\def\Eq#1{Eq.~(\ref{#1})}
\def\Eqst#1#2{Eqs.~(\ref{#1})--(\ref{#2})}
\def\Eqs#1#2{Eqs.~(\ref{#1}) and (\ref{#2})}
\def\Eqst#1#2{Eqs.~(\ref{#1})--(\ref{#2})}

\def\vev#1{\langle #1 \rangle}

\def\cbma{c_{\beta-\alpha}}

\def\sbma{s_{\beta-\alpha}}
\def\ctwob{c_{2\beta}}
\def\stwob{s_{2\beta}}

\def\phm{\phantom{-}}

\def\beq{\begin{equation}}
\def\eeq{\end{equation}}
\def\ifmath#1{\relax\ifmmode #1\else $#1$\fi}

\def\stwob  {s_{2\beta}}
\def\ctwob  {c_{2\beta}}

\def\hl{h}
\def\ha{A}
\def\hh{H}
\def\hpm{{H^\pm}}

\def\mha{m_{\ha}}
\def\mhl{m_{\hl}}
\def\mhh{m_{\hh}}
\def\mhpm{m_{\hpm}}

\def\ls#1{\ifmath{_{\lower1.5pt\hbox{$\scriptstyle #1$}}}}
\def\lss#1{\ifmath{^{\,\lower2.5pt\hbox{$\scriptstyle #1$}}}}

\def\half{\ifmath{{\textstyle{\frac{1}{2}}}}}

\def\pht{\phantom{i}}

\begin{document}

\preprint{CFTP/21-001 \cr
SCIPP-21/01\phantom{,} \cr
February, 2021\phantom{-}}

\vspace*{-3cm}

\title{Exceptional regions of the 2HDM parameter space}
\author{\mbox{Howard E.~Haber}}
 \email[E-mail: ]{haber@scipp.ucsc.edu}
 \affiliation{Santa Cruz Institute for Particle Physics,
   University of California, Santa Cruz, California 95064, USA}
  \author{Jo\~ao P.~Silva}
  \email[E-mail: ]{jpsilva@cftp.ist.utl.pt}
 \affiliation{CFTP, Departamento de F\'{\i}sica, Instituto Superior T\'{e}cnico, Universidade de Lisboa, Avenida Rovisco Pais 1, Lisboa 1049, Portugal}

\vspace*{4cm}

\begin{abstract}
  
The exceptional region of the parameter space (ERPS) of the two Higgs doublet model (2HDM) is defined to be the parameter regime where
the scalar potential takes on a very special form.   In the standard parametrization of the 2HDM scalar potential with squared mass parameters $m_{11}^2$, $m_{22}^2$, $m_{12}^2$, and
dimensionless couplings, $\lambda_1$, $\lambda_2$, $\ldots,\lambda_7$, the ERPS corresponds to $\lambda_1=\lambda_2$, $\lambda_7=-\lambda_6$, 
$m_{11}^2=m_{22}^2$ and $m_{12}^2=0$, corresponding to a scalar potential with an enhanced generalized CP symmetry called GCP2.   
Many special features persist if  $\lambda_1=\lambda_2$ and $\lambda_7=-\lambda_6$ are retained while allowing for $m_{11}^2\neq m_{22}^2$ and/or $m_{12}^2\neq 0$,
corresponding to a scalar potential with a softly broken GCP2 symmetry, which we designate as the ERPS4.
In this paper, we examine many of the special features
of the ERPS4, as well
as even more specialized cases within the ERPS4 framework in which additional constraints on the scalar potential parameters are imposed.
By surveying the landscape of the ERPS4,  we complete the classification of 2HDM scalar potentials that
exhibit an exact
Higgs alignment (where the tree-level couplings of one neutral scalar coincide with those of the Standard Model Higgs boson), due to a residual symmetry that is unbroken in the vacuum.
One surprising aspect of the ERPS4 is the possibility that the scalar sector is CP-conserving despite the presence of a complex parameter of the scalar potential
whose complex phase cannot be removed by separate rephasings of the two scalar doublet fields.
The significance of the ERPS4 regime for custodial symmetry is also discussed, and
the cases where a custodial symmetric 2HDM scalar potential preserves an exact Higgs alignment are elucidated.

\end{abstract}
\maketitle

\section{Introduction}

After the discovery of the Higgs boson at the LHC~\cite{Aad:2012tfa,Chatrchyan:2012ufa}, the ATLAS and CMS Collaborations have ascertained that
the observed properties of the Higgs boson are consistent with the corresponding predictions of the Standard Model (SM). 
Various production mechanisms and decay channels have been detected, and many of the observed signal strengths are consistent with SM
expectations given the current precision of the LHC Higgs data, typically in the range of 10\%--20\% depending on the final state observable~\cite{Aad:2019mbh,Sirunyan:2018koj,CMS:2020gsy,ATLAS:2020qdt}. 

Can it be true that the scalar sector of the SM consists of a single spin-0 boson?  In light of the nonminimal nature of the SM fermions (which
consist of three generations of quarks and leptons) as well as the nonminimal nature of the SM gauge group [which is the direct product of two nonabelian
groups and the weak hypercharge U(1)$_{\rm Y}$], it would be surprising if the scalar sector did not possess a nonminimal structure as well.  Extended Higgs sectors
have been proposed and explored in the literature since the birth of the Standard Model.   Indeed, an important part of the LHC Higgs program is to search for
the existence of new scalar states related to the observed Higgs boson, and to study their properties if found.

Of course, an arbitrary extended Higgs sector can in many cases be ruled out by current experimental data. The observed electroweak $\rho$ parameter \cite{Ross:1975fq,Veltman:1977kh,Chanowitz:1978uj,Toussaint:1978zm},
which is close to 1, and the absence of tree-level Higgs-mediated flavor changing neutral currents (FCNCs) that otherwise would lead to observable FCNC effects, in
conflict with current experimental bounds, impose significant constraints on any theory with an extended Higgs sector.
The two Higgs doublet model (2HDM), which is one of the simplest extensions of the SM, possesses two scalar doublet fields $\Phi_1$ and $\Phi_2$
\cite{Lee:1973iz,Gunion:1989we,Branco:2011iw}, each with the
same hypercharge $Y=1$ (in a convention where the electric charge is given by $Q=T_3+\half Y$).  
Nearly all of the new scalar physics phenomena expected in theories
of extended Higgs models can be found in the 2HDM---charged scalars, CP-odd scalars (in models with a CP-conserving scalar sector) and/or scalars of indefinite
CP quantum numbers (in models with a CP-violating scalar sector).   Moreover, the 2HDM predicts a tree-level value of $\rho=1$ and 
is also compatible with the absence of tree-level Higgs-mediated FCNCs with a suitably chosen Higgs-fermion Yukawa interaction \cite{Glashow:1976nt,Paschos:1976ay}.\footnote{Radiative corrections
to the predicted value of $\rho$ and the size of Higgs-mediated FCNCs impose some interesting constraints on the 2HDM parameter space (e.g., see Refs.~\cite{WahabElKaffas:2007xd,Arbey:2017gmh,Eberhardt:2020dat}).} 
Finally, the 2HDM has often been employed in theories that introduce physics beyond the SM to solve other conceptual problems 
of the Standard Model.  Two well-known examples are the minimal supersymmetric extension of the Standard Model (MSSM), which employs the Higgs sector of the
2HDM~\cite{Fayet:1974pd,Flores:1982pr,Haber:1984rc,Gunion:1984yn} and has been used to provide an explanation of the origin of the energy scale of electroweak symmetry breaking~\cite{Susskind:1982mw}, and the inert doublet model (IDM) \cite{Barbieri:2006dq,LopezHonorez:2006gr}, which is
a special case of the 2HDM that provides a candidate for dark matter~\cite{Ma:2006km,Arina:2009um,Goudelis:2013uca}.

Despite the simplicity of the 2HDM, the corresponding scalar sector in its most general form is governed by 11 independent parameters~\cite{Davidson:2005cw}.  However, additional theoretical
assumptions can be brought to bear to reduce this large number of parameters.  For example, to avoid the presence of tree-level Higgs mediated FCNCs in a natural
way (i.e.~without a fine-tuning of parameters in the Yukawa Lagrangian), one can introduce a $\mathbb{Z}_2$ discrete symmetry under which one of the Higgs doublet
fields is even and the other is odd.  An appropriate assignment of $\mathbb{Z}_2$ quantum numbers to the fermion fields then provides a symmetry explanation for the absence
of Higgs-mediated tree-level FCNCs \cite{Glashow:1976nt,Paschos:1976ay}.  In fact, this result is robust even in the presence of a soft breaking of the $\mathbb{Z}_2$ symmetry by squared mass parameters
appearing in the scalar potential.  The softly broken $\mathbb{Z}_2$-symmetric 2HDM is governed by nine independent parameters and is called the complex 2HDM (C2HDM)~\cite{Ginzburg:2004vp,Arhrib:2010ju,Barroso:2012wz,Inoue:2014nva,Fontes:2014xva,Grzadkowski:2014ada,Fontes:2017zfn,Boto:2020wyf,Low:2020iua}.   

There is some motivation to try to reduce the parameter count even further.   For example, imposing CP invariance on the scalar potential \cite{Lee:1973iz} would reduce the number of
independent parameters to eight, corresponding to the vacuum expectation value, $v\simeq 246$~GeV, four scalar masses, two real angles, and one scalar self-coupling.
As another example, consider the requirement that one of the scalar states of the 2HDM should resemble the SM Higgs boson.  One way of achieving
this result is to posit an additional symmetry of the scalar potential, which would further reduce the number of independent scalar sector parameters~\cite{Dev:2014yca,Dev:2017org}.

The exceptional region of the parameter space of the 2HDM, first introduced in Ref.~\cite{Davidson:2005cw},
and designated by the acronym ERPS in Ref.~\cite{Ferreira:2009wh}, corresponds to a special parameter regime in which 
the coefficients of $(\Phi_1^\dagger\Phi_1)^2$ and $(\Phi_2^\dagger\Phi_2)^2$ appearing in the 2HDM scalar potential
are set equal and the coefficient of $(\Phi_1^\dagger\Phi_2)(\Phi_1^\dagger\Phi_1)$ is the negative of the 
coefficient of $(\Phi_1^\dagger\Phi_2)(\Phi_2^\dagger\Phi_2)$.  In addition, the squared mass coefficients of $\Phi_1^\dagger\Phi_1$ and $\Phi_2^\dagger\Phi_2$ are set equal and
the squared mass coefficient of $\Phi_1^\dagger\Phi_2+{\rm h.c.}$ is set to zero.   The number of free parameters of the ERPS is five, consisting of $v$ and the four scalar masses.
The ERPS conditions can be enforced by a global symmetry.
Allowing for the conditions on the quadratic terms of the scalar potential to be relaxed, which would constitute a soft breaking of the global symmetry, still yields a rather exceptional region of the 2HDM
parameter space, which we shall henceforth denote as the ERPS4 in order to emphasize that the global symmetry of the ERPS is still respected by the dimension-four terms of the
scalar potential.  

The ERPS4 is governed by eight parameters in its most general form, and the corresponding scalar potential is explicitly CP-violating.  If in addition one imposes a CP symmetry on the scalar potential (which may or may not be violated by the vacuum), the number of parameters is reduced by one.  One can identify the seven parameters as $v$, four scalar masses, one real angle and one scalar self-coupling.   One may also impose additional softly broken symmetries within the class of the ERPS4 scalar potentials, which yields a subset of the ERPS4 with additional exceptional features.   All scalar potentials obtained in this way automatically possess a CP-conserving scalar potential and vacuum.  The 2HDM employed in the MSSM provides one such example.

It is worth highlighting a number of the exceptional features of scalar potentials that reside within the ERPS4.  
 First, in contrast to a generic 2HDM, if the conditions on the scalar potential parameters that define the ERPS4 hold in one scalar field basis,\footnote{One is always free to change the scalar field basis by redefining the scalar fields, $\Phi_a\to U_{a\bbar}\Phi_b$ (summed over $b=1,2$ following the index notation of Ref.~\cite{Davidson:2005cw}), where $U$ is an arbitrary U(2) matrix.   A particular choice for $\Phi_1$ and $\Phi_2$ is called a choice of scalar field basis.  In a generic 2HDM, the squared mass coefficients and dimensionless quartic coefficients that appear in the scalar potential will be transformed by a change of scalar field basis. Often, relations among parameters that are valid in one basis cease to be valid in a different basis. The ERPS4 conditions are notable in that
 they hold in \textit{all} scalar field bases.} 
 then they are satisfied in \textit{all} scalar field bases.  
 
 Second, in a softly broken $\mathbb{Z}_2$-symmetric 2HDM, there are two potentially complex coefficients of the scalar potential, denoted by $m_{12}^2$ and $\lambda_5$ in \eq{lambdapotential}, since the other two complex coefficients in \eq{lambdapotential} are $\lambda_6=\lambda_7=0$ as a consequence of the $\mathbb{Z}_2$ symmetry.   Generically, one finds that the scalar potential is explicitly CP conserving if and only if $\Im(\lambda_5^*[m_{12}^2]^2)=0$, since the latter condition implies that one can rephase the scalar fields $\Phi_1$ and $\Phi_2$ to remove the complex phases of $m_{12}^2$ and $\lambda_5$.  The resulting scalar potential is then invariant with respect to the CP transformation, $\Phi\to\Phi^*$.  
 Remarkably, the ``only if'' part of this statement is no longer true in the ERPS4.  We find that in the special case of $\vev{\Phi_1^0}=\vev{\Phi_2^0}$, the scalar potential is explicitly CP conserving despite the fact that $\Im(\lambda_5^*[m_{12}^2]^2)\neq 0$.  Indeed, we can identify the modified definition of CP that governs the ERPS4 in this special case.   Moreover, if we constrain the ERPS4 by adding an additional softly broken global symmetry, we find that the ``only if'' part of the original statement is no longer true independently of the scalar field vacuum expectation values.   Once again, we can understand this behavior by identifying an appropriately redefined generalized CP transformation law (as shown explicitly in Appendix~\ref{app:cp}).

Third, the Higgs alignment limit~\cite{Gunion:2002zf,Craig:2012vn,Craig:2013hca,Asner:2013psa,Carena:2013ooa,Haber:2013mia}
(in which the tree-level properties of one neutral scalar coincide with those of the SM Higgs boson)
can be achieved by imposing a symmetry on the scalar potential~\cite{Dev:2014yca,Dev:2017org}.    For example, exact 
Higgs alignment is realized in the IDM, where the
scalar potential and vacuum both respect a $\mathbb{Z}_2$ symmetry.
It is of interest to classify all possible symmetries of the scalar potential beyond the $\mathbb{Z}_2$ symmetry of the IDM in which the Higgs alignment is exact.   All scalar potentials of the ERPS fall within this class.  But, one can also maintain exact Higgs alignment in some cases in which the symmetry is softly broken, corresponding to the ERPS4.   Including these cases completes the classification of all symmetry based explanations for exact Higgs alignment in the 2HDM.

Fourth, it is known that custodial symmetry is an accidental symmetry of the SM Higgs potential~\cite{Sikivie:1980hm,Mannheim:1983ti}.  In the 2HDM, the custodial symmetry is an accidental symmetry of the scalar potential if an additional constraint is imposed~\cite{Pomarol:1993mu,Gerard:2007kn,Grzadkowski:2010dj,Haber:2010bw,Aiko:2020atr}.   A custodial symmetric 2HDM scalar potential is automatically CP conserving.   Additional accidental symmetries can arise in special regions of the parameter space.
Of particular interest is the case of a custodial symmetric scalar potential that preserves an exact Higgs alignment.  Indeed, with two exceptions, the resulting scalar potential is necessarily in the ERPS4 regime.

In Section~\ref{softzeetwo}, we introduce the 2HDM with a softly broken $\mathbb{Z}_2$-symmetric scalar potential.  
The possible enhanced global symmetries of the 2HDM scalar potential beyond the $\mathbb{Z}_2$ symmetry are summarized in Section~\ref{enhanced}, and their connections to the ERPS are exhibited in Section~\ref{sec:erps}.  In this section, we provide a set of basis-independent conditions that correspond to the ERPS4 and special subregions of the ERPS4 where additional global symmetries (perhaps softly broken) are imposed.

A convenient scalar field basis for the ERPS4 is one where the softly broken $\mathbb{Z}_2$ symmetry and a softly broken permutation symmetry (that interchanges $\Phi_1\leftrightarrow\Phi_2$) are simultaneously imposed.   We examine the properties of the resulting scalar sector in Section~\ref{zee2pi2}
and note that for generic choices of the parameters, the scalar potential is CP-violating.   If the corresponding scalar potential is explicitly CP conserving, then CP may or may not be spontaneously broken by the vacuum.   The CP-conserving ERPS4 is examined in detail and we exhibit the special parameter regime where CP is conserved, despite the fact that a simple rephasing of $\Phi_1$ and $\Phi_2$ is not sufficient to produce a scalar potential whose parameters are all real.  In Section~\ref{sec:UPibasis}, we extend the softly broken $\mathbb{Z}_2$ symmetry to U(1) and explore the properties of this special subregion of the ERPS4.  One can show that the corresponding scalar potential respects a generalized CP symmetry (denoted by GCP3) when expressed in a different scalar field basis.   The implications of the scalar potential when expressed in terms of the GCP3 basis of scalar fields are exhibited in Section~\ref{sec:GCP3basis} and the relations between the scalar potential parameters in the two different basis choices is made explicit in Section~\ref{transforming}.

As noted above, exact Higgs alignment is realized in the ERPS.  If soft-symmetry breaking squared mass terms are included, the resulting ERPS4 may or may not exhibit exact Higgs alignment.  In Section~\ref{alignment}, we provide a complete classification of the symmetries (which in some cases is softly broken) that naturally yield a neutral scalar mass eigenstate whose tree-level properties are identical to those of the SM Higgs boson.  In Section~\ref{custodial}, we combine exact Higgs alignment with the constraint of custodial symmetry and exhibit 
the implications for the ERPS4 regime.
Conclusions and future directions appear in Section~\ref{conclude}, followed by five appendices that provide additional details on the consequences of the ERPS4 for CP symmetry and other related matters.

\section{2HDM with a softly broken \texorpdfstring{$\mathbb{Z}_2$}{\blackboardZ\texttwoinferior}-symmetric scalar potential}
\label{softzeetwo}

Let $\Phi_1$ and $\Phi_2$ denote two complex $Y=1$, SU(2)$\ls{L}$ doublet scalar fields.
The most general gauge invariant renormalizable scalar potential (in the $\Phi$-basis) is given by
\beqa
\label{lambdapotential}
\mathcal{V}&=& m_{11}^2\Phi_1^\dagger\Phi_1+m_{22}^2\Phi_2^\dagger\Phi_2
-[m_{12}^2\Phi_1^\dagger\Phi_2+{\rm h.c.}]
 +\half\lambda_1(\Phi_1^\dagger\Phi_1)^2
+\half\lambda_2(\Phi_2^\dagger\Phi_2)^2
+\lambda_3(\Phi_1^\dagger\Phi_1)(\Phi_2^\dagger\Phi_2)\nonumber\\[8pt]
&&\qquad\quad\,\,
+\lambda_4(\Phi_1^\dagger\Phi_2)(\Phi_2^\dagger\Phi_1)
+\left\{\half\lambda_5(\Phi_1^\dagger\Phi_2)^2
+\big[\lambda_6(\Phi_1^\dagger\Phi_1)
+\lambda_7(\Phi_2^\dagger\Phi_2)\big]
\Phi_1^\dagger\Phi_2+{\rm h.c.}\right\}\,. \label{pot}
\eeqa
In general, $m_{12}^2$, $\lambda_5$, $\lambda_6$ and $\lambda_7$ can be complex.
To avoid tree-level Higgs-mediated FCNCs, we shall impose a softly broken discrete
$\mathbb{Z}_2$ symmetry, $\Phi_1\to +\Phi_1$ and $\Phi_2\to -\Phi_2$
on the quartic terms of \eq{pot}, which implies that $\lambda_6=\lambda_7=0$, whereas $m_{12}^2\neq 0$ is allowed.  
In this basis of scalar doublet fields (denoted as the $\mathbb{Z}_2$ basis), 
the discrete $\mathbb{Z}_2$ symmetry of the quartic terms of \eq{lambdapotential} is manifestly realized.
In the $\mathbb{Z}_2$ basis, it is convenient to rephase the scalar fields such that
$\lambda_5$ is real.   Then, the requirement that $\mathcal{V}$ is bounded from below yields the following conditions~\cite{Deshpande:1977rw,Maniatis:2006fs},\footnote{If \eq{bb} is satisfied then the softly broken $\mathbb{Z}_2$-symmetric tree-level scalar potential is said to be stable in the strong sense.   If we replace $>$ with $\geq$ in one of the above inequalities (corresponding to a particular direction in field space) and then impose the condition that the sum of the quadratic terms in the same field direction is strictly positive, then the scalar potential is said to be stable in the weak sense~\cite{Maniatis:2006fs}. \label{fnstable}}
\beq \label{bb}
\lambda_1>0\,,\qquad \lambda_2>0\,,\qquad \lambda_3>-(\lambda_1\lambda_2)^{1/2}\,,\qquad \lambda_3+\lambda_4\pm\lambda_5>-(\lambda_1\lambda_2)^{1/2}\,.
\eeq

The scalar fields will
develop non-zero vacuum expectation values (vevs) if the Higgs mass matrix
$m_{ij}^2$ has at least one negative eigenvalue.
We assume that the
parameters of the scalar potential are chosen such that
the minimum of the scalar potential respects the
U(1)$\ls{\rm EM}$ gauge symmetry.  Then, the scalar field
vevs are of the form
\beq
\langle \Phi_1 \rangle=\frac{v}{\sqrt{2}} \left(
\begin{array}{c} 0\\ c_\beta \end{array}\right), \qquad \langle
\Phi_2\rangle=
\frac{v}{\sqrt{2}}\left(\begin{array}{c}0\\  e^{i\xi} s_\beta
\end{array}\right)\,,\label{potmin}
\eeq
where $c_\beta\equiv\cos\beta=v_1/v$ and $s_\beta\equiv\sin\beta=v_2/v$ with $v\equiv (v_1^2+v_2^2)^{1/2}\simeq 246~{\rm GeV}$.   By convention,
$0\leq\beta\leq\half\pi$ and $0\leq\xi<2\pi$.   

The parameters 
$v_1$, $v_2$ and $\xi$ are determined by minimizing the scalar potential.   The resulting minimization conditions in the case of $\lambda_6=\lambda_7=0$ and real $\lambda_5$ are given by,
\beqa
m_{11}^2 v_1&=& \Re(m_{12}^2 e^{i\xi})v_2-\half\lambda_1 v_1^3-\half\lambda_{345} v_1 v_2^2\,,\label{minone} \\
m_{22}^2 v_2&=& \Re(m_{12}^2 e^{i\xi})v_1-\half\lambda_2 v_2^3-\half\lambda_{345} v_2 v_1^2\,,\label{mintwo} \\
\Im(m_{12}^2 e^{i\xi})v_1&=& \half\lambda_5 v_1^2 v_2\sin 2\xi\,,  \label{minthree}\\
\Im(m_{12}^2 e^{i\xi})v_2&=& \half\lambda_5 v_2^2 v_1\sin 2\xi\,, \label{minfour}
\eeqa
where $\lambda_{345}\equiv\lambda_3+\lambda_4+\lambda_5\cos 2\xi$.
Note that both \eqs{minthree}{minfour} are provided in case one of the vevs vanishes.
If both $v_1\neq 0$ and $v_2\neq 0$, then the minimization conditions simplify,
\beqa
m_{11}^2 &=& \Re(m_{12}^2 e^{i\xi})\tan\beta-\half\lambda_1 v^2 c^2_\beta-\half\lambda_{345} v^2 s^2_\beta\,,\label{min1} \\
m_{22}^2 &=& \Re(m_{12}^2 e^{i\xi})\cot\beta-\half\lambda_2 v^2  s_\beta^2-\half\lambda_{345} v^2 c_\beta^2\,, \label{min2} \\
\Im(m_{12}^2 e^{i\xi})&=& \half\lambda_5 v^2 s_\beta c_\beta\sin 2\xi\,.\label{min3}
\eeqa
The value of the potential at the minimum is given by,
\begin{align}
V_{\rm min}=&\half v^2\bigl[m_{11}^2 c_\beta^2+m_{22}^2 s_\beta^2-2\Re(m_{12}^2 e^{i\xi}) s_{\beta}c_{\beta}+\tfrac14 \lambda_1 v^2 c_\beta^4+\tfrac14\lambda_2 v^2 s_\beta^4 +\half \lambda_{345} v^2 s_\beta^2 c_\beta^2\bigr]\nonumber \\
=& -\tfrac18 v^4\bigl[\lambda_1 c_\beta^4+ \lambda_2 s_\beta^4+2\lambda_{345}s_\beta^2 c_\beta^2\bigr]\,, \label{minV}
\end{align}
after making use of \eqs{min1}{min2}.  
In light of \eq{bb}, $V_{\rm min}<0$, which means that the extremum with $v_1=v_2=0$ always is less favorable than the asymmetric minimum, assuming that there is a solution to \eqst{minone}{minfour} with nonvanishing vevs.

If one of the two vevs vanishes, then the minimization conditions are given by
\beqa
m_{12}^2 &=&0\,, \qquad m_{22}^2=-\half \lambda_2 v^2\,,\qquad \text{if $v_1=0$ and $v_2=v$}, \label{inert1} \\
m_{12}^2 &=&0\,, \qquad m_{11}^2=-\half \lambda_1 v^2\,,\qquad \text{if $v_2=0$ and $v_1=v$}.\label{inert2}
\eeqa
This corresponds to an inert phase in which there exists a $\mathbb{Z}_2$ symmetry that is respected both by the scalar potential and the vacuum.  This phase exists if and only if $m_{12}^2=0$ and $m_{22}^2<0$ [$m_{11}^2<0$] in the case of $v_1=0$ [$v_2=0$].   These two cases are physically equivalent, as they are related by a basis change where $\Phi_1\leftrightarrow\Phi_2$.   The inert phase is stable if all the physical scalar squared masses are non-negative.

\section{Enhanced global symmetries of the scalar potential}
\label{enhanced}

The possible global symmetries of the 2HDM scalar potential have been classified in Refs.~\cite{Ivanov:2007de,Ferreira:2009wh,Ferreira:2010yh,Battye:2011jj}.
Starting from a generic $\Phi$-basis, 
these symmetries fall into two separate categories: (i) Higgs family symmetries of the form $\Phi_a\to U_{ab}\Phi_b$, and (ii) Generalized CP (GCP) symmetries of the form  $\Phi_a\to U_{ab}\Phi_b^*$, where $U$ resides in a subgroup (either discrete or continuous) of U(2).  Although it might appear that the number of possible symmetries is quite large, it turns out that different choices of $U$ often yield the same constraints on the 2HDM scalar potential parameters.  

The full global U(2) Higgs family symmetry transformation is the largest global symmetry group under which the gauge covariant kinetic terms of the scalar fields are invariant.
Moreover, the scalar potential is invariant under a global hypercharge transformation, U(1)$_{\rm Y}$, which is a subgroup of U(2).  Thus, any enhanced Higgs family symmetries that are respected by the scalar potential would be a subset of the U(2) transformations that do not contain U(1)$_{\rm Y}$ as a subgroup.   We summarize below possible discrete and continuous Higgs family symmetries modulo the
U(1)$_{\rm Y}$ hypercharge symmetry that can impose constraints on the 2HDM scalar potential in Tables~\ref{tab:symm1} and \ref{tab:symm2}.

\begin{table}[t!]
\begin{tabular}{|ll|}
\hline
\pht  symmetry &\phantom{xxxxx} transformation law \\
\hline
\pht $\mathbb{Z}_2$ &
$\Phi_1 \rightarrow \Phi_1$,
\hspace{7.9ex}
$\Phi_2 \rightarrow  -\Phi_2$ \\
\pht  $\Pi_2$\,\,\, ({\rm mirror symmetry})
&
$\Phi_1 \longleftrightarrow \Phi_2$ \\
\pht U(1)$_{\rm PQ}$\,\,\,({\rm Peccei-Quinn symmetry~\cite{Peccei:1977hh}}) & 
$\Phi_1 \rightarrow e^{-i \theta} \Phi_1$,
\hspace{4ex}
$\Phi_2 \rightarrow e^{i \theta} \Phi_2$, \hspace{4ex} for $-\half\pi<\theta\leq\half\pi$ \pht \\
\pht SO(3)\,\,\,({\rm maximal Higgs flavor symmetry)} \pht\pht& $\Phi_a\to U_{ab}\Phi_b$\,,\qquad 
\,\, where $U\in{\rm U(2)}/{\rm U(1)}_{\rm Y}$ \\
\hline
\end{tabular}
\caption{\small \baselineskip=15pt
Classification of the Higgs family symmetries of the scalar potential in a generic $\Phi$-basis where the symmetries are manifestly realized~\cite{Ivanov:2007de,Ferreira:2009wh,Ferreira:2010yh,Battye:2011jj}.  Note that $\mathbb{Z}_2$ is a subgroup of U(1)$_{\rm PQ}$.  The corresponding constraints on the 2HDM scalar potential parameters are shown in Table~\ref{tab:class}. 
\label{tab:symm1}}
\end{table}
\begin{table}[ht!]
\begin{tabular}{|ll|}
\hline
\pht symmetry & \phantom{xxxxx}transformation law \\
\hline
\pht GCP1 &
$\Phi_1 \rightarrow \Phi_1^*$,
\hspace{7.5ex}
$\Phi_2 \rightarrow \Phi_2^*$ \\
\pht GCP2 &
$\Phi_1 \rightarrow \Phi_2^*$,
\hspace{7.5ex}
$\Phi_2 \rightarrow -\Phi_1^*$ \\
\pht GCP3 &
$\begin{cases} \Phi_1 \rightarrow \Phi_1^*\cos\theta+\Phi_2^*\sin\theta & \\
\Phi_2 \rightarrow -\Phi_1^*\sin\theta+\Phi_2^*\cos\theta \end{cases}$\qquad \text{for $0<\theta<\half\pi$} \\
\hline
\end{tabular}
\caption{\small \baselineskip=15pt
Classification of the generalized CP (GCP) symmetries of the scalar potential in the $\Phi$-basis~\cite{Ivanov:2007de,Ferreira:2009wh,Ferreira:2010yh,Battye:2011jj}.  Note that a GCP3 symmetry transformation with \text{any} value of $\theta$ that lies between $0$ and $\half\pi$ yields the same constrained 2HDM scalar potential.   The corresponding constraints on the 2HDM scalar potential parameters are shown in Table~\ref{tab:class}. 
\label{tab:symm2}}
\end{table}
\begin{table}[ht!]
\begin{tabular}{|cccccccccccc|}
\hline
\pht symmetry &  $\phantom{m_{11}^2}$ & $m_{22}^2$ &\quad $m_{12}^2$ \quad & $\phantom{\lambda_1}$ &
 $\lambda_2$ & $\phantom{\lambda_3}$ & $\lambda_4$ &
$\Re\lambda_5$ &\pht  $\Im\lambda_5$  \pht & $\lambda_6$\quad  & \quad $\lambda_7$ \qquad \\
\hline
$\mathbb{Z}_2$ & &   & $0$
   &  &  &  & & &
   & $0$ & $\phm 0$ \\
$\Pi_2$  &  &$ m_{11}^2$ &\pht  real \pht &&
    $ \lambda_1$ & &  &  
 & $0$ &  & $\phm\lambda_6^\ast$
\\
$\mathbb{Z}_2\otimes\Pi_2$ & & $m_{11}^2$ & $0$ && $\lambda_1$ &&&  & $0$ &  $0$ & $\phm 0$
\\
U(1) & &  & $0$ 
 &  & &  & &
$0$ & $0$ & $0$ &  $\phm 0$ \\
U(1)$\otimes\Pi_2$  & & $m_{11}^2$ & $0$ && $\lambda_1$ &&&  $0$ & $0$ & $0$ & $\phm 0$
\\
SO(3)  && $ m_{11}^2$ & $0$
   && $\lambda_1$ &  & \pht $\lambda_1 - \lambda_3$\pht  &
$0$ & $0$ & $0$ & $\phm 0$ \\
GCP1   & & & real
 & &  &  &&
& $0$ &\pht  real \pht & \pht real \pht \\
GCP2   && $m_{11}^2$ & $0$
  && $\lambda_1$ &  &  &
&   &  & $- \lambda_6\pht$ \\
GCP3   && $m_{11}^2$ & $0$
   && $\lambda_1$ &  &  &
\pht  $\lambda_1 - \lambda_3 - \lambda_4$\pht  & $0$ & $0$ & $\phm 0$ \\
\hline
\end{tabular}
\caption{\small \baselineskip=15pt
Classification of 2HDM scalar potential symmetries and their impact on the parameters of the scalar potential [cf.~\eq{pot}] in the $\Phi$-basis~\cite{Ivanov:2007de,Ferreira:2009wh,Ferreira:2010yh,Battye:2011jj}.
Empty entries correspond to a lack of constraints on the corresponding parameters. Note that $\Pi_2$, $\mathbb{Z}_2\otimes\Pi_2$ and
U(1)$\otimes\Pi_2$ are
 not independent from other symmetry conditions, since a change of scalar field basis can be performed in each case to yield a new basis in which the $\mathbb{Z}_2$, GCP2 and GCP3 symmetries, respectively, are manifestly realized. 
\label{tab:class}}
\end{table}

Note that the list of symmetries in Table~\ref{tab:symm1} contains a redundancy.  Although it might appear that the $\mathbb{Z}_2$ and $\Pi_2$ discrete symmetries are distinct (as they yield different constraints on the 2HDM scalar potential parameters in the $\Phi$-basis), one can show that 
starting from a $\Pi_2$-symmetric scalar potential, one can find a different basis of scalar fields in which the corresponding scalar potential manifestly exhibits the $\mathbb{Z}_2$ symmetry, and vice versa~\cite{Davidson:2005cw}.  In Table~\ref{tab:class},  the constraints of the various possible Higgs family symmetries and GCP symmetries on the 2HDM scalar potential in a 
generic $\Phi$-basis are exhibited.  In the list of symmetries, U(1) corresponds to U(1)$_{\rm PQ}$ (henceforth, we shall suppress the PQ subscript).

One can also consider applying two of the symmetries listed above simultaneously in the same basis.  It was shown in Refs.~\cite{Ivanov:2007de,Ferreira:2009wh} that no new independent models arise in this way.  For example, applying $\mathbb{Z}_2$ and $\Pi_2$ in the same basis yields a $\mathbb{Z}_2\otimes \Pi_2$-symmetric scalar potential that is equivalent to GCP2 when expressed in a different basis.
Similarly, applying U(1)
and~$\Pi_2$ in the same basis yields a U(1)$\otimes\Pi_2$-symmetric scalar potential that is equivalent to a GCP3-symmetric scalar potential when expressed in a different basis.  This equivalence of GCP3 and U(1)$\otimes\Pi_2$ is explicitly demonstrated in Section~\ref{transforming}.

There are a number of additional Higgs family symmetries and generalized CP symmetries that are closely related to the ones displayed in Tables~\ref{tab:symm1} and \ref{tab:symm2}
that will be useful in our work.   In Tables~\ref{tab:symm3} and \ref{tab:symm4}, we have listed three additional Higgs family symmetries and two additional GCP symmetries that can be used to constrain the parameters of the 2HDM scalar potential.
The corresponding constraints are exhibited in Table~\ref{tab:class2}.  Given scalar potentials where $\Pi_2$, U(1)$^\prime$ and GCP3 symmetries are manifestly realized, the basis change 
\beq \label{simplebasischange}
\Phi_1\to\Phi_1, \qquad \Phi_2\to i\Phi_2\,,
\eeq
yields $m_{12}^2\to im_{12}^2$, $\lambda_5\to -\lambda_5$, $\lambda_6\to i\lambda_6$ and $\lambda_7\to i\lambda_7$, and
produces a scalar potential where $\Pi_2^\prime$, U(1)$^{\prime\prime}$ and GCP3$^\prime$ symmetries, respectively, are manifestly realized.

The origin of U(1)$^\prime$ is slightly more subtle and is derived in Section~\ref{transforming}.  It arises in the following way.   We have noted above that the $\ug\otimes\Pi_2$ and GCP3 symmetries are equivalent in the sense that the scalar field bases in which these symmetries are manifestly realized are related by a change in basis.    Moreover, 
as shown in Section~\ref{transforming}, by transforming from the $\ug\otimes\Pi_2$ basis to the GCP3 basis, the U(1) symmetry constraints are mapped onto the U(1)$^\prime$ symmetry constraints.

\begin{table}[t!]
\begin{tabular}{|llc|}
\hline
\pht symmetry\phantom{xxxxx}  &\phantom{xxxxxxxxxxxxx} transformation law & related symmetry \pht \\
\hline
\pht $\Pi^\prime_2$ &
$\Phi_1 \rightarrow \Phi_2$,
\hspace{20ex}
$\Phi_2 \rightarrow  -\Phi_1$ & $\Pi_2$ \\
\pht U(1)$^\prime$ & 
$\Phi_1 \rightarrow \Phi_1\cos\theta +\Phi_2\sin\theta$,
\hspace{4.7ex}
$\Phi_2 \rightarrow -\Phi_1\sin\theta+\Phi_2\cos\theta$  & U(1)$_{\rm PQ}$\\
\pht U(1)$^{\prime\prime}$ & 
$\Phi_1 \rightarrow \Phi_1\cos\theta+ i\Phi_2\sin\theta$,
\hspace{4ex}
$\Phi_2 \rightarrow i\Phi_1\sin\theta+\Phi_2\cos\theta$  & U(1)$_{\rm PQ}$\\
\hline
\end{tabular}
\caption{\small \baselineskip=15pt
2HDM scalar potential Higgs family symmetries in a generic $\Phi$-basis that are related by a simple change of basis to the family symmetries of
Table~\ref{tab:symm1}.  As in the case of the Peccei-Quinn symmetry, a scalar potential that respects the U(1) symmetries above must be invariant for any value
of $-\half\pi<\theta\leq\half\pi$.  Note that $\Pi_2^\prime$ [$\Pi_2$] is a subgroup of U(1)$^\prime$ [U(1)$^{\prime\prime}$], respectively.
The corresponding constraints on the 2HDM scalar potential parameters are shown in Table~\ref{tab:class2}.
\label{tab:symm3}}
\end{table}
\begin{table}[t!]
\begin{tabular}{|llc|}
\hline
symmetry & \phantom{xxxxxxxxxxxxx}transformation law  & related symmetry \pht \\
\hline
GCP1$^\prime$ &
$\Phi_1 \rightarrow \Phi_2^*$,
\hspace{10ex}
$\Phi_2 \rightarrow  \Phi_1^*$ & GCP1 \pht \\[5pt]
GCP3$^\prime$ &
$\begin{cases} \Phi_1 \rightarrow \Phi_1^*\cos\theta-i\Phi_2^*\sin\theta & \\
\Phi_2 \rightarrow i\Phi_1^*\sin\theta-\Phi_2^*\cos\theta \end{cases}$\qquad \text{for $0<\theta<\half\pi$}\phantom{xx} & GCP3 \pht \\
\hline
\end{tabular}
\caption{\small \baselineskip=15pt
Generalized CP (GCP) symmetries of the scalar potential in the $\Phi$-basis that are related by a change of basis to the GCP symmetries of
Table~\ref{tab:symm2}.   Note that a GCP3$^\prime$ symmetry with \text{any} value of $\theta$ that lies between $0$ and $\half\pi$ yields the same constrained 2HDM scalar potential.   The corresponding constraints on the 2HDM scalar potential parameters are shown in Table~\ref{tab:class2}.
\label{tab:symm4}}
\end{table}
\begin{table}[ht!]
\begin{tabular}{|ccccccccccc|}
\hline
symmetry &  $\phantom{m_{11}^2}$ & $m_{22}^2$ &\quad $m_{12}^2$ \quad & $\phantom{\lambda_1}$ &
 $\lambda_2$ & $\phantom{\lambda_3}$ &
$\Re\lambda_5$ & \pht $\Im\lambda_5$ \pht  &\pht  $\lambda_6$ \pht & \pht $\phm\lambda_7$ \pht \\
\hline
$\Pi^\prime_2$  &  &$ m_{11}^2$ & \pht purely imaginary &&
    $ \lambda_1$ & &    
& $0$ & &  $\pht -\lambda_6^\ast\pht $
\\
$\Pi_2\otimes\Pi_2^\prime$  &  & $m_{11}^2$ &0 &&
    $ \lambda_1$ & &    
& $0$ &  0 &  $\phm 0$
\\
U(1)$^\prime$ & & $m_{11}^2$ &\pht  purely imaginary
 &  & $\lambda_1$&  & 
$\lambda_1-\lambda_3-\lambda_4$ & $0$ & $0$ & $\phm 0$ \\
U(1)$^{\prime\prime}$ & & $m_{11}^2$ & real
 &  &$ \lambda_1$&  & 
$\lambda_3+\lambda_4-\lambda_1$ & $0$ & $0$ & $\phm 0$ \\
U(1)$^\prime\otimes\mathbb{Z}_2$  && $m_{11}^2$ & $0$
   && $\lambda_1$ &  &  
$\lambda_1 - \lambda_3 - \lambda_4$  & $0$ & $0$ & $\phm 0$ \\
U(1)$^{\prime\prime}\otimes\mathbb{Z}_2$  && $m_{11}^2$ & $0$
   && $\lambda_1$ &  &  
$\lambda_3 +\lambda_4-\lambda_1$ & $0$ & $0$ & $\phm 0$ \\
GCP1$^\prime$   && $m_{11}^2$ & 
   && $\lambda_1$ &  &  
 &&  &  $\phm \lambda_6$ \\
GCP3$^\prime$   && $m_{11}^2$ & $0$
   && $\lambda_1$ &  &  
$\lambda_3 +\lambda_4-\lambda_1$ & $0$ &\pht \phm  $0$\pht\phm   & $\phm 0$ \\
\hline
\end{tabular}
\caption{\small \baselineskip=15pt
The impact of the 2HDM scalar potential symmetries listed in Tables~\ref{tab:symm3} and \ref{tab:symm4} 
 on the parameters of the scalar potential [cf.~\eq{pot}] in the $\Phi$-basis.
Empty entries correspond to a lack of constraints on the corresponding parameters. Note that the constraints on the scalar potential parameters due to the $\mathbb{Z}_2\otimes\Pi_2$, GCP3 and
GCP3$^\prime$ symmetries coincide with those of the  $\Pi_2\otimes\Pi_2^\prime$,
U(1)$^\prime\otimes\mathbb{Z}_2$ and U(1)$^{\prime\prime}\otimes\mathbb{Z}_2$ symmetries, respectively.
\label{tab:class2}}
\end{table}

Starting from a GCP1 symmetry transformation in the $\Phi$-basis, consider an arbitrary basis change, $\Phi\to\Phi^\prime=U\Phi$.  Then, \eqs{GCP}{XprimeGCP} yield the corresponding GCP transformation in the $\Phi^\prime$-basis, $\Phi\to V\Phi^*$, where $V\equiv UU^{\T}$ is a symmetric unitary matrix.   The choice of $V=\left(\begin{smallmatrix} 0 & \phm 1 \\ 1 & \phm 0\end{smallmatrix}\right)$ corresponds to the definition of GCP1$^\prime$ exhibited in Tables~\ref{tab:symm4} and \ref{tab:class2}.  In contrast to the GCP1 symmetry, the GCP1$^\prime$ symmetry transformation is especially noteworthy in that it does not enforce reality conditions on the potentially complex parameters $m_{12}^2$, $\lambda_5$, $\lambda_6$ and $\lambda_7$. 

Finally, it should be noted that the constraints on the scalar potential in a scalar field basis where the GCP3 symmetry is manifestly realized are precisely the same as the constraints due to the U(1)$^\prime\otimes\mathbb{Z}_2$ family symmetry when imposed \textit{in the same basis of scalar fields}.
This should be contrasted with the $\ug\otimes\Pi_2$-symmetric scalar potential, which is equivalent to the GCP3-symmetric scalar potential when expressed with respect to a different scalar field basis.  Likewise the parameter constraints in a basis where the GCP3$^\prime$ symmetry is manifestly realized coincide with those that arise from the U(1)$^{\prime\prime}\otimes\mathbb{Z}_2$ family symmetry.

\section{An exceptional region of the 2HDM parameter space}
\label{sec:erps}

The exceptional region of the parameter space (ERPS) of the 2HDM corresponds to a regime in which the parameters of the scalar potential satisfy the following conditions: $m_{11}^2=m_{22}^2$, $m_{12}^2=0$, $\lambda_1=\lambda_2$ and $\lambda_7=-\lambda_6$.   These conditions can be imposed by a GCP2 symmetry,
\beq
\Phi_1 \rightarrow \Phi_2^*,
\qquad\quad
\Phi_2 \rightarrow -\Phi_1^*\,.
\eeq
 However, in the case of a softly broken GCP2 symmetry, the conditions on the $m_{ij}^2$ are relaxed.  In general, one can take $m_{11}^2\neq m_{22}^2$ and allow for nonzero complex values of~$m_{12}^2$.  The resulting parameter regime maintains many of the exceptional characteristics of the ERPS and will henceforth be designated as the ERPS4.

If the relations, $\lambda_1=\lambda_2$ and $\lambda_7=-\lambda_6$, hold in one scalar field basis, then they hold in all choices of the scalar field basis.  
Indeed, one can construct a quantity, $\mathcal{Z}$, which is explicitly given in \eq{cp2invariant}, which is manifestly basis invariant under a change of scalar field basis.  Evaluating this invariant in a generic $\Phi$-basis, we obtain
\beq \label{zee}
\mathcal{Z}=\tfrac14 (\lambda_1-\lambda_2)^2+|\lambda_6+\lambda_7|^2\,.
\eeq
Thus, the invariant condition for the ERPS4 is $\mathcal{Z}=0$, which yields $\lambda_1=\lambda_2$ and $\lambda_7=-\lambda_6$ for any choice of the scalar field basis.
Moreover as first noted in Ref.~\cite{Davidson:2005cw}, there exists a choice of basis 
such that $\lambda_6=\lambda_7=0$
and $\lambda_5$ is real.  This basis corresponds to an enhanced symmetry obtained by simultaneously imposing a $\mathbb{Z}_2$ and $\Pi_2$ symmetry,
   \beqa
   \mathbb{Z}_2:\quad  &&  \Phi_1 \rightarrow \Phi_1,\qquad \Phi_2 \rightarrow -\Phi_2\,,\\
   \Pi_2:\quad && \Phi_1 \longleftrightarrow \Phi_2\,,
   \eeqa 
on the quartic terms of the scalar potential.
Indeed, this symmetry adds the constraints, $\lambda_6=\lambda_7=0$ and $\lambda_5\in\mathbb{R}$ to the ERPS4 conditions.  
That is, a softly broken GCP2-symmetric scalar potential can be realized as a softly broken $\mathbb{Z}_2\otimes\Pi_2$-symmetric scalar potential in a different scalar field basis.  
A simple proof of this result is given below \eq{Zinvariant}.

One can impose an additional constraint on the 
ERPS4 by imposing a GCP3 symmetry,
\beq
\Phi_1 \rightarrow \Phi_1^*\cos\theta+\Phi_2^*\sin\theta,\qquad\quad \Phi_2 \rightarrow -\Phi_1^*\sin\theta+\Phi_2^*\cos\theta,
\eeq
for all  $0<\theta<\half\pi$.
   This symmetry adds the additional constraint, $\lambda_5=\lambda_1-\lambda_3-\lambda_4$ (which implies that $\lambda_5$ is real).  We will allow for a general soft breaking of the GCP3 symmetry so that one can again take $m_{11}^2\neq m_{22}^2$ and allow for nonzero complex values of~$m_{12}^2$.   Another possible choice for an enhanced symmetry is to impose simultaneously a U(1) and $\Pi_2$ symmetry~\cite{Ferreira:2009wh},
   \beqa
   \rm{U}(1):\quad  &&  \Phi_1 \rightarrow e^{-i \theta} \Phi_1,\qquad \Phi_2 \rightarrow e^{i \theta} \Phi_2\,,\\
   \Pi_2:\quad && \Phi_1 \longleftrightarrow \Phi_2\,,
   \eeqa
   for any $0<\theta<\half\pi$.
   This symmetry adds the constraint $\lambda_5=0$ to the $\mathbb{Z}_2\otimes\Pi_2$ symmetry.   
  As shown explicitly in Section~\ref{transforming}, if the 2HDM scalar potential respects a GCP3 symmetry, then there exists a basis of scalar fields in which the symmetry can be identified as U(1)$\otimes\Pi_2$.
 
 A basis-invariant condition can be found that corresponds to the case in which the quartic terms of the scalar potential respect the GCP3 symmetry in some basis.   
 The invariant was first constructed in Ref.~\cite{Ferreira:2009wh}\footnote{The published version of Ref.~\cite{Ferreira:2009wh} contains some typographical errors---in eq.~(39),
$\det\tilde\Lambda$ should be replaced by $-\det\tilde\Lambda$ and in eq.~(44), $\half$ should be replaced by $\tfrac13$.   All other equations in Section II.E of Ref.~\cite{Ferreira:2009wh} are correct.}  
and then rederived using a different technique
 in Appendix~B of Ref.~\cite{Haber:2012np}.  Below, we shall review the method employed in Ref.~\cite{Haber:2012np} while providing additional details 
of the derivation.
 
 First, we make use of the notation of \eq{ZZ} to assemble the 2HDM scalar potential couplings into a rank four tensor denoted by $Z_{ab,cd}$.   
 It is also convenient to introduce a related rank four tensor,
 \beq \label{ZoverZ}
 \overline{Z}_{ab,cd}\equiv Z_{ba,cd}=Z_{ab,dc}\,,
 \eeq
 where the two expressions for $\overline{Z}_{ab,cd}$ given above are equivalent in light of \eqs{herm}{ZZ}.  
Next, we define a three-vector whose components $P_B$ (for $B=1,2,3$) are given by,
\beq \label{PB}
P_B=\tfrac14 (Z_{ab,cd}+\overline{Z}_{ab,cd})\delta_{ca}\sigma^B_{db}\,,
\eeq
and a $3\times 3$ real symmetric
 matrix whose elements $D_{AB}$ are given by~\cite{Ivanov:2005hg,Haber:2012np},\footnote{Quantities that are invariant with respect to scalar field basis transformations can be constructed out of
 objects such as $D_{AB}$.  Although $D_{AB}$ is not an invariant, related objects such as $\Tr D$, $\det D$ and the eigenvalues of $D$ can be used to construct invariant quantities.
 Ivanov published the first paper that presented this strategy in Ref.~\cite{Ivanov:2005hg}.}
 \beq \label{DAB}
 D_{AB}=\tfrac14 (Z_{ab,cd}+\overline{Z}_{ab,cd})\sigma^A_{ca}\sigma^B_{db}-\tfrac{1}{12} (Z_{ab,ab}+\overline{Z}_{ab,ab})\delta^{AB}\,,
 \eeq
 where the $\sigma^A$ are the Pauli matrices and there is an implicit sum over repeated indices.  
 Under a change of scalar field basis, $\Phi\to\Phi^\prime=U\Phi$, \eq{Z-transf} yields,
 \beq
 P_B\to P^\prime_B=\mathcal{R}_{BD}P_D\,,\qquad\quad
 D_{AB}\to D^\prime_{AB}=\mathcal{R}_{AC}\mathcal{R}_{BD}D_{CD}=(\mathcal{R}D\mathcal{R}^{\T})_{AB}\,,\label{deeprimeAB}
 \eeq
 after employing the identity $U^\dagger\sigma^A U=\mathcal{R}_{AB}\sigma^B$,
 where $\mathcal{R}$ is a real orthogonal matrix that is explicitly given by 
 \beq \label{areAB}
 \mathcal{R}_{AB}=\half\Tr(U^\dagger\sigma^A U\sigma^B)\,.
 \eeq

 Using the Kronecker product notation introduced in \eqs{AotimesB}{Kproduct}, we
 can rewrite \eq{DAB} in a convenient form in terms of two $4\times 4$ matrices, $Z$ and $\overline{Z}$, where $Z$ is defined in \eq{Zmatrix} and $\overline{Z}$ is obtained from $Z$ by interchanging $\lambda_3\leftrightarrow\lambda_4$.
 Then, the equivalent forms of \eqs{PB}{DAB} are given by,
 \beq \label{peeB}
 P_B=\tfrac14\Tr\bigl[(\mathds{1}_{3\times 3}\otimes\sigma^B)(Z+\overline{Z})\bigr]\,,
 \eeq
 where $\mathds{1}_{3\times 3}$ is the $3\times 3$ identity matrix, and
 \beq \label{deeAB}
 D_{AB}=\tfrac14\Tr\bigl[(\sigma^A\otimes\sigma^B)(Z+\overline{Z})\bigr]-\tfrac{1}{12}\Tr(Z+\overline{Z})\delta^{AB}\,.
 \eeq
Using $Z_{ab,cd}=Z_{ba,dc}$ [cf.~\eq{ZoverZ}], it follows that
\beq
\Tr\bigl[(\sigma^A\otimes\sigma^B)(Z+\overline{Z})\bigr]=\Tr\bigl[(\sigma^B\otimes\sigma^A)(Z+\overline{Z})\bigr]\,,
\eeq
which shows that $D$ is a symmetric matrix.   Moreover,
\beq
\sum_C\Tr\bigl[(\sigma^C\otimes\sigma^C)(Z+\overline{Z})\bigr]=\Tr(Z+\overline{Z})=2(\lambda_1+\lambda_2+\lambda_3+\lambda_4)\,,
\eeq
which implies that $D$ is a traceless matrix.
 Indeed, a straightforward calculation yields,
 \beq \label{pee}
 P=\begin{pmatrix} \Re(\lambda_6+\lambda_7) & \quad -\Im(\lambda_6+\lambda_7) & \quad \half(\lambda_1-\lambda_2)\end{pmatrix},
 \eeq
 and~\!\footnote{The matrix $D$ is related to $\tilde\Lambda$ employed in Ref.~\cite{Ferreira:2009wh} by $D=\tilde\Lambda-\frac13(\Tr\tilde\Lambda)\mathds{1}_{3\times 3}$.
Thus, if $\lambda$ is an eigenvalue of $\tilde\Lambda$ then $\lambda-\frac13\Tr\tilde\Lambda$ is the corresponding eigenvalue of $D$.  Consequently, the condition for degenerate eigenvalues is the same if applied to either $D$ or $\tilde{\Lambda}$.   There are some advantages to 
employing \eq{dab}, as the condition $\Tr D=0$ simplifies the algebraic manipulations.}  
  \beq \label{dab}
   D
   =\begin{pmatrix}
     -\frac{1}{3}\Delta+\Re\,\lambda_5 &\,\,\,
     -\Im\,\lambda_5 & \,\,\,\phm\Re\,\lt(\lambda_6-\lambda_7\rt)\\
     -\Im\,\lambda_5 &\,\,\,
     -\frac{1}{3}\Delta-\Re\,\lambda_5 &\,\,\, -\Im\,\lt(\lambda_6\!-\!\lambda_7\rt)\\
    \phm \Re\,\lt(\lambda_6\!-\!\lambda_7\rt) &\,\,\,
     -\Im\,\lt(\lambda_6\!-\!\lambda_7\rt)
     & \,\,\,\frac{2}{3}\Delta\end{pmatrix}\!\!,
\eeq
where 
\beq
\Delta\equiv \half(\lambda_1+\lambda_2)-\lambda_3-\lambda_4\,.
\eeq
In particular, the following condition for the ERPS4, which makes use of the vector $P_B$ [cf.~\eq{pee}], reproduces the invariant previously given in \eq{zee}, 
\beq \label{Zinvariant}
\mathcal{Z}\equiv\sum_B P_B P_B=\tfrac14(\lambda_1-\lambda_2)^2+G|\lambda_6+\lambda_7|^2=0\,.
\eeq

We now prove that if $\lambda_1=\lambda_2$ and $\lambda_7=-\lambda_6$ in the $\Phi$-basis, then there exists a $\Phi^\prime$-basis, defined by $\Phi^\prime=U\Phi$, in which $\lambda^\prime_1=\lambda^\prime_2$ and $\Im\lambda_5^\prime=\lambda^\prime_6=\lambda^\prime_7=0$.  In light of \eq{pee}, if $\lambda_1=\lambda_2$ and $\lambda_7=-\lambda_6$ in the $\Phi$-basis then it follows that $P=0$.  Moreover, $D$ is a real traceless symmetric matrix [cf.~\eq{dab}], which can always be 
transformed into a real diagonal matrix via an orthogonal similarity transformation.
Thus, there exists a real orthogonal matrix $\mathcal{R}$ [explicitly given in terms of $U$ in \eq{areAB}] such that $P^\prime=\mathcal{R}P=0$ and $D^\prime=\mathcal{R}D\mathcal{R}^{\T}$ is diagonal.
Noting the explicit forms of $P$ and $D$ given above, it follows that $\lambda^\prime_1=\lambda^\prime_2$ and $\Im\lambda_5^\prime=\lambda^\prime_6=\lambda^\prime_7=0$ in the $\Phi^\prime$-basis, as previously asserted.

Next, we demonstrate, following Ref.~\cite{Haber:2012np},
that there exists a basis in which the Peccei-Quinn U(1) symmetry of the quartic terms of the scalar potential is manifestly realized if and only if $P_B$ and $D_{AB}$ can be written in the following forms,
\beq \label{PD}
P_B=c_2 q_B\,,\qquad\quad D_{AB}=c_3\left(q_A q_B-\tfrac13 \delta_{AB}\right),
\eeq
where the $q_B$ are components of a real three-vector of unit length and $c_2$ and $c_3$ are constants.   It then follows that
\beq
\Tr D=0\,,\qquad\quad \Tr(D^2)=\tfrac23 c_3^2\,,\qquad\quad \det D=\tfrac13\Tr(D^3)=\tfrac{2}{27}c_3^3\,,
\eeq
which yields a characteristic equation for the eigenvalues of $D$ [cf.~\eqst{secular}{traceless}], 
\beq
x^3-\tfrac13 c_3 ^2 x-\tfrac{2}{27}c_3^3=\bigl(x+\tfrac13 c_3\bigr)^2\bigl(x-\tfrac23 c_3\bigr)=0\,.
\eeq
Hence, the eigenvalues of $D$ are $-\frac13 c_3$, $-\frac13 c_3$, and $\frac23 c_3$.  That is, if $D\neq 0$ then two of the eigenvalues of $D$ are degenerate.
Moreover, in light of the eigenvalue equation,
\beq \label{Deigen}
D_{AB}q_B=c_3\left(q_A q_B-\tfrac13 \delta_{AB}\right)q_B=\tfrac23 c_3 q_A\,,
\eeq
it follows that $q_A$ is the eigenvector of the nondegenerate eigenvalue.
Thus, in a scalar field basis in which $D$ as defined by \eq{dab} is diagonal with two degenerate diagonal elements, it follows
that $\lambda_5=0$ and $\lambda_6=\lambda_7$, in which case we can identify
$c_3=\Delta$ and the unit vector $q=(0\quad 0\quad 1)$.   Applying this result for $q_B$ in \eq{PD} and comparing with \eq{pee} then yields 
$c_2=\half(\lambda_1-\lambda_2)$ and $\lambda_7=-\lambda_6$.  Hence, we conclude that
$\lambda_5=\lambda_6=\lambda_7=0$ in the $D$-diagonal basis, corresponding to a softly broken U(1)$_{\rm PQ}$-symmetric scalar potential.

If we now impose the ERPS4 condition on the softly broken U(1)$_{\rm PQ}$-symmetric scalar potential then $\lambda_1=\lambda_2$, which implies that $P=0$.   In this case,
the quartic terms of the scalar potential respect a U(1)$\otimes\Pi_2$ symmetry.  Given a softly broken U(1)$\otimes\Pi_2$-symmetric scalar potential in the $\Phi$-basis, one can perform
a basis change $\Phi^\prime=U\Phi$ such that\footnote{Using the explicit form for $U$ given by \eq{you} in \eq{areAB} yields the result exhibited in \eq{R}.}
\beq \label{R}
\mathcal{R}=\begin{pmatrix}  1 & \,\,\,\phm 0 & \,\,\,\phm 0 \\  0 & \,\,\,\phm 0 & \,\,\,-1 \\  0 & \,\,\,\phm 1& \,\,\,\phm 0\end{pmatrix}\,.
\eeq
Then, \eq{deeprimeAB} yields $P^\prime=0$, $\Im\lambda_5^\prime=\lambda_6^\prime=\lambda_7^\prime=0$ and
\beq
-\tfrac13\Delta^\prime+\Re\lambda_5^\prime=\tfrac23\Delta^\prime=-\tfrac13\Delta\,.
\eeq
\clearpage

\noindent
It follows that $\Delta^\prime=\Re\lambda_5^\prime$.  That is,
$\lambda_1^\prime=\lambda_2^\prime=\lambda_3^\prime+\lambda_4^\prime+\lambda_5^\prime$, and $\Im\lambda_5^\prime=\lambda_6^\prime=\lambda_7^\prime=0$, which are the conditions for a softly broken GCP3-symmetric scalar potential in the $\Phi^\prime$-basis.

Finally, if $D=P=0$ then
$\lambda_1=\lambda_2=\lambda_3+\lambda_4$ and
$\lambda_5=\lambda_6=\lambda_7=0$,  corresponding to a scalar potential whose quartic terms
respect an SO(3) symmetry.   In summary, we have successfully provided simple basis-invariant conditions for the 2HDM with a softly broken U(1), GCP2 [or $\mathbb{Z}_2\otimes\Pi_2$], GCP3 [or U(1)$\otimes\Pi_2$] and SO(3) symmetry, respectively.

Thus, we  seek a condition that guarantees that the matrix $D$ given in \eq{dab} possesses two degenerate eigenvalues.
In general, the characteristic equation of a generic $3\times 3$ matrix~$D$ is of the form,
\begin{equation} \label{secular}
x^3 + a_2 x^2 + a_1 x + a_0 = 0\,,
\end{equation}
where
\beqa
a_0&=&-\det D=-\tfrac16(\Tr D)^3+\half \Tr D\Tr(D^2)-\tfrac13\Tr(D^3) \,,\\
a_1&=& \half(\Tr D)^2-\half\Tr(D^2)\,,\\
a_2&=& -\Tr D\,.\label{traceless}
\eeqa
The cubic equation given in \eq{secular} has exactly two
degenerate roots if the following two conditions are satisfied~\cite{AbrSte,Beyond},
\begin{equation} \label{dd}
\mathcal{D} \equiv
\tfrac92 a_0 a_1 a_2-a_0 a_2^3+\tfrac14 a_1^2 a_2^2-a_1^3-\tfrac{27}{4}a_0^2=0
\quad \text{and} \quad a_2^2\neq 3a_1\,.
\end{equation}
Since the matrix $D$ given in \eq{dab} is symmetric and traceless, the conditions that $D$ possesses exactly two degenerate eigenvalues simplify to, 
\beq \label{scriptdee}
\mathcal{D}=-a_1^3-\tfrac{27}{4} a_0^2=\tfrac18\bigl[\Tr(D^2)]^3-\tfrac{3}{4}\bigl[\Tr(D^3)\bigr]^2=0\quad \text{and} \quad D\neq 0\,.
\eeq

If the quartic terms of the scalar potential exhibit a U(1)$\otimes\Pi_2$ symmetry, then it follows that $\mathcal{Z}=\mathcal{D}=0$.   Thus, we conclude that the basis-invariant condition,
$\mathcal{I}\equiv \mathcal{Z}+\mathcal{D}^2=0$, is satisfied
if and only if the quartic terms of the scalar potential exhibit a U(1)$\otimes\Pi_2$ symmetry in some basis (which implies that the quartic terms of the scalar potential exhibits a GCP3 symmetry in some other basis).  One can determine this condition explicitly by setting $\lambda\equiv\lambda_1=\lambda_2$ and $\lambda_7=-\lambda_6$
when evaluating the characteristic equation of the matrix $D$, which yields
\beqa
\Tr(D^3)&=&\tfrac{2}{9}\Delta^3+2\Delta\bigl(2|\lambda_6|^2-|\lambda_5|^2\bigr)+12\Re(\lambda_5^*\lambda_6^2)\,,\\
\Tr(D^2)&=&\tfrac23\Delta^2+2|\lambda_5|^2+8|\lambda_6|^2\,.
\eeqa
\clearpage
Inserting the above results into \eq{scriptdee} yields an expression for
$\mathcal{D}$.   First, we assume that $\lambda_6\neq 0$, in which case the end result is,
\beq \label{Dinvariant} 
 \mathcal{D}=\bigl[2|\lambda_6|^2-\mathcal{R}_{56}(\Delta+\mathcal{R}_{56})\bigr]^2 
  \left[(\Delta-\mathcal{R}_{56})^2+16|\lambda_6|^2\right]+CI_{56}^2\,,
  \eeq
where $\Delta\equiv \lambda-\lambda_3-\lambda_4$,
\beq \label{C}
C\equiv (\Delta^2-|\lambda_5|^2-\mathcal{R}_{56}^2)^2-\mathcal{R}_{56}^2|\lambda_5|^2+ 2|\lambda_6|^2\bigl[\Delta^2+9(\Delta+\mathcal{R}_{56})^2+3\mathcal{I}_{56}^2+3|\lambda_5|^2+24|\lambda_6|^2\bigr]\,,
\eeq  
and  
\beq
\mathcal{R}_{56}\equiv\frac{\Re(\lambda_5^*\lambda_6^2)}{|\lambda_6|^2}\,,\qquad\quad \mathcal{I}_{56}\equiv \frac{\Im(\lambda_5^*\lambda_6^2)}{|\lambda_6|^2}\,.
\eeq
Note that $\mathcal{R}^2_{56}+\mathcal{I}^2_{56}=|\lambda_5|^2$.

The product of the first two factors on the right-hand side of \eq{Dinvariant} is non-negative definite.  Thus, one
solution to the equation $\mathcal{D}=0$ can be obtained by setting 
\beq \label{lam6sq}
|\lambda_6|^2=\half \mathcal{R}_{56}(\Delta+\mathcal{R}_{56})>0\,, 
\eeq
which implies that $\Delta/\mathcal{R}_{56}>-1$ [after dividing by~$\mathcal{R}_{56}^2$].\footnote{Note that $\mathcal{R}_{56}$ and/or $\Delta$ can be zero.  If these quantities are nonvanishing, then 
their individual signs can be either positive or negative.}
It then follows from \eq{Dinvariant} that either
$\mathcal{I}_{56}=0$ or $C=0$.  We now demonstrate that the latter possibility is never realized.  
After inserting \eq{lam6sq} into the expression for~$C$ given in \eq{C}, we obtain,
\beq \label{see}
C=\mathcal{I}_{56}^4+(9\mathcal{R}_{56}^2+6\Delta \mathcal{R}_{56}-2\Delta^2)\mathcal{I}_{56}^2+(\Delta+\mathcal{R}_{56})(\Delta+3\mathcal{R}_{56})^3\,,
\eeq
which is a quadratic equation whose discriminant is given by,
\beq
{\rm Disc}=(9\mathcal{R}_{56}^2+6\Delta \mathcal{R}_{56}-2\Delta^2)^2-4(\Delta+\mathcal{R}_{56})(\Delta+3\mathcal{R}_{56})^3=-\mathcal{R}_{56}(4\Delta+3\mathcal{R}_{56})^3\,.
\eeq
If $\Delta/\mathcal{R}_{56}> -\frac34$ then ${\rm Disc}<0$ and it follows that $C>0$ for all values of $\mathcal{I}_{56}$.  Finally, if $-1<\Delta/\mathcal{R}_{56}\leq-\frac34$, then \eq{see} yields
$C>0$ for any nonzero value of $\mathcal{I}_{56}$.   Thus, we have shown that for $\lambda_6\neq 0$, if \eq{lam6sq} is satisfied then
$\mathcal{D}=0$ if and only if
\beq \label{GCP3invar}
|\lambda_6|^2=\half \mathcal{R}_{56}(\Delta+\mathcal{R}_{56})> 0\quad \text{and}\quad \mathcal{I}_{56}=0\,.
\eeq
One can rewrite the two conditions given in \eq{GCP3invar} as a single complex equation, 
\beq \label{GCP3invcond}
\lambda^2_5\lambda_6^*+\lambda_5\lambda_6(\lambda_1-\lambda_3-\lambda_4)-2\lambda_6^3=0\,,
\eeq
which must hold true for any choice of scalar field basis.

If $C>0$ were valid for all nonzero values of $\lambda_6$, then it would immediately follow that \eq{GCP3invar} is the unique solution of the equation $\mathcal{D}=0$.   However one can verify
that regions of the parameter space exist in which $C<0$.   
This seems to leave open the possibility that if $\lambda_6\neq 0$ then $\mathcal{D}=0$ can be satisfied with a nonzero value of $\mathcal{I}_{56}$ due to a cancellation between terms in \eq{Dinvariant}.\footnote{If such a solution existed, it would not be continuously connected to the solution given by \eq{GCP3invar}, since any small perturbation of the scalar potential parameters from \eq{GCP3invar} would still yield $C>0$.
We have numerically checked in Mathematica using graphical techniques that in the region of parameter space where $C<0$,
there are no solutions to $\mathcal{D}=0$ for $\mathcal{I}_{56}\neq 0$ and $\lambda_6\neq 0$.
However, it is disappointing that we are unable to analytically establish the condition $\mathcal{I}_{56}=0$ directly from $\mathcal{D}=0$ when $\lambda_6\neq 0$.}

Nevertheless, we shall now argue that under the assumption that $\lambda_6\neq 0$, the condition $\mathcal{D}=0$ holds if and only if \eq{GCP3invar} is satisfied.
Recall that \eq{scriptdee} states that the $3\times 3$ traceless  real symmetric matrix~$D$ (assumed to be nonzero) possesses a doubly degenerate eigenvalue  if and only if
$\mathcal{D}=0$.  
Moreover,
any $3\times 3$ traceless real symmetric matrix that possesses a doubly degenerate eigenvalue must have the form specified in \eq{PD}.\footnote{Given a 
$3\times 3$ traceless real symmetric matrix $D$ with eigenvalues $-c$, $-c$ and $2c$ (where $c\in\mathbb{R}$),  it then follows that there exists a real orthogonal matrix $\mathcal{R}$ such that
$D=\mathcal{R}\,{\rm diag}(-c,-c,2c)\mathcal{R}^{\T}$.  However in this case one can write ${\rm diag}(-c,-c,2c)_{AB}=c_3(q_A q_B-\tfrac13\delta_{AB})$ with $q=(0,0,1)$ and $c_3=3c$.  Hence, we conclude that $D=c_3(q'_A q'_B-\tfrac13\delta_{AB})$ with unit vector $q^\prime_A=\mathcal{R}_{AB}q_B$.}
We can then use the discussion below \eq{Deigen} to conclude that in the $D$-diagonal basis, $\lambda_5=\lambda_6=\lambda_7=0$.   Performing a basis transformation to an arbitrary basis
(e.g., cf.~eqs.~(A9) and (A10) of Ref.~\cite{Boto:2020wyf}), it follows that $\Im(\lambda_5^*\lambda_6^2)=0$ in any scalar field basis.   Thus, we are justified in setting $\mathcal{I}_{56}=0$
in \eq{Dinvariant}, in which case \eqs{GCP3invar}{GCP3invcond} must be valid for any choice of scalar field basis.

In the case of $\lambda_6=0$, one can either evaluate $\mathcal{D}$ directly using \eq{scriptdee} or simply set $|\lambda_6|=0$ in \eq{Dinvariant} while keeping $\mathcal{R}_{56}$ and $\mathcal{I}_{56}$ fixed.  Both procedures yield the same result,
\beq
\mathcal{D}=|\lambda_5|^2\bigl(\Delta^2-|\lambda_5|^2\bigr)^2\,.
\eeq
In particular, if $\lambda_6=0$ then we can rephase $\Phi_2$ such that $\lambda_5$ is real, in which case either
\beq \label{fiveconds}
\text{$\lambda_5=0$ \quad or \quad $\lambda_5=\pm (\lambda_1-\lambda_3-\lambda_4)$},
\eeq
corresponding to the manifest realization of U(1)$\otimes\Pi_2$ and GCP3/GCP3$^\prime$, respectively, as indicated by the quartic coupling relations exhibited in Tables~\ref{tab:class} and \ref{tab:class2}.  
\clearpage

\section{The \texorpdfstring{$\mathbb{Z}_2\otimes\Pi_2$}{\blackboardZ\texttwoinferior\directprod\uppercasePi\texttwoinferior} scalar field basis}
\label{zee2pi2}

Since the softly broken GCP2-symmetric  scalar potential is equivalent to a softly broken $\mathbb{Z}_2\otimes\Pi_2$-symmetric scalar potential in a different scalar field basis, we henceforth focus on the $\mathbb{Z}_2\otimes\Pi_2$ basis,
where $\lambda\equiv \lambda_1=\lambda_2$, $\lambda_5\neq 0$ is real and $\lambda_6=\lambda_7=0$.  The softly broken parameters, $m_{11}^2$, $m_{22}^2$ and $m_{12}^2$, are arbitrary with $m_{12}^2$ potentially complex.    
If we demand that the potential is bounded from below, the following conditions must be satisfied,
\beq \label{ineqcp2}
\lambda>0\,,\qquad \lambda+\lambda_3>0\,,\qquad \lambda+\lambda_3+\lambda_4-|\lambda_5|>0\,,
\eeq
modulo the remarks of footnote~\ref{fnstable}.
It is convenient to introduce the parameter,
\beq \label{aredef}
R\equiv \frac{\lambda_3+\lambda_4+\lambda_5}{\lambda}\,.
\eeq
Using the definition of $\lambda_{345}$ given below \eq{minfour}, it follows that
\beq
\lambda_{345}=\lambda R-2\lambda_5\sin^2\xi\,.
\eeq

\subsection{The softly broken \texorpdfstring{$\mathbb{Z}_2\otimes\Pi_2$}{\blackboardZ\texttwoinferior\directprod\uppercasePi\texttwoinferior}-symmetric scalar potential with \texorpdfstring{$v_1\neq 0$}{v\textoneinferior \textneq 0} and \texorpdfstring{$v_2\neq 0$}{v\texttwoinferior \textneq 0}}

We shall first assume that $v_1$ and $v_2$ are both nonzero, or equivalently, $\sin 2\beta\neq 0$.  We then use \eqst{min1}{min3} [with \mbox{$\lambda\equiv\lambda_1=\lambda_2$]} to fix the values of $\beta$ and $\xi$.  In particular,
\beqa 
c_{2\beta}&=&\frac{m_{22}^2-m_{11}^2}{m_{11}^2+m_{22}^2+\lambda v^2}\,, \label{betacp2eq} \\
\cos\xi&=&\frac{2\Re m_{12}^{2}}{s_{2\beta}\bigl[m_{11}^{2}+m_{22}^{2}+\half\lambda(1+R)v^2\bigr]}\,,\label{cosxi}\\
\sin\xi&=&\frac{-2\Im m_{12}^{2}}{s_{2\beta}\bigl[m_{11}^{2}+m_{22}^{2}+\bigl(\half\lambda(1+R)-\lambda_5\bigr) v^2\bigr]}\,,\label{sinxi}
\eeqa
where $s_{2\beta}\equiv \sin 2\beta$ and $c_{2\beta}\equiv \cos 2\beta$.
Writing $m_{12}^2=|m_{12}^2|e^{i\theta_{12}}$ in \eqs{cosxi}{sinxi} and imposing $\cos^2\xi+\sin^2\xi=1$ yields an equation that determines the phase
$\theta_{12}$ in terms of~$\xi$ and the other scalar potential parameters.   Thus, the ERPS4 is governed by eight real parameters: $\lambda$, $\lambda_3$, $\lambda_4$, $\lambda_5$,
$|m_{12}^2|$, $v$, $\beta$ and $\xi$.

It is convenient to introduce the Higgs basis as follows~\cite{Donoghue:1978cj,Georgi:1978ri,Botella:1994cs,Branco:1999fs,Davidson:2005cw,Haber:2006ue,Boto:2020wyf}.   The Higgs basis fields $\mathcal{H}_1$ and $\mathcal{H}_2$ are defined 
by the linear combinations of $\Phi_1$ and $\Phi_2$ such that $\langle \mathcal{H}_1^0\rangle=v/\sqrt{2}$ and $\langle \mathcal{H}_2^0\rangle=0$.  That is,
\beq \label{invhiggs}
\mathcal{H}_1\equiv c_\beta\Phi_1+s_\beta e^{-i\xi}\Phi_2\,,\qquad\quad \mathcal{H}_2=e^{i\eta}\bigl[-s_\beta e^{i\xi}\Phi_1+c_\beta \Phi_2\bigr]\,,
\eeq
where we have introduced (following Ref.~\cite{Boto:2020wyf}) the complex phase factor $e^{i\eta}$ to account for the non-uniqueness of the Higgs basis, since one is always free to rephase the Higgs basis field whose vacuum expectation value vanishes.  In particular,  $e^{i\eta}$ is a pseudo-invariant quantity~\cite{Boto:2020wyf} that is rephased under the unitary basis transformation,
$\Phi_a\to U_{a\bbar}\Phi_b$, as
\beq \label{etatrans}
e^{i\eta}\to (\det~U)^{-1} e^{i\eta}\,,
\eeq
where $\det U$ is a complex number of unit modulus.
In terms of the Higgs basis fields defined in \eq{invhiggs}, the scalar potential is given by, 
 \beqa
 \mathcal{V}&=& Y_1 \mathcal{H}_1^\dagger \mathcal{H}_1+ Y_2 \mathcal{H}_2^\dagger \mathcal{H}_2 +[Y_3 e^{-i\eta}
\mathcal{H}_1^\dagger \mathcal{H}_2+{\rm h.c.}]
\nn\\
&&\quad 
+\half Z_1(\mathcal{H}_1^\dagger \mathcal{H}_1)^2+\half Z_2(\mathcal{H}_2^\dagger \mathcal{H}_2)^2
+Z_3(\mathcal{H}_1^\dagger \mathcal{H}_1)(\mathcal{H}_2^\dagger \mathcal{H}_2)
+Z_4(\mathcal{H}_1^\dagger \mathcal{H}_2)(\mathcal{H}_2^\dagger \mathcal{H}_1) \nn \\
&&\quad
+\left\{\half Z_5 e^{-2i\eta}(\mathcal{H}_1^\dagger \mathcal{H}_2)^2 +\big[Z_6 e^{-i\eta} (\mathcal{H}_1^\dagger
\mathcal{H}_1) +Z_7 e^{-i\eta} \mathcal{H}_2^\dagger \mathcal{H}_2)\big] \mathcal{H}_1^\dagger \mathcal{H}_2+{\rm
h.c.}\right\}\,.\label{higgspot}
\eeqa
The coefficients of the quadratic and quartic terms of the scalar potential in \eq{higgspot} are independent of the initial choice of the $\Phi$-basis.  It then follows that $Y_3$, $Z_5$, $Z_6$ and $Z_7$ are also pseudo-invariant quantities~\cite{Haber:2006ue} that are rephased under $\Phi_a\to U_{a\bbar}\Phi_b$ as
\beq \label{rephasing}
 [Y_3, Z_6, Z_7]\to (\det~U)^{-1}[Y_3, Z_6, Z_7] \quad{\rm and}\quad
Z_5\to  (\det~U)^{-2} Z_5\,.
\eeq

It is straightforward to compute the corresponding Higgs basis parameters.
The $Y_i$ are given by,
\beqa
Y_1&=& m_{11}^2 c_\beta^2+m_{22}^2 s_\beta^2-\Re(m_{12}^2 e^{i\xi})s_{2\beta}\,,\\
Y_2&=& m_{11}^2 s_\beta^2+m_{22}^2 c_\beta^2+\Re(m_{12}^2 e^{i\xi})s_{2\beta}\,,\\
Y_3&=&\bigl[\half(m_{22}^2-m_{11}^2)s_{2\beta}-\Re(m_{12}^2 e^{i\xi})c_{2\beta}-i\Im(m_{12}^2 e^{i\xi})\bigr]e^{-i\xi}\,.\label{why3}
\eeqa
Employing \eqst{min1}{min3} [with $\lambda\equiv\lambda_1=\lambda_2$] to eliminate $m_{11}^2$, $m_{22}^2$ and $\Im(m_{12}^2 e^{i\xi})$, it follows that,
\beqa
&& Y_2=\frac{2\Re(m_{12}^2 e^{i\xi})}{s_{2\beta}}-\half\lambda v^2+\half v^2\bigl[\lambda(1-R)+2\lambda_5 \sin^2\xi\bigr](1-\half s_{2\beta}^2)\,,\label{whytwocp2}\\
&& Z_1=Z_2=\lambda-\half\bigl[\lambda(1-R)+2\lambda_5 \sin^2\xi\bigr]s_{2\beta}^2\,,\label{zeeonecp2}\\
&& Z_3=\lambda_3+\half\bigl[\lambda(1-R)+2\lambda_5 \sin^2\xi\bigr]s_{2\beta}^2\,, \label{zeethreecp2}\\
&& Z_4=\lambda_4+\half\bigl[\lambda(1-R)+2\lambda_5 \sin^2\xi\bigr]s_{2\beta}^2\,, \label{zeefourcp2}\\
&& Z_5=\bigl\{\half\bigl[\lambda(1-R)+2\lambda_5 \sin^2\xi\bigr]s_{2\beta}^2+\lambda_5(\cos 2\xi+ic_{2\beta}\sin 2\xi)\bigr\} e^{-2i\xi}\,,\label{zeefivecp2} \\
&& Z_6=-Z_7=\bigl\{-\half\bigl[\lambda(1-R)+2\lambda_5 \sin^2\xi\bigr] c_{2\beta}+\half i\lambda_5\sin 2\xi\bigr\}s_{2\beta} e^{-i\xi}\,.\label{zeesevencp2}
\eeqa
One can also check that the scalar potential minimization conditions in the Higgs basis,
\beq \label{YZ}
Y_1=-\half Z_1 v^2\,,\qquad\quad Y_3=-\half Z_6 v^2\,,
\eeq
are satisfied.   The eight parameters that specify the ERPS4 can now be identified as $v$,~$Y_2$, $Z_1$, $Z_3$, $Z_4$, $\Re Z_5$, $\Im Z_5$, and $|Z_6|$ after using the freedom to rephase the Higgs basis field~$\mathcal{H}_2$ to remove the complex phase from $Z_6$ and $Z_7$ [cf.~\eq{rephasing}].

The subregion of the ERPS4 where $Z_6=0$ is worthy of special attention.   The ERPS4 condition, $Z_6=-Z_7$, along with \eq{YZ} yields $Y_3=Z_6=Z_7=0$, which signals the presence of a $\mathbb{Z}_2$ symmetry that is manifestly realized in the Higgs basis and is unbroken by the vacuum.  We recognize this scenario as a special case of the IDM, and hence we shall refer to this  parameter regime as the \textit{inert limit} of the softly broken $\mathbb{Z}_2\otimes\Pi_2$-symmetric scalar potential.  Moreover, Higgs alignment is exact in the inert limit, as discussed in Section~\ref{alignment}.  The conditions for achieving the inert limit will be elucidated below \eq{subs}.

Three additional limiting cases are noteworthy.  
First, if $\lambda_5=0$, then the quartic terms of the scalar potential exhibit a U(1)$\otimes\Pi_2$ symmetry, which will be discussed in more detail in Section~\ref{sec:UPibasis}.   Second, if $R=1$, then the quartic terms of the scalar potential exhibit a GCP3 symmetry, which will be discussed in more detail in Section~\ref{sec:GCP3basis}.  Both these limits yield the same physical scalar sector, since they correspond to the softly broken GCP3-symmetric scalar potential expressed in two different choices of the scalar field basis.  Finally, if $\lambda_5=0$ \textit{and} $R=1$, then the quartic terms of the scalar potential exhibit an SO(3) symmetry.

The charged Higgs mass is given by,
\beq \label{chargedHmass}
m^2_{H^\pm}=Y_2+\half Z_3 v^2=
\frac{2\Re(m_{12}^{2}e^{i\xi})}{s_{2\beta}}
-\half v^2(\lambda_4+\lambda_5\cos 2\xi)\,.
\eeq
The squared masses of the neutral Higgs bosons are given by the eigenvalues of
the neutral scalar squared-mass matrix, 
\beq  \label{matrix33}
\mathcal{M}^2=v^2\left( \begin{array}{ccc}
Z_1&\quad \Re(Z_6 e^{-i\eta}) &\quad -\Im(Z_6 e^{-i\eta})\\
\Re(Z_6 e^{-i\eta})  &\quad \half\bigl[Z_{34}+\Re(Z_5 e^{-2i\eta})\bigr]+Y_2/v^2 & \quad
- \half \Im(Z_5  e^{-2i\eta})\\ -\Im(Z_6  e^{-i\eta}) &\quad - \half \Im(Z_5  e^{-2i\eta}) &\quad
\half\bigl[Z_{34}-\Re(Z_5 e^{-2i\eta})\bigr]+Y_2/v^2\end{array}\right), 
\eeq
which is expressed with respect to the $\{\sqrt{2}\,\Re \mathcal{H}_1^0-v,\sqrt{2}\,\Re \mathcal{H}_2^0,\sqrt{2}\,\Im \mathcal{H}_2^0\}$ basis, where
\beq \label{zeethreefour}
Z_{34}\equiv Z_3+Z_4=\lambda\bigl[R+ s^2_{2\beta}(1-R)\bigr]-\lambda_5(1-2 s_{2\beta}^2\sin^2\xi)\,,
\eeq
after making use of \eqs{zeethreecp2}{zeefourcp2}.\footnote{The expressions given for $m_{H^\pm}^2$ in \eq{chargedHmass} and for $\mathcal{M}^2$ in \eqs{matrix33}{zeethreefour}  in terms of the Higgs basis parameters are valid for the most general 2HDM scalar potential.}
The eigenvalues of $\mathcal{M}^2$ are independent of the choice of $\eta$, since these cannot depend on the phase choice used in the definition of the Higgs basis field $\mathcal{H}_2$.  Hence, in practical calculations, one can choose $\eta$ to facilitate the analysis.  

For example, if we choose $\eta=-\xi$, then the neutral scalar squared-mass matrix is given by,
\beq \label{massN}
\mathcal{M}^2=v^2 \begin{pmatrix} \lambda-L s_{2\beta}^2& \quad -L s_{2\beta}c_{2\beta}
& \quad -\half\lambda_5   s_{2\beta} \sin 2\xi \\ -L s_{2\beta}c_{2\beta} & \quad \frac{2\Re(m_{12}^2 e^{i\xi})}{v^2 s_{2\beta}}+L s_{2\beta}^2 & \quad -\half \lambda_5 c_{2\beta}\sin 2\xi \\  -\half\lambda_5   s_{2\beta} \sin 2\xi  & \quad -\half \lambda_5 c_{2\beta}\sin 2\xi  & \quad  \frac{2\Re(m_{12}^2 e^{i\xi})}{v^2 s_{2\beta}} -\lambda_5\cos 2\xi\end{pmatrix},
\eeq
where
\beq \label{elldef}
L\equiv \half\lambda(1-R)+\lambda_5\sin^2\xi\,.
\eeq

If $Z_6\neq 0$ and/or $Z_7\neq 0$ then the neutral scalar squared-mass matrix has a block diagonal form consisting of a $2\times 2$ block and a $1\times 1$ block if and only if
$\Im(Z_5 e^{-2i\eta})=0$ and $\Re(Z_6 e^{-i\eta})\Im(Z_6 e^{-i\eta})=0$.
In such cases, the scalar potential and vacuum are CP conserving, and
we shall employ the following convention for the names of the neutral scalar mass eigenstates:  the CP-even scalars whose squared masses are the eigenvalues of the $2\times 2$ block will be denoted by $h$ and $H$ where $m_h\leq m_H$, and the $1\times 1$ block will be identified with the squared mass of the CP-odd scalar, $A$.  

\subsection{The softly broken \texorpdfstring{$\mathbb{Z}_2\otimes\Pi_2$}{\blackboardZ\texttwoinferior\directprod\uppercasePi\texttwoinferior}-symmetric scalar potential with one vanishing vev}

In the case where one of the vevs vanishes (i.e., $s_{2\beta}=0$), \eqs{inert1}{inert2} imply that $m_{12}^2=0$.
For example, if $v_1=v$ and $v_2=0$ then $Y_2$ is a free parameter, $Z_i=\lambda_i$, $c_{2\beta}=1$, and $\xi$ is indeterminate.  In particular, $Y_3=Z_6=Z_7=0$, which signals the presence of a $\mathbb{Z}_2$ symmetry, $\mathcal{H}_1\to +\mathcal{H}_1$, $\mathcal{H}_2\to -\mathcal{H}_2$, that is not broken by the vacuum.  This is a special case of the IDM and 
corresponds to the inert limit of the softly broken $\mathbb{Z}_2\otimes\Pi_2$-symmetric scalar potential.  In particular, 
$Y_2=m_{22}^2$ is a~free~parameter that is generically not equal to $Y_1=m_{11}^2=-\half \lambda v^2$.
To obtain the neutral scalar squared-mass matrix from \eq{matrix33}, we must make a choice of $\eta$.
For reasons discussed below \eq{etapr}, 
we shall choose $e^{-2i\eta}=-1$.
The neutral scalar squared-mass matrix is then diagonal and
we may identify,
\beqa
m_A^2&=&Y_2+\half\lambda  v^2 R\,,\label{inertmass1}\\
m_{H^\pm}^2&=&Y_2+\half Z_3 v^2=m_A^2-\half(\lambda_4+\lambda_5) v^2\,,\label{inertmass2}\\\
m^2_h&=& \lambda v^2\,,\label{inertmass3}\\\
m^2_H&=&m_A^2-\lambda_5 v^2\,,\label{inertmass4}
\eeqa
where we have denoted the mass eigenstate neutral scalar fields in the inert limit by
\beq \label{inertnames}
h\equiv \sqrt{2}\,\Re \mathcal{H}_1^0-v\,,\qquad H\equiv\sqrt{2}\,\Re \mathcal{H}_2^0\,,\qquad A\equiv \sqrt{2}\,\Im \mathcal{H}_2^0\,.
\eeq
This nomenclature (where no mass ordering is implied) will be employed in all subsequent occurrences of the inert limit in this work, and differs from the convention adopted in the paragraph following \eq{elldef} for the CP-conserving case where $Z_6\neq 0$ and/or $Z_7\neq 0$.  

If $v_1=0$ and $v_2=v$, then
one transforms to the Higgs basis via $\Phi\to U\Phi$ with $U=\left(\begin{smallmatrix} 0 & 1 \\ 1 & 0\end{smallmatrix}\right)$.   
In this case, $Y_2=m_{11}^2$ is a free parameter, $Y_1=m_{22}^2=-\half\lambda v^2$, $Y_3=Z_6=Z_7=0$, $Z_i=\lambda_i$, $c_{2\beta}=-1$, and $\xi=0$.  The scalar squared masses are again given by \eqst{inertmass1}{inertmass4}.

In the inert limit where $Y_3=Z_6=Z_7=0$, the scalar potential and vacuum are automatically CP-conserving.  In particular, in the inert limit the neutral scalars consist of a CP-even neutral scalar $h$ whose properties coincide with those of the SM Higgs boson and
two neutral scalars $H$ and $A$ with opposite sign CP quantum numbers.  However, one cannot separately assign unique CP quantum numbers to $H$ and $A$, respectively, based on the interactions of the scalars with the gauge bosons and the scalar self-interactions.\footnote{Indeed, the choice of $e^{-2i\eta}=1$ would have interchanged the identities of $H$ and $A$ in \eqst{inertmass1}{inertmass4}.}
CP-conserving interactions of the scalars with other sectors of the theory, if present,
will often resolve the ambiguity and identify $A$ as the neutral CP-odd scalar of the inert scalar doublet.   

For example, the most general form for CP-conserving neutral Higgs interactions with
one generation of fermions in the inert limit is obtained by setting $q_{11}=1$, $q_{22}=1$, $q_{32}=i$ (with all other $q_{kj}=0$), $\rho^{D*}=\rho^D$, and $\rho^{U*}=\rho^U$ in eq.~(58) of Ref.~\cite{Boto:2020wyf}, which yields,\footnote{Introducing Yukawa interactions constitutes a hard breaking of the symmetries responsible for the ERPS.  Thus, in this paper we shall not
entertain such terms further.} 
\beq \label{Yuk}
-\mathscr{L}_{\rm Y}=\frac{1}{v}\left(m_d\bar{d}d+m_u\bar{u}u\right)h+\frac{1}{\sqrt{2}}\left(\rho^D\bar{d}d+\rho^U\bar{u}u\right)H+\frac{i}{\sqrt{2}}
\left(\rho^D\bar{d}\gamma\ls{5}d-\rho^U\bar{u}\gamma\ls{5}u\right)A\,,
\eeq
indicating that $h$ behaves like the SM Higgs boson,
$H$ is CP-even and $A$ is CP-odd.  Note that $\rho^D=\rho^U=0$ in the IDM, since $\mathcal{H}_2$ is the only $\mathbb{Z}_2$-odd field of the model, in which case
the individual CP-quantum numbers of $H$ and $A$ are not resolved.   

\subsection{Non-coexistence of an inert phase and a mixed phase}

Let us examine more closely when a vacuum in which one of the two vevs vanishes can arise.   
Here, we shall extend the analysis of Ref.~\cite{Draper:2020tyq} to the case of $\lambda_5\neq 0$.
First, we require that $R> -1$ due to \eq{ineqcp2}.
If $v_1=v$ and $v_2=0$, then \eq{inert2} yields $m_{12}^2 =0$\ and $m_{11}^2=-\half \lambda v^2<0$.
The value of the scalar potential at the minimum is $V_{\rm min}=-(m_{11}^2)^2/(2\lambda^2)$.
The positivity of $m_A^2$ given in \eq{inertmass4} yields $m_{22}^2+\half\lambda Rv^2>0$.    Hence, it follows that
\beq \label{m22gtRm11}
m_{22}^2>Rm_{11}^2\,.
\eeq
The above inequality is equivalent to
\beq
(1+R)(m_{11}^2-m_{22}^2)<(1-R)(m_{11}^2+m_{22}^2)\,.
\eeq
Since $1+R$ is always positive, it follows that
\beq \label{Rineq2}
m_{22}^2-m_{11}^2> -\left(\frac{1-R}{1+R}\right)(m_{11}^2+m_{22}^2)\,.
\eeq

In the case of $v_1=0$ and $v_2=v$, the roles of $m_{11}^2$ and $m_{22}^2$ are interchanged.  That is,
\beq \label{Rineq3}
m_{22}^2-m_{11}^2< \left(\frac{1-R}{1+R}\right)(m_{11}^2+m_{22}^2)\,.
\eeq

Although the vanishing of one of the two vevs requires that $m_{12}^2=0$, the converse is not necessarily true.   If $m_{12}^2=0$ then two different phases of the 2HDM are possible---an inert phase where one of the two vevs vanishes and a mixed phase where both vevs are nonzero.  To analyze the latter possibility in more detail, 
 we again extend the analysis presented in Ref.~\cite{Draper:2020tyq} to the case of $\lambda_5\neq 0$.
If $m_{12}^2=0$ and $v_1$, $v_2\neq 0$, then 
\eqst{minone}{minfour} yield,
\beqa
m_{11}^2 &=&-\half\lambda\bigl(v_1^2+R v_2^2\bigr)\,,\label{minonemixed} \\
m_{22}^2 &=& -\half\lambda\bigl(v_2^2+Rv_1^2\bigr)\,,\label{mintwomixed} \\
0&=& \lambda_5\sin 2\xi\,.
\eeqa
Since $\lambda_5\neq 0$ by assumption, it follows that $\sin 2\xi=0$ and $\cos 2\xi=\pm 1$.   One is always free to rephase one of the scalar doublet fields so that $\xi=0$, since the only possible effect on the scalar potential parameters is a sign change of $\lambda_5$.  In the convention where $\xi=0$, \eq{massN} yields $m_A^2=-\lambda_5 v^2$, which implies that $\lambda_5<0$.  \Eq{minV} then yields
\beq \label{vminmixed}
V_{\rm min}=-\tfrac18\lambda(v_1^4+v_2^4+2Rv_1^2 v_2^2)\,.
\eeq
It is convenient to eliminate $v_1$ and $v_2$ in favor of the scalar potential parameters.
Using \eqs{minonemixed}{mintwomixed}, one easily obtains,
\beq \label{vineq}
v_1^2=\frac{2}{\lambda}\left(\frac{m_{22}^2 R-m_{11}^2}{1-R^2}\right)\,,\qquad\quad
v_2^2 =\frac{2}{\lambda}\left(\frac{m_{11}^2 R-m_{22}^2}{1-R^2}\right)\,.
\eeq
Plugging these values into \eq{vminmixed} yields,
\beqa
V_{\rm min}&=&-\frac{1}{2\lambda(1-R^2)}\bigl[m_{11}^4+m_{22}^4-2Rm_{11}^2 m_{22}^2\bigr]\nonumber \\[6pt]
&=&-\frac{1}{4\lambda}\left[\frac{(m_{11}^2+m_{22}^2)^2}{1+R}+\frac{(m_{11}^2-m_{22}^2)^2}{1-R}\right]\,.
\eeqa

One can work out a number of inequalities that must be satisfied if the mixed phase is stable.  We again require that $R> -1$ in light of \eq{bb}.
Using \eq{massN} with $m_{12}^2=\xi=0$, the trace and determinant of the $2\times 2$ neutral CP-even scalar squared-mass matrix yield,
\beq
m^2_h+m^2_H=\lambda v^2\,,\qquad\quad m_h^2 m_H^2=\tfrac14 \lambda^2 v^4 s_{2\beta}^2(1-R^2)\,.
\eeq
Hence, the positivity of the scalar squared masses implies that $|R|<1$.

Next, we employ \eqs{minonemixed}{mintwomixed} along with $|R|<1$ to obtain,
\beqa
m_{11}^2+m_{22}^2&=&-\half\lambda v^2(1+R)<0\,,\\
m_{11}^2+m_{22}^2+\lambda v^2&=&\half\lambda v^2(1-R)>0\,.\label{secondineq}
\eeqa
Using \eq{betacp2eq} and $|c_{2\beta}|\leq 1$, it follows that $m_{11}^2\geq -\half \lambda v^2$ and $m_{22}^2\geq -\half \lambda v^2$.  However, it is again more useful to provide inequalities that are independent of the vevs.   In light of \eq{vineq}, the requirement that $v_1^2$ and $v_2^2$ are strictly positive implies that
\beq \label{Rineq}
m_{22}^2 R>m_{11}^2\,,\qquad\quad m_{11}^2 R>m_{22}^2\,,
\eeq
The above inequalities are equivalent to
\beq \label{Rineq1}
\left(\frac{1-R}{1+R}\right)(m_{11}^2+m_{22}^2)< m_{22}^2-m_{11}^2< -\left(\frac{1-R}{1+R}\right)(m_{11}^2+m_{22}^2)\,.
\eeq
Comparing \eq{Rineq1} with \eqs{Rineq2}{Rineq3}, it follows that the mixed phase and the inert phase do not coexist~\cite{Barroso:2007rr,Ivanov:2007de,Draper:2020tyq}.

\subsection{CP properties of the softly broken \texorpdfstring{$\mathbb{Z}_2\otimes\Pi_2$}{\blackboardZ\texttwoinferior\directprod\uppercasePi\texttwoinferior}-symmetric scalar potential with \texorpdfstring{$m_{12}^2\neq 0$}{m\texttwosuperior\textoneinferior\texttwoinferior \textneq 0}}

Returning to the more general case where $m_{12}^2\neq 0$,
the scalar sector is CP conserving if and only if $\Im(Z_5^* Z_6^2)=0$.    A straightforward computation yields, 
\beqa
\Im(Z_5^* Z_6^2)&=&-\tfrac14\lambda\lambda_5(1-R)s_{2\beta}^2 c_{2\beta}\sin 2\xi\bigl[\lambda(1-R)+2\lambda_5\bigr] \nonumber \\
&=&-\tfrac14\lambda_5(\lambda-\lambda_3-\lambda_4-\lambda_5)(\lambda-\lambda_3-\lambda_4+\lambda_5)s_{2\beta}^2 c_{2\beta}\sin 2\xi\,. \label{nocpv}
\eeqa
The case of $s_{2\beta}=0$ corresponds to the inert limit, which has already been treated above.  In light of \eq{fiveconds}, the conditions $\lambda_5=0$ and
$\lambda-\lambda_3-\lambda_4=\pm\lambda_5$ correspond to the ERPS4 where a U(1) symmetry is manifestly realized in some basis.  In particular,
 $\lambda=\lambda_3+\lambda_4\pm\lambda_5$ correspond to GCP3 and GCP3$^\prime$, respectively, whereas $\lambda_5=0$ corresponds to U(1)$\otimes\Pi_2$, which is equivalent to GCP3 and GCP3$^\prime$ in different choices of the scalar field basis.  For example,
GCP3$^\prime$ is related to GCP3 via the basis change specified in \eq{simplebasischange}. These enhanced symmetry cases will be treated separately in Sections~\ref{sec:UPibasis}--\ref{transforming}.  

In this section, we shall assume that $s_{2\beta}\neq 0$, $\lambda\neq\lambda_3+\lambda_4\pm\lambda_5$ and $\lambda_5\neq 0$, in which case
CP is conserved if either (or both) of the following two conditions hold,
\beq \label{resCPconds}
 c_{2\beta}=0\quad \text{and/or}\quad  \sin 2\xi=0\,.
\eeq

\subsubsection{$\cos 2\texorpdfstring{\beta}{\textbeta}=0$}

In the case of $c_{2\beta}=0$, 
it follows that CP is conserved despite the fact that one cannot separately rephase $\Phi_1$ and $\Phi_2$ in the $\mathbb{Z}_2\otimes\Pi_2$ basis such that all the parameters of the scalar potential are real if $\Im\bigl(\lambda_5^* [m^2_{12}]^2]\bigr)\neq 0$, as was already noticed in Ref.~\cite{Boto:2020wyf}.  To understand the origin of this result, note that \eq{betacp2eq} implies that $m_{11}^2=m_{22}^2$ when $c_{2\beta}=0$.  Together with the ERPS4 conditions, it follows that the scalar potential is invariant with respect to a GCP1$^\prime$ transformation [cf.~Tables~\ref{tab:symm4} and \ref{tab:class2}].  Moreover, the condition of $c_{2\beta}=0$ ensures that the GCP1$^\prime$ symmetry is 
preserved by the vacuum.

When $c_{2\beta}=0$, \eq{massN} is rendered block diagonal, with the $2\times 2$ block identified as the squared-mass matrix of the neutral CP-even scalars.  It then follows that
\beqa
m_A^2&=&2\Re(m_{12}^2 e^{i\xi})+\half \lambda v^2(1-R)+\lambda_5 v^2\sin^2\xi\,, \label{Amassc} \\
m^2_{H^\pm}&=&m_A^2-\half v^2\bigl[\lambda(1-R)+\lambda_4+\lambda_5\bigr]\,, \label{Hpmmassc}
\eeqa
where $A=\varphi_2\equiv\sqrt{2}\,\Re \mathcal{H}_2^0$.  The squared-mass matrix of the neutral CP-even scalars is 
\beq \label{hHmassc}
\mathcal{M}_H^2=\begin{pmatrix} \half\lambda v^2(1+R)-\lambda_5 v^2\sin^2\xi & \quad -\lambda_5 v^2\sin\xi\cos\xi \\ -\lambda_5 v^2\sin\xi\cos\xi & \quad m_A^2-\half\lambda v^2(1-R)-\lambda_5 v^2\cos^2\xi\end{pmatrix},
\eeq
with respect to the $\{\varphi_1,\varphi_3\}$ basis, where $\varphi_1\equiv \sqrt{2}\,\Re \mathcal{H}_1^0-v$ and $\varphi_3\equiv\sqrt{2}\,\Im \mathcal{H}_2^0$.
The neutral CP-even scalar mass eigenstates are given by,
\beq \label{Hscalareigenstates}
\hh=\varphi_1\cbma-\varphi_3\sbma\,,\qquad\quad \hl=\varphi_1\sbma+\varphi_3\cbma\,,
\eeq
where $0\leq\beta-\alpha\leq\pi$, $\sbma\equiv\sin(\beta-\alpha)$ and $\cbma\equiv\cos(\beta-\alpha)$,
\beq \label{msquaredHh}
m^2_{H,h}=\half\biggl\{\mha^2+(\lambda R-\lambda_5) v^2\pm \sqrt{\bigl[\mha^2-v^2(\lambda+\lambda_5\cos 2\xi)\bigr]^2+\lambda_5^2 v^4\sin^2 2\xi}\,
\biggr\}\,,
\eeq
with $m_h\leq m_H$, and~\!\footnote{In obtaining \eq{cbmawhenc2bzero}, we have employed \eqs{CPmassmatrix}{ctozero}, where the real quantity $Z_6 v^2$ in these equations is to be identified with the off-diagonal element of $\mathcal{M}^2_H$ given in \eq{hHmassc}.}
\beq \label{cbmawhenc2bzero}
c_{\beta-\alpha}=\frac{\lambda_5 v^2\sin 2\xi}{2\sqrt{(m_H^2-m_h^2)\bigl[m_H^2-\half\lambda v^2(1+R)+\lambda_5 v^2\sin^2\xi\bigr]}}\,.
\eeq

\subsubsection{$\sin 2\texorpdfstring{\xi}{\textxi}=0$}

In the case of $\sin 2\xi=0$, \eqs{cosxi}{sinxi} imply that $\Im\bigl[(m_{12}^2)^2\bigr]\!=\!2\Re m_{12}^2 \Im m_{12}^2\!=\!0$.   If $\sin\xi=0$ then $\Im m_{12}^2=0$ and all scalar potential parameters are real, whereas if $\cos\xi=0$ then $\Re m_{12}^2=0$ and a rephasing $\Phi_2\to i\Phi_2$ changes the sign of the real parameter $\lambda_5$ while removing the complex phase of $m_{12}^2$.    Hence a real basis exists,\footnote{A real basis is defined to be a scalar field basis in which the scalar potential parameters and the vevs are simultaneously real.} 
which implies that the scalar potential and the vacuum are CP conserving.
The neutral scalar squared-mass matrix given in
\eq{massN} is block diagonal when $\sin 2\xi=0$, with the 33 element identified as 
the squared mass of the CP-odd scalar, $A=\varphi_3\equiv\sqrt{2}\,\Im \mathcal{H}_2^0$.

For $\sin\xi=0$, it follows from \eqs{chargedHmass}{massN} that 
\beq \label{Amass}
m_A^2=\pm\frac{2\Re m^2_{12}}{s_{2\beta}}-\lambda_5 v^2\,,\qquad\quad m^2_{H^\pm}=m_A^2+\half(\lambda_5-\lambda_4)v^2\,,
\eeq
where 
the choice of signs corresponds to $\cos\xi=\pm 1$.  
The upper $2\times 2$ block of \eq{massN} is identified as the
squared-mass matrix of the CP-even neutral scalars,

\beq \label{calmh}
\mathcal{M}^2_H=\begin{pmatrix} \lambda v^2\bigl[1-\half  s^2_{2\beta}(1-R)\bigr] & \quad -\half\lambda v^2 s_{2\beta}c_{2\beta}(1-R) \\
 -\half\lambda v^2 s_{2\beta}c_{2\beta}(1-R) & \quad m_A^2+\lambda_5 v^2+\half\lambda v^2  s^2_{2\beta}(1-R) \end{pmatrix},
 \eeq
 with respect to the $\{\varphi_1,\varphi_2\}$ basis, where $\varphi_1\equiv \sqrt{2}\,\Re \mathcal{H}_1^0-v$ and $\varphi_2\equiv\sqrt{2}\,\Re \mathcal{H}_2^0$.
Hence,
 \beq \label{hHmass}
 m_{H,h}^2=\half\biggl\{m_A^2+(\lambda+\lambda_5)v^2\pm\sqrt{\bigl[m_A^2+\lambda_5 v^2-\lambda v^2(c_{2\beta}^2+Rs_{2\beta}^2)\bigr]^2+\lambda^2 s_{2\beta}^2 c_{2\beta}^2(1-R)^2 v^4}\biggr\}\,,
 \eeq
 with $m_h\leq m_H$, and
 \beq
 c_{\beta-\alpha}=\frac{\lambda v^2 s_{2\beta} c_{2\beta}(1-R)}{2\sqrt{(m_H^2-m_h^2)\bigl[m_H^2-\lambda v^2\bigl(1-\half s_{2\beta}^2(1-R)\bigr)\bigr]}}\,.\label{cbmawhens2xizero}
 \eeq
 
For $\cos\xi=0$, the results of \eqst{Amass}{cbmawhens2xizero} are modified by the following substitutions,
\beq \label{subs}
\pm\Re m_{12}^2\to \mp\Im m_{12}^2\,,\qquad \lambda_5\to -\lambda_5\,,\qquad R\to \overline{R}\equiv (\lambda_3+\lambda_4-\lambda_5)/\lambda\,,
\eeq
where the choice of signs in front of $\Im m_{12}^2$ corresponds to $\sin\xi=\pm 1$.  Note that $\overline{R}\neq 1$ under the assumption specified above \eq{resCPconds}.   
If $\overline{R}=1$ then the (softly broken) $\mathbb{Z}_2\otimes\Pi_2$ symmetry of the scalar potential is promoted to GCP3$^\prime$, as discussed below \eq{nocpv}.

\subsubsection{\texorpdfstring{$\cos 2\beta=\sin 2\xi=0$}{cos 2\textbeta=sin 2\textxi=0}}

If $c_{2\beta}=\sin 2\xi=0$ then \eqs{zeesevencp2}{YZ} yield $Y_3=Z_6=Z_7=0$, corresponding to an inert limit
of the softly broken $\mathbb{Z}_2\otimes\Pi_2$-symmetric scalar potential.   If $\sin\xi=0$, then one can obtain the scalar squared masses either by taking the 
$\sin\xi=\Im m_{12}^2=0$ limit of \eqst{Amassc}{hHmassc} or by taking the $c_{2\beta}=0$ limit of 
\eqst{Amass}{hHmass}.  Recall that we have identified the neutral scalar mass eigenstates in the convention specified in \eq{inertnames}.
Taking into account that $\mathcal{M}^2_H$ is exhibited with respect to the $\{\varphi_1,\varphi_3\}$ basis in \eq{hHmassc} and with respect to the $\{\varphi_1,\varphi_2\}$ basis in \eq{calmh}, respectively, it follows that
\beqa 
m_h^2&=&\half\lambda v^2(1+R)\,,\qquad \qquad \qquad\qquad\quad\,\,\, m_A^2=\pm 2\Re m_{12}^2-\lambda_5 v^2\,, \nonumber \\
m_H^2&=& m_A^2+\lambda_5 v^2+\half\lambda v^2(1-R)\,,\qquad\quad 
m_{H^\pm}^2= m_A^2+\half(\lambda_5-\lambda_4)v^2\,. \label{spectrum}
\eeqa
If $\cos\xi=0$, then \eq{spectrum} is modified by applying the substitutions indicated in \eq{subs}.
In the inert limit, the vacuum preserves the $\Pi_2$ symmetry (whereas the $\mathbb{Z}_2$ symmetry remains softly broken since $m_{12}^2\neq 0$).

\subsubsection{Spontaneous vs.~Explicit CP violation}

Spontaneous CP violation can occur when $\Im\bigl[m_{12}^2\bigr]^2=0$ (with $m_{12}^2\neq 0$) and $\sin 2\xi\neq 0$.   In addition, as noted in Appendix~\ref{app:scpv} below \eq{H}, one must assume that $\lambda_5>0$ in order to guarantee that this CP-violating vacuum solution is a local minimum. 
If  $\Im m_{12}^2=0$ and $\sin 2\xi\neq 0$, 
then \eq{sinxi} implies that $m_{11}^2+m_{22}^2+\half\lambda(1+R)v^2=\lambda_5 v^2$.  Inserting this result into \eq{cosxi} yields
$\cos\xi=\Re m_{12}^2/(\lambda_5 v^2 s_\beta c_\beta)$;
i.e., spontaneous CP violation occurs if~\cite{Gunion:2002zf},
\beq \label{spcpv}
0<|m_{12}^2|<\lambda_5 v^2 s_\beta c_\beta\,.
\eeq
Likewise, if 
$\Re m_{12}^2=0$ and $\sin 2\xi\neq 0$, then \eq{cosxi} implies that $m_{11}^2+m_{22}^2+\half\lambda(1+R)v^2=0$. Inserting this result into \eq{sinxi} yields $\sin\xi=\Im m_{12}^2/(\lambda_5 v^2 s_\beta c_\beta)$.
Once again, spontaneous CP violation occurs if \eq{spcpv} is satisfied.

If $\Im(Z_5^* Z_6^2)\neq 0$ then the scalar potential is explicitly CP-violating.   In this case, one must diagonalize the $3\times 3$ neutral scalar squared-mass matrix given in \eq{massN} to determine the neutral scalar mass eigenstates.  The inert limit cannot be realized in this case, so the presence of scalar-mediated CP-violating effects necessarily implies that the tree-level properties of any of the three neutral scalar mass eigenstates must deviate from those of the SM Higgs boson.   This is a special case of the C2HDM that has been explored in Ref.~\cite{Boto:2020wyf}.

\subsection{Scalar potential with an unbroken \texorpdfstring{$\mathbb{Z}_2\otimes\Pi_2$}{\blackboardZ\texttwoinferior\directprod\uppercasePi\texttwoinferior}-symmetry}

The $\mathbb{Z}_2\otimes\Pi_2$ symmetry of the scalar potential is unbroken if $m_{11}^2=m_{22}^2$ and $m_{12}^2=0$.  First, we suppose that both vevs are nonzero.
Then \eq{why3} implies that
$Y_3=0$, which yields $Y_3=Z_6=Z_7=0$ in light of \eq{YZ} and the ERPS condition, 
corresponding to the inert limit of the $\mathbb{Z}_2\otimes\Pi_2$-symmetric scalar potential.
Moreover in light of \eqs{min3}{betacp2eq}, $c_{2\beta}=\sin 2\xi=0$ in the $\mathbb{Z}_2\otimes\Pi_2$ symmetry limit, and it follows that
the vacuum breaks $\mathbb{Z}_2$ but conserves $\Pi_2$.
The scalar squared masses in the limit of $c_{2\beta}=\sin\xi=0$ and $m_{12}^2=0$ are given by the $m^2_{12}=0$ limit of \eq{spectrum}.
Stability of the scalar potential requires that the squared masses of the scalars are positive, which yields $\lambda_5=-|\lambda_5|<0$, $\lambda_4<|\lambda_5|$ and $|R|<1$.
Likewise, for $c_{2\beta}=\cos\xi=0$, the scalar squared masses are given by $m_{12}^2=0$ limit of \eq{spectrum} after replacing $\lambda_5\to -\lambda_5$ and $R\to\overline{R}$ [cf.~\eq{subs}], in 
which case the stability requirement yields $\lambda_5>0$, $\lambda_4<\lambda_5$ and $|\overline{R}|<1$.

If one of the vevs vanishes (i.e., $s_{2\beta}=0$) then
the $\mathbb{Z}_2\otimes\Pi_2$ symmetry limit   \pagebreak
of \eq{inertmass1} corresponds
to setting $Y_2=Y_1=-\half\lambda v^2$ [cf.~\eqs{inert1}{inert2}], which yields
\beq \label{mhainert}
m_{A}^2=\half\lambda v^2(R-1)\,,
\eeq  
which requires that $R>1$.
The squared masses of $H^\pm$, $h$ and $H$ in terms of $m_A^2$ given in \eqst{inertmass2}{inertmass4} remain unchanged.
In this case, the vacuum breaks $\Pi_2$ but conserves $\mathbb{Z}_2$.

\subsection{The landscape of ERPS4--Part I: scalar potential with a softly broken or unbroken \texorpdfstring{$\mathbb{Z}_2\otimes\Pi_2$}{\blackboardZ\texttwoinferior\directprod\uppercasePi\texttwoinferior} symmetry}

The landscape of scalar potentials in the ERPS4 that respects a softly broken or exact $\mathbb{Z}_2\otimes\Pi_2$ symmetry (but no larger symmetry) is summarized in Table~\ref{tab:erps4}. 
\vskip 0.2in

\begin{table}[h!]
\begin{tabular}{|c|c|c|c|c|c|c|}
\hline
$\beta$ &  $\pht\sin 2\xi\pht$ & $m_{11}^2$, $m_{22}^2$ & $m_{12}^2$ & CP-violation? & Higgs alignment  & comment \\
\hline
$s_{2\beta}\neq 0$ & $\neq 0$ &  $m_{11}^2\neq m_{22}^2$ & complex & explicit &  no & $\Im\bigl[m_{12}^2\bigr]^2\neq 0$  \\
$s_{2\beta}\neq 0$ &  $\neq 0$ &  $m_{11}^2\neq m_{22}^2$  & $\pht\Im\bigl[m_{12}^2\bigr]^2=0\pht$ & spontaneous &  no & $0<|m_{12}^2|<\half\lambda_5 v^2 s_{2\beta}$   \\
$s_{2\beta}\neq 0$ & $\neq 0$ &  $m_{11}^2\neq m_{22}^2$  & $\Im\bigl[m_{12}^2\bigr]^2=0$& no &  no & $|m_{12}^2|>\half\lambda_5 v^2 s_{2\beta}$   \\
$c_{2\beta}= 0$ &  $\neq 0$ &  $\pht m_{11}^2= m_{22}^2\pht$  & complex & no &  no & $m_{12}^2\neq 0$ \\
$s_{2\beta}c_{2\beta}\neq 0$ & $0$ &  $m_{11}^2\neq m_{22}^2$  & $\Im\bigl[m_{12}^2\bigr]^2=0$ & no&  no &  \\
$s_{2\beta}=0$ &  &  $m_{11}^2\neq m_{22}^2$  & 0 & no &  yes &   \\
$c_{2\beta}=0$ & $0$ &  $m_{11}^2= m_{22}^2$  & $\Im\bigl[m_{12}^2\bigr]^2=0$ & no&  yes & $m_{12}^2\neq 0$ \\
$s_{2\beta}=0$ &  &  $m_{11}^2=m_{22}^2$  & 0 & no& yes& unbroken $\mathbb{Z}_2\otimes\Pi_2$ \\
$c_{2\beta}=0$ & $0$ &  $m_{11}^2= m_{22}^2$  & 0 & no&  yes &  unbroken $\mathbb{Z}_2\otimes\Pi_2$ \\
\hline
\end{tabular}
\caption{\small \baselineskip=15pt
Landscape of the ERPS4--Part I: Scalar potentials of the 2HDM with either an unbroken or softly broken $\mathbb{Z}_2\otimes\Pi_2$ symmetry that is manifestly realized in the $\Phi$-basis.   In all cases,
$\lambda\equiv\lambda_1=\lambda_2\neq \lambda_3+\lambda_4\pm \lambda_5$, where $\lambda_5$ is real and nonzero, $\lambda_6=\lambda_7=0$, and $\lambda$, $\lambda+\lambda_3$, and $\lambda+\lambda_3+\lambda_4-|\lambda_5|$ are all positive.  Note that if $m_{12}^2$ is purely imaginary, one can rephase $\Phi_2\to i\Phi_2$ to obtain a new basis where 
$m_{12}^2$ is real and~$\lambda_5$ flips sign.  An exact Higgs alignment in the ERPS4 is realized in the inert limit where $Y_3=Z_6=Z_7=0$. \\[-20pt]
\label{tab:erps4}}
\end{table}

\section{The \texorpdfstring{U(1)$\otimes\Pi_2$}{U(1)\directprod\uppercasePi\texttwoinferior} scalar field basis}
\label{sec:UPibasis}
\vskip -0.1in
  
Consider the softly broken U(1)$\otimes\Pi_2$-symmetric scalar potential where $\lambda\equiv \lambda_1=\lambda_2$ and $\lambda_5=\lambda_6=\lambda_7=0$.  The softly broken parameters, $m_{11}^2$, $m_{22}^2$ and $m_{12}^2$, are arbitrary with $m_{12}^2$ potentially complex.    \pagebreak
If we demand that the potential is bounded from below, the following conditions must be satisfied (modulo the remarks of footnote~\ref{fnstable}),
\beq \label{ineq}
\lambda>0\,,\qquad \lambda+\lambda_3>0\,,\qquad \lambda+\lambda_3+\lambda_4>0\,.
\eeq

\subsection{The softly broken \texorpdfstring{U(1)$\otimes\Pi_2$}{U(1)\directprod\uppercasePi\texttwoinferior}  or SO(3)-symmetric scalar potential with \texorpdfstring{$v_1\neq 0$}{v\textoneinferior \textneq 0} and \texorpdfstring{$v_2\neq 0$}{v\texttwoinferior \textneq 0}}

We shall first assume that $v_1$ and $v_2$ are both nonzero, or equivalently $\sin 2\beta\neq 0$.  We then use 
\eqst{min1}{min3} [with $\lambda\equiv\lambda_1=\lambda_2$] to obtain,
\beqa
m_{11}^2 &=& \Re(m_{12}^2 e^{i\xi})\tan\beta-\half\lambda v^2 c^2_\beta-\half(\lambda_3+\lambda_4) v^2 s^2_\beta\,,\label{min1a} \\
m_{22}^2 &=& \Re(m_{12}^2 e^{i\xi})\cot\beta-\half\lambda v^2  s_\beta^2-\half(\lambda_{3}+\lambda_4) v^2 c_\beta^2\,, \label{min2a} \\
\Im(m_{12}^2 e^{i\xi})&=&0\,.\label{min3a}
\eeqa
\Eqs{min1a}{min2a} fix the value of $\beta$.  In particular,
\beq \label{betaeq}
c_{2\beta}=\frac{m_{22}^2-m_{11}^2}{m_{11}^2+m_{22}^2+\lambda v^2}\,.
\eeq

Since $m_{12}^2$ is the only potentially complex parameter, one can assume without loss of generality that $m_{12}^2$ is real and non-negative after an appropriate rephasing of one of the two Higgs doublet fields.   
Hence, \eq{min3a} implies that the scalar sector is CP-conserving.
Nevertheless, in the analysis presented in this section, we find it convenient to retain all factors of $e^{i\xi}$ for later purposes, which simply means that 
$m_{12}^2 e^{i\xi}=\Re(m_{12}^2 e^{i\xi})\geq 0$,
in light of \eq{min3a} and the requirement that $m_A^2\geq 0$ [cf.~\eq{mha}].

The corresponding parameters of the Higgs basis are obtained by setting $\lambda_5=0$ in \eqst{whytwocp2}{zeesevencp2},
\beqa
&& Y_2=\frac{2\Re(m_{12}^2 e^{i\xi})}{s_{2\beta}}-\half\lambda v^2\bigl[R+\half s_{2\beta}^2(1-R)\bigr]\,,\label{whytwo}\\
&& Z_1=Z_2=\lambda\bigl[1-\half s^2_{2\beta}(1-R)\bigr]\,,\label{zeeone}\\
&& Z_3=\lambda_3+\half\lambda  s^2_{2\beta}(1-R)\,, \\
&& Z_4=\lambda_4+\half\lambda  s^2_{2\beta}(1-R)\,, \\
&& Z_5=\half\lambda s^2_{2\beta}(1-R) e^{-2i\xi}\,,\label{zeefive} \\
&& Z_6=-Z_7=-\half \lambda s_{2\beta}c_{2\beta}(1-R)e^{-i\xi}\,,\label{zeeseven}
\eeqa
where 
\beq \label{Rdef}
R\equiv \frac{\lambda_3+\lambda_4}{\lambda}\,.   
\eeq
\clearpage

\noindent
Note that $R> -1$ in light of \eq{ineq}.\footnote{If $R=-1$ then the quartic terms of the scalar potential exhibit a flat direction.   One can then ensure the stability of the scalar potential in the weak sense [cf.~footnote~\ref{fnstable}] if $m_{11}^2+m_{22}^2>2|m_{12}^2|$~\cite{Maniatis:2006fs,Martin:1997ns}. \label{fnminusone}}
The limit of $R=1$ corresponds to the softly broken SO(3)-symmetric scalar potential, where
the conditions $\lambda_1=\lambda_2=\lambda_3+\lambda_4$ and $\lambda_5=\lambda_6=\lambda_7=0$ hold for all choices of the scalar field basis.

The Higgs basis parameters $Y_1$ and $Y_3$ are fixed by the potential minimum conditions given in \eq{YZ}.
Note that $\Im(Z_5^* Z_6^2)=0$, which implies that a real Higgs basis
exists after an appropriate rephasing of the Higgs basis field $\mathcal{H}_2$.  That is,
there is no CP violation (neither explicit nor spontaneous) arising from a scalar potential that exhibits a
softly broken U(1)$\otimes\Pi_2$ symmetry.
Using \eqst{zeeone}{zeeseven}, it follows that the following conditions are satisfied,
\beq \label{invcond12}
[\Re(Z_5^* Z_6^2)]^2+\Re(Z_5^* Z_6^2)|Z_6|^2(Z_1-Z_{34})-2|Z_6|^6=0\quad \text{and} \quad
 \Im(Z_5^* Z_6^2)=0\,.
\eeq
We recognize these conditions as equivalent to \eq{GCP3invar} when applied in the Higgs basis. 

The squared masses of the neutral Higgs bosons are obtained by computing the eigenvalues of \eq{matrix33}.  In light of \eqs{zeefive}{zeeseven},
it is convenient to take $\eta=-\xi$ in \eq{matrix33}, since this choice yields $\Im(Z_5 e^{-2i\eta})=\Im(Z_6 e^{-i\eta})=0$.   One can then immediately identity the squared mass of the CP-odd neutral scalar $A=\varphi_3\equiv \sqrt{2}\,\Im \mathcal{H}_2^0$,
\beq \label{mha}
m_A^2=\half v^2\bigl[Z_{34}-\Re(Z_5 e^{2i\xi})\bigr]+Y_2=\frac{2\Re(m_{12}^2 e^{i\xi})}{s_{2\beta}}\,.
\eeq
Combining \eqss{min1a}{min2a}{mha} yields an alternative expression,
\beq \label{mha2}
m_A^2=m_{11}^2+m_{22}^2+\half \lambda v^2(1+R)\,.
\eeq
In light of \eq{betaeq} [which was obtained under the assumption that $s_{2\beta}\neq 0$] and \eq{mha2}, it follows that
\beq \label{m22m11diff}
m_{22}^2-m_{11}^2=\bigl[m_{11}^2+m_{22}^2+\lambda v^2\bigr]c_{2\beta}=\bigl[m_A^2+\half\lambda v^2(1-R)\bigr]c_{2\beta}\,.
\eeq
Hence, if $m_{11}^2=m_{22}^2$ then either $c_{2\beta}=0$ or $m_A^2=\half\lambda v^2(R-1)$.
If the latter is satisfied, then the stability of the vacuum requires that $R>1$.  
If in addition $m_{12}^2\neq 0$ and $c_{2\beta}\neq 0$, then 
$m_A^2\neq 0$ in light of \eq{mha2}, which eliminates the possibility of $R=1$.

Likewise, the charged Higgs squared mass is given by
\beq \label{mch}
m_{H^\pm}^2= Y_2+\half Z_3 v^2=m_A^2-\half \lambda_4 v^2\,,
\eeq
after making use of \eq{mha}.
Finally, the squared masses of the CP-even neutral scalars, denoted by $h$ and $H$, are the eigenvalues of the $2\times 2$ matrix,
\beqa
\mathcal{M}^2_H&=&\begin{pmatrix} Z_1 v^2 & \quad \Re(Z_6 e^{i\xi}) v^2 \\  \Re(Z_6 e^{i\xi}) v^2 & \quad m_A^2+\Re(Z_5 e^{2i\xi}) v^2\end{pmatrix} \nonumber \\[10pt]
&=&\begin{pmatrix}  \lambda v^2\bigl[1-\half s_{2\beta}^2(1-R)\bigr] & \quad -\half \lambda v^2 s_{2\beta}c_{2\beta}(1-R) \\ 
-\half \lambda v^2 s_{2\beta}c_{2\beta}(1-R)  & \quad m_A^2+\half \lambda v^2 s_{2\beta}^2 (1-R)\end{pmatrix}, \label{calmsqH}
\eeqa
with respect to the $\{\varphi_1,\varphi_2\}$ basis,\footnote{The computation of the squared-mass matrix of the CP-even neutral scalars in the $\Phi$-basis is given in Appendix~\ref{app:sqmass}.}
where $\varphi_1\equiv \sqrt{2}\,\Re \mathcal{H}_1^0-v$ and $\varphi_2\equiv\sqrt{2}\,\Re \mathcal{H}_2^0$.
The neutral CP-even scalar masses are given by,
\beq \label{Hhmasses2}
m^2_{H,h}=\half\biggl\{\mha^2+\lambda v^2\pm \sqrt{\bigl[\mha^2-\lambda v^2(c_{2\beta}^2+Rs_{2\beta}^2)\bigr]^2+\lambda^2 s_{2\beta}^2 c_{2\beta}^2(1-R)^2 v^4}\,
\biggr\}\,,
\eeq
with $\mhl\leq\mhh$, and
\beq
c_{\beta-\alpha}=\frac{\lambda v^2 s_{2\beta}c_{2\beta}(1-R)}{2\sqrt{(\mhh^2-\mhl^2)\bigl[\mhh^2-\lambda v^2\bigl(1-\half s_{2\beta}^2(1-R)\bigr)\bigr]}}\,.
\eeq

A stable minimum requires that the scalar squared masses should be non-negative.   This condition implies that 
\beq 
\Re(m_{12}^2 e^{i\xi})\geq 0\quad \text{and} \quad m_A^2\geq\half \lambda_4 v^2\,.
\eeq
In addition, we demand that
\beqa
\Tr\mathcal{M}_H^2&=&m_A^2+\lambda v^2\geq 0\,,\label{ineq1}\\
\frac{1}{v^2}\det \mathcal{M}_H^2&=&\tfrac14\lambda^2v^2 s_{2\beta}^2(1-R^2)+\lambda m_A^2\bigl[1-\half s_{2\beta}^2(1-R)\bigr]\geq 0\,.\label{ineq2}
\eeqa

Since $m_A^2\geq 0$ by assumption, \eq{ineq1} is automatically satisfied in light of \eq{ineq}.  On the other hand, \eq{ineq2} is satisfied only if $R$ lies below a critical positive value that depends on $\lambda$, $\beta$ and $m_A^2/v^2$,
\beq \label{bound}
-1< R\leq \frac{m_A^2}{\lambda v^2}+\sqrt{\left(\frac{m_A^2}{\lambda v^2}-1\right)^2+\frac{4m_A^2}{\lambda v^2s_{2\beta}^2}}\,,
\eeq
after employing \eq{ineq}.\footnote{Apart from the upper bound given in \eq{bound}, one can obtain an independent upper bound by imposing either tree-level unitarity~\cite{Weldon:1984wt,Ginzburg:2005dt,Horejsi:2005da,Kanemura:2015ska,Goodsell:2018fex} or a perturbativity constraint.   One would then expect $R/(4\pi)\lsim \mathcal{O}(1)$.}
It follows that \eq{ineq2} is satisfied for all values of $\beta$ if
\beq
-1< R\leq 1+\frac{2m_A^2}{\lambda v^2}\,.
\eeq

One can fix the parameter space of the softly broken U(1)$\otimes\Pi_2$ scalar potential by specifying the values of six real parameters: $\lambda$, $\lambda_4$, $R$, $\beta$, $m_A$ and $v=246$~GeV.   By replacing $\lambda$ with $m_h$ (see eq.~(131) of Ref.~\cite{Draper:2020tyq})
and $\lambda_4$ with $m_{H^\pm}$,  the independent parameters of the  softly broken U(1)$\otimes\Pi_2$ scalar potential can be taken as $m_h$, $m_A$, $m_{H^\pm}$, $v$, $R$ and $\beta$, in which case $m_H^2=m_A^2-m_h^2+\lambda v^2$ [cf.~\eq{ineq1}] is a derived quantity.

\subsection{The inert limit of the softly broken \texorpdfstring{U(1)$\otimes\Pi_2$}{U(1)\directprod\uppercasePi\texttwoinferior} or SO(3)-symmetric scalar potential}

The inert limit of the scalar potential, where $Y_3=Z_6=Z_7=0$, possesses
an exact $\mathbb{Z}_2$ symmetry 
despite the presence of squared mass parameters that softly break the $\ug\otimes\Pi_2$ symmetry.   The inert limit arises if either $v_1=0$ or $v_2=0$, but is more general.  Indeed, \eqs{bb}{zeeseven} imply that the inert limit arises if any one of the three conditions, $R=1$, $c_{2\beta}=0$, or $s_{2\beta}=0$, is satisfied. \\[-30pt]

\subsubsection{softly broken SO(3)-symmetric scalar potential ($R=1$)}

The case of $R=1$ corresponds to the softly broken SO(3) scalar potential as noted below \eq{Rdef}.  In light of \eqst{min1a}{min3a}, it follows that if $s_{2\beta}c_{2\beta}\neq0$ then $m_{11}^2\neq m_{22}^2$ and $\Re(m_{12}^2 e^{i\xi})\neq 0$.  
In this case, the squared masses of the Higgs bosons are given by
\beq \label{masssrelso3}
m_h^2=\lambda v^2\,,\qquad m_H^2=m_A^2\,,\qquad  m^2_{H^\pm}=m_A^2-\half\lambda_4 v^2\,,
\eeq
where $m_A^2=2\Re(m_{12}^2 e^{i\xi})/s_{2\beta}$.
The mass degeneracy of $H$ and~$A$ arises due to an unbroken U(1) symmetry of the scalar potential in the Higgs basis (since $Y_3=Z_5=Z_6=Z_7=0$) that is preserved by the vacuum.  
The $\Pi_2$ symmetry remains softly broken (since $Y_1\neq Y_2$). \\[-30pt]

\subsubsection{The softly broken \texorpdfstring{U(1)$\otimes\Pi_2$}{U(1)\directprod\uppercasePi\texttwoinferior}-symmetric scalar potential with $\cos 2\texorpdfstring{\beta}{\textbeta}=0$}

In the case of $c_{2\beta}=0$, \eq{betaeq} implies that $m_{11}^2=m_{22}^2$.  \Eqst{mha}{Hhmasses2} yield,
\beq \label{ctwobetamasses}
m_h^2=\half\lambda v^2(1+R)\,,\qquad m_H^2=m_A^2+\half\lambda v^2(1-R)\,,\qquad m^2_{H^\pm}=m_A^2-\half\lambda_4 v^2\,,
\eeq
where $m_A^2=2\Re(m_{12}^2 e^{i\xi})$, in agreement with the $\lambda_5=0$ limit of \eq{spectrum}.  
In this limiting case, after rephasing one of the two Higgs doublet fields to set $\xi=0$, the vacuum preserves the $\Pi_2$ symmetry [whereas the U(1) symmetry remains softly broken since $m_{12}^2\neq 0$].

\subsubsection{The softly broken \texorpdfstring{U(1)$\otimes\Pi_2$}{U(1)\directprod\uppercasePi\texttwoinferior}-symmetric scalar potential with one vanishing vev}

The case where one of the vevs vanishes (i.e., $s_{2\beta}=0$) should be treated separately and implies that $m_{12}^2=0$ in light of \eqs{inert1}{inert2}.  
One can check that \eqst{zeeone}{zeeseven} remain valid after setting $s_{2\beta}=0$.
In this case, the U(1) symmetry~of the scalar potential is unbroken, whereas the $\Pi_2$ symmetry is softly broken if $m_{11}^2\neq m_{22}^2$.

First, suppose that $v_2=0$ and $v_1=v$.   Then, \eq{inertmass1} yields,
\beq \label{inertA}
m_A^2=Y_2+\half \lambda v^2R\,,
\eeq
where $Y_2$ is a free parameter of the scalar potential that is no longer given by \eq{whytwo}.  Moreover, \eq{betaeq} is no longer valid since $Y_2=m_{22}^2$ is independent of the squared mass parameter~$m_{11}^2$; only the latter is fixed by the scalar potential minimum condition.
The squared masses of the other scalars are given by \eqst{inertmass2}{inertmass4} by setting $\lambda_5=0$,
\beq \label{inertB}
m_h^2=\lambda v^2\,,\qquad\quad m_H^2=m_A^2\,,\qquad\quad m_{H^\pm}^2=m_A^2-\half\lambda_4 v^2\,.
\eeq
Note that the U(1) symmetry is preserved by the vacuum, which results in the mass degeneracy of $H$ and $A$.
Second, if $v_1=0$ and $v_2=v$, then it follows that $Y_2=m_{11}^2$ is a free parameter and $Y_1=m_{22}^2=-\half \lambda v^2$.   
\Eqst{zeeone}{zeeseven} remain valid after setting $\beta=\half\pi$.   Moreover, the Higgs masses given by \eqs{inertA}{inertB} also remain valid.

\subsection{The mixed phase of the softly broken \texorpdfstring{U(1)$\otimes\Pi_2$}{U(1)\directprod\uppercasePi\texttwoinferior}-symmetric scalar potential with \texorpdfstring{$m_{12}^2=0$}{m\texttwosuperior\textoneinferior\texttwoinferior =0}}

Although the vanishing of one of the two vevs requires that $m_{12}^2=0$, the converse is not necessarily true, as previously noted.   That is, if $m_{12}^2=0$, then both an inert phase and a mixed phase of the 2HDM are possible.   The inequalities previously obtained that distinguish the inert and mixed phases in \eqss{Rineq2}{Rineq3}{Rineq1} still apply (after setting $\lambda_5=0$), and again imply that the inert and mixed phases do not coexist.  In the mixed phase with $m_{12}^2=0$, 
the scalar potential respects the U(1) symmetry, which is spontaneously broken by the vacuum.  Consequently, $m_A^2=0$ and the other scalar squared masses are given by,
\beq \label{mhhmixed}
 m^2_{H,h}=\half\lambda v^2\bigl[1\pm\sqrt{c_{2\beta}^2+R^2 s_{2\beta}^2}\,\bigr]\,, \qquad m_{H^\pm}^2=-\half\lambda_4 v^2\,,
\eeq
with $m_h\leq m_H$.  Stability of the vacuum requires that $\lambda_4<0$.

\subsection{Scalar potential with an unbroken \texorpdfstring{U(1)$\otimes\Pi_2$}{U(1)\directprod\uppercasePi\texttwoinferior} or SO(3)  symmetry}

The $\ug\otimes\Pi_2$ symmetry of the scalar potential is unbroken if
$m_{11}^2=m_{22}^2$ and $m_{12}^2=0$.
Then, as noted 
at the end of Section~\ref{zee2pi2}, the squared mass conditions yield 
$Y_3=Z_6=Z_7=0$, corresponding 
to the inert limit of the $\ug\otimes\Pi_2$-symmetric scalar potential.   

First, we suppose that both vevs are nonzero.  
Then in the $\ug\otimes\Pi_2$ symmetry limit, \eqst{min1a}{min3a} imply that $(R-1)c_{2\beta}=0$.  Hence,
the $\ug\otimes\Pi_2$ symmetry limit arises in two distinct cases.  If $m_{12}^2=c_{2\beta}=0$ and $R\neq 1$, then 
\eqst{mha}{calmsqH} yield,
 \beq \label{msymlimit2}
 m_h^2=\half \lambda v^2(1+R)\,,\qquad m_{H}^2= \half\lambda v^2(1-R)\,,\qquad m_A^2=0\,,\qquad m_{H^\pm}^2=-\half \lambda_4 v^2\,.
\eeq
Note that a stable minimum exists if $\lambda_4<0$ and $|R|<1$.
The $\Pi_2$ symmetry is preserved by the vacuum, whereas the $\ug$ symmetry is spontaneously broken by the vacuum and results in a massless Goldstone boson.

If  $m_{11}^2=m_{22}^2$ , $m_{12}^2=0$ and $R=1$, then an SO(3) symmetry is explicitly preserved by the scalar
potential and \eqst{mha}{calmsqH} yield,
\beq \label{sothreemasses}
m_h^2=\lambda v^2\,,\qquad m_H^2=m_A^2=0\,,\qquad  m_{H^\pm}^2=-\half \lambda_4 v^2\,.
\eeq
The SO(3) symmetry is spontaneously broken by the vacuum, leaving a residual unbroken U(1) symmetry, which results in two massless Goldstone bosons, $H$ and~$A$.

If one of the vevs vanishes (i.e., $s_{2\beta}=0$), 
then setting $\lambda_5=0$ in \eq{mhainert} and in \eqst{inertmass2}{inertmass4} yields,
 \beq \label{msymlimit3}
m_h^2=\lambda v^2\,,\qquad m_H^2=m_A^2=\half\lambda v^2(R-1)\,,\qquad m_{H^\pm}^2=m_A^2-\half \lambda_4 v^2\,,
\eeq
which corresponds to a stable minimum if $R>1$. 
Note that in this case the $\Pi_2$ symmetry is broken by the vacuum, whereas the
$\ug$ symmetry is preserved by the vacuum and results in the mass degeneracy of $H$ and $A$.  In the limit of $R=1$, corresponding to an SO(3)-symmetric scalar potential,
the resulting scalar masses are again given by \eq{sothreemasses}.

\subsection{The landscape of ERPS4--Part II(a): scalar potential with a softly broken or unbroken \texorpdfstring{U(1)$\otimes\Pi_2$}{U(1)\directprod\uppercasePi\texttwoinferior} or SO(3) symmetry}

Table~\ref{tab:erps4s} provides a summary of
the landscape of scalar potentials in the subspace of the ERPS4 regime where the $\ug\otimes\Pi_2$
or SO(3) symmetry of the scalar potential is either softly broken ($m_{11}^2\neq m_{22}^2$ and/or $m_{12}^2\neq 0$) or unbroken 
($m_{11}^2=m_{22}^2$ and $m_{12}^2=0$).

\begin{table}[h!]
\begin{tabular}{|c|c|c|c|c|c|}
\hline
$\beta$ &  $m_{11}^2$, $m_{22}^2$ & $\pht m_{12}^2 e^{i\xi}\pht$ & $R$ & Higgs alignment  & comment \\
\hline
$\pht s_{2\beta}c_{2\beta}\neq 0 \pht$ &  $\pht m_{11}^2\neq m_{22}^2\pht $ & $>0$ & $R\neq 1$ & no &  see \eq{bound} \\
$\pht s_{2\beta}c_{2\beta}\neq 0 \pht$ &  $\pht m_{11}^2= m_{22}^2\pht $ & $>0$ & $R>1$ & no &  $m_A^2=\half\lambda v^2(R-1)$ \\
$\pht s_{2\beta}c_{2\beta}\neq 0 \pht$ &  $\pht m_{11}^2\neq m_{22}^2\pht $ & $0$ & $|R|<1$ & no &  $m_A^2=0$\\
$c_{2\beta}=0$  &  $m_{11}^2=m_{22}^2$  &$>0$ & $R\neq 1$ &  yes &  $-1<R\leq 1+2m_A^2/(\lambda v^2)$   \\
$s_{2\beta}= 0$  &  $m_{11}^2\neq m_{22}^2$  & $0$ & $\pht R\neq 1 \pht $ &  yes &  $m^2_H=m^2_A>0$ \\
$c_{2\beta}=0$  &  $m_{11}^2= m_{22}^2$  & 0&  $|R|<1$  & yes & one massless scalar \\
$s_{2\beta}=0$  &  $m_{11}^2=m_{22}^2$  & 0 &  $R> 1$ & yes& $m^2_H=m^2_A>0$ \\
\hline
$s_{2\beta}c_{2\beta}\neq 0$ &  $m_{11}^2\neq m_{22}^2$  & $> 0$ &  $R=1$ & yes & $\pht m^2_H=m^2_A> 0\pht $ \\
$c_{2\beta}=0$  &  $m_{11}^2=m_{22}^2$  &$> 0$ & $R=1$ &  yes &   $m^2_H=m^2_A> 0$   \\
 $s_{2\beta}= 0$  &  $m_{11}^2\neq m_{22}^2$  & $0$ & $\pht R=1 \pht $ &  yes &   $m^2_H=m^2_A>0$ \\
 &  $m_{11}^2= m_{22}^2$  & 0 &  $R=1$  & yes &  $m^2_H=m^2_A=0$  \\
\hline
\end{tabular}
\caption{\small \baselineskip=14pt
Landscape of the ERPS4--Part II(a): Scalar potentials of the 2HDM with either an unbroken or softly broken $\ug\otimes\Pi_2$ symmetry that is manifestly realized in the $\Phi$-basis, where   
$\lambda\equiv\lambda_1=\lambda_2$, $\lambda_5=\lambda_6=\lambda_7=0$, and CP is conserved by the scalar potential and vacuum.
The parameter $m_{12}^2 e^{i\xi}$ is real and non-negative [as a consequence of \eqs{min3a}{mha}];
if $m^2_{12}=0$ and $s_{2\beta}\neq 0$ then a massless neutral scalar is present in the neutral scalar spectrum.
The parameter $R\equiv (\lambda_3+\lambda_4)/\lambda > -1$;  when $R=1$
the (softly broken) $\ug\otimes\Pi_2$ symmetry is promoted to a (softly broken) SO(3) symmetry.  
An exact Higgs alignment in the ERPS4 is realized in the inert limit where $Y_3=Z_6=Z_7=0$. \\[-15pt]
\label{tab:erps4s}}
\end{table}

It is noteworthy that the tree-level Higgs scalar potential of the MSSM exhibits a softly broken $\ug\otimes\Pi_2$ symmetry with
$m_{11}^2\neq m_{22}^2$, $m_{ 12}^2\neq 0$, $s_{2\beta}\neq 0$ and $R=-1$~\cite{Gunion:1984yn,Ferreira:2010jy}, corresponding to the first line of Table~\ref{tab:erps4s}.\footnote{Since $R=-1$, the scalar potential stability 
conditions require that $m_{11}^2+m_{22}^2>2|m_{12}^2|$ as noted in footnote~\ref{fnminusone}.  Moreover, $m_{11}^2 m_{22}^2<|m_{12}^2|^2$ in order to have electroweak symmetry breaking~\cite{Martin:1997ns}, thereby excluding $m_{11}^2=m_{22}^2$.}
Of course, radiative corrections to the scalar potential of the MSSM are significant~\cite{Draper:2016pys} and yield an effective 2HDM scalar potential below the energy scale of supersymmetry breaking that lies outside the domain of the ERPS4~\cite{Haber:1993an}.

\section{The GCP3 scalar field basis}
\label{sec:GCP3basis}

Consider a softly broken GCP3 symmetric scalar potential whose parameters (denoted with prime superscripts) satisfy the following conditions: $\lambda^\prime\equiv\lambda^\prime_1=\lambda^\prime_2=\lambda^\prime_3+\lambda^\prime_4+\lambda^\prime_5$ and $\Im\lambda_5^\prime=\lambda^\prime_6=\lambda^\prime_7=0$.
The softly broken parameters $m_{11}^{\prime\,2}$, $m_{12}^{\prime\,2}$ and $m_{12}^{\prime\,2}$ are arbitrary (with $m_{12}^{\prime\,2}$ potentially complex).   
If we demand that the potential is bounded from below, then
\beqa \label{ineq5}
\lambda^\prime>0\,,\qquad \lambda^\prime+\lambda^\prime_3>0\,,\qquad \lambda_5^\prime< \lambda^\prime\,.
\eeqa

\subsection{The softly broken GCP3 symmetric scalar potential with \texorpdfstring{$v_1\neq 0$}{v\textoneinferior \textneq 0} and \texorpdfstring{$v_2\neq 0$}{v\texttwoinferior \textneq 0}}
\vskip -0.1in

Assuming that $v^\prime_1$ and $v^\prime_2$ are both nonzero (or equivalently, $s_{2\beta^\prime}\neq 0$), \eqst{min1}{min3} yield,
\beqa
m_{11}^{\prime\,2} &=& \Re(m_{12}^{\prime\,2} e^{i\xi^\prime})\tan\beta^\prime-\half\lambda^\prime v^2+\lambda^\prime_5 v^2s^2_{\beta^\prime}\sin^2\xi^\prime \,,\label{min4a} \\
m_{22}^{\prime\,2} &=& \Re(m_{12}^{\prime\,2} e^{i\xi^\prime})\cot\beta^\prime-\half\lambda^\prime v^2 +\lambda^\prime_5 v^2 c^2_{\beta^\prime}\sin^2\xi^\prime \,, \label{min5a} \\
\Im(m_{12}^{\prime\,2} e^{i\xi^\prime})&=&\half\lambda^\prime_5 v^2 s_{\beta^\prime} c_{\beta^\prime}\sin 2\xi^\prime\,,\label{min6a}
\eeqa
which can be rewritten in the following forms,
\beqa 
c_{2\beta^\prime}(m_{11}^{\prime\,2}+m_{22}^{\prime\,2}+\lambda^\prime v^2)&=&m_{22}^{\prime\,2}-m_{11}^{\prime\,2}\,,\label{cosbp} \\
\half s_{2\beta^\prime} \cos\xi^\prime(m_{11}^{\prime\,2}+m_{22}^{\prime\,2}+\lambda^\prime v^2)&=&\Re m_{12}^{\prime\,2}\,,\label{cxi}\\
-\half  s_{2\beta^\prime} \sin\xi^\prime\bigl[m_{11}^{\prime\,2}+m_{22}^{\prime\,2}+(\lambda^\prime-\lambda_5^\prime) v^2\bigr]&=&\Im m_{12}^{\prime\,2}\,.\label{sxi}
\eeqa
\Eqst{cosbp}{sxi} can be used to fix the value of $\beta^\prime$ and $\xi^\prime$.  However,
 if $m_{11}^{\prime\,2}+m_{22}^{\prime\,2}+\lambda^\prime v^2=0$ then it follows that $m_{11}^{\prime\,2}=m_{22}^{\prime\,2}$ and $\Re m_{12}^{\prime\,2}=0$, in which case only $s_{2\beta^\prime}\sin\xi^\prime$ is determined.   
As noted below \eq{sinxi}, inserting
$m^{\prime\,2}_{12}=|m^{\prime\,2}_{12}|e^{i\theta^\prime_{12}}$ in \eqs{cxi}{sxi} and imposing $\cos^2\xi^\prime+\sin^2\xi^\prime=1$ yields an equation that determines the phase
$\theta^\prime_{12}$ in terms of $\xi^\prime$ and the other GCP3 scalar potential parameters.

The corresponding parameters of the Higgs basis are obtained by setting $R=1$ in \eqst{whytwocp2}{zeesevencp2},
\beqa
&& Y_2=\frac{2\Re(m_{12}^{\prime\,2}e^{i\xi^\prime})}{s_{2\beta^\prime}}-\half\lambda^\prime v^2+\lambda_5^\prime v^2\bigl(1-\half s_{2\beta^\prime}^2\bigr)\sin^2\xi^\prime\,, \label{why2cp3}\\
&&  Z_1=Z_2=\lambda^\prime-\lambda_5^\prime s_{2\beta^\prime}^2\sin^2\xi^\prime \,, \label{zee1} \\
&&  Z_3=\lambda^\prime_3+\lambda_5^\prime s_{2\beta^\prime}^2\sin^2\xi^\prime \,, \\
&&  Z_4=\lambda^\prime_4+\lambda_5^\prime s_{2\beta^\prime}^2\sin^2\xi^\prime\,, \\
&& Z_5=\lambda_5^\prime e^{-2i\xi^\prime}\bigl(\cos\xi^\prime +ic_{2\beta^\prime}\sin\xi^\prime \bigr)^2\,,\label{zee5}\\
&& Z_6=-Z_7=i\lambda_5^\prime s_{2\beta^\prime}\sin\xi^\prime e^{-i\xi^\prime}\bigl(\cos\xi^\prime+ic_{2\beta^\prime}\sin\xi^\prime\bigr)\,.\label{zee6}
\eeqa
The Higgs basis parameters $Y_1$ and $Y_3$ are fixed by the potential minimum conditions given in \eq{YZ}.
Note that \eq{invcond12} is satisfied, as expected.  
In addition, in the limit of $\lambda_5^\prime=0$, we recover the softly broken SO(3)-symmetric scalar potential, where
the conditions $\lambda'_1=\lambda'_2=\lambda'_3+\lambda'_4$ and $\lambda'_5=\lambda'_6=\lambda'_7=0$ hold for all choices of the scalar field basis.

One can check that CP is conserved in light of the relation,
\beq \label{fivesix}
Z^2_6=-\lambda_5^\prime s^2_{2\beta^\prime}\sin^2\xi^\prime Z_5\,,
\eeq
which implies that $\Im(Z_5^* Z_6^2)=0$.  Thus, there exists an appropriate rephasing of the Higgs basis such that $Z_5$, $Z_6$ and $Z_7$ are real.  This is remarkable in light of the fact that $\lambda_5$ is real but $m_{12}^2$ can be complex, which implies that one cannot perform a simple rephasing of the scalar doublet fields in the GCP3 basis to render all parameters real.  In light of the CP invariance of a softly broken GCP3-symmetric scalar potential, it must be possible to find a residual generalized CP transformation under which the scalar potential and the vacuum in the GCP3 basis is left invariant.   In Appendix~\ref{app:cp}, we provide an explicit construction of this residual generalized CP transformation.  Of course, the existence of such a transformation is a foregone conclusion given that the existence of the residual CP symmetry in the U(1)$\otimes\Pi_2$ basis can be established by inspection.

The scalar masses can now be evaluated.  First, \eq{chargedHmass} is still valid,
\beq \label{chiggs}
m^2_{H^\pm}=Y_2+\half Z_3 v^2=\frac{2\Re(m_{12}^{\prime\,2}e^{i\xi^\prime})}{s_{2\beta^\prime}}-\half v^2(\lambda^\prime_4+\lambda^\prime_5\cos 2\xi^\prime)\,.
\eeq
Next, consider the neutral scalar squared-mass matrix, which is given by \eq{matrix33}.   Noting that the complex number, $\cos\xi^\prime+ic_{2\beta^\prime}\sin\xi^\prime$ appears in both \eqs{zee5}{zee6}, it is convenient to define the complex phase $\psi$ via,
\beq \label{psidef}
\cos\xi^\prime+ic_{2\beta^\prime}\sin\xi^\prime=(1-s_{2\beta^\prime}^2\sin^2\xi^\prime)^{1/2} e^{i\psi}\,.
\eeq
In order to make use of \eq{matrix33}, we must choose a value for $\eta$.   Following \eq{etatrans}, we shall transform $\eta=-\xi$ [which was employed in the $\ug\otimes\Pi_2$ basis]
to the GCP3 basis.   The derivation is provided in Section~\ref{transforming} [cf.~\eqst{expetap}{etapr}] and instructs us to choose
\beq \label{etachoice}
\eta=\psi-\xi^\prime-\half\pi\,.
\eeq
Inserting this result into \eq{matrix33}, it then follows that 
\beqa
Z_5&=&-\lambda_5^\prime(1-s_{2\beta^\prime}^2\sin^2\xi^\prime)e^{2i\eta}\,, \\
Z_6&=&-\lambda_5^\prime s_{2\beta^\prime}\sin\xi^\prime(1-s_{2\beta^\prime}^2\sin^2\xi^\prime)^{1/2} e^{i\eta}\,.
\eeqa
In particular, $\Im(Z_5 e^{-2i\eta})=\Im(Z_6 e^{-i\eta})=0$.   Thus, we can immediately read off the squared mass of the CP-odd neutral scale from \eq{matrix33},
\beq \label{masq}
m_A^2=Y_2+\half v^2\bigl[Z_3+Z_4-\Re(Z_5 e^{-2i\eta})\bigr]=\frac{2\Re(m_{12}^{\prime\,2}e^{i\xi^\prime})}{s_{2\beta^\prime}}+\lambda_5^\prime v^2\sin^2\xi^\prime\,,
\eeq
where $A=\varphi_3\equiv\sqrt{2}\,\Im \mathcal{H}_2^0$.  
Combining the results of \eqss{min4a}{min5a}{masq} yields,
\beq \label{macp3}
m_A^2=m_{11}^{\prime\, 2}+m_{22}^{\prime\,2}+\lambda^\prime v^2\,,
\eeq
Hence, we can rewrite \eqst{cosbp}{sxi} as,
\beqa 
m_{22}^{\prime\,2}-m_{11}^{\prime\,2} &=& m_A^2 c_{2\beta^\prime}\,,\label{mam22minusm11} \\
\Re m^{\prime\,2}_{12}&=&\half s_{2\beta^\prime}\cos\xi^\prime m_A^2\,,\label{realmp12} \\
\Im m^{\prime\,2}_{12}&=&\half s_{2\beta^\prime}\sin\xi^\prime (\lambda_5^\prime v^2-m_A^2)\,.
\label{imaginarymp12}
\eeqa
For example, suppose that $s_{2\beta^\prime}c_{2\beta^\prime}\neq 0$, $\sin \xi^\prime\neq 0$, $m_{11}^{\prime\,2}=m_{22}^{\prime\,2}$ and $\lambda_5^\prime\neq 0$.   
Then, \eqst{mam22minusm11}{imaginarymp12} imply
that  $m^{\prime\,2}_{12}$ is purely imaginary.  Consulting Tables~\ref{tab:symm3} and \ref{tab:class2}, it follows that the scalar potential respects a U(1)$^\prime$
symmetry that is spontaneously broken.  Hence, a massless Goldstone boson exists that can be identifed as the CP-odd scalar $A$.   

Likewise, the charged Higgs squared mass is given by,
\beq \label{chcp3}
m^2_{H^\pm}=m_A^2-\half (\lambda_4^\prime+\lambda_5^\prime)v^2\,,
\eeq
after making use of \eqs{chiggs}{masq}.  Finally,
the squared masses of the CP-even neutral scalars, $h$ and $H$ are the eigenvalues of the $2\times 2$ matrix exhibited below,\footnote{The computation of the squared-mass matrix of the CP-even neutral scalars starting from the $\Phi^\prime$-basis is much more difficult.  Details are provided in Appendix~\ref{app:sqmass}.}
\beqa
\mathcal{M}^2_H&=&\begin{pmatrix} Z_1 v^2 & \quad -\Im(Z_6 e^{i(\xi^\prime-\psi)}) v^2 \\  -\Im(Z_6 e^{i(\xi^\prime-\psi)}) v^2 & \quad m_A^2-\Re(Z_5 e^{2i(\xi^\prime-\psi)}) v^2\end{pmatrix} \nonumber \\[10pt]
&=&\begin{pmatrix} \bigl(\lambda^\prime-\lambda_5^\prime s_{2\beta^\prime}^2\sin^2\xi^\prime\bigr)v^2 & \quad -\lambda_5^\prime v^2  s_{2\beta^\prime}\sin\xi^\prime(1-s_{2\beta^\prime}^2\sin^2\xi^\prime)^{1/2}  \\ -\lambda_5^\prime v^2 s_{2\beta^\prime}\sin\xi^\prime (1-s_{2\beta^\prime}^2\sin^2\xi^\prime)^{1/2}  
 & \quad m_A^2-\lambda_5^\prime v^2(1-s_{2\beta^\prime}^2\sin^2\xi^\prime)\end{pmatrix}, \nonumber \\
 &&\phantom{line} \label{emhtwo}
\eeqa
with respect to the $\{\varphi_1,\varphi_2\}$ basis, where $\varphi_1\equiv \sqrt{2}\,\Re \mathcal{H}_1^0-v$ and $\varphi_2\equiv\sqrt{2}\,\Re \mathcal{H}_2^0$.
The neutral CP-even scalar masses are given by
\beq \label{HL}
m^2_{H,h}=\half\biggl\{m_A^2+(\lambda^\prime-\lambda_5^\prime)v^2\pm\sqrt{\bigl[m_A^2-(\lambda^\prime+\lambda_5^\prime)v^2\bigr]^2+4\lambda_5^\prime v^2(m_A^2-\lambda^\prime v^2)s_{2\beta^\prime}^2\sin^2\xi^\prime}\biggr\}\,,
\eeq
where $\mhl\leq\mhh$, and
\beq
c_{\beta-\alpha}=\frac{\lambda_5^\prime v^2 s_{2\beta^\prime}\sin\xi^\prime (1-s_{2\beta^\prime}^2\sin^2\xi^\prime)^{1/2} }{\sqrt{(m_H^2-m_h^2)\bigl[m_H^2- (\lambda^\prime-\lambda_5^\prime s_{2\beta^\prime}^2\sin^2\xi^\prime\bigr)v^2\bigr]}}\,.
\eeq

A stable minimum requires that the scalar squared masses should be non-negative.   This condition implies that 
\beq \label{positive}
\Re(m_{12}^2 e^{i\xi^\prime})+\half\lambda_5^\prime v^2 s_{2\beta}\sin^2\xi^\prime\geq 0\quad \text{and} \quad m_A^2\geq\half (\lambda^\prime_4+\lambda^\prime_5) v^2\,.
\eeq
In addition, we demand that
\beqa
\Tr\mathcal{M}^2_H&=&m_A^2+(\lambda^\prime-\lambda_5^\prime)v^2\geq 0\,, \label{trcp3}\\
\frac{1}{v^2}\det\mathcal{M}^2_H&=&m_A^2(\lambda^\prime-\lambda_5^\prime s_{2\beta^\prime}^2\sin^2\xi^\prime)-\lambda^\prime\lambda_5^\prime v^2(1-s_{2\beta^\prime}^2\sin^2\xi^\prime)\geq 0\,.\label{detcp3}
\eeqa
Since $m_A^2\geq 0$ by assumption, \eq{trcp3} is automatically satisfied in light of \eq{ineq5}.  On the other hand, \eq{detcp3} is satisfied if and only if
\beq
\lambda^\prime_5\leq {\rm min}\left\{\lambda^\prime\,,\,\frac{\lambda^\prime m_A^2}{\lambda^\prime v^2+s_{2\beta^\prime}^2 \sin^2\xi^\prime(m_A^2-\lambda^\prime v^2)}\right\}.
\eeq

One can fix the parameter space of the softly broken GCP3 scalar potential by specifying the values of $\lambda^\prime$, $\lambda_4^\prime$, $\lambda_5^\prime$, $s_{2\beta^\prime}\sin\xi^\prime$, $m_A$ and $v$.   In particular, once $m_A$ is fixed, we see that $\beta^\prime$ and~$\xi^\prime$ do not appear independently in any 2HDM observable.
By replacing $\lambda$ with $m_h$ and $\lambda_4$ with $m_{H^\pm}$, the independent parameters of the  softly broken GCP3 scalar potential can be taken to be $m_h$, $m_A$, $m_{H^\pm}$, $v$, $\lambda_5^\prime$ and $s_{2\beta^\prime}\sin\xi^\prime$.  That is, just as in the case of the softly broken U(1)$\otimes\Pi_2$ scalar potential, the parameter space is fixed by six real parameters.

\subsection{The inert limit of the softly broken GCP3-symmetric scalar potential}

The inert limit ($Y_3=Z_6=Z_7=0$) of the softly broken GCP3-symmetric scalar potential
arises 
if either $v_1=0$ or $v_2=0$, but is more general.  
In the inert limit [using \eq{zee6}], 
one of the following three conditions, $s_{2\beta^\prime}\sin\xi^\prime=0$, $c_{2\beta^\prime}=\cos\xi^\prime=0$, or $\lambda_5^\prime=0$, is satisfied.   

\subsubsection{The softly broken GCP3-symmetric scalar potential with one vanishing vev}

The case where one of the vevs vanishes (i.e., $s_{2\beta^\prime}=0$) 
implies that $m_{12}^{\prime\,2}=0$ in light of \eqs{inert1}{inert2}.
Then, setting $R=1$ in \eqst{inertmass1}{inertmass4} yields,
 \beq \label{inertCP3}
 m^2_h=\lambda^\prime v^2\,,\qquad m^2_H=m_A^2-\lambda^\prime_5 v^2\,,\qquad m^2_{H^\pm}=m_A^2-\half(\lambda^\prime_4+\lambda^\prime_5)v^2\,,
 \eeq
where $m_A^2=Y_2+\half\lambda^\prime v^2$ and $Y_2$ is a free parameter.    Likewise, if $\sin\xi^\prime=0$, then \eqst{masq}{emhtwo} also yield \eq{inertCP3}, where
$m_A^2=2|\Re m_{12}^{\prime\, 2}|/s_{2\beta^\prime}$.\footnote{If $\sin\xi^\prime =0$ then $\Re(m_{12}^{\prime\,2} e^{i\xi^\prime })=\pm\Re m_{12}^{\prime\,2}=|\Re m_{12}^{\prime\,2}|$ after choosing the sign that yields $m_A^2\geq 0$.}
That is, if $s_{2\beta^\prime}\sin\xi^\prime=0$ then \eq{inertCP3} is satisfied where
 \beq \label{firstcaseMA}
 m_A^2=\begin{cases} \displaystyle\frac{2|\Re m_{12}^{\prime\, 2}|}{s_{2\beta^\prime}}\,, & \quad \text{if $\sin\xi^\prime =0$ and $s_{2\beta^\prime }\neq 0$}, \\
 Y_2+\half\lambda^\prime v^2\,, & \quad \text{if $s_{2\beta^\prime }=0$}. \end{cases}
 \eeq
 Note that if $\sin\xi^\prime=0$, $s_{2\beta^\prime}c_{2\beta^\prime}\neq 0$ and $m_{11}^{\prime\,2}\neq m_{22}^{\prime\,2}$, then 
 $\Re m_{12}^{\prime\,2}\neq 0$ due to \eqst{min4a}{min6a}.

 \subsubsection{The softly broken GCP3-symmetric scalar potential with $\cos 2\texorpdfstring{\beta^\prime}{\textbeta\textprime}=\cos \texorpdfstring{\xi^\prime}{\textxi\textprime}=0$}

Second, if $c_{2\beta^\prime}=\cos\xi^\prime=0$, then it follows from \eqst{masq}{emhtwo} that,
\beq \label{secondcase}
m_h^2=(\lambda^\prime -\lambda^\prime_5)v^2\,,\qquad m_H^2=m_A^2=\pm 2\Im m_{12}^{\prime\,2}+\lambda^\prime_5 v^2\,,\qquad m^2_{H^\pm}=m_A^2-\half(\lambda^\prime _4+\lambda^\prime _5)v^2\,.
\eeq
Using the results of Section~\ref{transforming}, this case corresponds to $s_{2\beta}=0$ in the $\ug\otimes\Pi_2$ basis.  Then the choice of plus [minus] sign in the expression for $m^2_{H,A}$ in \eq{secondcase} corresponds to $\beta=0$ [$\beta=\half\pi$], respectively.
Moreover, recall that an unbroken GCP3 symmetry is equivalent to U(1)$^\prime\otimes\mathbb{Z}_2$ [cf.~Table~\ref{tab:class2}].
Although the $\mathbb{Z}_2$ symmetry is explicitly broken (due to $m^{\prime\,2}_{12}\neq 0$), 
a residual U(1)$^\prime$ symmetry survives that is preserved by the vacuum if $c_{2\beta^\prime}=\cos\xi^\prime=0$ since,
\beq \label{youprime}
\begin{pmatrix} \phm\cos\theta & \quad \sin\theta \\ -\sin\theta & \quad \cos\theta\end{pmatrix}\begin{pmatrix} 1 \\ \pm i\end{pmatrix}=e^{\pm i\theta}\begin{pmatrix} 1 \\ \pm i\end{pmatrix}\,,
\eeq
which results in the mass degeneracy of $H$ and $A$. 

\subsubsection{The softly broken SO(3)-symmetric scalar potential (\texorpdfstring{$\lambda_5^\prime=0$}{\textlambda\textprime\textfiveinferior=0})}

Third, if $\lambda_5^\prime=0$, then the softly broken GCP3 symmetry is promoted to a softly broken SO(3) symmetry.    In light of \eqst{min4a}{min6a}, it follows that if $s_{2\beta^\prime}c_{2\beta^\prime}\neq 0$ then $m_{11}^{\prime\,2}\neq m_{22}^{\prime\,2}$ and $\Re(m_{12}^{\prime\,2}\ e^{i\xi^\prime})\neq 0$.  We then obtain,
\beq \label{thirdcase}
m_h^2=\lambda^\prime v^2\,,\qquad m_H^2=m_A^2=\frac{2\Re(m_{12}^{\prime\,2}e^{i\xi^\prime})}{s_{2\beta^\prime}}\,,\qquad  m^2_{H^\pm}=m_A^2-\half\lambda^\prime_4 v^2\,,
\eeq
in agreement with \eq{masssrelso3}, as expected.  Using the results of Section~\ref{transforming}, this case corresponds to $R=1$ in the $\ug\otimes\Pi_2$ basis.

\subsection{Scalar potential with an unbroken GCP3 symmetry}

Finally, if $m_{11}^{\prime\,2}=m_{22}^{\prime\,2}$ and $m_{12}^{\prime\,2}=0$ then
the GCP3 symmetry is explicitly preserved by the scalar potential.   
In light of \eqs{why3}{YZ} and the ERPS condition, it follows that $Y_3=Z_6=Z_7=0$, corresponding
to the inert limit of the scalar potential.  
If both vevs are nonzero, then it follows from \eqst{min4a}{min6a} that 
$\lambda_5^\prime c_{2\beta^\prime}\sin^2\xi^\prime=\lambda_5^\prime\sin \xi^\prime\cos\xi^\prime=0$.
Consequently, the GCP3 symmetry limit arises in the following three distinct cases.

First, if $\lambda_5^\prime\neq 0$ and $\sin\xi^\prime=0$, then \eqs{inertCP3}{firstcaseMA} yield
 \beq \label{firstcase2}
 m^2_h=\lambda^\prime v^2\,,\qquad m^2_H=-\lambda^\prime_5 v^2\,,\qquad m_A^2=0\,, \qquad m^2_{H^\pm}=-\half(\lambda^\prime_4+\lambda^\prime_5)v^2\,,
 \eeq
 which corresponds to a stable minimum if $\lambda^\prime_5<0$ and $\lambda^\prime_4<-\lambda^\prime_5$.  
 The GCP3 symmetry is spontaneously broken by the vacuum, resulting in a massless scalar.
 
Second, if $\lambda_5^\prime\neq 0$, $c_{2\beta^\prime}=0$ and $\cos\xi^\prime=0$, then \eq{secondcase} yields,
\beq 
m_h^2=(\lambda^\prime -\lambda^\prime_5)v^2\,,\qquad m_H^2=m_A^2=\lambda^\prime_5 v^2\,,\qquad m^2_{H^\pm}=\half(\lambda^\prime _5-\lambda^\prime _4)v^2\,,
\eeq
which corresponds to a stable minimum if $0<\lambda_5^\prime < \lambda^\prime$ and $\lambda_4^\prime<\lambda_5^\prime$.
The mass degeneracy of $H$ and~$A$ is again a result of a residual U(1)$^\prime$ symmetry that is preserved by the vacuum. 

Third, if $\lambda^\prime_5=0$, then the GCP3 symmetry of the scalar potential is promoted to an SO(3) symmetry.   In this case, 
\beq \label{thirdcase2}
m_h^2=\lambda^\prime v^2\,,\qquad m_H^2=m_A^2=0\,,\qquad  m^2_{H^\pm}=-\half\lambda^\prime_4 v^2\,,
\eeq
corresponding to a stable minimum if $\lambda^\prime_4<0$.  In particular, the SO(3) symmetry is spontaneously broken down to U(1), which yields two massless scalars $H$ and $A$.

If only one of the two vevs is nonzero (i.e., $s_{2\beta}=0$), then \eqs{inert1}{inert2} yield $m_{12}^{\prime\,2}=0$.   Setting 
$R=1$ and $Y_1=Y_2=-\half \lambda^\prime v^2$ in \eqst{inertmass1}{inertmass4}, we end up with 
 \beq 
 m_h^2=\lambda^\prime v^2\,,\qquad m_H^2=-\lambda^\prime_5 v^2\,,\qquad m_{A}^2=0\,,\qquad m_{H^\pm}^2=-\half (\lambda^\prime_4+\lambda^\prime_5) v^2\,,
\eeq
which coincides with the mass spectrum given in \eq{firstcase2}.  If in addition we set $\lambda_5^\prime=0$, then we obtain the mass spectrum of \eq{thirdcase2}.

\subsection{The landscape of ERPS4--Part II(b) and (c): scalar potential with a softly broken or unbroken GCP3 or SO(3) symmetry}

Tables~\ref{tab:erps4gcp3} and \ref{tab:erps4so3} provide summaries of
the landscape of possible scalar potentials in the subspace of the ERPS4 regime where the GCP3 or SO(3) symmetry of the scalar potential is either softly broken ($m_{11}^{\prime\,2}\neq m_{22}^{\prime\,2}$ and/or $m_{12}^{\prime\,2}\neq 0$) or unbroken ($m_{11}^{\prime\,2}=m_{22}^{\prime\,2}$ and $m_{12}^{\prime\,2}=0$).    
Using the results Section~\ref{transforming}, one can check that each entry (or entries) of Tables~\ref{tab:erps4gcp3} and \ref{tab:erps4so3} can be matched up with a corresponding entry (or entries) of Table~\ref{tab:erps4s}, and vice versa.

\enlargethispage{3\baselineskip}
The analysis presented in this section can be repeated for the closely related GCP3$^\prime$ basis, where $\lambda_1^\prime=\lambda_2^\prime=\lambda_3^\prime+\lambda_4^\prime-\lambda_5^\prime$ and $\Im\lambda_5^\prime=\lambda_6^\prime=\lambda_7^\prime=0$.  Details are left for the reader.
\vskip 0.1in

\begin{table}[b!]
\begin{tabular}{|c|c|c|c|c|c|c|}
\hline
$\beta^\prime$ & $\xi^\prime$ & $m_{11}^{\prime\,2}$, $m_{22}^{\prime\,2}$ & $\pht m_{12}^{\prime\,2}\pht$  & Higgs alignment  & comment \\
\hline
$\pht s_{2\beta^\prime}c_{2\beta^\prime}\neq 0 \pht$ & $\pht\sin 2\xi^\prime\neq 0\pht$  &  $\pht m_{11}^{\prime\,2}\neq m_{22}^{\prime\,2}\pht$ &  \pht complex ($\neq 0$)\pht  & no &  \\
$\pht s_{2\beta^\prime}c_{2\beta^\prime}\neq 0 \pht$ & $\pht\sin 2\xi^\prime\neq 0\pht$  &  $\pht m_{11}^{\prime\,2}\neq m_{22}^{\prime\,2}\pht$ &  \pht real ($\neq 0$)\pht  & no &  $m_A^2=\lambda_5^\prime v^2$  \\
$s_{2\beta^\prime}c_{2\beta^\prime}\neq 0$  & $\cos \xi^\prime=0$ & $m_{11}^{\prime\,2}\neq m_{22}^{\prime\,2}$  &\pht purely imaginary  ($\neq 0$) \pht&   no &  \\
$s_{2\beta^\prime}c_{2\beta^\prime}\neq 0$  & $\cos \xi^\prime=0$ & $m_{11}^{\prime\,2}\neq m_{22}^{\prime\,2}$  &\pht 0 \pht&   no &  $m_A^2=\lambda_5^\prime v^2$  \\
$s_{2\beta^\prime}c_{2\beta^\prime}\neq 0$  & $\sin \xi^\prime\neq 0$ & $m_{11}^{\prime\,2}=m_{22}^{\prime\,2}$  &\pht purely imaginary ($\neq 0$) \pht&  no &  $m^2_A=0\phantom{_5 v^2}$  \\
$c_{2\beta^\prime}=0$  & $\sin 2\xi^\prime\neq 0$ & $m_{11}^{\prime\,2}=m_{22}^{\prime\,2}$  &\pht purely imaginary ($\neq 0$) \pht&  no &   $m^2_A=0\phantom{_5 v^2}$  \\
$c_{2\beta^\prime}=0$  & $\sin 2\xi^\prime\neq 0$ & $m_{11}^{\prime\,2}=m_{22}^{\prime\,2}$  &\pht real ($\neq 0$) \pht&  no &   $m_A^2=\lambda_5^\prime v^2$ \\
$c_{2\beta^\prime}=0$  & $\sin 2\xi^\prime\neq 0$ & $m_{11}^{\prime\,2}=m_{22}^{\prime\,2}$  &\pht complex ($\neq 0$) \pht&  no &  $m^2_A\neq 0$, $\lambda_5^\prime v^2$ \\
$c_{2\beta^\prime}=0$  & $\sin\xi^\prime=0$ & $m_{11}^{\prime\,2} = m_{22}^{\prime\,2}$  &  real ($\neq 0$) & yes&  \\
$c_{2\beta^\prime}= 0$  & $\cos \xi^\prime= 0$ & $m_{11}^{\prime\,2}= m_{22}^{\prime\,2}$  &purely imaginary ($\neq 0$)& yes & \pht   $m^2_H=m^2_A\neq 0$, $\lambda^\prime_5 v^2$\pht \\
$s_{2\beta^\prime}c_{2\beta^\prime}\neq 0$  & $\sin\xi^\prime=0$ & $m_{11}^{\prime\,2} \neq m_{22}^{\prime\,2}$  &  real  ($\neq 0$) & yes&  \\
$s_{2\beta^\prime}= 0$  && $m_{11}^{\prime\,2}\neq m_{22}^{\prime\,2}$  &  0  & yes&   \\
$s_{2\beta^\prime}\neq 0$  & $\sin\xi^\prime=0$ & $m_{11}^{\prime\,2}=m_{22}^{\prime\,2}$  &  0  & yes& one massless scalar  \\
$c_{2\beta^\prime}=0$ & $\cos\xi^\prime=0$ &  $m_{11}^{\prime\,2}= m_{22}^{\prime\,2}$  & 0 &   yes &  $m^2_H=m^2_A=\lambda_5^\prime v^2$  \\
$s_{2\beta^\prime}= 0$  && $m_{11}^{\prime\,2}=m_{22}^{\prime\,2}$  &  0  & yes& one massless scalar  \\
\hline
\end{tabular}
\caption{\small \baselineskip=15pt
Landscape of the ERPS4--Part II(b): Scalar potentials of the 2HDM with either an unbroken or softly broken GCP3 symmetry that is manifestly realized in the $\Phi$-basis.   
In all cases,
$\lambda\equiv\lambda^\prime_1=\lambda^\prime_2=\lambda^\prime_3+\lambda^\prime_4+\lambda^\prime_5$, (with $\lambda^\prime_5$ real and nonzero) and $\lambda^\prime_6=\lambda_7^\prime=0$, and CP is conserved by the scalar potential and vacuum.
The results shown in the fourth column for $m^{\prime\,2}_{12}$ have been obtained using \eqst{mam22minusm11} {imaginarymp12}.
The term ``complex'' means neither real nor purely imaginary.
An exact Higgs alignment in the ERPS4 is realized in the inert limit where $Y_3=Z_6=Z_7=0$.
\label{tab:erps4gcp3}}
\end{table}
\begin{table}[ht!]
\begin{tabular}{|c|c|c|c|c|c|}
\hline
$\beta^\prime$  & $m_{11}^{\prime\,2}$, $m_{22}^{\prime\,2}$ & $\pht m_{12}^{\prime\,2}e^{i\xi^\prime}\pht$  & Higgs alignment  & comment \\
\hline
$\pht s_{2\beta^\prime}c_{2\beta^\prime}\neq 0\pht $ &  $\pht m_{11}^{\prime\,2}\neq m_{22}^{\prime\,2}\pht $  & \pht real ($\neq 0$) \pht & yes & $\pht m^2_H=m^2_A>0\pht $ \\
$c_{2\beta^\prime}=0$  &  $m_{11}^{\prime\,2}= m_{22}^{\prime\,2}$  &\pht real ($\neq 0$) \pht & yes &   $m^2_H=m^2_A> 0$   \\
$s_{2\beta^\prime}=0$  &   $m_{11}^{\prime\,2}\neq m_{22}^{\prime\,2}$  &$ 0$ &  yes &   $m^2_H=m^2_A> 0$   \\
  &   $m_{11}^{\prime\,2}=m_{22}^{\prime\,2}$  &  0 & yes& $m^2_H=m^2_A=0$ \\
\hline
\end{tabular}
\caption{\small \baselineskip=15pt
Landscape of the ERPS4--Part II(c): Scalar potentials of the 2HDM with either an unbroken or softly broken SO(3) symmetry that is manifestly realized in the $\Phi$-basis.   
In all cases,
$\lambda\equiv\lambda^\prime_1=\lambda^\prime_2=\lambda^\prime_3+\lambda^\prime_4$ and $\lambda^\prime_5=\lambda^\prime_6=\lambda_7^\prime=0$, and CP is conserved by the scalar potential and vacuum. 
In all cases of an unbroken or softly broken SO(3) symmetric scalar potential, an exact Higgs alignment is realized as a consequence of $Y_3=Z_6=Z_7=0$. 
\label{tab:erps4so3}}
\thispagestyle{empty}
\end{table}

\section{Transforming between the \texorpdfstring{U(1)$\otimes\Pi_2$}{U(1)\directprod\uppercasePi\texttwoinferior} basis and GCP3 basis}
\label{transforming}
\vskip -0.1in

In Ref.~\cite{Ferreira:2009wh}, it was shown that the U(1)$\otimes\Pi_2$ and GCP3-symmetric scalar potentials are in fact the same scalar potential expressed in different scalar field bases.
In this section, we extend this result to the softly broken U(1)$\otimes\Pi_2$ and GCP3-symmetric scalar potentials by
providing an explicit mapping between the corresponding scalar potential parameters.   

Consider the following unitary transformation,
\beq \label{you}
U=\frac{e^{i\phi}}{\sqrt{2}}\begin{pmatrix} \phm 1 &\quad -i \\ -i& \quad\phm 1\end{pmatrix}\,,
\eeq
where the phase $\phi$ is determined in \eq{phasephi}.
Starting from the $\ug\otimes\Pi_2$ basis defined in Section~\ref{sec:UPibasis}, it then follows that (independently of the choice of $\phi$),
\beqa
\lambda^\prime&=& \lambda_1^\prime=\lambda_2^\prime =\half\lambda(1+R)\,,\label{lamp}\\
\lambda_3^\prime&=&\lambda_3+\half\lambda (1-R)\,,\\
\lambda_4^\prime&=&\lambda_4+\half\lambda(1-R)\,,\label{lam4p}\\
\lambda_5^\prime&=&-\half\lambda(1-R)\,,\label{lam5p}\\
\lambda_6^\prime&=&-\lambda_7^\prime=0\,,
\eeqa
where $R\equiv (\lambda_3+\lambda_4)/\lambda$.
In particular, 
$\lambda_5^\prime=\lambda^\prime-\lambda_3^\prime-\lambda_4^\prime$ is real and $\lambda_6^\prime=\lambda_7^\prime=0$, corresponding to the GCP3 basis defined in Section~\ref{sec:GCP3basis}.
In addition, the corresponding soft-breaking squared mass parameters are,
\beqa
&& m^{\prime\,2}_{11}=\half (m_{11}^2+m_{22}^2)+\Im m_{12}^2\,, \label{m11p} \\
&& m^{\prime\,2}_{22}=\half (m_{11}^2+m_{22}^2)-\Im m_{12}^2\,,\label{m22p} \\
&& m^{\prime\,2}_{12}=\Re m_{12}^2+\half i (m_{22}^2-m_{11}^2)\,.\label{m12p}
\eeqa

Finally, the vevs in the GCP3 basis are given by
\beq \label{vprimes}
v_1^\prime=\frac{e^{i\phi}}{\sqrt{2}}\bigl(v_1-iv_2e^{i\xi}\bigr)\,,\qquad\quad v_2^\prime e^{i\xi^\prime}=-e^{i\phi}\frac{i}{\sqrt{2}}\bigl(v_1+iv_2 e^{i\xi}\bigr)\,,
\eeq
where $v_1^\prime\equiv vc_{\beta^\prime}$ and $v_2^\prime\equiv vs_{\beta^\prime}$ are real and positive.    
Hence,
\beq \label{mags}
c_{\beta^\prime}=\frac{1}{\sqrt{2}}\bigl(1+s_{2\beta}\sin\xi\bigr)^{1/2}\,,\qquad\quad s_{\beta^\prime} =\frac{1}{\sqrt{2}}\bigl(1-s_{2\beta}\sin\xi\bigr)^{1/2}\,,
\eeq
and it immediately follows that
\beq \label{sintwob}
s^2_{2\beta^\prime}=1-s_{2\beta}^2\sin^2\xi\,.
\eeq
By convention, $0\leq\beta^\prime\leq \half\pi$ (or equivalently, $\sin 2\beta^\prime\geq 0$).

The phase $\phi$ is determined by the positivity of~$v_1^\prime$.  Hence, it follows that
\beq \label{phasephi}
e^{i\phi}=\frac{c_\beta+is_{\beta}e^{-i\xi}}{(1+s_{2\beta}\sin\xi)^{1/2}}\,.
\eeq
Then, \eq{vprimes} yields,
\beq
 e^{i\xi^\prime} s_{\beta^\prime}=-\frac{i}{\sqrt{2}}\frac{c_{2\beta}+is_{2\beta}\cos\xi}{\bigl(1+s_{2\beta}\sin\xi\bigr)^{1/2}}\,.
\eeq
Likewise, the relative phase $\xi^\prime$ is given by,
\beq \label{expp}
e^{i\xi^\prime}=\frac{s_{2\beta}\cos\xi-ic_{2\beta}}{(1-s_{2\beta}^2\sin^2\xi)^{1/2}}\,.
\eeq
That is,
\beq \label{sin2xi}
\sin\xi^\prime=\frac{-c_{2\beta}}{(1-s_{2\beta}^2\sin^2\xi)^{1/2}}\,,\qquad\quad \cos\xi^\prime=\frac{s_{2\beta}\cos\xi}{(1-s_{2\beta}^2\sin^2\xi)^{1/2}}\,.
\eeq
Consequently, \eqs{sintwob}{sin2xi} yield,
\beq \label{cs}
s_{2\beta^\prime}\sin\xi^\prime=-c_{2\beta}\,.
\eeq
Finally, if $\beta=\tfrac14\pi$ and $\sin\xi=\pm 1$, then one of the vevs vanishes.  It then follows that $s_{2\beta^\prime}=0$, in which case $\xi^\prime$ is indeterminate if $s_{\beta^\prime}=0$
and $\xi^\prime=0$ if $c_{\beta^\prime}=0$.

Using \eqss{m12p}{sintwob}{expp}, it is instructive to note that
\beq \label{rem12p}
\frac{2\Re(m_{12}^{\prime\,2}e^{i\xi^\prime})}{s_{2\beta^\prime}}=\frac{2\Re(m_{12}^2)s_{2\beta}\cos\xi+c_{2\beta}(m_{22}^2-m_{11}^2)}{1-s_{2\beta}^2\sin^2\xi}\,.
\eeq
In light of \eq{min3a}, it follows that 
\beq \label{realm12}
\Re m_{12}^2= \Re(m_{12}^2e^{i\xi})\cos\xi+\Im(m_{12}^2e^{i\xi})\sin\xi=\Re(m_{12}^2e^{i\xi})\cos\xi\,.
\eeq
Hence, after using \eqs{min1a}{min2a} for $m_{22}^2-m_{11}^2$ and \eq{realm12} for $\Re m_{12}^2$, it then follows that \eq{rem12p} yields,
\beq
\frac{2\Re(m_{12}^{\prime\,2}e^{i\xi^\prime})}{s_{2\beta^\prime}}=\frac{2\Re(m_{12}^2e^{i\xi})}{s_{2\beta}}+\frac{\lambda v^2(1-R)c_{2\beta}^2}{2(1-s_{2\beta}^2\sin^2\xi)}\,.
\eeq
In particular, \eq{masq} yields,
\beq
m_A^2=\frac{2\Re(m_{12}^{\prime\,2}e^{i\xi^\prime})}{s_{2\beta^\prime}}+\lambda_5^\prime v^2\sin^2\xi^\prime=\frac{2\Re(m_{12}^2e^{i\xi})}{s_{2\beta}}\,,
\eeq
after employing \eqs{lam5p}{sin2xi}.
Comparing with \eq{mha}, we see that one obtains the same result for $m_A^2$ in the GCP3 basis and the U(1)$\otimes\Pi_2$ basis respectively, as required.   Note that the same conclusion can be drawn by plugging the results of
\eqss{lamp}{m11p}{m22p} into \eq{macp3}, which reproduces the result of \eq{mha2}. 

For completeness, we check that the scalar mass spectrum derived from the softly broken U(1)$\otimes\Pi_2$ and GCP3-symmetric scalar potentials coincide, as required. 
Plugging in the results of \eqs{lam4p}{lam5p} into \eq{chcp3} reproduces the result of 
\eq{mch} for $m^2_{H^\pm}$.   To check the squared masses of the neutral CP-even scalars, we 
plug the results of \eqs{lamp}{lam5p} into \eq{trcp3}, which reproduces the result of \eq{ineq1}.   Finally, we plug in the results of \eqss{lamp}{lam5p}{cs} into \eq{detcp3}, which reproduces the result of \eq{ineq2}.

As a final check of our computations, one can verify that the invariant quantities, $Y_2$, $Z_1,\ldots,Z_4$, $|Z_5|$, $|Z_6|$ and $Z_5^* Z_6^2$ are independent of the choice of basis.   
For example, starting from the GCP3 basis,
\beq
Z_5^* Z_6^2=-\lambda_5^\prime s_{2\beta^\prime}^2\sin^2\xi^\prime |Z_5|^2=-\lambda_5^{\prime\,3} s_{2\beta^\prime}^2\sin^2\xi^\prime(1-s_{2\beta^\prime}^2\sin^2\xi^\prime)^2
=\tfrac18\lambda^3(1-R)^3 c_{2\beta}^2 s^4_{2\beta}\,,
\eeq
in agreement with \eqs{zeefive}{zeeseven}.  One can also check that all the other invariants yield the same values in the GCP3 and U(1)$\otimes\Pi_2$ bases.

Given a softly broken \mbox{$\ug\otimes\Pi_2$}-symmetric scalar potential that is displayed in the \mbox{$\ug\otimes\Pi_2$} basis where the softly broken symmetry is manifestly realized, it may turn out that the scalar potential is invariant under some discrete or continuous subgroup of $\ug\otimes\Pi_2$.  It is of interest to determine implications of this invariance for the scalar potential parameters when expressed in the GCP3 basis.   We proceed by assuming that the scalar potential, which is specified in \eq{VH2} in terms of squared mass parameters $Y_{a\bbar}$ and dimensionless parameters $Z_{ab,cd}$ [defined in \eq{ZZ}], is invariant under the transformations, $\Phi_a \to X_{a\bbar}\Phi_b$ and $\Phi^\dagger_{\abar} \to\Phi^\dagger_{\bbar}X^\dagger_{b\abar}$.
That is [cf.~\eqs{Y-transf}{Z-transf}],
\beq \label{XYX}
Y=X YX^\dagger \,,\qquad\quad Z= (X\otimes X) Z(X^\dagger\otimes X^\dagger)\,.
\eeq

Consider a change of scalar field basis specified by $U$ [e.g., $U$ specified in \eq{you} transforms the $\ug\otimes\Pi_2$ basis into the GCP3 basis].  In light of \eqs{Y-transf}{Z-transf},
$Y^\prime=UYU^\dagger$ yields the squared mass parameters in the new basis, and $Z^\prime= (U\otimes U) Z(U^\dagger\otimes U^\dagger)$ yields the dimensionless parameters in the new basis.   In the new basis, the symmetry transformation matrix $X$ will be denoted by $X^\prime$.  That is, $Y'=X'Y'X^{\prime\dagger}$, since the parameters in the new basis are invariant with respect to the transformations induced by $X'$.  It then follows that $UYU^\dagger=X^\prime(UYU^\dagger)X^{\prime\dagger}$, which yields $Y=(U^\dagger X^\prime U) Y(U^\dagger X^\prime U)^\dagger$.
Comparing this result with \eq{XYX}, we can conclude that
\beq \label{XprimeUU}
X'= e^{i\zeta}UXU^\dagger\,,
\eeq
where the complex phase factor $e^{i\zeta}$ is arbitrary and can be chosen for convenience as it corresponds to an additional hypercharge U(1)$_{\rm Y}$ transformation, which has no effect on the scalar potential parameters.
One can now check that $Z'= (X'\otimes X') Z'(X^{\prime\dagger}\otimes X^{\prime\dagger})$ by inserting \eq{XprimeUU} for $X'$ and evaluating the products and hermitian conjugates according to \eqs{kprop1}{kprop2}.

We shall now employ \eq{XprimeUU} in several examples.  First, suppose that the softly broken $\ug\otimes\Pi_2$-symmetric scalar potential is invariant with respect to $\mathbb{Z}_2$.   Using Table~\ref{tab:symm1}, it follows that
$X=\left(\begin{smallmatrix} 1 & \phm 0 \\ 0 & -1\end{smallmatrix}\right)$.  Hence, when transformed to the GCP3 basis using $U$ specified in \eq{you}, and choosing $e^{i\zeta}=-i$, it follows that,
\beq
X^\prime=\begin{pmatrix} \phm 0 & \,\,\, 1 \\ -1& \,\,\, 0\end{pmatrix}\,,
\eeq
which corresponds to the $\Pi^\prime_2$ symmetry defined in Table~\ref{tab:symm3}.  This is easily checked using \eqst{m11p}{m12p}.   In particular, if $\Im m_{12}^2=0$ and $m_{11}^2\neq m_{22}^2$, then it follows that $m^{\prime\,2}_{11}=m^{\prime\,2}_{22}$ and $\Re m^{\prime\,2}_{12}=0$ [cf.~Table~\ref{tab:class2}].

Second, suppose that the softly broken $\ug\otimes\Pi_2$-symmetric scalar potential is invariant with respect to $\Pi_2$. Using Table~\ref{tab:symm1}, it follows that
$X=\left(\begin{smallmatrix} 0 & \phm 1 \\ 1 & \phm 0\end{smallmatrix}\right)$.  Hence, when transformed to the GCP3 basis using $U$ specified in \eq{you}, and choosing $e^{i\zeta}=1$, it follows that $X'=X$.  That is, the scalar potential in the GCP3 basis also exhibits a $\Pi_2$ symmetry.  This is easily checked using \eqst{m11p}{m12p}.   Namely, if $m_{11}^2=m_{22}^2$ and $\Im m_{12}^2=0$ [cf. Table~\ref{tab:class}], then the same relations also hold for the primed parameters.

Third, suppose that the softly broken $\ug\otimes\Pi_2$-symmetric scalar potential is invariant with respect to U(1). Using Table~\ref{tab:symm1}, it follows that
$X=\left(\begin{smallmatrix} e^{-i\theta} & \phm 0 \\ 0 & \phm e^{i\theta}\end{smallmatrix}\right)$, where $-\half\pi<\theta\leq\half\pi$.  Transforming to the GCP3 basis using $U$ given in \eq{you}, and choosing $e^{i\zeta}=1$, 
we obtain
\beq
X^\prime=\begin{pmatrix} \phm\cos\theta & \,\,\, \sin\theta \\ -\sin\theta & \,\,\,\cos\theta\end{pmatrix},\qquad \text{for $-\half\pi<\theta\leq\half\pi$},
\eeq
which defines the U(1)$^\prime$ symmetry transformation.
In particular, if $\lambda_1=\lambda_2$ and $m_{12}^2=\lambda_5=\lambda_6=\lambda_7=0$ then \eqst{lamp}{m12p} yield,
  $m^{\prime\,2}_{11}=m^{\prime\,2}_{22}$, $\Re m^{\prime\,2}_{12}=0$, $\lambda_5^\prime=\lambda^\prime-\lambda_3^\prime-\lambda_4^\prime$ is real and $\lambda_6^\prime=\lambda_7^\prime=0$, which are the constraints due to U(1)$^\prime$ as indicated in Table~\ref{tab:class2}.
  
As our final example, we reconsider the residual unbroken symmetry in the case of a softly broken $\ug\otimes\Pi_2$-symmetric scalar potential when $c_{2\beta}=0$.  Below \eq{ctwobetamasses}, we noted that in this limiting case, after performing a rephasing to set $\xi=0$, the residual symmetry of the scalar potential and vacuum was $\Pi_2$.   
However, in this example, we shall keep $\xi$ arbitrary.  In order to accommodate $\xi\neq 0$ we define a new discrete symmetry,
\beq \label{pi2alphadef}
\Pi^{(\alpha)}_2:\quad \Phi_1\to e^{-i\alpha}\Phi_2\,,\qquad \Phi_2\to e^{i\alpha}\Phi_1\,,\quad \text{where $\alpha$ is a fixed real parameter}.
\eeq
Note that when $\Pi^{(\alpha)}_2$ is applied twice, one obtains the identity.  This means that for any fixed value of $\alpha$, the $\Pi^{(\alpha)}_2$ symmetry is equivalent to a $\mathbb{Z}_2$ symmetry that is manifestly realized in a different scalar field basis.

Of course, for $\alpha=0$, we regain the $\Pi_2$ symmetry.   Moreover, up to an overall hypercharge U(1)$_{\rm Y}$ transformation, the $\Pi^\prime_2$ symmetry corresponds to $\alpha=\half\pi$.  If the scalar potential in the $\Phi$-basis is invariant under $\Pi^{(\alpha)}_2$, then it follows that,
\beq \label{pi2xief}
m_{11}^2=m_{22}^2\,,\qquad \Im(m_{12}^2 e^{i\alpha})=0\,,\qquad \lambda_1=\lambda_2\,,\qquad \Im(\lambda_5e^{2i\alpha})=0\,,\qquad \lambda_7=\lambda_6^* e^{-2i\alpha}\,.
\eeq
In light of \eq{min3a}, it follows that for $c_{2\beta}=0$, the residual symmetry of the softly broken $\ug\otimes\Pi_2$-symmetric scalar potential and vacuum is $\Pi_2^{(\xi)}$ (for any fixed value of $\xi$).

Suppose that the $\Pi^{(\alpha)}_2$ symmetry is unbroken by the scalar potential in the $\ug\otimes\Pi_2$ basis.  Then, we can
deduce the corresponding symmetry in the GCP3 basis.  In this example, $X=\left(\begin{smallmatrix} 0 & \phm e^{-i\alpha} \\ e^{i\alpha} & \phm 0 \end{smallmatrix}\right)$, where
$\alpha$ is a fixed real parameter.  Hence, when transformed to the GCP3 basis using $U$ specified in \eq{you}, and choosing $e^{i\zeta}=1$, it follows that 
\beq \label{exprime}
X^\prime=\begin{pmatrix} \sin\alpha & \,\,\, \phm \cos\alpha \\ \cos\alpha & \,\,\,-\sin\alpha\end{pmatrix},\qquad \text{for a fixed real value of $\alpha$}.
\eeq
Without loss of generality, we may take $-\half\pi<\alpha\leq\half \pi$ (since $\alpha\to\alpha+\pi$ yields a hypercharge U(1)$_{\rm Y}$ transformation).
We shall denote this symmetry by,
\beq \label{barpidef}
\overline\Pi^{(\alpha)}_2:\quad \Phi_1\to \Phi_1\sin\alpha+\Phi_2\cos\alpha\,,\qquad \Phi_2\to \Phi_1\cos\alpha-\Phi_2\sin\alpha\,.
\eeq 
Note that for $\alpha=0$ [$\alpha=\half\pi$], the $\overline\Pi^{(\alpha)}_2$ symmetry coincides with $\Pi_2$ [$\mathbb{Z}_2$].
That is, $\overline\Pi^{(\alpha)}_2$ provides 
an interpolation from the $\Pi_2$ to the $\mathbb{Z}_2$ symmetry.

Imposing the $\overline\Pi^{(\alpha)}_2$ symmetry on the parameters of the scalar potential in the $\Phi$-basis for a fixed value of $\alpha\neq 0$, $\half\pi$, it follows that, 
\beqa
 \Im m_{12}^2&=&\Im \lambda_5=\Im\lambda_6=\Im\lambda_7=0\,, \label{pialpha1} \\
 m_{22}^2-m_{11}^2 &=& 2\tan\alpha\Re m_{12}^2\,, \label{pialpha2} \\
  \lambda_1-\lambda_2&=& 2\tan\alpha\Re(\lambda_6+\lambda_7)\,, \label{pialpha3} \\
 \lambda_1+\lambda_2-2(\lambda_3+\lambda_4+\Re\lambda_5)&=& 4\cot 2\alpha\Re(\lambda_6-\lambda_7)\,.\label{pialpha4} 
 \eeqa
Applying the above results to the softly broken GCP3-symmetric scalar potential, we set $\lambda^\prime\equiv\lambda_1^\prime=\lambda_2^\prime=\lambda_3^\prime+\lambda_4^\prime+\lambda_5^\prime$ and $\Im\lambda_5^\prime=\lambda_6^\prime=\lambda_7^\prime=0$.  Note that \eqst{pialpha1}{pialpha4} are consistent with these constraints.  In addition, the softly broken GCP3-symmetric scalar potential preserves the $\overline\Pi^{(\alpha)}_2$ symmetry in two cases: (i) if $m_{12}^{\prime\,2}$ is real and nonzero then $m_{11}^{\prime\,2}\neq m_{22}^{\prime\,2}$ [in which case, $\alpha$ is determined from \eq{pialpha2}]; or 
(ii)  if $m_{12}^{\prime\,2}=0$ then $m_{11}^{\prime\,2}= m_{22}^{\prime\,2}$ (in which case $\overline\Pi^{(\alpha)}_2$\!, which is a symmetry of the scalar potential for all values of $\alpha$, is promoted to an unbroken  GCP3 symmetry).   

For example, in the inert limit of the softly broken GCP3-symmetric scalar potential, where $s_{2\beta^\prime}c_{2\beta^\prime}\neq 0$,
$\sin\xi^\prime=0$, $m_{11}^{\prime\,2}\neq m_{22}^{\prime\,2}$ and $m_{12}^{\prime\,2}e^{i\xi^\prime}$ is real [See Table~\ref{tab:erps4gcp3}], we see that the conditions for the $\overline\Pi^{(\alpha)}_2$ symmetry are satisfied.  Moreover, one can show that the $\overline\Pi^{(\alpha)}_2$ symmetry is unbroken by the vacuum as follows.
Using \eqst{min4a}{min6a} under the assumption that $\sin\xi^\prime=0$, it follows that $m_{12}^{\prime\,2}e^{i\xi^\prime}=m_{12}^{\prime\,2}\cos\xi^\prime=\pm m_{12}^{\prime\,2}$ is real (where $\pm$ corresponds to $\xi^\prime=0$ or $\xi^\prime=\pi$, respectively), and
\beq
m_{22}^{\prime\,2}-m_{11}^{\prime\,2}=\pm\frac{2m_{12}^{\prime\,2} c_{2\beta^\prime}}{s_{2\beta^\prime}}\,.
\eeq
Hence, \eq{pialpha2} yields $\tan\alpha=\pm\cot 2\beta^\prime$.  In the convention where $-\half\pi<\alpha\leq\half\pi$ and $0\leq\beta^\prime\leq\half\pi$, it follows that $\sin\alpha=\pm\cos 2\beta^\prime$ and $\cos\alpha=\sin 2\beta^\prime$, or equivalently
\beq \label{sincos}
\alpha=\pm\left(\half\pi-2\beta^\prime\right)\,.
\eeq
  The $\overline\Pi^{(\alpha)}_2$ symmetry is unbroken by the vacuum if $ X^\prime\left(\begin{smallmatrix} v_1^\prime \\ v_2^\prime e^{i\xi^\prime}\end{smallmatrix}\right)=\pm \left(\begin{smallmatrix} v_1^\prime \\ v_2^\prime e^{i\xi^\prime}\end{smallmatrix}\right)$,
  where the $\pm$ sign reflects the fact that $\bigl\{\vev{\Phi^0_1},\vev{\Phi^0_2}\bigr\}$ is equivalent to $\bigl\{-\vev{\Phi^0_1},-\vev{\Phi^0_2}\bigr\}$, as the two are related by a hypercharge U(1)$_{\rm Y}$ transformation. 
After using the value of $\alpha$ obtained in
\eq{sincos} to determine $X'$ [cf.~\eq{exprime}], it follows that the $\overline\Pi^{(\alpha)}_2$ symmetry is unbroken 
by the vacuum since the following equation is an identity,
\beq 
\begin{pmatrix} \pm c_{2\beta^\prime}& \,\,\,\phm s_{2\beta^\prime} \\ \phm s_{2\beta^\prime}& \,\,\,\mp c_{2\beta^\prime}\end{pmatrix}\begin{pmatrix} \phm c_{\beta^\prime} \\ \pm s_{\beta^\prime}\end{pmatrix}=\pm\begin{pmatrix} \phm c_{\beta^\prime} \\ \pm s_{\beta^\prime}\end{pmatrix}\,.
\eeq
Consequently, one can conclude that the $\overline\Pi^{(\alpha)}_2$ symmetry is responsible for the inert limit (and the attendant exact Higgs alignment) in this case.

In some applications, it is useful to invert the relations obtained in \eqst{lamp}{m12p}.   This can be achieved by 
starting
from the GCP3 basis and employing the 
unitary transformation
\beq
U^{-1}=\frac{e^{-i\phi}}{\sqrt{2}}\begin{pmatrix} 1 &\quad i \\ i& \quad 1\end{pmatrix}\,.
\eeq
The resulting U(1)$\otimes\Pi_2$ basis parameters are,
\beqa
\lambda&=& \lambda^\prime -\lambda_5^\prime\,, \label{inv1}\\
\lambda_3&=& \lambda_3^\prime +\lambda_5^\prime\,, \label{inv2}\ \\
\lambda_4&=& \lambda_4^\prime +\lambda_5^\prime\,,\label{inv3}\ \\
\lambda R&=&\lambda^\prime+\lambda_5^\prime\,,\label{inv4}\ \\
\lambda_5&=&\lambda_6=\lambda_7=0\,. \label{inv5}\
\eeqa
In addition, the corresponding soft-breaking squared mass parameters are:
\beqa
&& m^2_{11}=\half (m_{11}^{\prime\,2}+m_{22}^{\prime\,2})-\Im m_{12}^{\prime\,2}\,,  \\
&& m^2_{22}=\half (m_{11}^{\prime\,2}+m_{22}^{\prime\,2})+\Im m_{12}^{\prime\,2}\,, \\
&& m^2_{12}=\Re m_{12}^{\prime\,2}-\half i (m_{22}^{\prime\,2}-m_{11}^{\prime\,2})\,.\label{unprimed}
\eeqa

Finally, the vevs in the U(1)$\otimes\Pi_2$ basis are given by
\beq \label{vprimes2}
v_1=\frac{e^{-i\phi}}{\sqrt{2}}\bigl(v'_1+iv'_2e^{i\xi^\prime}\bigr)\,,\qquad\quad v_2 e^{i\xi}=e^{-i\phi}\frac{i}{\sqrt{2}}\bigl(v'_1-iv'_2 e^{i\xi^\prime}\bigr)\,,
\eeq
where $v_1$ and $v_2$ are real and positive.    
Hence,
\beq \label{mags2}
c_\beta=\frac{1}{\sqrt{2}}\bigl(1-s_{2\beta^\prime}\sin\xi^\prime\bigr)^{1/2}\,,\qquad\quad s_\beta=\frac{1}{\sqrt{2}}\bigl(1+s_{2\beta^\prime}\sin\xi^\prime\bigr)^{1/2}\,,
\eeq
and it immediately follows that
\beq \label{sintwob2}
s^2_{2\beta}=1-s_{2\beta^\prime}^2\sin^2\xi^\prime\,.
\eeq
By convention, $0\leq\beta\leq \half\pi$ (or equivalently, $\sin 2\beta\geq 0$).

The phase $\phi$ is again fixed by the positivity of~$v_1$, which yields
\beq \label{phasephi2}
e^{-i\phi}=\frac{c_{\beta^\prime}-is_{\beta^\prime}e^{-i\xi^\prime}}{(1-s_{2\beta^\prime}\sin\xi^\prime)^{1/2}}\,,
\eeq
and is consistent with \eq{phasephi} after employing \eqs{mags2}{expxi}.
Then, \eq{vprimes2} yields,
\beq
e^{i\xi} s_\beta =\frac{i}{\sqrt{2}}\frac{c_{2\beta^\prime}-is_{2\beta^\prime}\cos\xi^\prime}{\bigl(1-s_{2\beta^\prime}\sin\xi^\prime\bigr)^{1/2}}\,.
\eeq
Likewise, 
$\xi$ is given by,
\beq \label{expxi}
e^{i\xi}=\frac{s_{2\beta^\prime}\cos\xi^\prime +ic_{2\beta^\prime}}{(1-s_{2\beta^\prime}^2\sin^2\xi^\prime)^{1/2}}\,.
\eeq
That is,
\beq \label{sin2xi2}
\sin\xi=\frac{c_{2\beta^\prime}}{(1-s_{2\beta^\prime}^2\sin^2\xi^\prime)^{1/2}}\,,\qquad\quad \cos\xi=\frac{s_{2\beta^\prime}\cos\xi^\prime}{(1-s_{2\beta^\prime}^2\sin^2\xi^\prime)^{1/2}}\,.\qquad\quad 
\eeq
Hence \eqs{sintwob2}{sin2xi2} yield,
\beq
s_{2\beta}\sin\xi=c_{2\beta^\prime}\,.
\eeq
Once the U(1)$\otimes\Pi_2$ basis parameters have been derived, one can perform one further rephasing to remove the phase $\xi$ (which is unphysical).
Finally, if $\beta^\prime=\tfrac14\pi$ and $\sin\xi^\prime=\pm 1$, then one of the vevs vanishes.  It then follows that $s_{2\beta}=0$, in which case $\xi$ is indeterminate if $s_{\beta}=0$
and $\xi=0$ if $c_{\beta}=0$.

The scalar masses obtained in the U(1)$\otimes\Pi_2$ basis and the GCP3 basis were derived by applying \eq{matrix33}.   In employing this equation, a specific value of $\eta$ was chosen.
The eigenvalues of $\mathcal{M}^2$ are independent of this choice.  However, the identification of $H$ and $A$ depend on this choice in the inert limit.   For a consistent treatment of the two basis choices, one should also transform $\eta$ when changing the scalar field basis according to \eq{etatrans}.   
In particular, given the choice of $\eta=-\xi$ that was employed in the $\ug\otimes\Pi_2$ basis, the corresponding $\eta^\prime$ in the GCP3 basis is given by
\beq \label{expetap}
e^{-i\eta^\prime}=(\det U) e^{-i\eta}=e^{2i\phi}e^{i\xi}= \frac{(c_{\beta^\prime}+is_{\beta^\prime}e^{i\xi^\prime})^2(s_{2\beta^\prime}\cos\xi^\prime+ic_{2\beta^\prime})}{(1-s_{2\beta^\prime}\sin\xi^\prime)(1-s_{2\beta^\prime}^2\sin^2\xi^\prime)^{1/2}}\,,
\eeq
after employing \eq{you} to obtain $\det U=e^{2i\phi}$ and making use of \eqs{phasephi2}{expxi}.  

The numerator of \eq{expetap}  can be simplified with a little algebra,
\beqa
(c_{\beta^\prime}+is_{\beta^\prime}e^{i\xi^\prime})^2(s_{2\beta^\prime}\cos\xi^\prime+ic_{2\beta^\prime})&=&
e^{i\xi^\prime}\bigl[c_{2\beta^\prime}\cos\xi^\prime+i(s_{2\beta^\prime}-\sin\xi^\prime)\bigr](s_{2\beta^\prime}\cos\xi^\prime+ic_{2\beta^\prime}) \nonumber \\
&=& e^{i\xi^\prime}(c_{2\beta^\prime}\sin\xi^\prime+i\cos\xi^\prime)(1-s_{2\beta^\prime}\sin\xi^\prime)\,. \nonumber \\
&=& ie^{i(\xi^\prime-\psi)}(1-s_{2\beta^\prime}\sin\xi^\prime)(1-s_{2\beta^\prime}^2\sin^2\xi^\prime)^{1/2}\,,
\eeqa
where we have used \eq{psidef} in the final step.  Inserting this result back into \eq{expetap} yields
\beq \label{etapr}
\eta^\prime=\psi-\xi^\prime-\half\pi\,,
\eeq
which justifies the choice of $\eta$ that was employed in \eq{etachoice}.
Note that if $s_{2\beta^\prime}=0$, then one of the two vevs in the GCP3 basis vanishes.   If $c_{2\beta^\prime}=1$, then \eq{psidef} yields $\psi=\xi^\prime$ and we conclude that $\eta^\prime=-\half\pi$.   If $c_{2\beta^\prime}=-1$, then $\psi=\xi^\prime=0$ (since in this case $\vev{\Phi_2^0}=v/\sqrt{2}$ is real and positive) and we again find that $\eta^\prime=-\half\pi$.  That is, $e^{-2i\eta^\prime}=-1$, which is the motivation for the choice of $\eta$ employed in obtaining \eqst{inertmass1}{inertmass4}.\footnote{If $s_{2\beta}=0$ in the $\ug\otimes\Pi_2$ basis then $Z_5=0$, in which case $m_H=m_A$ and the results of  \eqst{inertmass1}{inertmass4} do not depend on the choice of $\eta$.} 

For a satisfying check of \eq{etapr}, one can compute the value of $Z_5$ in the GCP3 basis starting from its value in the $\ug\otimes\Pi_2$ basis given in \eq{zeefive}.
In performing this computation, one must remember to rephase $Z_5$ as indicated in \eq{rephasing}.  
After making use of eqs.~(\ref{inv1}), (\ref{inv4}), (\ref{etapr}) and (\ref{psidef}), it then follows that
\beqa
Z_5&=& \half \lambda s_{2\beta}^2(1-R)e^{-2i\xi}e^{-4i\phi}=-\lambda_5^\prime(1-s_{2\beta^\prime}^2\sin^2\xi^\prime)e^{2i\eta^\prime} \nonumber \\
&=& \lambda_5^\prime e^{-2i\xi^\prime} (1-s_{2\beta^\prime}^2\sin^2\xi^\prime)e^{2i\psi}=\lambda_5^\prime e^{-2i\xi^\prime}(\cos\xi^\prime+ic_{2\beta^\prime}\sin\xi^\prime)^2\,,
\eeqa
in agreement with \eq{zee5}.

Although cases in which one of the two vevs vanish appear to be isolated from the parameter regimes in which both vevs are nonvanishing, in fact the two parameter regimes can be regarded as being continuously connected.  For example, starting from the GCP3 basis, the parameter regime in which one of the two vevs vanishes (i.e., $s_{2\beta^\prime}=0$) implies that $m^{\prime\,2}_{12}=0$ due to \eqs{inert1}{inert2}.\footnote{Note that it follows that the ratio $\Re(m_{12}^{\prime\,2} e^{i\xi^\prime})/s_{2\beta^\prime}$ that appears in \eq{masq} is indeterminate, in which case it can be replaced by $m_A^2-\lambda_5^\prime v^2\sin^2\xi^\prime$, with $m_A^2$ being regarded as a free parameter.}   In light of  \eq{m12p}, it follows that in terms of the U(1)$\otimes\Pi_2$ basis parameters, $\Re m^2_{12}=0$ and $m_{11}^2=m_{22}^2$, but in general $\Im m_{12}^2\neq 0$. Moreover, \eq{betaeq} then yields $\beta=\tfrac14\pi$.  As expected, this parameter regime can be identified as the inert limit, independently of the basis choice.  Nevertheless, it is clear that $\beta=\tfrac14\pi$ is continuously connected to other regions of the parameter space in the U(1)$\otimes\Pi_2$ basis.  Similarly, the case of one vanishing vev in the U(1)$\otimes\Pi_2$ basis corresponds to $s_{2\beta^\prime}\sin\xi=\pm 1$, which implies that $\beta^\prime=\tfrac14\pi$ in the GCP3 basis.   
Finally, the inert limit in the U(1)$\otimes\Pi_2$ basis when $R=1$ corresponds to the inert limit in the GCP3 basis when $\lambda_5^\prime=0$.

\section{The Higgs alignment limit}
\label{alignment}

The neutral scalar squared-mass matrix, $\mathcal{M}^2$, given in \eq{matrix33} is expressed with respect to a basis of neutral scalar interaction eigenstates,
$\{\sqrt{2}\,{\rm Re}~\!\mathcal{H}_1^0-v\,,$ ${\rm Re}\!~\mathcal{H}_2^0\,,\,{\rm Im}\!~\mathcal{H}_2^0\}$, where $\mathcal{H}_1^0$ and $\mathcal{H}_2^0$ are the neutral components of the Higgs basis fields.
The neutral scalar interaction eigenstate, $\varphi_1 \equiv \sqrt{2}\,{\rm Re}~\!\mathcal{H}_1^0-v$, possesses tree-level couplings to SM particles that coincide precisely with those of
the SM Higgs boson.  Consequently, if the mixing of $\varphi_1$ with ${\rm Re}\!~\mathcal{H}_2^0$ and ${\rm Im}\!~\mathcal{H}_2^0$ were to vanish exactly, then $\varphi_1$ would be a mass eigenstate
with tree-level properties that are indistinguishable from those of the SM Higgs boson.  In this case, the direction of $\varphi_1$ in field space is exactly aligned with the direction of the vacuum expectation value $v$.  Hence, the limit of zero mixing described above is called the Higgs alignment limit~\cite{Gunion:2002zf,Craig:2012vn,Craig:2013hca,Asner:2013psa,Carena:2013ooa,Haber:2013mia}.

In light of \eq{matrix33}, it follows that the Higgs alignment is realized exactly if and only if $Z_6=0$, in which case we can identify $m^2_{\varphi_1}=Z_1 v^2$.   In this limit, the masses of the two other neutral scalars are not immediately constrained (beyond experimental bounds based on the absence of any newly discovered scalar states at the LHC).   Although the observed Higgs boson at the LHC is SM-like, the precision of the current data allows for $10\%$--$20\%$ deviations from SM behavior of the Higgs boson couplings to vector bosons and third generation quarks and charged leptons~\cite{Aad:2019mbh,Sirunyan:2018koj,CMS:2020gsy,ATLAS:2020qdt}.   Thus, the present Higgs data requires only an approximate Higgs alignment, which allows for a small mixing of $\varphi_1$ with the other two neutral scalar interaction eigenstates.   This small mixing can be achieved in one of two ways---either $|Z_6|\ll 1$ or $Y_2\gg v^2$.   The latter corresponds to the decoupling limit of the 2HDM~\cite{Haber:1989xc,Gunion:2002zf}.

For example, if CP is conserved then the squared-mass matrix, $\mathcal{M}^2$, of the neutral scalars, expressed with respect to Higgs basis fields, breaks up into a $2\times 2$ block and a $1\times 1$ block, 
\beq \label{CPmassmatrix}
\mathcal{M}^2=\begin{pmatrix} Z_1 v^2 & \quad Z_6 v^2 & \quad 0 \\ Z_6v^2 & \quad m_A^2+Z_5 v^2 & \quad 0 \\  0 & \quad 0 & \quad m_A^2\end{pmatrix}\,.
\eeq
Once again, we see that $Z_6=0$ corresponds to exact Higgs alignment, whereas approximate 
Higgs alignment is realized when $|Z_6|\ll 1$ and/or $m_A\gg v$.

Diagonalizing the $2\times 2$ block yields the CP-even Higgs mass eigenstates $H$ and $h$, 
\beq \label{Hh}
\begin{pmatrix} H\\ h\end{pmatrix}=\begin{pmatrix} \cbma & \,\,\, -\sbma \\
\sbma & \,\,\,\phantom{-}\cbma\end{pmatrix}\,\begin{pmatrix} \sqrt{2}\,\,{\rm Re}~\mathcal{H}_1^0-v \\ 
\sqrt{2}\,{\rm Re}~\mathcal{H}_2^0
\end{pmatrix}\,,
\eeq
where $\mhl\leq\mhh$, $\cbma\equiv\cos(\beta-\alpha)$ and $\sbma\equiv\sin(\beta-\alpha)$ in a convention where \mbox{$0\leq\beta-\alpha\leq\pi$.} In a real $\Phi$-basis with real non-negative vevs, $\vev{\Phi^0_a}=v_a/\sqrt{2}$ ($a=1,2$),
$\tan\beta=v_2/v_1$ 
 and $\alpha$ is the mixing angle that diagonalizes the CP-even Higgs squared-mass matrix when expressed with respect to the $\{\sqrt{2}\,{\rm Re}~\Phi_1^0-v_1\,,\,\sqrt{2}\,{\rm Re}~\Phi_2^0-v_2\}$ basis.  Nevertheless, the quantity $s_{\beta-\alpha}$ is independent of the choice of the scalar field basis.

After diagonalizing the matrix $\mathcal{M}_H^2$, the neutral CP-even scalar masses are given by,
\beq \label{cpevenmasses}
m^2_{H,h}=\half\biggl\{\mha^2+(Z_1+Z_5)v^2\pm \sqrt{\bigl[\mha^2-(Z_1-Z_5)v^2\bigr]^2+|Z_6|^2 v^4}\,
\biggr\}\,,
\eeq
where $\mhl\leq\mhh$, and
\beq \label{sc}
s_{\beta-\alpha}c_{\beta-\alpha}=\frac{-Z_6 v^2}{m_H^2-m_h^2}\,,\qquad\qquad
c^2_{\beta-\alpha}-s^2_{\beta-\alpha}=\frac{\mha^2-(Z_1-Z_5)v^2}{m_H^2-m_h^2}\,.
\eeq
We shall henceforth assume that 
$h\simeq\sqrt{2}\,\Re \mathcal{H}_1^0-v$ is SM-like and thus should be identified with the observed Higgs boson with $m_h\simeq 125$~GeV.  Under this assumption, it follows from \eq{Hh} that $\cbma\to 0$ in the Higgs alignment limit.  Indeed, one can use \eq{sc} to derive~\cite{Bernon:2015qea,Haber:2015pua},
\beq \label{ctozero}
\cbma=\frac{-Z_6 v^2}{\sqrt{(m_H^2-m_h^2)(m_H^2-Z_1 v^2)}}\,,
\eeq
in a convention where $\sbma\geq 0$.  Having identified $h$ as SM-like, it follows that $m^2_H> Z_1 v^2$, which confirms that $\cbma\to 0$ in the Higgs alignment limit.\footnote{If $H$ were SM-like (with $m_H\geq m_h$) then $m_H^2\to Z_1 v^2$ in the Higgs alignment limit, in which case \eq{ctozero} would not be very useful.  Indeed, in this case $\sbma\to 0$ in the Higgs alignment limit, and a more useful formula to replace \eq{ctozero} would be $\sbma=-Z_6 v^2/\sqrt{(m_H^2-m_h^2)(Z_1 v^2-m_h^2)}$ in an alternative convention where $\cbma\geq 0$~\cite{Bernon:2015wef}.
However, in the conventions adopted in this paper $h$ is always SM-like in the Higgs alignment limit in light of \eq{inertnames}, irrespective of the mass ordering of $h$ and $H$.}

One can now ask the following question---is there a symmetry that can be imposed on the 2HDM scalar potential such that the Higgs alignment limit is exact, corresponding to the condition that $Z_6=0$.  In fact, it is straightforward to identify the complete list of all possible symmetries that enforce the $Z_6=0$ condition in the 2HDM by noting that
the scalar potential minimum condition [cf.~\eq{YZ}] would then imply that $Y_3=0$.  Thus, in light of \eq{why3}, the Higgs alignment  limit is exact if 
\beq \label{why30}
Y_3=\bigl[\half(m_{22}^2-m_{11}^2)s_{2\beta}-\Re(m_{12}^2 e^{i\xi})c_{2\beta}-i\Im(m_{12}^2 e^{i\xi})\bigr]e^{-i\xi}=0\,.
\eeq
A sufficient (but not necessary) condition for satisfying \eq{why30} can be obtained by setting $m_{12}^2=0$, in which case either $s_{2\beta}=0$ or $m_{11}^2=m_{22}^2$.

If $s_{2\beta}=0$, then the $\mathbb{Z}_2$ symmetry is unbroken by the vacuum.  
This corresponds to the IDM, where an unbroken $\mathbb{Z}_2$ symmetry is present in the Higgs basis, which implies that $Y_3=Z_6=Z_7=0$.  In the IDM, the Higgs basis field $\mathcal{H}_2$ is odd under the $\mathbb{Z}_2$ symmetry, whereas $\mathcal{H}_1$ along with all other SM fields are $\mathbb{Z}_2$-even.   If in addition, one imposes the condition $Z_5=0$, then the IDM scalar potential will exhibit an unbroken U(1) symmetry that is preserved by the vacuum (resulting in a mass-degenerate pair of inert neutral scalars, $m_H=m_A$).
In both cases, one can identify $\varphi_1=\sqrt{2}\,\Re \mathcal{H}_1^0-v$ as the SM Higgs boson at tree level.  Deviations of the properties of $\varphi_1$ from that of the SM Higgs boson can arise due to the other Higgs fields (beyond $\varphi_1$) contributing to radiative loop corrections to physical observables (e.g., the charged Higgs boson loop that contributes to $\varphi_1\to\gamma\gamma$ decay).  

Note that if $m_{11}^2=m_{22}^2$ and $m_{12}^2\neq 0$ then one can achieve $Y_3=0$ by simultaneously imposing a $\Pi_2$ symmetry and a GCP1 symmetry in the $\Phi$-basis.  Consulting Table~\ref{tab:class}, these symmetries taken together yield the following constraints on the scalar potential,
\beq \label{IDMconstraints}   
m_{11}^2=m_{22}^2\,,\qquad m^2_{12}, \lambda_5 \in\mathbb{R}\,, \qquad \lambda_1=\lambda_2\,,\qquad \lambda_6=\lambda_7\in\mathbb{R}\,.
\eeq
Such a scalar potential does not lie within the ERPS4.   One can now determine the scalar potential minimum conditions (see eq.~(E3) of Ref.~\cite{Boto:2020wyf}), which yield 
\beq \label{IDMbasis}
c_{2\beta}=0 \quad \text{and} \quad \sin\xi=0\,.
\eeq
Hence, a scalar potential whose parameters satisfy \eqs{IDMconstraints}{IDMbasis} yields $Y_3=0$ [cf.~\eq{why30}], corresponding to an exact Higgs alignment~\cite{Darvishi:2020teg}.   Moreover,  \eqst{idminert1}{idminert6} are satisfied after employing \eqs{IDMconstraints}{IDMbasis}, which implies that $Y_3=Z_6=Z_7=0$.
That is, the scalar potential parameters of \eq{IDMconstraints} corresponds to the IDM in a particular scalar field basis.   
Indeed, one can adopt a definition of the IDM scalar potential as corresponding to the existence of a scalar field basis in which
the $\Pi_2$ and GCP1 symmetries are both manifestly preserved by the 2HDM scalar potential and vacuum.

If $s_{2\beta}\neq 0$ then \eq{why30} is automatically satisfied if $m_{12}^2=0$ and 
$m_{11}^2=m_{22}^2$.  Consulting the results of Table~\ref{tab:class}, it follows that exact Higgs alignment is automatically implemented if the 2HDM scalar potential respects one of the following symmetries: $\mathbb{Z}_2\otimes\Pi_2$, U(1)$\otimes\Pi_2$, SO(3), GCP2, or GCP3.   Of course, given that GCP2 is equivalent to  $\mathbb{Z}_2\otimes\Pi_2$ in a different scalar field basis and GCP3 is equivalent to U(1)$\otimes\Pi_2$ in a different scalar field basis, it follows that there are three inequivalent symmetries of the 2HDM scalar potential beyond the IDM that yield exact Higgs alignment.  What is common to these three inequivalent symmetries is that they all reside in the ERPS.

The conditions that $m_{11}^2=m_{22}^2$ and $m_{12}^2=0$ imply that the symmetries identified above are preserved by the scalar potential.   
In particular, in the ERPS (where $\lambda\equiv\lambda_1=\lambda_2$ and $\lambda_7=-\lambda_6$), eq.~(A10) of Ref.~\cite{Boto:2020wyf} yields,
\beq \label{zsix}
Z_6  =e^{-i\xi}\biggl\{
-\half\stwob\ctwob(\lambda-\lambda_{345})+\half is_{2\beta}\Im(\lambda_5 e^{2i\xi})+c_{4\beta}\Re(\lambda_6 e^{i\xi})+i\ctwob\Im(\lambda_6 e^{i\xi})\biggr\}\,,
\eeq
where $\lambda_{345}\equiv \lambda_3+\lambda_4+\Re(\lambda_5 e^{2i\xi})$.   We have already noted that the inert limit of the scalar potential in the ERPS regime, where exact Higgs alignment is achieved, corresponds to $Y_3=0$, which then implies
that $Z_6=0$ via \eq{YZ} and $Z_7=0$ due to the ERPS conditions.   For example, in the case of a $\mathbb{Z}_2\otimes\Pi_2$-symmetric scalar potential, applying the conditions exhibited in Table~\ref{tab:class} yields the expression for $Z_6$ given in \eq{zeesevencp2}.
Because $m_{11}^2=m_{22}^2$ and $m_{12}^2=0$, it automatically follows that $Y_3=0$ which implies that $Z_6=0$.  The vanishing of $Z_6$ [although not immediately evident from \eq{zsix}] is a consequence of the scalar potential minimum conditions of the ERPS which yield,
\beq \label{erps4inertconds} 
\lambda (1-R)c_{2\beta}=\lambda_5 s_{2\beta}\sin 2\xi=0\quad \text{or} \quad s_{2\beta}=0\,,
\eeq
where $R\equiv (\lambda_3+\lambda_4+\lambda_5)/\lambda$.  Since $R\neq 1$ and $\lambda_5\neq 0$ (otherwise, the symmetry group of the scalar potential is larger than $\mathbb{Z}_2\otimes\Pi_2$), it then follows that either $c_{2\beta}=\sin 2\xi=0$ or $s_{2\beta}=0$.    Inserting these conditions in \eq{zsix} along with the $\mathbb{Z}_2\otimes\Pi_2$ symmetry conditions, $\Im\lambda_5=\lambda_6=\lambda_7=0$, yields $Z_6=0$ as expected.

In the case of an unbroken $\ug\otimes\Pi_2$ symmetry, we simply add one additional condition, $\lambda_5=0$ to the scalar potential parameters (while maintaining $R\neq 1$).  In this case, \eq{erps4inertconds} implies that either $c_{2\beta}=0$ or $s_{2\beta}=0$ (with no restriction on $\xi$, which is an unphysical phase that can be rephased away), and again \eq{zsix} yields $Z_6=0$.

It is instructive to consider the GCP3-symmetric scalar potential, which can be obtained from the $\mathbb{Z}_2\otimes\Pi_2$-symmetric scalar potential by adding one additional constraint, $R=1$ (while maintaining $\lambda_5\neq 0$).   The case of an unbroken GCP3 symmetry
is not physically distinct from the previous case since it is equivalent to a $\ug\otimes\Pi_2$ symmetry in a different scalar field basis.  The minimum conditions of the GCP3-symmetric scalar potential (where we now employ primed parameters) yield,
\beq \label{erps4inertconds2} 
\lambda_5^\prime c_{2\beta^\prime}\sin^2\xi^\prime=\lambda_5^\prime s_{2\beta^\prime}\sin 2\xi^\prime=0\quad \text{or} \quad  s_{2\beta^\prime}=0\,.
\eeq
These conditions guarantee that $Z_6=0$ [cf~\eq{zee6}], independently of the parameters of the GCP3-symmetric scalar potential.  In particular, 
exact Higgs alignment is achieved for all values 
of $\beta^\prime$ in cases of an unbroken and some softly broken GCP3-symmetric scalar potentials, 
in contrast to the cases of $\mathbb{Z}_2\otimes\Pi_2$ and $\ug\otimes\Pi_2$ where exact Higgs alignment is satisfied only when $\beta=0$, $\frac14\pi$ or~$\half\pi$.\footnote{The reader might wonder how it is possible that exact alignment can be achieved for all values of $\beta^\prime$ but only special values of $\beta$ in light of the fact that the $\ug\otimes\Pi_2$ and GCP3-symmetric scalar potentials can be transformed into one another by an appropriate change of basis.  The answer can be seen by examining \eq{sintwob2}.    Employing \eq{erps4inertconds2} with $\lambda_5^\prime\neq 0$, it follows that either $\sin\xi^\prime=0$, in which case all values of~$\beta^\prime$ are permitted, or $c_{2\beta^\prime}=\cos\xi^\prime=0$.  Using \eq{sintwob2}, it follows that the possible values of $\beta^\prime$ correspond to either $s_{2\beta}=0$ or  $c_{2\beta}=0$.}

In Refs.~\cite{Dev:2014yca,Dev:2017org}, Higgs alignment enforced by a symmetry is defined to be ``natural'' if $Z_6=0$ is achieved 
independently of the value of~$\beta$.  Based on the discussion above, this definition
eliminates the $\mathbb{Z}_2\otimes\Pi_2$, GCP2 and $\ug\otimes\Pi_2$-symmetric scalar potentials from the list of potentials that exhibit a ``natural'' Higgs alignment.\footnote{In Refs.~\cite{Dev:2014yca,Dev:2017org}, the maximal symmetry groups associated with the GCP3 and SO(3)-symmetric scalar potentials, which  exhibit a ``natural'' Higgs alignment, are identified as 
$\mathbb{Z}_2 \otimes{\rm O}(2)\otimes{\rm O}(2)$ and O(3)$ \otimes $O(2), respectively.  In addition, 
if the U(1)$_{\rm Y}$ hypercharge gauge coupling $g^\prime=0$ in the gauge covariant kinetic terms of the scalar fields, then an SO(3)-symmetric scalar potential with $\lambda_4=\lambda_5=0$
(cf.~Table~\ref{tab:cust}) yields an SO(5)-symmetric scalar Lagrangian that also exhibits a ``natural'' Higgs alignment. \label{fnnatural}}
In particular, the GCP3-symmetric scalar potential exhibits a ``natural'' alignment in the sense of Refs.~\cite{Dev:2014yca,Dev:2017org}, whereas the $\ug\otimes\Pi_2$-scalar potential does not.  
However, this distinction is problematical given that the GCP3-symmetric and U(1)$\otimes\Pi_2$-symmetric scalar potentials are physically equivalent and can be transformed into each other by an appropriate change of scalar field basis (as shown explicitly in Section~\ref{transforming}).   To avoid such an undesirable feature, a better definition of  ``natural'' alignment 
in the spirit of Refs.~\cite{Dev:2014yca,Dev:2017org} would be to require that the conditions, $Y_3=Z_6=0$, should be independent of the scalar potential minimum conditions.   
Under this stricter definition, the GCP3-symmetric scalar potential would not exhibit a ``natural'' Higgs alignment.  
In contrast, the SO(3)-symmetric scalar potential with \hbox{$m_{11}^2=m_{22}^2$,} $m_{12}^2=0$, $\lambda_1=\lambda_2=\lambda_3+\lambda_4$ and $\lambda_5=\lambda_6=\lambda_7=0$
does satisfy $Y_3=Z_6=0$ independently of the scalar potential minimum conditions [cf.~\eq{zsix}] and thus would exhibit a ``natural'' Higgs alignment according to the stricter definition proposed above.

However, our preference is to employ
 the concept of naturalness as introduced by `t Hooft in Ref.~\cite{tHooft:1979rat}, which implies that a small parameter of a theory should be considered natural if the symmetry of the Lagrangian is increased by setting the parameter to zero.  In the present context, the small parameters are the potentially soft-breaking  parameters, $m_{11}^2-m_{22}^2$ and $m_{12}^2$, of the ERPS4 that could potentially generate departures from exact Higgs alignment.  
 All symmetry groups of the ERPS---GCP2, $\mathbb{Z}_2\otimes\Pi_2$, GCP3, $\ug\otimes\Pi_2$ and SO(3)---yield an exact Higgs alignment naturally in the sense of `t Hooft.  Indeed, exact Higgs alignment realized in this way is stable under renormalization group running, which is further evidence that the symmetry based approach that we have adopted is correct.\footnote{In general, renormalization group running does not preserve the scalar field basis.  However, the group theoretic properties of the symmetries of the scalar potential, whose specific realization may change in different choices of the scalar field basis, do not depend on the basis choice.} 

The conditions that $m_{11}^2=m_{22}^2$ and $m_{12}^2=0$ are sufficient but not necessary for exact Higgs alignment.  In particular, exact Higgs alignment arises in any inert limit of the 2HDM.    Thus to obtain a complete classification of 2HDM scalar potentials that yield an exact Higgs alignment due to a symmetry, it suffices to enumerate the inert limits of the softly broken $\mathbb{Z}_2\otimes\Pi_2$, U(1)$\otimes\Pi_2$ or GCP3, and SO(3)-symmetric scalar potentials.\footnote{The possibility of natural Higgs alignment in the presence of soft symmetry-breaking squared mass terms has also been treated in Ref.~\cite{Darvishi:2020teg}.}  
These results can be found in Tables~\ref{tab:erps4}, \ref{tab:erps4s}, \ref{tab:erps4gcp3} and \ref{tab:erps4so3}.   
We proceed to list all the relevant subcases below.

Given a softly broken $\mathbb{Z}_2\otimes\Pi_2$-symmetric scalar potential, 
exact Higgs alignment arises in two subcases, as shown in Section~\ref{zee2pi2}: (i) $s_{2\beta}=\sin 2\xi=0$, $m_{11}^2\neq m_{22}^2$ and $m_{12}^2=0$, which preserves a $\mathbb{Z}_2$ symmetry that is unbroken in the vacuum, and (ii) $c_{2\beta}=\sin 2\xi=0$,  $m_{11}^2=m_{22}^2$ and $\Im\bigl[m_{12}^2\bigr]^2=2\Re m_{12}^2 \Im m_{12}^2=0$, which preserves a $\Pi_2$~[$\Pi^\prime_2$] symmetry if $\Im m_{12}^2=0$ [$\Re m_{12}^2=0$] that is unbroken in the vacuum. 
In the absence of soft breaking, the constraints on the scalar potential parameters due to $\Pi_2$ and $\Pi_2^\prime$ are identical.
The $\mathbb{Z}_2$, $\Pi_2$ or~$\Pi_2^\prime$ residual symmetries are responsible for maintaining the exact Higgs alignment.

Given a softly broken $\ug\otimes\Pi_2$-symmetric scalar potential, exact Higgs alignment arises in two subcases, as shown in Section~\ref{sec:UPibasis}: (i) $s_{2\beta}=0$, $m_{11}^2\neq m_{12}^2$ and $m_{12}^2=0$, which preserves a U(1) symmetry that is unbroken in the vacuum, and (ii) $c_{2\beta}=0$, $m_{11}^2=m_{22}^2$ and $m_{12}^2\neq 0$.
In light of \eq{min3a}, one can rephase $\Phi_2\to e^{-i\xi}\Phi_2$
to achieve a real basis, in which case
the scalar potential in subcase (ii)
preserves a $\Pi_2$ symmetry that is unbroken in the vacuum.  If one does not remove
the (unphysical) parameter $\xi$, then \eq{pi2xief} can be used to identify the unbroken vacuum symmetry as $\Pi^{(\xi)}_2$, which is $\Pi_2$ in the rephased scalar field basis. 
In the case of an unbroken $\ug\otimes\Pi_2$-symmetric scalar potential, the $\ug\otimes\Pi_2$ symmetry is 
spontaneously broken down to U(1) if $s_{2\beta}=0$ or to $\Pi_2$ if $c_{2\beta}=0$.  

Although a softly broken GCP3-symmetric scalar potential is equivalent to a softly broken $\ug\otimes\Pi_2$-symmetric scalar potential in a different basis, it is instructive to enumerate the cases in which a softly broken GCP3-symmetric scalar potential exhibits exact Higgs alignment.   Using the results of Section~\ref{sec:GCP3basis}, exact Higgs alignment arises in four subcases in terms of the primed GCP3 basis parameters:
(i) $s_{2\beta^\prime}=0$, $m_{11}^{\prime\,2}\neq m_{22}^{\prime\,2}$ and $m_{12}^{\prime\,2}=0$, which preserves a $\mathbb{Z}_2$ symmetry that is unbroken in the vacuum;
(ii) $c_{2\beta^\prime}=\cos\xi^\prime=0$, $m_{11}^{\prime\,2}= m_{22}^{\prime\,2}$ and $\Im m_{12}^{\prime\,2}\neq 0$, which preserves a U(1)$^\prime$ symmetry that is unbroken in the vacuum; (iii) $c_{2\beta^\prime}=\sin\xi^\prime=0$, $m_{11}^{\prime\,2}= m_{22}^{\prime\,2}$ and $\Re m_{12}^{\prime\,2}\neq 0$, which preserves a $\Pi_2$ symmetry that is unbroken by the vacuum; and
(iv) $s_{2\beta^\prime}c_{2\beta^\prime}\neq 0$, $\sin\xi^\prime=0$, $m_{11}^{\prime\,2}\neq m_{22}^{\prime\,2}$ and $\Re m_{12}^{\prime\,2}\neq 0$, which preserves a $\overline\Pi^{(\alpha)}_2$ vacuum symmetry, where $\alpha=\left(\half\pi-2\beta^\prime\right)\cos\xi^\prime=\pm \left(\half\pi-2\beta^\prime\right)$.  This result is derived in Section~\ref{transforming}, where the $\overline\Pi^{(\alpha)}_2$ symmetry is introduced in \eq{barpidef} and the relation that yields $\alpha$ in terms of $\beta^\prime$ is obtained in \eq{sincos}.
Finally, in the case of an unbroken GCP3-symmetric scalar potential, the GCP3 symmetry, which is equivalent to a U(1)$^\prime\otimes\mathbb{Z}_2$ symmetry, is spontaneously broken down to U(1)$^\prime$ [$\mathbb{Z}_2$] if $c_{2\beta}=\cos\xi^\prime=0$ [$s_{2\beta^\prime}=0$], or to $\overline\Pi^{(\alpha)}_2$ with $\alpha=\pm\left(\half\pi-2\beta^\prime\right)$ if $s_{2\beta^\prime}\neq 0$ and $\cos\xi^\prime=\pm 1$.  

Finally, we examine the case of a softly broken SO(3)-symmetric scalar potential.   We noted above that in this case $Z_6=0$ independently of the scalar potential minimum conditions.  This means that all softly broken SO(3)-symmetric scalar potentials exhibit exact Higgs alignment, since the scalar potential minimum conditions will guarantee that $Y_3=0$ even when $m_{11}^2\neq m_{22}^2$ and/or $m_{12}^2\neq 0$. 

Below \eq{masssrelso3}, we noted the presence of 
mass-degenerate scalars, $H$ and $A$, which was attributed to a Peccei-Quinn U(1) symmetry in the Higgs basis, $\mathcal{H}_1\to \mathcal{H}_1$, $\mathcal{H}_2\to e^{2i\theta}\mathcal{H}_2$ (for any value of $0\leq\theta<\pi$), which is unbroken by the vacuum.\footnote{The version of the Peccei-Quinn symmetry transformation that is used here corresponds to U(1)$_{\rm PQ}$ given in Table~\ref{tab:symm1} followed by a hypercharge U(1)$_{\rm Y}$ transformation, which is also a symmetry of the vacuum in the Higgs basis.}
It is instructive to ascertain the precise form of the U(1) symmetry in the $\Phi$-basis.  To accomplish this, we employ \eq{XprimeUU}, where the unitary matrix
\beq \label{U}
U=\begin{pmatrix} c_\beta & \quad -e^{-i(\xi+\eta)}s_\beta \\ e^{i\xi}s_\beta &  \quad e^{-i\eta}c_\beta\end{pmatrix}\,,
\eeq
transforms the Higgs basis into the $\Phi$-basis.  The phase $e^{i\eta}$, which appears in \eq{higgspot} and reflects the freedom to rephase the Higgs basis field $\mathcal{H}_2$, cancels exactly when \eq{XprimeUU} is applied.  Starting with $X=\left(\begin{smallmatrix} 1 & \,\,\,0 \\ 0  &\,\,\, e^{2i\theta}\end{smallmatrix}\right)$, we make use of \eq{XprimeUU} with $\zeta=-\theta$ to obtain
\beq
X^\prime=\begin{pmatrix} \cos\theta-ic_{2\beta}\sin\theta & \quad -ie^{-i\xi}s_{2\beta}\sin\theta \\ 
-ie^{i\xi}s_{2\beta}\sin\theta & \quad \cos\theta+ic_{2\beta}\sin\theta\end{pmatrix}\,.
\eeq
Thus, in the $\Phi$-basis characterized by $\tan\beta=|\vev{\Phi_2^0}/\vev{\Phi_1^0}|$ and $\xi=\arg\bigl[\vev{\Phi_1^0}^*\vev{\Phi_2^0}\bigr]$, the Peccei-Quinn symmetry, which we designate by U(1)$_{\rm H}$
(to remind the reader that it has been first applied in the Higgs basis), is given by
\beqa
U(1)_{\rm H}: & \quad & \Phi_1\longrightarrow (\cos\theta-ic_{2\beta}\sin\theta)\Phi_1-ie^{-i\xi}s_{2\beta}\sin\theta\,\Phi_2\,,\nonumber \\
&& \Phi_2\longrightarrow -ie^{i\xi}s_{2\beta}\sin\theta\, \Phi_1+ (\cos\theta+ic_{2\beta}\sin\theta)\Phi_2\,. \label{youoneH}
\eeqa

Imposing the U(1)$_{\rm H}$ symmetry on the parameters of a
general 2HDM scalar potential in the $\Phi$-basis yields the following constraints,
\beqa
&& \Im(m_{12}^2 e^{i\xi})=\lambda_5=\lambda_6=\lambda_7=0\,,\\
&& \lambda\equiv\lambda_1=\lambda_2=\lambda_3+\lambda_4\,, \\
&& m_{22}^2-m_{11}^2=2\cot 2\beta\,\Re(m_{12}^2 e^{i\xi})\,.
\eeqa
These constraints correspond to a softly broken SO(3)-symmetric scalar potential and scalar potential minimum conditions [cf.~\eqst{min4a}{min6a} with $\lambda=\lambda_3+\lambda_4$ and $\lambda_5=0$].  Moreover, 
\beq
X'\begin{pmatrix} c_\beta \\ s_\beta e^{i\xi}\end{pmatrix}= e^{-i\theta} \begin{pmatrix}c_\beta \\ s_\beta e^{i\xi}\end{pmatrix}\,,
\eeq
which confirms that the vacuum is invariant under the U(1)$_{\rm H}$ transformation (up to an overall hypercharge U(1)$_{\rm Y}$ transformation that has no effect on the scalar potential parameters).

We conclude that for a generic softly broken SO(3)-symmetric scalar potential, a U(1)$_{\rm H}$  subgroup remains unbroken and is responsible for the mass degeneracy of $H$ and $A$ as well as the exact Higgs alignment.\footnote{In the case of $s_{2\beta}=0$, the $\Phi$-basis coincides with the Higgs basis (up to a possible discrete $\Pi_2$ transformation), in which case U(1)$_{\rm H}$ reduces to the standard U(1)$_{\rm PQ}$ symmetry.}  
In the case of an unbroken SO(3)-symmetric scalar potential, the SO(3) symmetry is spontaneously broken down to U(1)$_{\rm H}$, in which case both $H$ and $A$ can be identified as massless Goldstone bosons (of opposite CP quantum numbers).

This completes 
the classification of all unbroken or softly broken symmetries of the 2HDM scalar potential that yield an exact Higgs alignment.   This classification is summarized in
Table~\ref{tab:align}.  
Many aspects of this table can be easily understood by employing the results of Appendix~\ref{app:IDM}.  Applying the ERPS4 conditions
($\lambda_1=\lambda_2$ and $\lambda_7=-\lambda_6$) in \eqst{idminert1}{idminert6}, the parameters 
of the scalar potential in the ERPS4 regime in the $\Phi$-basis 
satisfy, 
\beqa
 \Im(m_{12}^2 e^{i\xi})&=&0\,, \label{gcp2inert1}\\[-1.5pt] 
(m_{22}^2-m_{11}^2)s_{2\beta} &=& 2\Re(m_{12}^2 e^{i\xi})c_{2\beta}\,,\label{gcp2inert2}\\[-1.5pt] 
c_{4\beta}\Re(\lambda_6 e^{i\xi})&=&\half s_{2\beta}c_{2\beta}\bigl[\lambda-\lambda_3-\lambda_4-\Re(\lambda_5 e^{2i\xi})\bigr]\,,\label{gcp2inert3}\\[-1.5pt] 
c_{2\beta}\Im(\lambda_6 e^{i\xi})&=&-\half s_{2\beta}\Im(\lambda_5 e^{2i\xi})\,.\label{gcp2inert4} 
\eeqa
\Eqst{gcp2inert1}{gcp2inert4} provide the necessary and sufficient conditions for the inert limit of the scalar potential in the ERPS4 regime, 
thereby producing an exact Higgs alignment.  
\clearpage

\begin{table}[t!]
\begin{tabular}{|c|c|c|c|c|}
\hline
\Lower{\pht Symmetry\pht } &   \Lower{soft breaking} & \Lower{parameter}  & \multicolumn{2}{c|}{\pht  residual unbroken symmetry of  \pht}  \\
&& constraints & \pht scalar potential \hspace{1pt}& vacuum \\
\hline
$\mathbb{Z}_2$  & none & \pht $s_{2\beta}=0$\pht  &$\mathbb{Z}_2$ & $\mathbb{Z}_2$ \\
U(1) & none & \pht $s_{2\beta}=0$\pht  & U(1) & U(1) \\
$\mathbb{Z}_2\otimes\Pi_2$ &  $m_{11}^2\neq m_{22}^2$  & $s_{2\beta}=0$ &  $\mathbb{Z}_2$ & $\mathbb{Z}_2$ \\
$\mathbb{Z}_2\otimes\Pi_2$ & $\Re m_{12}^2\neq 0$  & $c_{2\beta}=\sin \xi=0$ & $\Pi_2$ & $\Pi_2$ \\
$\mathbb{Z}_2\otimes\Pi_2$ & $\Im m_{12}^2\neq 0$   & $c_{2\beta}=\cos \xi=0$ & $\Pi^\prime_2$ & $\Pi^\prime_2$ \\
$\mathbb{Z}_2\otimes\Pi_2$ & none  & $s_{2\beta}=0$ &$\mathbb{Z}_2\otimes\Pi_2$ & $\mathbb{Z}_2$ \\
$\mathbb{Z}_2\otimes\Pi_2$ &  none  & $c_{2\beta}=\sin 2\xi=0$  &  $\mathbb{Z}_2\otimes\Pi_2$ & $\Pi_2$ \\
U(1)$\otimes\Pi_2$  &   $m_{11}^2\neq m_{22}^2$  & $s_{2\beta}= 0$ & U(1) & U(1) \\
U(1)$\otimes\Pi_2$  &  $\Re(m_{12}^2 e^{i\xi})\neq  0$  & $c_{2\beta}= 0$ & $\Pi^{(\xi)}_{2}$ & $\Pi^{(\xi)}_2$ \\
U(1)$\otimes\Pi_2$  & none  & $s_{2\beta}=0$ &U(1)$\otimes\Pi_2$  & U(1) \\
U(1)$\otimes\Pi_2$  & none   & $c_{2\beta}=0$ &U(1)$\otimes\Pi_2$   & $\Pi_2$ \\
GCP3  &  $m_{11}^{\prime\,2}\neq m^{\prime\,2}_{22}$, $\Re m_{12}^{\prime\,2}\neq 0$  &  $s_{2\beta^\prime}c_{2\beta^\prime}\neq 0$, $\sin\xi^\prime=0$ & $\overline\Pi^{(\alpha)}_2$  & $\overline\Pi^{(\alpha)}_2$ \\
GCP3  &  $m_{11}^{\prime\,2}\neq m^{\prime\,2}_{22}$ &  $s_{2\beta^\prime}=0$ &  $\mathbb{Z}_2$ & $\mathbb{Z}_2$ \\
GCP3  &  $\Re m_{12}^{\prime\,2}\neq 0$  &  $c_{2\beta^\prime}=0$, $\sin\xi^\prime=0$  & $\Pi_2$ & $\Pi_2$ \\
GCP3 &  $\Im m_{12}^{\prime\,2}\neq 0$  & $c_{2\beta^\prime}=0$, $\cos\xi^\prime=0$ & U(1)$^\prime$ & U(1)$^\prime$ \\
GCP3  & none & $s_{2\beta^\prime}=0$ & U(1)$^\prime\otimes\mathbb{Z}_2$ & $\mathbb{Z}_2$ \\
GCP3  & none& $s_{2\beta^\prime}\neq 0$, $\sin\xi^\prime=0$ & U(1)$^\prime\otimes\mathbb{Z}_2$ &  $\overline\Pi^{(\alpha)}_2$\\
GCP3  &none &  $c_{2\beta^\prime}=0$, $\cos\xi^\prime=0$ & U(1)$^\prime\otimes\mathbb{Z}_2$ & U(1)$^\prime$ \\
SO(3)  &   $m_{11}^{\prime\,2}\neq m^{\prime\,2}_{22}$,  $\Re(m_{12}^{\prime\,2}e^{i\xi^\prime})\neq 0$ & $s_{2\beta^\prime}c_{2\beta^\prime}\neq 0$  & U(1)$_{\rm H}$ &  U(1)$_{\rm H}$ \\
SO(3)  &  $\Re(m_{12}^{\prime\,2}e^{i\xi^\prime})\neq 0$ &  $c_{2\beta^\prime}=0$ & U(1)$_{\rm H}$ &  U(1)$_{\rm H}$ \\
SO(3)  &   $m_{11}^{\prime\,2}\neq m^{\prime\,2}_{22}$ &  $s_{2\beta^\prime}=0$ &  U(1)  & U(1) \\
SO(3)  &  none & none & SO(3) & U(1)$_{\rm H}$  \\
\hline
\end{tabular}
\caption{\small  \baselineskip=15pt 
Classification of 
symmetries  of the 2HDM scalar potential that yield exact Higgs alignment, where the tree-level properties of one of the neutral scalar mass eigenstates coincide with those of the SM Higgs boson.  
Note that $m_{11}^2=m_{22}^2$ and $\Re(m_{12}^2 e^{i\xi})=\Im(m_{12}^2 e^{i\xi})=0$ (and similarly for the primed parameters) unless otherwise indicated.  
The unprimed parameters correspond to the $\mathbb{Z}_2\otimes\Pi_2$ or $\ug\otimes\Pi_2$ basis, whereas the primed parameters correspond to the GCP3 basis.   
All such basis choices are consistent with the ERPS4 with $\lambda_6=\lambda_7=0$ and real $\lambda_5$; the corresponding parameter constraints for the softly broken GCP2-symmetric scalar potential are given in \eqst{gcp2inert1}{gcp2inert4}.   
In cases where the residual symmetry is given by $\overline\Pi_2^{(\alpha)}$, the value of $\alpha=\left(\half\pi-2\beta^\prime\right)\cos\xi^\prime$, where $\cos\xi^\prime=\pm 1$.  
Although separate listings are provided for scalar potentials that exhibit the 
$\ug\otimes\Pi_2$ or GCP3 symmetry (either of which may be softly broken), they represent the same scalar potential expressed in two different choices of the scalar field basis.
\label{tab:align}}
\end{table}
\thispagestyle{empty}
\clearpage

 One can check that all the entries listed in Table~\ref{tab:align} (including the first two lines, which correspond respectively to the $\mathbb{Z}_2$-symmetric IDM and the U(1)-symmetric IDM, outside of the domain of the ERPS4) satisfy the four conditions specified in
\eqst{gcp2inert1}{gcp2inert4}.
For example, starting from the softly broken or unbroken GCP2-symmetric scalar potential transformed to the basis in which $\lambda_5$ is real and $\lambda_6=\lambda_7=0$, one 
easily obtains the following correlations of the parameters $\beta$ and $\xi$ for the symmetry cases listed in 
Table~\ref{tab:align}, 
\beqa
\mathbb{Z}_2\otimes\Pi_2: & \qquad  & s_{2\beta}s_{2\xi}=0\quad  \text{or} \quad  s_{2\beta}c_{2\beta}=0\,,\\
 \ug\otimes\Pi_2: & \qquad  & s_{2\beta}c_{2\beta}=0\,,\\
 \text{GCP3}: & \qquad  & s_{2\beta^\prime}s_{2\xi^\prime}=0\,,\\
 \text{SO(3)}: & \qquad & \text{no constraints.}
 \eeqa
 One can then employ \eqs{gcp2inert1}{gcp2inert2} to determine the allowed soft breaking due to $m_{11}^2\neq m_{22}^2$ and/or $m_{12}^2\neq 0$ that is consistent with exact Higgs alignment.

\section{Implications of custodial symmetry}
\label{custodial}

One of the possible symmetries of the scalar potential that does not appear in Table~\ref{tab:class} is custodial symmetry.   This symmetry is necessarily violated by the hypercharge U(1)$_{\rm Y}$ gauge interactions.\footnote{That is, the custodial symmetry is violated by the gauge covariant kinetic term of the scalar fields that is proportional to $g'$.} 
Nevertheless, in the limit of $g'\to 0$ this is a potential symmetry of the bosonic sector of the 2HDM.   Given that custodial symmetry is an approximate symmetry of nature in light of the small deviation of the electroweak $\rho$-parameter from its custodial symmetric value of $\rho=1$, it is of interest to consider the possibility that the 2HDM scalar potential respects the custodial symmetry.  In more detail, if the 2HDM scalar potential is symmetric under SU(2)$_{L}\otimes$SU(2)$_{R}$ transformations where SU(2)$_L$ is identified with the SU(2) part of the electroweak gauge group and SU(2)$_R$ is a global symmetry group, then after the symmetry breaking of SU(2)$_{L}$ the residual custodial symmetry can be identified with an unbroken
diagonal SU(2)$_{L+R}$ global symmetry.

Details of the SU(2)$_{L}\otimes$SU(2)$_{R}$ transformation laws 
are given in Ref.~\cite{Pomarol:1993mu,Grzadkowski:2010dj,Haber:2010bw,Aiko:2020atr}.  
As~shown in Ref.~\cite{Grzadkowski:2010dj}, a~$\Phi$-basis exists for any 2HDM custodial symmetric scalar potential 
such that,\footnote{If $\lambda_4=\lambda_5$ in the $\Phi$-basis where $\lambda_5\in\mathbb{R}$, then $\lambda_4=\lambda_5$ in any real basis, $\Phi^\prime=U\Phi$, such that $U$ is a real orthogonal matrix.   However, if $m_{12}^2=\lambda_6=\lambda_7=0$, then one can perform a basis transformation where $U$ is unitary but not real orthogonal that still preserves the reality of the basis.   For such basis transformations the relation $\lambda_4=\lambda_5$ is no longer preserved.  \Eq{custard2} provides a trivial example of this.  A more interesting example is provided by the basis transformation that converts \eq{oh2oh3} into \eq{oh2oh3prime}.  Of course, if $U$ is not real orthogonal, then the basis transformation will not preserve the reality of the vevs.} 
\beq \label{custard}
\lambda_4=\Re \lambda_5\quad \text{and \quad $m_{12}^2$, $\lambda_5$, $\lambda_6$, $\lambda_7\in\mathbb{R}$}.
\eeq
Hence, all 2HDM custodial symmetric scalar potentials are explicitly CP conserving.  In the case of an unbroken $\mathbb{Z}_2$-symmetric scalar potential where $m_{12}^2=\lambda_6=\lambda_7=0$, one is always free to rephase $\Phi_2\to i\Phi_2$,\footnote{Although this rephasing maintains the reality of the scalar potential parameters, it introduces
a relative phase, $\xi=\half\pi$, in the vevs.}
in which case the custodial symmetry condition of \eq{custard} specializes to
\beq \label{custard2}
\lambda_4=\pm\Re \lambda_5\quad \text{and \quad $m_{12}^2=\Im\lambda_5=\lambda_6=\lambda_7=0$}.
\eeq

Additional information is provided by minimizing the scalar potential and determining the Higgs basis.  Then, as shown in
Ref.~\cite{Haber:2010bw}, the scalar potential respects the custodial symmetry if the Higgs basis parameters satisfy,
\beq \label{custconds}
Z_4=Z_5 e^{-2i\eta}\in\mathbb{R}\,,\qquad Y_3 e^{-i\eta}=-\half Z_6 e^{-i\eta}v^2\in\mathbb{R}\,,\qquad  Z_7 e^{-i\eta}\in\mathbb{R}\,,
\eeq
where the phase $\eta$ represents the freedom to rephase the Higgs basis field $\mathcal{H}_2$.   It follows that one can choose $\eta$ such that the parameters of the scalar potential in the Higgs basis are all real, which implies that GCP1 is a symmetry of the scalar potential and vacuum. 
In particular, in a real Higgs basis, \eq{custconds} yields two possible solutions,
\beqa
Z_4&=&Z_5\,, \label{Aplus}\\[-5pt]
&\text{or} \nonumber  \\[-5pt]
Z_4&=&\pm Z_5\,,\quad \text{and}\quad Y_3=Z_6=Z_7=0\,.\label{Aminus}
\eeqa

\Eq{Aplus} is a consequence of choosing $\eta= 0$~(mod~$\pi$).   In the case of $Y_3=Z_6=Z_7=0$, the condition $Z_4=-Z_5$ is now possible by choosing $\eta=\half\pi$~(mod~$\pi$), as indicated in \eq{Aminus}.   Note that if the Yukawa interactions are neglected then
the sign of $Z_5$ in a real Higgs basis is not physical since one can always redefine $\mathcal{H}_2\to i\mathcal{H}_2$ while maintaining the reality of the Higgs basis.   
Thus, \eq{Aminus} can be understood to mean that $Z_4=\pm |Z_5|$ in a real Higgs basis, which corresponds to two physically inequivalent conditions. 

Moreover, employing the results of \eq{custconds} in \eqst{chargedHmass}{zeethreefour}, it follows that if $Z_6\neq 0$ then we can identify the squared mass of the CP-odd mass eigenstate as corresponding to the $33$ element of the squared-mass matrix $\mathcal{M}^2$ given in \eq{matrix33}, namely $m_A^2=m_{H^\pm}^2$.   If $Z_6=0$, then $\mathcal{M}^2$ is diagonal; nevertheless, one can determine the CP properties of the neutral Higgs mass eigenstates via the three-scalar and four-scalar interaction terms assuming that $Z_7\neq 0$~\cite{Haber:2006ue}.  One can again confirm that the CP-odd mass eigenstate corresponds to the $33$ element of $\mathcal{M}^2$, in which case we also conclude that $m_A^2=m_{H^\pm}^2$.  

Finally, in the case of a custodially symmetric scalar potential with
$Y_3=Z_6=Z_7=0$, an exact Higgs alignment is realized and
we can identify $m_{h}^2=Z_1 v^2$ following the convention of \eq{inertnames}, and
$m^2_{H,A}=m_{H^\pm}^2+\half(Z_4\pm Z_5)v^2$.  Although the CP-quantum numbers of $H$ and $A$ are of opposite sign, there are no bosonic interactions that can uniquely identify which of the two states $H$ and $A$ is CP-even and which is CP-odd, as previously noted in Section~\ref{zee2pi2}.  Ultimately, the CP-quantum numbers of $H$ and $A$ may be fixed by the Higgs-fermion Yukawa couplings (if these interactions are CP conserving), except in special cases where the ambiguity persists [cf.~\eq{Yuk}].   Assuming that the CP-quantum numbers of $H$ and $A$ are unambiguously determined by the Yukawa couplings, then the sign of $Z_5$ is promoted to a physical parameter in the case of $Y_3=Z_6=Z_7=0$.  It then follows that~\cite{Haber:2010bw},
\beq \label{equalmasses2}
m_{H^\pm}^2=\begin{cases} m_{A}^2 & \quad \text{if}~Z_4
=Z_5 \quad\text{and}\quad Z_6=Z_7=0\,,\\
m_{H}^2~ &
\quad \text{if}~Z_4=-Z_5
\quad\text{and}\quad Z_6=Z_7=0\,,\end{cases}
\eeq
in a real Higgs basis.  In particular,
\beqa
&& m_h<m_H\quad \text{if $Z_4=-Z_5$, $Z_6=Z_7=0$ and $\mhpm^2>Z_1 v^2$}\,,\nonumber \\
&& m_h>m_H\quad \text{if $Z_4=-Z_5$, $Z_6=Z_7=0$ and $\mhpm^2<Z_1 v^2$}\,.
\eeqa
 Indeed, the transformation $\mathcal{H}_2\to i\mathcal{H}_2$ changes the sign of $Z_5$ while also changing the scalar Yukawa coupling into a pseudoscalar Yukawa coupling and vice versa.
 
In this section, we propose to classify all 2HDM custodial-symmetric scalar potentials that exhibit exact Higgs alignment due to an unbroken or softly broken symmetry.  All such scalar potentials will satisfy the Higgs basis conditions given in \eq{Aminus}.  If the parameters of the corresponding Higgs potential lie in the ERPS4 regime, then this classification amounts to supplementing the results of Table~\ref{tab:align} with the conditions of custodial symmetry.   

As a first step, we review the classification of custodial symmetric 2HDM scalar potentials first obtained in Refs.~\cite{Battye:2011jj,Pilaftsis:2011ed} (and recently reproduced in Ref.~\cite{Darvishi:2019dbh}).  
If the scalar potential respects a custodial symmetry, then the Higgs Lagrangian can exhibit seven additional global symmetries in the limit of $g'=0$ beyond the symmetries listed in Tables~\ref{tab:symm1} and \ref{tab:symm2}.   
Three of the seven symmetries correspond to GCP1, the Peccei-Quinn U(1) and $\Pi_2$, which when combined with the custodial symmetry yield
maximal symmetry groups of SO(3), SO(4) and $\mathbb{Z}_2\otimes{\rm O}(3)$, respectively~\cite{Pilaftsis:2011ed}.    The case of GCP1 corresponds to the minimal implementation of custodial symmetry in the most general 2HDM scalar potential.  Indeed, the custodial-symmetric scalar potential of the 2HDM must be CP conserving as noted below \eq{custard2}.   
A custodial symmetric, $\Pi_2$-symmetric scalar potential is equivalent to a custodial symmetric, $\mathbb{Z}_2$ symmetric scalar potential in another scalar field basis.   
To validate this remark, we first combine \eq{custard} with the constraints of the $\Pi_2$ symmetry shown in Table~\ref{tab:class} to obtain,
\beqa
&& m_{11}^2=m_{22}^2\,,\qquad\quad \lambda_1=\lambda_2\,,\qquad\quad \lambda_4=\Re\lambda_5\,,\qquad\quad  \lambda_6=\lambda_7\,,\,\nonumber \\
&&\Im m_{12}^2=\Im\lambda_5=\Im\lambda_6=\Im\lambda_7=0\,,
\eeqa
in the $\Phi$-basis.  We now transform to a new basis by defining $\Phi_1^\prime=(\Phi_1+\Phi_2)/\sqrt{2}$ and $\Phi_2^\prime=(\Phi_2-\Phi_1)/\sqrt{2}$.  In this new basis, the corresponding scalar potential parameters are 
\beqa
&& \hspace{-0.6in}  m_{11}^{\prime\,2}=m_{11}^2-\Re m_{12}^2\,,\qquad\qquad  m_{22}^{\prime\,2}=m_{11}^2+\Re m_{12}^2\,,\qquad\qquad\quad\, m_{12}^{\prime\,2}=0\,,\nonumber \\
&& \hspace{-0.6in}\lambda_1^\prime=\half(\lambda_1+\lambda_3+2\lambda_4+4\lambda_6)\,,\qquad\qquad\qquad\!\! \lambda_2^\prime=\half(\lambda_1+\lambda_3+2\lambda_4-4\lambda_6)\,,\nonumber \\
&& \hspace{-0.6in} \lambda_3^\prime=\half(\lambda_1+\lambda_3-2\lambda_4)\,,\qquad\quad \lambda_4^\prime=\Re\lambda_5^\prime=\half(\lambda_1-\lambda_3)\,,\qquad\quad \Im\lambda_5^\prime=\lambda_6^\prime=\lambda_7^\prime=0\,,\,
\eeqa
which combines the constraints of  \eq{custard} with the constraints of the $\mathbb{Z}_2$ symmetry. 

The remaining four symmetry cases of Ref.~\cite{Pilaftsis:2011ed} correspond to $\mathbb{Z}_2\otimes\Pi_2$, $\ug\otimes\Pi_2$, GCP3 and SO(3), which when combined with custodial symmetry [cf.~\eq{custard2}] yields a maximal symmetry group of $\mathbb{Z}_2\otimes\mathbb{Z}_2\otimes{\rm SO}(3)$, $\rm{O}(2)\otimes{\rm O}(3)$, $\mathbb{Z}_2\otimes{\rm O}(4)$ and SO(5), respectively.\footnote{Since a GCP3-symmetric scalar potential is a $\ug\otimes\Pi_2$-symmetric scalar potential in a different scalar field basis, one cannot unambiguously associate  $\rm{O}(2)\otimes{\rm O}(3)$ and $\mathbb{Z}_2\otimes{\rm O}(4)$ with either ERPS symmetry.   A physical criterion for distinguishing these two maximal symmetry groups is provided by the two Higgs basis conditions specified in \eq{zonetwothree} and exhibited in Table~\ref{tab:cust} and in the discussion that follows.}  
In light of \eq{why30}, each of these symmetry cases corresponds to the inert limit in the ERPS, and thus satisfy \eq{Aminus} in a real Higgs basis.  The case of $\rm{O}(2)\otimes{\rm O}(3)$ requires some clarification.  In Table 1 of Ref.~\cite{Pilaftsis:2011ed},
the constraints on the scalar potential parameters corresponding to the $\rm{O}(2)\otimes{\rm O}(3)$ symmetry are,
\beq \label{oh2oh3}
m_{11}^2=m_{22}^2\,, \qquad\quad m_{12}^2=0\,,\qquad\quad \lambda_1=\lambda_2=\lambda_3\,,\qquad\quad \lambda_5=\lambda_6=\lambda_7=0\,.
\eeq
This is a $\ug\otimes\Pi_2$-symmetric scalar potential, but it does not satisfy \eq{custard}.  However, if we transform to the GCP3 basis then \eqst{lamp}{m12p} yield,
\beqa
&& m_{11}^{\prime\,2}=m_{22}^{\prime\,2}\,,\qquad\quad m_{12}^{\prime\,2}=0\,,\qquad\quad  \lambda^\prime_1=\lambda^\prime_2=\lambda^\prime_3+\lambda^\prime_4+\Re\lambda^\prime_5\,,\qquad\quad  \lambda^\prime _4=\Re \lambda^\prime_5\,,\nonumber \\
&& \Im\lambda_5^\prime=\lambda^\prime_6=\lambda^\prime_7=0\,, \label{oh2oh3prime}
\eeqa
which corresponds to custodial-symmetric, GCP3-symmetric scalar potential.  

We are now ready to present the classification of custodial-symmetric 2HDM scalar potentials that satisfy exact Higgs alignment due to an unbroken or softly broken symmetry.  Exact Higgs alignment requires $Y_3=Z_6=0$, and then to achieve Higgs alignment via a symmetry also requires $Z_7=0$.    Two immediate examples are the IDM with either an unbroken $\mathbb{Z}_2$ or U(1) symmetry in the Higgs basis.  Supplementing these two examples with the condition that $Z_4=\pm Z_5$ yields a custodial symmetric scalar potential with exact Higgs alignment.  

In light of the classification of custodial symmetric scalar potentials discussed above, we now consider cases in which additional unbroken or softly broken symmetries are present.
It is clear that at minimum, a softly broken $\mathbb{Z}_2$ symmetry that is manifestly realized in the $\Phi$-basis must be present.
Consequently, let us consider a softly broken $\mathbb{Z}_2$-symmetric scalar potential in the $\Phi$-basis that is distinct from the Higgs basis
which satisfies $Y_3=Z_6=Z_7=0$.
Since $\lambda_6=\lambda_7=0$  holds in a $\Phi$-basis such that $s_{2\beta}\neq 0$, the
ERPS4 conditions must be satisfied as we now demonstrate.   

First, we shall employ eqs.~(A.26)--(A.28) of Ref.~\cite{Boto:2020wyf} with $s_{2\beta}\neq 0$, $\lambda_6=\lambda_7=0$ 
and $Z_6=Z_7=0$ to obtain,
\beqa
s_{2\beta}\bigl[Z_1 c^2_{\beta}-Z_2s_{\beta}^2-Z_{34}c_{2\beta}-\Re(Z_5 e^{2i\xi})c_{2\beta}-i\Im(Z_5 e^{2i\xi})\bigr]&=&0\,,\\
s_{2\beta}\bigl[Z_1 s^2_{\beta}-Z_2c_{\beta}^2+Z_{34}c_{2\beta}+\Re(Z_5 e^{2i\xi})c_{2\beta}+i\Im(Z_5 e^{2i\xi})\bigr]&=&0\,,
\eeqa
where $Z_{34}\equiv Z_3+Z_4$.  Adding and subtracting these two equations yields,
\beqa
s_{2\beta}(Z_1-Z_2)&=&0\,,\label{z1m2} \\
s_{2\beta}\bigl\{c_{2\beta}\bigl[Z_1+Z_2-2Z_{34}-2\Re(Z_5 e^{2i\xi})\bigr]-2i\Im(Z_5 e^{2i\xi})\bigr\}&=&0\label{secondcondition} \,.
\eeqa
Moreover, we can use eqs.~(A20) of Ref.~\cite{Boto:2020wyf} along with \eq{YZ} to obtain,
\beq \label{phibasisparms}
m_{12}^2 e^{i\xi}=\half(Y_2-Y_1)s_{2\beta}\,.
\eeq
It follows that 
$\Im(m_{12}^2 e^{i\xi})=0$.
Imposing the scalar potential minimum condition given by \eq{min3}, it follows that either $\sin 2\xi=0$ or $\lambda_5=0$.  If $\lambda_5=0$ then the only remaining complex parameter of the scalar potential, $m_{12}^2$, can be arbitrarily rephased.  Thus without loss of generality, we may assume that $\sin 2\xi=0$ holds in all cases, or equivalently $e^{2i\xi}=\pm 1$.
Moreover, having assumed that $s_{2\beta}\neq 0$,
we see that \eqs{z1m2}{secondcondition} yields $Z_1=Z_2$ and $\Im Z_5=0$.  
That is, the Higgs basis parameters satisfy the ERPS4 conditions!\footnote{As expected, if $s_{2\beta}=0$ then \eqs{z1m2}{secondcondition} are automatically satisfied,  in which case no enhanced symmetry beyond $\mathbb{Z}_2$ is present for generic parameters of the scalar potential.} 

In the case of $s_{2\beta}c_{2\beta}\neq 0$ and $e^{2i\xi}=\pm 1$ (and $Z_6=Z_7=0$), \eqs{z1m2}{secondcondition} yield,
\beq \label{z12345}
Z_1=Z_2=Z_3+Z_4\pm Z_5 \quad \text{and}\quad \Im Z_5=0\,.
\eeq
Hence, the quartic terms of the scalar potential 
satisfy the GCP3 or GCP3$^\prime$ symmetry conditions in the Higgs basis.  If we now also impose the custodial symmetry condition, $Z_4=\pm Z_5$, 
where the sign choice is uncorrelated with the $\pm$ sign appearing in \eq{z12345}, 
then it follows that two relations are possible that are physically distinguishable,
\beq \label{zonetwothree}
Z_1=Z_2=Z_3+2Z_4\quad \text{or} \quad Z_1=Z_2=Z_3\,.
\eeq

Moreover, we can use eqs.~(A21)--(A28) of Ref.~\cite{Boto:2020wyf} along with \eq{z12345} to obtain the scalar potential parameters in the $\Phi$-basis,
\beq \label{phibasisparms2}
\lambda_i=Z_i\quad \text{for $i=1,2,\ldots,7$}.
\eeq
When $Y_3=Z_6=Z_7=0$, it is always possible to rephase the Higgs basis field,~$\mathcal{H}_2\to i\mathcal{H}_2$, such that $Z_1=Z_2=Z_3+Z_4+Z_5$.
Then, in a GCP3 basis (where parameters in the $\Phi$-basis are indicated with prime superscripts), \eqs{zonetwothree}{phibasisparms2} respectively yield,
\beqa
\lambda^\prime_1=\lambda^\prime_2=\lambda^\prime_3+\lambda^\prime_4+\lambda^\prime_5\,,& \qquad  &\text{where $\lambda^\prime_4=\lambda^\prime_5$}\,, \label{prime1} \\[-5pt]
&\text{or} \nonumber \\[-5pt]
\lambda^\prime_1=\lambda^\prime_2=\lambda^\prime_3\,,& \qquad  &\text{where $\lambda^\prime_4=-\lambda^\prime_5$}\,.\label{prime2}
\eeqa

In both cases, the GCP3 conditions are manifestly realized  by the quartic terms of the scalar potential in the $\Phi$-basis.  
In light of \eq{oh2oh3prime}, in the case of unbroken GCP3 and custodial symmetry, \eq{prime1} yields a maximal symmetry group of ${\rm O}(2)\otimes{\rm O}(3)$ in the classification of Ref.~\cite{Pilaftsis:2011ed}.   To determine the maximal symmetry group of the scalar potential whose parameters satisfy \eq{prime2}, we transform to the $\ug\otimes\Pi_2$ basis.   Then
\eqst{inv1}{inv5} yield $\lambda_1=\lambda_2\neq\lambda_3$ and $\lambda_4=\lambda_5=\lambda_6=\lambda_7=0$, corresponding to a maximal symmetry group of $\mathbb{Z}_2\otimes {\rm O}(4)$ in the classification of Ref.~\cite{Pilaftsis:2011ed}. 
Note that the custodial symmetry is preserved in the presence of 
soft breaking of the GCP3 symmetry by allowing for $m_{12}^{\prime\,2}\neq 0$ subject to the condition, $m_{22}^{\prime\,2}-m_{11}^{\prime\,2}=2m_{12}^{\prime\,2} c_{2\beta^\prime}/s_{2\beta^\prime}$ [cf.~\eqs{min4a}{min5a}].

If $Z_5=0$, then
\eq{phibasisparms2} together with the custodial symmetry condition $Z_4=\pm Z_5$ imply that $\lambda^\prime_4=\lambda^\prime_5=0$.  
In light of \eqs{prime1}{prime2},
the SO(3) conditions are manifestly realized by the quartic terms of the scalar potential.  Indeed, in this case the quartic terms are given by $\mathcal{V}\ni \half\lambda\bigl(\Phi_1^\dagger\Phi_1+\Phi_2^\dagger\Phi_2)^2$, which corresponds to the maximally symmetric SO(5) limit of the 2HDM (after including the gauge covariant kinetic terms of the scalar fields with $g'=0$) analyzed in Ref.~\cite{Dev:2014yca}.
Soft breaking of the SO(3) symmetry due to $m_{12}^{\prime\,2}\neq 0$ is again allowed subject to the condition, $m_{22}^{\prime\,2}-m_{11}^{\prime\,2}=2m_{12}^{\prime\,2} c_{2\beta^\prime}/s_{2\beta^\prime}$.

In the case of $c_{2\beta}=\sin 2\xi=0$, 
\eqs{z1m2}{secondcondition} yield 
$Z_1=Z_2$ and $\Im Z_5=0$.  Hence, the quartic terms of the scalar potential in the real Higgs basis satisfy the $\mathbb{Z}_2\otimes\Pi_2$
symmetry conditions.
Using eqs.~(A21)--(A25) of Ref.~\cite{Boto:2020wyf},  we obtain
  \beqa 
&&\hspace{-0.5in}  \lambda_1=\lambda_2=Z_1-\half(Z_1-Z_{345})\,,\qquad \lambda_{i}=Z_i+\half(Z_1-Z_{345})\,,\quad \text{for $i=3,4$}, \nonumber \\
&&\hspace{-0.5in}   \lambda_{5}=Z_5\pm \half(Z_1-Z_{345})\,, \qquad\qquad\,\,  \tilde{\lambda}_{345}=Z_1+\half(Z_1-Z_{345})\,,\qquad\quad
\lambda_6=\lambda_7=0\,,
\label{lam12345}
\eeqa
 where $Z_{345}\equiv Z_3+Z_4\pm Z_5$ and $\tilde{\lambda}_{345}\equiv\lambda_3+\lambda_4\pm\lambda_5$.    Assuming that $Z_1\neq Z_{345}$ (otherwise, \eq{z12345} is satisfied and we return to the previous case), it follows that $\lambda_1\neq\tilde{\lambda}_{345}$. 
That is, the $\mathbb{Z}_2\otimes\Pi_2$ symmetry of the 
 quartic terms of the scalar potential of the $\Phi$-basis is manifestly realized.  Soft breaking of the $\mathbb{Z}_2\otimes\Pi_2$ symmetry 
due to $m_{12}^2\neq 0$, where $m_{12}^2$ is real [purely imaginary] if $\sin\xi=0$ [$\cos\xi=0$], is allowed subject to the condition $m_{11}^2=m_{22}^2$.

\begin{table}[t!]
\begin{tabular}{|l|c|c|c|c|}
\hline
\pht Higgs basis conditions & custodial &additional&scalar  &maximal \\[-8pt]
\quad  (all cases satisfy \quad  &  
symmetry  &\pht real $\Phi$-basis \pht  & Lagrangian   &symmetry \\[-8pt]
\quad $Y_3=Z_6=Z_7=0$) \pht  & conditions & \pht  constraints \pht & symmetry  & group\tnote{*} \\
\hline
  & $Z_4=\pm Z_5\neq 0$ &  $s_{2\beta}=0$ & $\mathbb{Z}_2$  & $\mathbb{Z}_2\otimes{\rm O}(3)$ \\
 & $Z_4= Z_5=0$ &  $s_{2\beta}=0$ & U(1)  & SO(4)\\
\pht $Z_1=Z_2\neq Z_{345}$ &\pht  $Z_4=\pm Z_5\neq 0$ \pht  & $c_{2\beta}\sin 2\xi =0$, $\lambda=\lambda_3$ or $\lambda_4=\pm\lambda_5$ &  $\mathbb{Z}_2\otimes\Pi_2$ & \pht $\mathbb{Z}_2\otimes\mathbb{Z}_2\otimes {\rm SO}(3)$ \pht  \\
\pht $Z_1=Z_2\neq Z_{345}$ &\pht  $Z_4=\pm Z_5\neq 0$ \pht  & $s_{2\beta}=0$, $\lambda_4=\pm\lambda_5$ &  $\mathbb{Z}_2\otimes\Pi_2$ & \pht $\mathbb{Z}_2\otimes\mathbb{Z}_2\otimes {\rm SO}(3)$ \pht  \\
  \pht $Z_1=Z_2=Z_3+2Z_4$\pht& $Z_4=\pm Z_5 \neq 0$  \pht  & $c_{2\beta}=0$, $\lambda=\lambda_3$, $\lambda_4\neq 0$ &  \pht $\ug\otimes\Pi_2$  \pht & ${\rm O}(2)\otimes{\rm O}(3)$ \\
\pht $Z_1=Z_2=Z_3$ \pht & $Z_4=\pm Z_5 \neq 0 \pht  $  & $c_{2\beta}=0$, $\lambda\neq\lambda_3$, $\lambda_4= 0$ &  $\ug\otimes\Pi_2$ & $\mathbb{Z}_2\otimes {\rm O}(4)$  \\
\pht  $Z_1=Z_2\neq Z_3$& $Z_4=Z_5=0$ & $s_{2\beta}=0$, $\lambda\neq\lambda_3$, $\lambda_4=0$ & \pht  $\ug\otimes\Pi_2$ \pht & $\mathbb{Z}_2\otimes {\rm O}(4)$  \\
\pht $Z_1=Z_2=Z_3+2Z_4$ \pht & $Z_4=Z_5 \neq 0 \pht  $  & $s_{2\beta^\prime}\sin\xi^\prime=0$, $\lambda_4^\prime=\lambda_5^\prime\neq 0$ & GCP3 & ${\rm O}(2)\otimes{\rm O}(3)$  \\
\pht $Z_1=Z_2=Z_3$ \pht & $Z_4=-Z_5 \neq 0 \pht  $  & $s_{2\beta^\prime}\sin\xi^\prime=0$, $\lambda_4^\prime=-\lambda_5^\prime\neq 0$ & GCP3 & $\mathbb{Z}_2\otimes {\rm O}(4)$  \\
 \pht  $Z_1=Z_2\neq Z_3$ & $Z_4=Z_5=0$ &\pht  $c_{2\beta^\prime}=\cos\xi^\prime=0$, $\lambda^\prime_4=-\lambda^\prime_5\neq 0$ \pht & \pht GCP3& $\mathbb{Z}_2\otimes {\rm O}(4)$  \\
 \pht $Z_1=Z_2=Z_3$ & $Z_4=Z_5=0$  & $\lambda_4=\lambda_5=0$  &  ${\rm SO}(3)$  & ${\rm SO}(5)$ \\
\hline
\end{tabular}
\caption{\small \baselineskip=15pt
Classification of 2HDM scalar potentials that possess an unbroken custodial symmetry and satisfy the inert conditions, $Y_3=Z_6=Z_7=0$, thereby exhibiting exact Higgs alignment. 
The Higgs basis field $\mathcal{H}_2$ has been rephased such that $Z_5$ is real.  
In the symmetry limit, the scalar Lagrangian symmetry that is manifestly realized in the $\Phi$-basis is shown along with the corresponding maximal symmetry group (that includes the custodial symmetry in the limit of $g'\to 0$) according to the classification provided in Ref.~\cite{Pilaftsis:2011ed}.   
Excluding the first two lines of the table, all entries correspond to the ERPS4 regime.
The corresponding ERPS symmetry may be softly broken if $m_{11}^2\neq m_{22}^2$ and/or $m_{12}^2\neq 0$ as indicated in Table~\ref{tab:align}.  Since GCP3 is equivalent to $\ug\otimes\Pi_2$ when expressed in a different scalar field basis, there is a one-to-one mapping between their corresponding entries above that is consistent with the results of Section~\ref{transforming}.  \\[-18pt]
\label{tab:cust}}
\end{table}

If we now impose the custodial symmetry condition, $Z_4=\pm Z_5$, then it follows that two parameter relations are possible,
 \beqa
\lambda_1=\lambda_2\neq \tilde{\lambda}_{345}\,,& \qquad  &\text{where $\lambda_4=\pm\lambda_5$}\,, \label{cust1cp2}\\[-5pt]
&\text{or} \nonumber  \\[-5pt]
\lambda_1=\lambda_2=\lambda_3\,,& \qquad  &\text{where $\lambda_4\neq \pm \lambda_5$}\,.\label{cust2cp2}
\eeqa
\Eqs{cust1cp2}{cust2cp2} are related by a change of scalar field basis.  For example, if $\sin\xi=0$ then we replace the $\pm$ sign with a plus sign in the above expressions.
Starting from \eq{cust2cp2} and employing \eq{you} to transform $\Phi\to \overline{\Phi}=U\Phi$, 
the scalar potential parameters in the new basis satisfy $\overline{m}\llsup{\,2}_{11}=\overline{m}\llsup{\,2}_{22}$, $\overline{m}\llsup{\,2}_{12}\neq 0$ and
\beq
\overline\lambda_1=\overline\lambda_2=\lambda_3+\half(\lambda_4-\lambda_5)\,,\qquad\quad \overline{\lambda}_3=\lambda_3-\half(\lambda_4-\lambda_5)\,,\quad\qquad
\overline{\lambda}_4=\overline\lambda_5=\half(\lambda_4+\lambda_5)\,.
\eeq
Since $\overline\lambda_3+\overline\lambda_4+\overline\lambda_5-\overline\lambda_1=2\lambda_5$, 
\eq{cust1cp2} is satisfied in the $\overline\Phi$-basis, assuming that $\lambda_5\neq 0$.  However, if $\lambda_5=0$, then $\overline\lambda_1=\overline\lambda_2=\overline\lambda_3+\overline\lambda_4+\overline\lambda_5$ is satisfied, corresponding to a \mbox{softly broken} GCP3-symmetric scalar potential in the $\overline\Phi$-basis. 
The case of $\cos\xi=0$ can be similarly treated by rephasing $\Phi_2\to i\Phi_2$, and yields
a softly broken GCP3$^\prime$-symmetric scalar potential.

Finally, if $s_{2\beta}=0$, then one can impose the unbroken or softly broken symmetries of the ERPS4 and the custodial symmetry condition directly in the Higgs basis.
In Table~\ref{tab:cust}, we provide a complete classification of the 2HDM scalar potentials that possess an unbroken custodial symmetry and exhibit exact Higgs alignment.  
For convenience, we have included entries corresponding to both $\ug\otimes\Pi_2$ and GCP3, which correspond to physically equivalent points in the ERPS4 in light of 
the results of Section~\ref{transforming}. 

It is noteworthy that two maximal symmetry groups are associated with both $\ug\otimes\Pi_2$ and GCP3.  These two cases are distinguished by the corresponding Higgs basis conditions.
Indeed, one can check that ${\rm O}(2)\otimes{\rm O}(3)$ is physically distinguished from $\mathbb{Z}_2\otimes{\rm O}(4)$.  In particular, in the cases of softly broken or unbroken $\ug\otimes\Pi_2$ and GCP3-symmetric scalar potentials, ${\rm O}(2)\otimes{\rm O}(3)$ is associated with the mass relation, $m_{H^\pm}=m_H\neq m_A$.  In contrast, $\mathbb{Z}_2\otimes{\rm O}(4)$ is associated with the mass relation $m_{H^\pm}=m_A$, which includes the possibility of $m_{H^\pm}=m_A=m_H$ if in addition $Z_4=Z_5=0$.   The latter is an example of the more general result that
any custodial symmetric 2HDM scalar potential with $Z_4=Z_5=Z_6=Z_7=0$ exhibits a Peccei-Quinn U(1) symmetry that is unbroken by the scalar potential and vacuum and thus 
possesses a scalar spectrum where $H^\pm$ is degenerate in mass with both $H$ and $A$.

\section{Conclusions and Future Directions}
\label{conclude}

There is a fascinating region of parameter space of the 2HDM that can be implemented by imposing the generalized CP symmetry, 
GCP2, on the quartic terms of the scalar potential, which enforces the relations, $\lambda_1=\lambda_2$
and $\lambda_7=-\lambda_6$.
We call this region the ERPS4, generalizing the exceptional region of the parameter space (ERPS) of the 2HDM introduced in
Refs.~\cite{Davidson:2005cw,Ferreira:2009wh}, where the GCP2 symmetry is also respected by the quadratic
terms of the scalar potential and yields
the additional constraints, $m_{11}^2=m_{22}^2$ and $m_{12}^2=0$.
That is, the ERPS4 is the parameter space of a softly broken GCP2-symmetric scalar potential.
In this paper, we have provided a comprehensive study of the many interesting properties of this 2HDM parameter regime,
including limiting cases of the ERPS4 parameters that extend the unbroken or softly broken symmetries of the scalar potential
beyond GCP2.

We have enumerated the basis invariant conditions that characterize the softly broken GCP2 symmetries and their extensions 
and evaluated the scalar squared masses and neutral scalar mixing matrices
in each of these cases.
We have discussed intricacies that arise when scalar potentials originating from two different symmetry conditions yield physically equivalent results.
In such cases, although the parameter constraints imposed by the symmetry conditions may differ,
one can show that a unitary transformation relates the scalar field bases in which each of the symmetries is manifestly realized.
Indeed, a GCP2-symmetric scalar potential is related by a unitary basis transformation to a scalar potential that is
invariant with respect to $\mathbb{Z}_2\otimes\Pi_2$, and a GCP3-symmetric scalar potential is related by a unitary basis transformation to a scalar potential that is
invariant with respect to $\ug\otimes\Pi_2$.  These considerations persist even if the corresponding symmetries are softly broken.

The equivalence of the softly broken GCP2 and $\mathbb{Z}_2\otimes\Pi_2$ symmetries, as demonstrated below \eq{Zinvariant},
does not yield a simple analytic formula that
relates the parameters of the GCP2 basis and the $\mathbb{Z}_2\otimes\Pi_2$ basis.
In this work, 
our analysis of the ERPS4 always starts from the $\mathbb{Z}_2\otimes\Pi_2$ basis, from which all subsequent special cases can be analyzed.
In contrast, the translation between the GCP3 basis and the $\ug\otimes\Pi_2$ basis can be made explicit, and a translation between the parameters defined in each of the two basis choices has been provided in Section~\ref{transforming}.   The results for the softly broken SO(3)-symmetric scalar potential, which are different limiting cases of the GCP3 and $\ug\otimes\Pi_2$ basis choices, ultimately yield identical results given that the form of the softly broken SO(3)-symmetric scalar potential  is invariant with respect to an arbitrary unitary transformation of the scalar field basis.

In examining the CP-invariance properties of scalar potentials in the ERPS4, we encountered
an interesting feature that runs contrary to a statement
usually found in the literature.  In a softly broken $\mathbb{Z}_2\otimes\Pi_2$-symmetric scalar potential
where the magnitudes of the two neutral scalar field vevs are equal, 
we originally noticed
in Ref.~\cite{Boto:2020wyf} that the scalar potential and vacuum were both CP conserving
even though the relative phase between the potentially complex parameters $m_{12}^2$ and $\lambda_5$  could 
not be removed by separate rephasings of the scalar fields $\Phi_1$ and $\Phi_2$.
In contrast, outside of the ERPS4 regime, it is straightforward to show that if $\lambda_6=\lambda_7=0$ and $\Im\bigl(\lambda_5^\ast  [m_{12}^2]^2 \bigr)\neq 0$,
then the corresponding scalar potential is explicitly CP violating.  
We were able to identify an alternative definition of CP, denoted by GCP1$^\prime$ in Tables~\ref{tab:symm4} and \ref{tab:class2}, which
provides an explanation for why the softly broken $\mathbb{Z}_2\otimes\Pi_2$-symmetric scalar potential with $v_1=v_2$ always preserves a CP symmetry.

Perhaps even more astonishing was that in a softly broken GCP3-symmetric scalar potential with $\Im\bigl(\lambda_5^\ast  [m_{12}^2]^2 \bigr)\neq 0$,
the scalar potential and vacuum are always CP-invariant independently of the vevs.   In this case, the identification of the relevant CP transformation
law is more obscure (see Appendix~\ref{app:cp}).  Of course, this result becomes almost trivial by transforming to the $\ug\otimes\Pi_2$ basis, where
a simple rephasing can be performed to remove all potential complex phases from the corresponding scalar potential parameters.
Moreover, in both CP-conserving examples above where $\Im\bigl(\lambda_5^\ast  [m_{12}^2]^2 \bigr)\neq 0$, a more general unitary transformation
of the scalar fields exists that can transform directly to a real scalar field basis in which the CP invariance of the scalar potential is manifest.
Because the scalar potentials of the ERPS4 are quite constrained, such
a unitary transformation is still consistent with the
parameter constraints imposed by the softly broken $\mathbb{Z}_2\otimes\Pi_2$  and GCP3 symmetries.

A very important subset of the ERPS4 is the so-called inert limit where the Higgs basis parameters satisfy $Y_3=Z_6=Z_7=0$.
In this parameter regime Higgs alignment is exact, which means that there exists a neutral scalar whose tree-level properties
coincide with those of the SM Higgs boson.   Indeed, the LHC Higgs data have already confirmed at the 10\%--20\% level
that the properties of the observed Higgs boson (of mass 125 GeV) are consistent with the predictions of the Standard Model.  Consequently, any phenomenologically
viable extended Higgs sector must exhibit at least an approximate Higgs alignment.   One can achieve an approximate Higgs alignment
automatically in the decoupling limit where the masses of all additional scalars are significantly larger than 125 GeV.  However, it is of interest to
consider the possibility of approximate Higgs alignment without decoupling, as this scenario would provide more options for potential discoveries
of new scalars of an extended Higgs sector in future LHC runs.  Higgs alignment without decoupling can be achieved without a fine-tuning of
scalar sector parameters if a symmetry is present that can enforce the Higgs alignment.  Thus, in the 2HDM it is especially useful to provide a
complete classification of all such symmetries.  The simplest example of a 2HDM with this property is the IDM which possesses an unbroken
$\mathbb{Z}_2$ or U(1) symmetry in the Higgs basis.  All other 2HDM scenarios that provide a natural explanation for exact Higgs alignment lie in 
the ERPS4 regime.   The complete classification has been provided in Table~\ref{tab:align}.

A phenomenologically viable extended Higgs sector must also be consistent with precision electroweak constraints.  The observation that
the electroweak $\rho$-parameter is approximately equal to one strongly suggests that the 
scalar potential should be invariant under a custodial symmetry.  The ERPS4 enters in these considerations as well, since one of the two
ways to satisfy the requirements of custodial symmetry is provided by the inert limit.   Thus, combining the requirements of exact
Higgs alignment and custodial symmetry yields a classification of 2HDM scenarios that is exhibited in Table~\ref{tab:cust}.

Finally, there is one aspect of the 2HDM that has been almost completely ignored in our comprehensive study of the ERPS4---namely, the
Higgs-fermion Yukawa interactions.  There is a reason for this neglect.   In a 2HDM with one generation of quarks and leptons, 
it is not possible to construct a Yukawa Lagrangian that respects a GCP2 symmetry or any of its symmetry extensions.   If three generations
of quarks and leptons are present, then it is possible to construct a set of Yukawa interactions that respect a GCP2 or GCP3 symmetry by
positing transformation laws that involve fermions of different generations. However, all such constructions are inconsistent
with the constraints of experimental observations with the possible exception of one very special implementation of the GCP3 symmetry in Ref.~\cite{Ferreira:2010bm}.  It is not clear whether these remarks also
hold if one were to construct a Yukawa Lagrangian that respects a $\mathbb{Z}_2\otimes\Pi_2$ or $\ug\otimes\Pi_2$ symmetry.
Scalar Lagrangians that are constrained by symmetries that are physically the same when only the scalar sector is
considered might be different once fermions are included.  This possibility is presently under study~\cite{future}.

If there is no phenomenologically successful 2HDM Yukawa Lagrangian consistent with the ERPS4 regime, then there are two
possible approaches.  
In one approach advocated in Refs.~\cite{Dev:2014yca,Darvishi:2019ltl}, the ERPS4 conditions are imposed at the Planck scale.
The Yukawa interactions represent a hard breaking of the symmetries responsible for the ERPS4 regime.  Hence, renormalization
group evolution down to the electroweak scale will generate an effective 2HDM that deviates from the ERPS4 but might
retain some of its best features (e.g., approximate Higgs alignment and approximate custodial symmetry).   The second approach
follows the proposals of Ref.~\cite{Draper:2016cag,Draper:2020tyq}, where vectorlike quark and lepton partners are introduced in an extended
Yukawa Lagrangian.   In this case, one can construct a Yukawa Lagrangian that is consistent with the ERPS4 regime (even
in a one generation model of fermions and the vectorlike partners).   To ensure that the vectorlike fermions are sufficiently heavy to avoid
current LHC search limits, one can introduce explicit mass terms 
for the vectorlike fermions, which then generate the soft-breaking squared mass terms of the ERPS4.

In either of these two approaches, one can 
determine parameter regimes that are consistent with 
observed Higgs boson phenomena, while setting useful
targets for precision Higgs studies at the LHC and future Higgs factories now under development.
It is also of theoretical interest to seek out ultraviolet complete models that include the Yukawa sector to ultimately explain the origin of the fundamental symmetries that underlie
the approximate symmetries governing the 2HDM at the electroweak scale.\footnote{For example, attempts to construct ultraviolet complete models that yield
Higgs alignment without decoupling in the 2HDM can be found in Refs.~\cite{Antoniadis:2006uj,Ellis:2016gxa,Benakli:2018vqz,Benakli:2018vjk,Lane:2018ycs,Eichten:2021qbm}.}
We shall defer such matters to future studies.  Given that the ERPS4 provides a simple framework
for the scalar potential of an extended Higgs sector with a reduced number of free parameters, we would anticipate that useful correlations
could emerge, such as relations among various three-scalar couplings, if deviations from SM Higgs properties are detected
and/or new scalar states are discovered.

\bigskip
\noindent
{\bf Acknowledgments}
\bigskip

H.E.H. is supported in part by the U.S. Department of Energy Grant
No.~\uppercase{DE-SC}0010107.  H.E.H. is grateful for the hospitality and support during his visit to the Instituto Superior T\'{e}cnico, Universidade de Lisboa,
and he
also acknowledges fruitful discussions that
took place at the University of
Warsaw during visits supported by the HARMONIA project of the National Science Centre,
Poland, under contract UMO-2015/18/M/ST2/00518 (2016--2021). 
The work of J.P.S. is supported in part by the Portuguese
Funda\c{c}\~{a}o para a Ci\^{e}ncia~e
Tecnologia (FCT) under Contracts
No.~CERN/FIS-PAR/0008/2019,
No.~PTDC/FIS-PAR/29436/2017,
No.~UIDB/00777/2020,
and No.~UIDP/00777/2020.

\begin{appendices}

\section{\texorpdfstring{The possibility of spontaneous CP violation}{The possibility of spontaneous CP violation}}
\label{app:scpv}
\renewcommand{\theequation}{A.\arabic{equation}}
\setcounter{equation}{0}

Consider the case of an explicitly CP-conserving, softly broken $\mathbb{Z}_2$-symmetric scalar potential written in a real scalar field basis, where $\lambda_6=\lambda_7=0$ and
the two potentially complex 
scalar potential parameters,
$m_{12}^2$ and $\lambda_5$, are real and nonzero.  In this case, spontaneous CP violation is possible \cite{Branco:1980sz,Branco:1985aq,Branco:1999fs}.\footnote{Spontaneous CP violation is also possible if $m_{12}^2$ is purely imaginary and $\lambda_5$ is real.  In this case, one can redefine $\Phi_2\to i\Phi_2$, which renders $m_{12}^2$ real while 
transforming $\lambda_5\to -\lambda_5$ and $\xi\to \xi+\half\pi$.} 
 It is instructive to examine the minimum and stability conditions under the assumption that $\langle\Phi_1^0\rangle=v_1/\sqrt{2}$ and $\langle\Phi_2^0\rangle=v_2 e^{i\xi}/\sqrt{2}$, where $v_1$ and $v_2$ are real and positive.  Following the analysis of Appendix B of Ref.~\cite{Gunion:2002zf}, 
 the vacuum value of the scalar potential is 
\beq
V_{\rm vac}=\half m_{11}^2 v_1^2+\half m_{22}^2 v_2^2-m_{12}^2 v_1 v_2\cos\xi+\tfrac18 (\lambda_1 v_1^4+\lambda_2 v_2^4)+\tfrac14(\lambda_3+\lambda_4-\lambda_5)v_1^2 v_2^2
+\half\lambda_5 v_1^2 v_2^2\cos^2\xi\,.
\eeq
The scalar potential minimum conditions are
\beqa
0&=&\frac{\partial  V_{\rm vac}}{\partial v_1}= m_{11}^2 v_1-m_{12}^2 v_2\cos\xi+\half \lambda_1 v_1^3+\half\lambda_{345} v_1 v_2^2\,, \label{mincond1}\\
0&=&\frac{\partial V_{\rm vac}}{\partial v_2}= m_{22}^2 v_2-m_{12}^2 v_1\cos\xi+\half \lambda_2 v_2^3+\half\lambda_{345} v_1^2 v_2\,,\label{mincond2}\\
0&=&\frac{1}{v}\frac{\partial V_{\rm vac}}{\partial\xi}= \frac{v_1 v_2}{v}\left(m_{12}^2-\lambda_5 v_1 v_2\cos\xi\right)\sin\xi\,,\label{mincond3}
\eeqa
where $\lambda_{345}$ is defined below \eq{minfour}.  The vacuum is CP conserving 
if $\sin 2\xi=0$~\cite{Branco:1980sz,Lin:1993pa},\footnote{If $\cos\xi=0$ then \eq{mincond3} yields $m_{12}^2=0$.  In this case the $\mathbb{Z}_2$ symmetry of the scalar potential is explicitly preserved and spontaneous CP violation does not occur~\cite{Branco:1980sz} (see also Theorem 23.3 of Ref.~\cite{Branco:1999fs}). \label{fnscpv}} 
whereas the vacuum is potentially CP violating if $\sin 2\xi\neq 0$.

First consider the case of $\sin 2\xi=0$.  Having excluded $m_{12}^2=0$ from consideration (cf.~footnote~\ref{fnscpv}), it follows that $\sin\xi=0$.   Without loss of generality, we may take $\cos\xi=1$ by rephasing $\Phi_2\to -\Phi_2$ (which also changes the sign of $m_{12}^2$ but otherwise has no effect on the other scalar potential parameters).   Then, \eqs{mincond1}{mincond2} yield, 
\beqa
m_{11}^2&=&m_{12}^2\frac{v_2}{v_1}-\half\lambda_1 v_1^2-\half(\lambda_3+\lambda_4+\lambda_5)v_2^2\,, \label{em11}\\
m_{22}^2&=&m_{12}^2\frac{v_1}{v_2}-\half\lambda_2 v_2^2-\half(\lambda_3+\lambda_4+\lambda_5)v_1^2\,.\label{em22}
\eeqa

The stability conditions can be discerned from the Hessian.  Computing the relevant second derivatives,
\beqa
\frac{\partial^2 V _{\rm vac}}{\partial v^2_1}&=& m_{11}^2+\tfrac32\lambda_1 v_1^2+\half(\lambda_3+\lambda_4+\lambda_5)v_2^2=m_{12}^2\frac{v_2}{v_1}+\lambda_1 v_1^2\,, \\
\frac{\partial^2 V _{\rm vac}}{\partial v^2_2}&=& m_{22}^2+\tfrac32\lambda_2 v_2^2+\half(\lambda_3+\lambda_4+\lambda_5)v_1^2=m_{12}^2\frac{v_1}{v_2}+\lambda_2 v_2^2\,,\\
\frac{\partial^2 V _{\rm vac}}{\partial v_1 \partial v_2}&=& -m_{12}^2+(\lambda_3+\lambda_4+\lambda_5)v_1 v_2\,,
\eeqa
after applying the results of \eqs{em11}{em22}.
Thus, the Hessian matrix is given by
\beq
H=\begin{pmatrix}  m_{12}^2\frac{v_2}{v_1}+\lambda_1 v_1^2 & \quad -m_{12}^2+(\lambda_3+\lambda_4+\lambda_5)v_1 v_2\\
-m_{12}^2+(\lambda_3+\lambda_4+\lambda_5)v_1 v_2 & \quad m_{12}^2\frac{v_1}{v_2}+\lambda_2 v_2^2\end{pmatrix}\,.
\eeq
Stability requires that $\Tr H>0$ and $\det H>0$.   In addition, we demand that the squared masses of the neutral Higgs bosons should be positive.   
Using the results of Ref.~\cite{Gunion:2002zf}, the following quantities all must be positive,
\beqa
m_A^2&=&\left( \frac{m_{12}^2 }{v_1 v_2}-\lambda_5\right) v^2\,, \\
m_h^2+m_H^2&=& m_A^2+\lambda_1 v_1^2+\lambda_2 v_2^2\,,\\
m_h^2 m_H^2 &=& \frac{m_A^2}{v^2}\bigl[\lambda_1 v_1^4+\lambda_2 v_2^4+2(\lambda_3+\lambda_4)v_1^2 v_2^2\bigr]+\bigl[\lambda_1\lambda_2-(\lambda_3+\lambda_4)^2\bigr]v_1^2 v_2^2\,.
\eeqa
Note that the trace and determinant of the Hessian matrix are related to the squared masses of the neutral scalars,
\beqa
\Tr H&=&m_h^2+m_H^2+\lambda_5 v^2\,,\\
\det H&=&m_h^2 m_H^2+\lambda_5\left[\left(\lambda_5+\frac{2 m_A^2}{v^2}\right)v_1^2 v_2^2+\lambda_1 v_1^4+\lambda_2 v_2^4\right]\,.
\eeqa

Next, consider the case of $\sin 2\xi\neq 0$.   In this case, it is convenient to replace \eq{mincond3} with
\beq
0=\frac{1}{v}\frac{\partial V_{\rm vac}}{\partial\cos\xi}=\frac{v_1 v_2}{v}\left(-m_{12}^2+\lambda_5 v_1 v_2\cos\xi\right)\,,\label{mincond3alt}
\eeq
which yields $m_{12}^2=\lambda_5 v_1 v_2\cos\xi$.
Inserting this result into \eqst{mincond1}{mincond2}, it follows that
\beqa
m_{11}^2&=&-\half\lambda_1 v_1^2-\half(\lambda_3+\lambda_4-\lambda_5)v_2^2\,, \label{emv11}\\
m_{22}^2&=&-\half\lambda_2 v_2^2-\half(\lambda_3+\lambda_4-\lambda_5)v_1^2\,.\label{emv22}
\eeqa

The elements of the $3\times 3$ Hessian matrix are given by the following second derivatives,
\beqa
 \frac{\partial^2 V _{\rm vac}}{\partial v^2_1}&=& m_{11}^2+\tfrac32\lambda_1 v_1^2+\half(\lambda_3+\lambda_4+\lambda_5\cos 2\xi)v_2^2=\lambda_1 v_1^2+\lambda_5v_2^2\cos^2\xi\,, \\
\frac{\partial^2 V _{\rm vac}}{\partial v^2_2}&=& m_{22}^2+\tfrac32\lambda_2 v_2^2+\half(\lambda_3+\lambda_4+\lambda_5\cos 2\xi)v_1^2=\lambda_2 v_2^2+\lambda_5v_1^2\cos^2\xi\,,\\
\frac{\partial^2 V _{\rm vac}}{\partial v_1 \partial v_2}&=& -m_{12}^2\cos\xi+(\lambda_3+\lambda_4+\lambda_5\cos 2\xi)v_1 v_2=(\lambda_3+\lambda_4-\lambda_5\sin^2\xi)v_1 v_2\,,
\phantom{xxx}\\
\frac{1}{v^2}\frac{\partial^2 V _{\rm vac}}{\partial (\cos\xi)^2}&=&\frac{\lambda_5 v_1^2 v_2^2}{v^2}\,,\\
\frac{1}{v}\frac{\partial^2 V _{\rm vac}}{\partial v_1\partial\cos\xi}&=&\frac{\lambda_5 v_1 v_2^2\cos\xi}{v}\,,\\
\frac{1}{v}\frac{\partial^2 V _{\rm vac}}{\partial v_2\partial\cos\xi}&=&\frac{\lambda_5 v_1^2 v_2\cos\xi}{v}\,.
\eeqa
Thus, the Hessian matrix is given by,
\beq \label{H}
H=v^2 \begin{pmatrix} \lambda_1 c_\beta^2+\lambda_5  s_\beta^2\cos^2\xi & \quad (\lambda_3+\lambda_4-\lambda_5\sin^2\xi)  s_\beta c_\beta & \quad  \lambda_5 v^2 s_\beta^2 c_\beta\cos\xi \\   (\lambda_3+\lambda_4-\lambda_5\sin^2\xi) s_\beta c_\beta  & \quad \lambda_2 s_\beta^2+\lambda_5 c_\beta^2\cos^2\xi & \quad \lambda_5  s_\beta c^2_\beta\cos\xi  \\  \lambda_5 s_\beta^2 c_\beta\cos\xi & \quad  \lambda_5  s_\beta c^2_\beta\cos\xi & \quad \lambda_5  s_\beta^2 c_\beta^2\end{pmatrix},
\eeq
where $s_\beta\equiv v_2/v$ and $c_\beta\equiv v_1/v$.
Stability requires that $H$ is positive definite.   By Sylvester's criterion~\cite{Gilbert}, it follows that the principal minors must all be positive.   A necessary (although not sufficient) condition is that all diagonal elements of $H$ must be positive.  In light of \eq{bb}, we conclude that $\lambda_5>0$ is a necessary condition for spontaneous CP violation.  

\section{\texorpdfstring{An invariant characterization of the ERPS4 and consequences for CP symmetry}{An invariant characterization of the ERPS4 and consequences for CP symmetry}}
\label{app:erps}
\renewcommand{\theequation}{B.\arabic{equation}}
\setcounter{equation}{0}

In \eq{Zinvariant}, we provided an 
invariant characterization of the ERPS4 that is defined by $\lambda_1=\lambda_2$ and $\lambda_7=-\lambda_6$,
which if realized in one scalar field basis is then satisfied in all scalar field bases.  Using \eq{PB} and employing the identity, $\sigma^B_{ab}\sigma^B_{cd}=2\delta_{ad}\delta_{bc}-\delta_{ab}\delta_{cd}$,
it follows that the ERPS4 invariant can be rewritten in terms of the quartic coefficients of the scalar potential,
\beq \label{zeeinv}
\mathcal{Z}=\tfrac18\Tr\bigl([Z^{(1)}+Z^{(2)}]^2\bigr)-\tfrac{1}{16}\bigl(\Tr Z^{(1)}+\Tr Z^{(2)}\bigr)^2\,,
\eeq
where, following Ref.~\cite{Davidson:2005cw}, we have defined
\beqa
Z_{ad}^{(1)} & \equiv & \delta_{bc}Z_{ab,cd}=Z_{ab,bd}=\begin{pmatrix} \lambda_1+\lambda_4 & \quad \lambda_6+\lambda_7 \\   \lambda^*_6+\lambda^*_7 & \quad  \lambda_2+\lambda_4 \end{pmatrix}\,, 
\eeqa
\beqa
Z_{bd}^{(2)} & \equiv & \delta_{ac}Z_{ab,cd}=Z_{ab,ad}=\begin{pmatrix} \lambda_1+\lambda_3 & \quad \lambda_6+\lambda_7 \\   \lambda^*_6+\lambda^*_7 & \quad  \lambda_2+\lambda_3 \end{pmatrix}\,,
\eeqa
and the $Z_{ab,cd}$ are  defined in terms of the quartic couplings of the scalar potential in \eq{ZZ}.   One can simplify \eq{zeeinv} using the symmetry properties of the
$Z_{ab,cd}$ to obtain a slightly more compact form than was originally obtained in Ref.~\cite{Davidson:2005cw},
\beq \label{cp2invariant}
\mathcal{Z}\equiv \half \Tr\bigl\{(Z^{(1)})^2\bigr\}-\tfrac14\bigl[\Tr Z^{(1)}\bigr]^2=\tfrac14(\lambda_1-\lambda_2)^2+|\lambda_6+\lambda_7|^2\,.
\eeq
As noted below \eq{Zinvariant},
the ERPS4 corresponds to the invariant condition, $\mathcal{Z}=0$, which implies that $\lambda_1=\lambda_2$ and $\lambda_7=-\lambda_6$.  This condition is invariant with respect to changes in the scalar field basis, $\Phi_a\to U_{a\bbar}\Phi_b$ for any $U\in{\rm U}(2)$.  Thus, if $\lambda_1=\lambda_2$ and $\lambda_7=-\lambda_6$ in one basis, then the same relation holds in \textit{any} scalar field basis.  In particular, it holds in the Higgs basis, which implies that $Z_1=Z_2$ and $Z_7=-Z_6$.

One can make an even stronger statement that 
for any scalar potential of the ERPS4, there exists a scalar field basis where $\lambda_6=\lambda_7=0$ and $\Im\lambda_5=0$ (the latter after an appropriate rephasing of $\Phi_2$), 
which defines the $\mathbb{Z}_2\otimes\Pi_2$ basis of Section~\ref{zee2pi2}.
A simple proof of this statement was provided below \eq{Zinvariant}.   
Moreover, if CP is conserved then the scalar field basis in which $\lambda_6=\lambda_7=0$ can be explicitly identified,
as shown in Appendix~B of Ref.~\cite{Boto:2020wyf} and summarized below.

The first step is to go to the Higgs basis of the ERPS4 where $Z_1=Z_2$ and $Z_7=-Z_6$.  If $Z_6=0$ then it trivially follows that $\lambda_6=\lambda_7=0$ in the Higgs basis.   
If $Z_6\neq 0$, then consider a U(2) transformation from the Higgs basis to the $\Phi$-basis with neutral vevs $v_1/\sqrt{2}$ and $v_2 e^{i\xi}/\sqrt{2}$ where $v_1$ and $v_2$ are positive, $\tan\beta\equiv v_2/v_1$, and $0\leq\xi<2\pi$.  It is straightforward to derive a $\Phi$-basis expression for $\lambda_6=-\lambda_7$ in terms of Higgs basis parameters,
\beq \label{lam6zero}
\lam_6 e^{i\xi}=
\half\stwob c_{2\beta}\bigl[Z_1-Z_{34}-\Re(Z_5 e^{2i\xi})\bigr]-\half i\stwob\Im(Z_5 e^{2i\xi})+c_{4\beta}\Re(Z_6 e^{i\xi})+ic_{2\beta}\Im(Z_6 e^{i\xi})\,,
\eeq
where $Z_{34}\equiv Z_3+Z_4$.
We now search for values of  $\beta$ and $\xi$ such that $\lambda_6=0$.  Moreover, since $\lambda_7=-\lambda_6$ in the ERPS4, it follows that if $(\beta,\xi)$ yield $\lambda_6=0$ then so does $(\half\pi-\beta,\xi+\pi)$, corresponding to the interchange of the scalar doublet fields, $\Phi_1\leftrightarrow\Phi_2$. 

Since $Z_6\neq 0$, we may write $Z_6=|Z_6|e^{i\theta_6}$.  It is then convenient to define 
\beq \label{xibardef}
\bar{\xi}\equiv \xi+\theta_6\,,
\eeq
where $\bar{\xi}$ is defined modulo $\pi$.   In the case where CP is preserved by the scalar potential and vacuum,
$\Im(Z_5^* Z_6^2)=0$, as noted above \eq{nocpv}.\footnote{If CP is violated, then $\Im(Z_5^* Z_6^2)\neq 0$ and the existence of the scalar field basis where $\lambda_6=\lambda_7=0$ can be identified numerically, although no simple analytic expressions exist for $\beta$ and $\xi$~\cite{Gunion:2005ja,Boto:2020wyf}.}  
Inserting $e^{i\xi}=e^{i\xi^{\prime}}Z_6^*/|Z_6|$ and $\Im(Z_5^* Z_6^2)=0$ into \eq{lam6zero}, we search for values of $\beta$ and $\xi$ 
such that,
\beqa
s_{2\beta}\Re(Z_5^* Z_6^2)\sin 2\bar{\xi}&=&2c_{2\beta}|Z_6|^3\sin\bar{\xi}\,, \label{except3} \\
\hspace{-0.4in}
\stwob\ctwob\bigl[|Z_6|^2(Z_1-Z_{34})-\Re(Z_5^* Z_6^2)\cos 2\bar{\xi}\bigr]&=&-2c_{4\beta}|Z_6|^3\cos\bar{\xi}\,.\label{except4}
\eeqa
We can immediately obtain one solution to \eq{except3}, $\sin\bar{\xi}=0$ or equivalently $\cos\bar{\xi}=\pm 1$.  The twofold ambiguity was anticipated in the remarks below \eq{lam6zero}.   
Inserting $\cos\bar{\xi}=\pm 1$ into \eq{except4} yields a quadratic equation for $\cot 2\beta$, 
\beq \label{tantwobeta}
2|Z_6|\cot^2 2\beta\pm\left(Z_1-Z_{34}-\frac{\Re(Z_5^* Z_6^2)}{|Z_6|^2}\right)\cot 2\beta-2|Z_6|=0\,,
\eeq
which possesses two real roots whose product is equal to $-1$.  As a result, one ends up with four choices of $(\beta,\xi)$, where $0\leq\beta\leq\half\pi$ and $\cos\bar{\xi}=\pm 1$, for which
\eqs{except3}{except4} are satisfied. 

Moreover, additional solutions can be found if $\sin\bar{\xi}\neq 0$, in which case one can divide \eq{except3} by $\sin\bar{\xi}$.   Solving 
\eq{except3} for $c_{2\beta}/s_{2\beta}$ and inserting this result into \eq{except4} yields
\beq \label{cosbarxi}
\cos\bar{\xi}\biggl\{[\Re(Z_5^* Z_6^2)]^2+\Re(Z_5^* Z_6^2)|Z_6|^2(Z_1-Z_{34})-2|Z_6|^6\biggr\}=0\,.
\eeq 
One immediate solution to this equation is $\cos\bar{\xi}=0$, which we can then plug back into \eq{except3} to obtain $\cos 2\beta=0$. Thus, we learn that
$(\beta=\tfrac14\pi\,,\,\bar{\xi}=\half\pi)$ and $(\beta=\tfrac14\pi\,,\,\bar{\xi}=\tfrac32\pi)$ are also solutions to \eqs{except3}{except4}.

The above results are consistent with the result of Section~\ref{zee2pi2}.  
\Eq{zeesevencp2} provides a relation between $Z_6$ and the scalar potential parameters of
the $\mathbb{Z}_2\otimes\Pi_2$ basis, which is of the form $Z_6=(x+iy)e^{-i\xi}$.   Thus, $x+iy=|Z_6|e^{i\bar{\xi}}$ and we can identify,
\beq
\tan\bar{\xi}=\frac{y}{x}=-\frac{\lambda_5\sin 2\xi}{\bigl[\lambda(1-R)+2\lambda_5\sin^2\xi\bigr]c_{2\beta}}\,.
\eeq
Taking into account the values of $(\beta,\bar{\xi})$ obtained above that provide solutions to  \eqs{except3}{except4}, we see that $\sin\bar{\xi}=0$ corresponds to $\sin 2\xi=0$
and $\cos\bar{\xi}=0$ corresponds to $c_{2\beta}=0$.  Since this analysis was based on the assumption that $\Im(Z_5^* Z_6^2)=0$, we have reproduced the result of \eq{resCPconds}.

It is now instructive to compute $\Im(\lambda_5^*[m_{12}^2]^2)$ in the $\mathbb{Z}_2\otimes\Pi_2$ basis.  Using eq.~(B11) of  Ref.~\cite{Boto:2020wyf}, written in a slightly different form under the assumption that $c_{4\beta}\neq 0$, we obtain\footnote{If $c_{4\beta}=0$, then the expression given in eq.~(B11) of Ref.~\cite{Boto:2020wyf} is more useful.} 

\beqa  \label{imlam5m12}
\Im\bigl(\lambda_5^*[m_{12}^2]^2\bigr)&=&-\frac{|Z_6|v^4s_{2\beta}\sin\bar{\xi}}{8c_{4\beta}}\Biggl\{c_{4\beta}\left(\frac{2Y_2}{v^2}\right)^2-\frac{2Y_2}{v^2}\bigl[s^2_{2\beta}Z_1+(1-3c_{2\beta}^2)Z_{34}\bigr]-c_{2\beta}^2\mathcal{R}_5^2 \nonumber \\
&& \quad
+\left(Z_1+\frac{2Y_2}{v^2}\right)\mathcal{R}_5
 +Z_{34}(Z_{34}c_{2\beta}^2-Z_1 s^2_{2\beta})-\frac{4c_{4\beta}}{s^2_{2\beta}}|Z_6|^2\sin^2\bar{\xi}\Biggr\}\,,
\eeqa
where
\beq
\mathcal{R}_5\equiv \Re(Z_5 e^{2i\xi})=\frac{\Re(Z_5^* Z_6^2)\cos 2\bar{\xi}+\Im(Z_5^* Z_6^2)\sin 2\bar{\xi}}{|Z_6|^2}\,.
\eeq

Under the assumption that $\Im(Z_5^* Z_6^2)=0$, the results obtained above imply that either $\sin\bar{\xi}=0$ or $\cos\bar{\xi}=0$.  If $\sin\bar{\xi}=0$ then it immediately follows that $\Im(\lambda_5^*[m_{12}^2]^2)=0$.  That is, one can rephase the scalar doublet fields in the GCP2 basis to obtain a scalar potential whose coefficients are all real in a scalar field basis where the vevs are also real.  
In contrast, if $\cos\bar{\xi}=0$, which implies that $c_{2\beta}=0$ as noted below \eq{cosbarxi}, then \eq{imlam5m12} yields,
\beq \label{nonzero}
\Im\bigl(\lambda_5^*[m^2_{12}]^2\bigr) =\pm \frac{v^4|Z_6|}{8}\Biggl\{4|Z_6|^2-\left(Z_1+\frac{2Y_2}{v^2}\right)^2 
+\left(Z_1+\frac{2Y_2}{v^2}\right)\left(Z_1-Z_{34}-\frac{\Re(Z_5^* Z_6^2)}{|Z_6|^2}\right)\Biggr\}\,, 
\eeq
which is generically nonzero.   Thus, it follows that if $\beta=\tfrac14\pi$ and $\Im m_{12}^2\neq 0$ in the GCP2 basis, then one cannot remove all complex phases from the scalar potential with a simple rephasing of the Higgs doublet fields.  Nevertheless, the scalar potential and vacuum are CP conserving since $\Im(Z_5^* Z_6^2)=0$ implies that a real Higgs basis exists (i.e.~all Higgs~basis scalar potential parameters are real after an appropriate rephasing of the Higgs basis field~$\mathcal{H}_2$).

In the analysis presented above, we used \eq{cosbarxi} to conclude that $\cos\bar{\xi}=0$.   However, there is an alternative solution to \eq{cosbarxi} where the coefficient of $\cos\bar{\xi}$ vanishes.  Indeed, this alternative solution corresponds to the case of the softly broken GCP3-symmetric 2HDM where \eq{invcond12} is satisfied.  In this case, division of \eq{except3} by $\sin\bar{\xi}$  is permitted when $\sin\bar{\xi}\neq 0$, which yields
\beq
\cos\bar{\xi}=\frac{|Z_6|^3\cot 2\beta}{\Re(Z_5^* Z_6^2)}\,.
\eeq
Plugging this result into \eq{imlam5m12} yields $\Im\bigl(\lambda_5^{\prime\,*}[m_{12}^{\prime\,2}]^2\bigr)\neq 0$ for generic values of the scalar potential parameters and $\beta$.
In particular, in the GCP3 basis (where the scalar potential parameters are designated with prime superscripts), there exists a residual CP invariance despite the fact that $\Im\bigl(\lambda_5^{\prime\,*}[m_{12}^{\prime\,2}]^2\bigr)\neq 0$ when $\sin\bar{\xi}^\prime\neq 0$, independently of the value of $\beta^\prime$.  In contrast, $\lambda_5=0$ and $Z_6=|Z_6|e^{i\theta_6}=\pm |Z_6|e^{-i\xi}$ [cf.~\eq{zeeseven}] in the U(1)$\otimes\Pi_2$ basis, in which case \eq{xibardef} yields $\sin\bar{\xi}=0$.

\section{\texorpdfstring{CP invariance of the softly broken GCP3-symmetric scalar potential}{CP invariance of the softly broken GCP3-symmetric scalar potential}}
\label{app:cp}
\renewcommand{\theequation}{C.\arabic{equation}}
\setcounter{equation}{0}

The softly broken GCP3 scalar potential contains a complex parameter, $m_{12}^{\prime\,2}$, in a basis in which the only other potentially complex parameter, $\lambda^\prime_5$, is taken to be real.   Moreover, there is a relative phase between the two vevs.   Thus, naively one would conjecture that the scalar sector of the softly broken GCP3-symmetric 2HDM is CP violating.  However, we have demonstrated that by changing the scalar field basis, this scalar potential can be transformed into a softly broken U(1)$\otimes\Pi_2$ scalar potential in which $\lambda_5=0$.   Then, one can rephase either $\Phi_1$ or $\Phi_2$ to remove the phase of $m_{12}^2$, which yields an explicitly CP-conserving scalar potential.  Moreover, one can show that the scalar potential minimum condition in the explicitly CP-conserving basis yields two vevs with no relative complex phase.  Hence, it follows that the softly broken GCP3-symmetric 2HDM is explicitly CP conserving, and the vacuum also preserves the CP symmetry.
These observations imply that in the original GCP3 basis, one should be able to identify a residual generalized CP transformation under which the GCP3 scalar potential and vacuum are invariant.   The purpose of this Appendix is to provide the explicit construction of this generalized CP transformation.

We begin by rewriting \eq{lambdapotential} following the notation of Ref.~\cite{Davidson:2005cw},
\beq \label{VH2}
\mathcal{V}(\Phi)=Y_{a\bbar} (\Phi_{\abar}^\dagger \Phi_b) + \half Z_{a\bbar c\dbar} (\Phi_{\abar}^\dagger \Phi_b) (\Phi_{\bar c}^\dagger \Phi_d),
\eeq
where the indices $a$, $\bbar$, $c$ and $\dbar$ can take one of two values 1, 2 (with an implicit sum over barred and unbarred index pairs of the same letter), and
hermiticity and symmetry under the interchange of barred and unbarred indices imply that 
\beq \label{herm}
Y_{a\bbar} = (Y_{b\abar})^\ast\,,\qquad\quad Z_{a\bbar c\dbar} \equiv Z_{c\dbar a\bbar} = (Z_{b\abar d\cbar})^\ast\,. 
\eeq
Note that as a matrix,
\beq \label{whymatrix}
Y=\begin{pmatrix} Y_{11} & \,\,\, Y_{12} \\ Y_{12}^* & \,\,\, Y_{22}\end{pmatrix}=\begin{pmatrix} \phm m_{11}^2& \,\,\, -m_{12}^2\phm  \\ -(m_{12}^2)^* & \,\,\, \phm m_{22}^2\phm \end{pmatrix},
\eeq
where the minus sign in the definition of $m_{12}^2$ is conventional.
It is convenient to assemble the elements of the tensor $Z_{abcd}$ into a $4\times 4$ hermitian matrix (denoted by $Z$) as follows.  First, we introduce a slightly different notation for the components of $Z$,
\beq \label{ZZ}
Z_{ac,bd}\equiv Z_{a\bbar c\dbar}\,,
\eeq
where the first pair of indices of $Z_{ac,bd}$ consists of unbarred indices and the second pair consists of barred indices.  In this notation, it is conventional to omit the bars in the second pair of indices.\footnote{The reader is cautioned that in contrast to \eq{ZZ}, the symbol $Z_{ab,cd}$ employed in Ref.~\cite{Ferreira:2009wh} is equivalent to $Z_{a\bbar c\dbar}$ without an interchange of the indices $b$ and $c$.}
With this notation, each element of a row of the matrix $Z$ is denoted by a pair of subscripts.  These index pairs arranged in the order $11$, $12$, $21$ and $22$, and similarly for the pair of subscripts denoting each element of a column of $Z$.   The matrix $Z$ is then given by, 
\beq \label{Zmatrix}
Z=\begin{pmatrix} Z_{11,11} & \,\,\, Z_{11,12} & \,\,\, Z_{11,21} \,\,\, Z_{11,22} \\
  Z_{12,11} & \,\,\, Z_{12,12} & \,\,\, Z_{12,21} \,\,\, Z_{12,22} \\  Z_{21,11} & \,\,\, Z_{21,12} & \,\,\, Z_{21,21} \,\,\, Z_{21,22} \\  Z_{22,11} & \,\,\, Z_{22,12} & \,\,\, Z_{22,21} \,\,\, Z_{22,22} \end{pmatrix} =\begin{pmatrix} \lambda_1 & \,\,\, \lambda_6 & \,\,\, \lambda_6 & \,\,\, \lambda_5\\
  \lambda_6^* & \,\,\, \lambda_3 & \,\,\,\lambda_4  & \,\,\, \lambda_7 \\  \lambda_6^* & \,\,\, \lambda_4 & \,\,\, \lambda_3 &  \,\,\, \lambda_7 \\  \lambda_5^* & \,\,\,\lambda_7^* & \,\,\, \lambda_7^* &  \,\,\, \lambda_2\end{pmatrix}.
  \eeq

Under a basis transformation,
\beq
\Phi_a \rightarrow \Phi_a^\prime = U_{a\bbar} \Phi_b\,,\qquad\quad  \Phi^\dagger_{\abar} \rightarrow \Phi^{\prime\dagger}_{\abar}=\Phi^\dagger_{\bbar}U^\dagger_{b\abar}\,,
\label{basis-transf}
\eeq
where $U\!\in\!$ U(2) is a $2 \times 2$ unitary matrix ($U^\dagger_{b\abar}U_{a\cbar}=\delta_{b\cbar}$).
Under this unitary basis transformation,
the vevs are transformed as $\vev{\Phi_a^0} \rightarrow \vev{\Phi_a^{\prime\,0}}= U_{a \bbar} \vev{\Phi_b^0}$.
Moreover, the gauge covariant kinetic terms of the scalar fields are invariant under a unitary basis transformation,
whereas the coefficients $Y_{a\bbar}$ and $Z_{ab,cd}$ transform covariantly with respect to U(2) transformations as
\beqa
Y_{a\bbar} & \rightarrow &
Y^\prime_{a\bbar} =
U_{a \bbar}\, Y_{c\dbar}\, U^\dagger_{d\bbar}=(UYU^\dagger)_{a\bbar} ,
\label{Y-transf}
\\
Z_{ab,cd} & \rightarrow &
Z^\prime_{ab,cd} =
U_{a\ebar}\, U_{b\gbar}\,
Z_{eg,fh}\, U_{f\cbar}^\dagger \, U_{h\dbar}^\dagger = \bigl[(U\otimes U) Z(U^\dagger\otimes U^\dagger)\bigr]_{ab,cd}\,,
\label{Z-transf}
\eeqa
where the Kronecker product of two $2\times 2$ matrices is a $4\times 4$ matrix 
given in block matrix form by,
\beq \label{AotimesB}
A\otimes B=\begin{pmatrix} A_{11}B & \,\,\, A_{12}B \\ A_{21}B & \,\,\, A_{22}B\end{pmatrix}.
\eeq
The Kronecker product of two matrices satisfies the following properties~\cite{Kronecker}:
\beqa
&&(A\otimes B)(C\otimes D)=AC\otimes BD\,,  \label{kprop1} \\  
&&(A\otimes B)^\dagger=A^\dagger\otimes B^\dagger\,,  \label{kprop2} \\
&&(A\otimes B)^\top=A^\top\otimes B^\top\,,  \label{kprop3}\\
&&(A\otimes B)^{-1}=A^{-1}\otimes B^{-1}\,,\quad \text{if $A^{-1}$ and $B^{-1}$ exist}. \label{kprop4}
\eeqa
In particular, if $A$ and $B$ are unitary then so is $A\otimes B$.  The Kronecker product $A\otimes B$ can be represented by a rank four tensor whose components are given by,
\beq \label{Kproduct}
(A\otimes B)_{ab,cd}\equiv A_{ac}B_{bd}\,,
\eeq
where a row of the matrix $A\otimes B$ is denoted by a pair of subscripts that are arranged in the order $11$, $12$, $21$ and $22$ (and similarly for the pair of subscripts denoting the columns of $A\otimes B$). This convention yields the $4\times 4$ matrix representation of $A\otimes B$ given in \eq{AotimesB}.  Hence, it follows that,
\beq
U_{a\ebar}\, U_{b\gbar}\,Z_{eg,fh}\, U_{f\cbar}^\dagger \, U_{h\dbar}^\dagger =(U\otimes U)_{ab,eg} Z_{eg,fh}(U^\dagger\otimes U^\dagger)_{fh,cd}=
\bigl[(U\otimes U) Z(U^\dagger\otimes U^\dagger)\bigr]_{ab,cd}\,,
\eeq
as indicated in \eq{Z-transf}.  That is, \eqs{Y-transf}{Z-transf} are equivalent to the matrix equations,
\beq \label{YZ-transf}
Y^\prime= UYU^\dagger\,,\qquad\quad Z^\prime=(U\otimes U) Z(U^\dagger\otimes U^\dagger)\,.
\eeq

It is common to consider the standard CP transformation
of the scalar fields as
\begin{equation}
\Phi_a (t; \boldsymbol{\vec{x}}) \rightarrow
\Phi^{\textrm{CP}}_a (t;\boldsymbol{\vec{x}}) = \Phi_a^\ast (t; - \boldsymbol{\vec{x}}),
\label{StandardCP}
\end{equation}
where we shall no longer distinguish between barred and unbarred indices, and
the reference to the time ($t$) and space ($\boldsymbol{\vec{x}}$)
coordinates will henceforth be suppressed.
However,
in the presence of several scalars with the same quantum numbers,
U(2) basis transformations can be included in the definition of the
CP transformation.
This yields the generalized CP transformation (GCP)~\cite{Ecker:1987qp,Branco:1999fs},\footnote{For early work on generalized CP transformations, see Refs.~\cite{Ecker:1981wv,Ecker:1983hz,Neufeld:1987wa}.  
Generalized CP transformations in the context of the 2HDM have been treated in Refs.~\cite{Botella:1994cs,Branco:2005em,Ginzburg:2004vp,Maniatis:2007vn,Maniatis:2007de,Ferreira:2009wh,Maniatis:2009vp,Maniatis:2009by,Ferreira:2010yh,Ferreira:2010bm,Battye:2011jj,Pilaftsis:2011ed}.}
\beqa
\Phi^{\textrm{GCP}}_a
&=& e^{i\gamma}X_{ab} \Phi_b^\ast
\equiv e^{i\gamma} X_{ab} (\Phi_b^\dagger)^\top,\label{GCP} \\
\Phi^{\dagger \textrm{GCP}}_a
&=& e^{-i\gamma}X_{ab}^\ast \Phi_b^\top
\equiv e^{-i\gamma} X_{a b}^\ast (\Phi_b^\dagger)^\ast,\label{GCPdag}
\eeqa
where $X$ is an arbitrary unitary
matrix of unit determinant and $\gamma\in\mathbb{R}$.  We will indicate below \eq{vevcond} how the complex phase factor $e^{i\gamma}$ is determined.

Note that the transformation
$\Phi_a\to\Phi^{\rm GCP}_a$, where $\Phi^{\rm GCP}_a$ is given
by \eq{GCP},
leaves invariant the gauge covariant kinetic terms of the scalar fields.
The GCP transformation of a scalar field bilinear yields
\begin{equation}
\Phi^{\dagger \textrm{GCP}}_a
\Phi^{\textrm{GCP}}_b
=
X_{a c}^\ast X_{bd}
(\Phi_c \Phi_d^\dagger)^\top,
\end{equation}
which does not depend on the complex phase factor $e^{i\gamma}$.
Under this GCP transformation,
the quadratic terms of the potential may be written as
\beqa
Y_{ab} \Phi^{\dagger \textrm{GCP}}_a
\Phi^{\textrm{GCP}}_b
&=&
Y_{ab} X_{a c}^\ast X_{bd}
\Phi_d^\dagger \Phi_c
=X_{b d} Y_{ba}^\ast X_{a c}^\ast
\Phi_d^\dagger \Phi_c
=
X_{c a} Y_{cd}^\ast X_{d b}^\ast
\Phi_a^\dagger \Phi_b \nonumber \\
&=&
( X^\dagger YX )^\ast_{ab}
\Phi_a^\dagger \Phi_b,
\eeqa
after making use of the hermiticity of $Y$ [cf.~\eq{herm}] and appropriately relabeling the indices.
A similar argument can be made for the quartic terms, by employing the properties of the Kronecker product given in \eqst{kprop1}{kprop4}.
We conclude that the scalar potential is invariant
under the GCP transformation exhibited in
\eq{GCP} if and only if the scalar potential coefficients obey
\beqa
Y_{ab}^\ast
&=&
X_{c a}^\ast Y_{cd} X_{d b}
= ( X^\dagger YX )_{ab}\,,
\label{Y-CPtransf}\\
Z_{ab,cd}^\ast &=&X_{e a}^\ast X_{g b}^\ast
Z_{eg,fh} X_{f c} X_{h d}\,,
\label{Z-CPtransf}
\eeqa
or equivalently,
\beq \label{YZ-CPtransf}
Y^*=X^\dagger YX\,,\qquad\quad  Z^*=(X^\dagger\otimes X^\dagger)Z(X\otimes X)\,.
\eeq

Finally, we must check to see whether the GCP symmetry is preserved by the vacuum, in which case the following condition must be satisfied,
\beq \label{vevcond}
\vev{\Phi^0_a}=e^{i\gamma}X_{ab} \vev{\Phi_b^{\dagger\,0}}\,.
\eeq
The complex phase factor $e^{i\gamma}$ will be chosen subject to the convention where $\vev{\Phi_1^0}$ is real and non-negative.  The latter can always be arranged by performing an appropriate hypercharge U(1)$_{\rm Y}$ transformation on the scalar doublet fields, which has no effect on the coefficients of the scalar potential.

So far, we have assumed that all statements apply in the $\Phi$-basis.   If we now perform a basis transformation to the $\Phi^\prime$-basis
as indicated by \eq{basis-transf}, then we can express the scalar potential in terms of the $\Phi^\prime$-basis scalar potential parameters,
\beq
\mathcal{V}(\Phi^\prime) =
Y_{ab}^\prime (\Phi_a^{\prime \dagger} \Phi^\prime_b) +
\tfrac{1}{2}
Z^\prime_{ac,bd} (\Phi_a^{\prime \dagger} \Phi^\prime_b)
(\Phi_c^{\prime \dagger} \Phi^\prime_d),
\eeq
where $Y_{ab}^\prime$ and $Z^\prime_{ac,bd} $
are given by
\eqs{Y-transf}{Z-transf}, respectively.

Suppose that $V(\Phi)$ is invariant under 
the GCP transformation of Eq.~(\ref{GCP}) with the matrix $X$.
\Eq{Y-CPtransf} guarantees that $Y^\ast = X^\dagger Y X$.
Now,
\eq{Y-transf} relates the coefficients in the two
bases through $Y = U^\dagger Y^\prime U$.
It then follows that
$U^\top Y^{\prime \ast} U^\ast
= X^\dagger (U^\dagger Y^\prime U) X$,
which implies that
\begin{equation} \label{XprimeGCP}
Y^{\prime \ast}
= (U^\ast X^\dagger U^\dagger) Y^\prime (U X U^\top)
= X^{\prime \dagger} Y^\prime X^\prime\,,
\end{equation}
where $X^\prime = U X U^\top$.  
A similar argument can be made for the quartic terms, by employing the properties of the Kronecker product given in \eqst{kprop1}{kprop4}.
Thus, we conclude that 
$V (\Phi^\prime)$ is invariant under a new GCP transformation
with matrix
\beq 
e^{i\gamma^\prime}X^\prime = e^{i\gamma^\prime} U X U^\top.
\label{X-prime}
\eeq
The phase $\gamma^\prime$ is not fixed by this computation and must instead be determined by examining \eq{vevcond} in the $\Phi^\prime$-basis in a convention where $\vev{\Phi_1^{\prime\,0}}$ is real and non-negative.

To construct the residual GCP transformation that is a symmetry of the softly broken GCP3-symmetric scalar potential, we begin our analysis in the U(1)$\otimes\Pi_2$ basis.   The parameters of the quadratic part of the scalar potential
are specified in \eq{whymatrix}, 
where $Y_{12}\equiv  |Y_{12}|e^{i\theta_{12}}$ is potentially complex.\footnote{Here, we are assuming that $Y_{12}\neq 0$.  In the case of $Y_{12}=0$, one can choose $\gamma=\xi$ and replace $\theta_{12}$ with $-\xi$ in the definition of $X$ given in \eq{ex} to ensure that \eqss{firstcond}{secondcond}{thirdcond} are all satisfied, thereby establishing invariance of the scalar potential and the vacuum under the residual generalized CP transformation. \label{fnwhy}}
In addition, the parameters of the quartic part of the scalar potential satisfy [cf.~\eq{Zmatrix}],
\beq \label{zees}
\lambda=Z_{11,11}=Z_{22,22}\,,\qquad\quad \lambda_3=Z_{12,12}=Z_{21,21}\,,\qquad\quad \lambda_4=Z_{12,21}=Z_{21,12}\,,
\eeq
and all the other $Z_{ab,cd}$ vanish.  The softly broken U(1)$\otimes\Pi_2$-symmetric scalar potential is invariant with respect to a GCP transformation with matrix,

\beq \label{ex}
e^{i\gamma} X=\begin{pmatrix} 1 & \quad 0 \\ 0 & \quad e^{-2i\theta_{12}}\end{pmatrix}.
\eeq
The phase $\gamma=-\theta_{12}$ has been chosen in anticipation of \eq{thirdcond} below.

To establish the presence of the residual GCP symmetry, we first verify that \eq{Y-CPtransf} is satisfied,
\beq \label{firstcond}
\begin{pmatrix} Y_{11} & \quad |Y_{12}|e^{-i\theta_{12}} \\  |Y_{12}|e^{i\theta_{12}} & \quad Y_{22}\end{pmatrix}=
\begin{pmatrix} 1 & \quad 0 \\ 0 & \quad e^{2i\theta_{12}}\end{pmatrix} 
\begin{pmatrix} Y_{11} & \quad |Y_{12}|e^{i\theta_{12}} \\  |Y_{12}|e^{-i\theta_{12}} & \quad Y_{22}\end{pmatrix}
 \begin{pmatrix} 1 & \quad 0 \\ 0 &\quad e^{-2i\theta_{12}}\end{pmatrix} .
 \eeq
 Next, we verify that \eq{Z-CPtransf} is satisfied,
 \beq \label{secondcond}
\begin{pmatrix} \lambda & \,\,\,0 & \,\,\, 0 & \,\,\, 0\\
 0 & \,\,\, \lambda_3 & \,\,\,\lambda_4 & \,\,\, 0 \\ 0 & \,\,\, \lambda_4 & \,\,\, \lambda_3 &  \,\,\,0 \\ 0 & \,\,\, 0 & \,\,\, 0 &  \,\,\, \lambda\end{pmatrix}=
 \begin{pmatrix} 1 & 0 &0 & \,\,\, 0 \\ 0 & e^{2i\theta_{12}} &0 &0 \\ 0 & 0 & e^{2i\theta_{12}} & 0  \\  0 &0 &0 &e^{4i\theta_{12} }\end{pmatrix} \begin{pmatrix} \lambda & \,\,\,0 & \,\,\, 0 & \,\,\, 0\\
 0 & \,\,\, \lambda_3 & \,\,\,\lambda_4 & \,\,\, 0 \\ 0 & \,\,\, \lambda_4 & \,\,\, \lambda_3 &  \,\,\,0 \\ 0 & \,\,\,0 & \,\,\, 0 &  \,\,\, \lambda\end{pmatrix}
 \begin{pmatrix} 1 &0 & 0 & 0 \\ 0 & e^{-2i\theta_{12}} & 0 &  0 \\ 0 &  0 & \,\,\, e^{-2i\theta_{12}} & 0  \\  0 & 0 & 0 &  e^{-4i\theta_{12} }\end{pmatrix}.
 \eeq  

Finally, the GCP symmetry is preserved by the vacuum if \eq{vevcond} is satisfied.   The vevs are given by $v_a=(v_1,v_2 e^{i\xi})$, where $v_1$ and $v_2$ are positive
and $\xi$ is determined by the scalar potential minimum condition [cf.~\eq{min3a}], $\Im(Y_{12}e^{i\xi})=0$, which yields $\sin(\theta_{12}+\xi)=0$, under the assumption
that $Y_{12}\neq 0$ [cf.~footnote~\ref{fnwhy}].
It then follows that $\theta_{12}+\xi=0~\text{mod}~\pi$,
which demonstrates that 
\eq{vevcond} is indeed satisfied.  That is,
\beq \label{thirdcond}
\begin{pmatrix} v_1 \\ v_2 e^{i\xi}\end{pmatrix}=\begin{pmatrix} 1 & \quad 0 \\ 0 & \quad e^{-2i\theta_{12}}\end{pmatrix}\begin{pmatrix} v_1 \\ v_2 e^{-i\xi}\end{pmatrix}\,.
\eeq

In the GCP3 basis, $\lambda^\prime_5=\lambda^\prime_1-\lambda^\prime_3-\lambda^\prime_4$ is real and nonzero.  Now, it is not immediately obvious that the softly broken GCP3-symmetric 2HDM preserves a CP symmetry, since $\Im(\lambda_5^*[m_{12}^2]^2)\neq 0$, which implies that one cannot rephase the scalar doublet fields to remove the phase of $m_{12}^2$.  Nevertheless, we know that CP is a symmetry of the softly broken GCP3-symmetric scalar potential and vacuum since it corresponds to a
softly broken symmetric U(1)$\otimes\Pi_2$ scalar potential expressed in a different basis.  Thus, it should be possible to explicitly construct the residual generalized CP transformation that preserves the
softly broken GCP3-symmetric 
scalar potential and vacuum by employing \eq{X-prime}, with $U$~given
by \eq{you}.  Indeed, we will construct $X^\prime$ below and explicitly verify that  \eqs{YZ-CPtransf}{vevcond} are satisfied when expressed in terms of the GCP3 basis parameters.

\Eqs{you}{X-prime} yield,\footnote{In obtaining \eq{eX}, we have absorbed the phase $\phi$ [cf.~\eq{you}] into the definition of $\gamma^\prime$.} 
\beq \label{eX}
e^{i\gamma^\prime}X^\prime=-ie^{i\gamma^\prime}\begin{pmatrix} -s_{12} & \,\,\, c_{12} \\ \phm c_{12} & \,\,\, s_{12}\end{pmatrix},
\eeq
where $s_{12}\equiv\sin\theta_{12}$ and $c_{12}\equiv\cos\theta_{12}$.   Using \eq{unprimed}, which we can rewrite as
\beq
Y_{12}=\Re Y^\prime_{12}+\half i(Y^\prime_{22}-Y_{11}^\prime)\,,
\eeq
it follows that
\beq \label{csonetwo}
c_{12}=\frac{\Re Y^\prime_{12}}{\sqrt{\bigl(\Re Y^\prime_{12}\bigr)^2+\tfrac14(Y_{22}^\prime-Y_{11}^\prime)^2}}\,,
\qquad\quad
s_{12}=\frac{Y^\prime_{22}-Y^\prime_{11}}{2\sqrt{\bigl(\Re Y^\prime_{12}\bigr)^2+\tfrac14(Y_{22}^\prime-Y_{11}^\prime)^2}}\,.
\eeq
One can check that \eq{Y-CPtransf} is satisfied in the GCP3 basis by rewriting this equation as,
\beq \label{check}
\begin{pmatrix} Y^\prime_{11} & \quad Y_{12}^{\prime\ast} \\  Y_{12}^\prime & \quad Y^\prime_{22}\end{pmatrix}-
\begin{pmatrix} -s_{12} & \,\,\, c_{12} \\ \phm\ c_{12} & \,\,\, s_{12}\end{pmatrix}
\begin{pmatrix} Y^\prime_{11} & \quad Y_{12}^\prime\\  Y_{12}^{\prime\ast} & \quad Y^\prime_{22}\end{pmatrix}
\begin{pmatrix} -s_{12} & \,\,\, c_{12} \\ \phm c_{12} & \,\,\, s_{12}\end{pmatrix}=0\,.
\eeq
To verify \eq{check}, we multiply out the left-hand side above to obtain,
\beq \label{check2}
\bigl[2s_{12}\Re Y^\prime_{12}-c_{12}(Y^\prime_{22}-Y^\prime_{11})\bigr]\begin{pmatrix} c_{12} & \,\,\,\phm s_{12} \\ s_{12} & \,\,\, -c_{12}\end{pmatrix}=0,
\eeq
which is equal to the zero matrix after making use of \eq{csonetwo}.

Next, we check the validity of \eq{Z-CPtransf} in the GCP3 basis.  The explicit form of $X^\prime\otimes X^\prime$ 
is given by
\beq
X^\prime\otimes X^\prime= \begin{pmatrix} s_{12}^2 & \,\,\, -s_{12}c_{12} & \,\,\, -s_{12}c_{12} & \,\,\, c_{12}^2 \\
-s_{12}c_{12} & \,\,\, -s_{12}^2  & \,\,\, \phm c_{12}^2 & \,\,\, s_{12} c_{12} \\
-s_{12}c_{12} & \,\,\, \phm c_{12}^2  & \,\,\, -s_{12}^2 & \,\,\, s_{12} c_{12} \\
 c_{12}^2 & \,\,\, \phm s_{12}c_{12} & \,\,\, \phm s_{12}c_{12} & \,\,\, s_{12}^2 \end{pmatrix}.
 \eeq
 In the GCP3 basis,
 \beq
 Z^\prime=\begin{pmatrix} \lambda^\prime & \,\,\, 0& \qquad\,\,\, 0& \,\,\, \lambda^\prime-\lambda_3^\prime-\lambda_4^\prime\\
0 & \,\,\, \lambda^\prime_3 & \qquad\,\,\,\lambda^\prime_4  & \,\,\, 0\\  0 & \,\,\, \lambda^\prime_4 & \qquad \,\,\, \lambda^\prime_3 &  \,\,\, 0 \\  \lambda^\prime-\lambda^\prime_3-\lambda^\prime_4 & \,\,\,0 & \qquad \,\,\, 0 &  \,\,\, \lambda^\prime\end{pmatrix}.
\eeq
Indeed, \eq{Z-CPtransf} is satisfied in the GCP3 basis, independently of the value of~$\theta_{12}$.

Our final check involves confirming the validity of \eq{vevcond} in the GCP3 basis.   This computation will then determine the phase $\gamma^\prime$. Before performing the computation, we shall record an important result that is a consequence of the scalar potential minimum condition that is used to fix $\xi^\prime$.  In light of \eqs{cosbp}{cxi}, it follows that\footnote{\Eq{cxprime}  can also be deduced from \eq{sin2xi2} after making use of $\tan\xi=-\tan\theta_{12}$, which is a consequence of the scalar potential minimum condition, $\sin(\theta_{12}+\xi)=0$, obtained above \eq{thirdcond}.}
\beq \label{cxprime}
\cos\xi^\prime=-\frac{c_{12}c_{2\beta^\prime}}{s_{12}s_{2\beta^\prime}}\,,
\eeq
after employing \eq{csonetwo}.

Thus, we must verify
\beq \label{etap}
\begin{pmatrix} c_{\beta^\prime} \\ e^{i\xi^\prime} s_{\beta^\prime} \end{pmatrix}=-i e^{i\gamma^\prime}\begin{pmatrix} -s_{12} & \,\,\, c_{12} \\ \phm c_{12} & \,\,\, s_{12}\end{pmatrix}\begin{pmatrix} c_{\beta^\prime} \\ e^{-i\xi^\prime} s_{\beta^\prime} \end{pmatrix}
=\begin{pmatrix} -i e^{i\gamma^\prime}\bigl(-s_{12}c_{\beta^\prime}+c_{12}s_{\beta^\prime}e^{-i\xi^\prime}\bigr) \\
-i e^{i\gamma^\prime}\bigl(c_{12} c_{\beta^\prime}+s_{12}s_{\beta^\prime}e^{-i\xi^\prime}\bigr) \end{pmatrix}.
\eeq
We can immediately determine $\gamma^\prime$ from the equation for $c_{\beta^\prime}$ in \eq{etap},
\beq \label{etaprime}
-i e^{i\gamma^\prime}=-s_{12}+c_{12} e^{i\xi^\prime}\tan\beta^\prime\,.
\eeq
One can verify that $-s_{12}+c_{12} e^{i\xi^\prime}\tan\beta^\prime$ is a complex number of unit modulus after making use of \eq{cxprime}, which provides one independent check of the validity of \eq{etap}.
The explicit form of the residual GCP symmetry in the GCP3 basis has now been fixed.

Finally, we must verify the second complex equation for $e^{i\xi^\prime} s_{\beta^\prime}$ given in \eq{etap},
\beq \label{seq}
s_{\beta^\prime}=-i e^{i(\gamma^\prime-\xi^\prime)}\bigl(c_{12} c_{\beta^\prime}+s_{12}s_{\beta^\prime}e^{-i\xi^\prime}\bigr) \,.
\eeq
Straightforward algebra shows that \eq{seq} is an identify after making use of \eqs{cxprime}{etaprime} to eliminate $\cos {\xi^\prime}$ and $e^{i\gamma^\prime}$. 

Thus, we have verified that the scalar potential and vacuum of the softly broken GCP3-symmetric 2HDM are invariant with respect to a residual GCP transformation with matrix
\beq
e^{i\gamma^\prime}X^\prime=\bigl(c_{12}e^{i\xi^\prime}\tan\beta^\prime-s_{12}\bigr)\begin{pmatrix} -s_{12} & \,\,\, c_{12} \\ \phm c_{12} & \,\,\, s_{12}\end{pmatrix},
\eeq
where $s_{12}$ and $c_{12}$ are given by \eq{csonetwo} and $\beta^\prime$ and $\xi^\prime$ are determined by the GCP3 scalar potential parameters as indicated in \eqst{cosbp}{sxi}.
Note that although the form of $e^{i\gamma^\prime}X^\prime$ depends on the parameters of the softly broken GCP3 scalar potential, our calculation demonstrates that for any choice of the parameters (and in particular for any choice of the parameters that softly breaks the GCP3 symmetry), there exists a residual GCP symmetry characterized by the matrix $e^{i\gamma^\prime}X^\prime$.

These results are not surprising given that we knew from the beginning that the softly broken GCP3-symmetric scalar potential is equivalent to a softly broken U(1)$\otimes\Pi_2$-symmetric scalar potential where the residual CP symmetry transformation law is identified as GCP1, after removing all complex phases from the coefficients of the scalar potential parameters by an appropriate rephasing of the scalar doublet fields.   Nevertheless, it is satisfying to explicitly identify the residual GCP symmetry of the softly broken GCP3-symmetric scalar potential independently of the relations between U(1)$\otimes\Pi_2$ basis and the GCP3 basis obtained in Section~\ref{transforming}.

\section{\texorpdfstring{Scalar squared mass matrices in the \texorpdfstring{$\Phi$}--basis}{Scalar squared mass matrices in the \uppercasePhi--basis}}
\label{app:sqmass}
\renewcommand{\theequation}{D.\arabic{equation}}
\setcounter{equation}{0}

In Sections~\ref{sec:UPibasis} and \ref{sec:GCP3basis}, the neutral scalar squared mass matrices were evaluated in the Higgs basis.   Of course, the same scalar squared masses can be obtained by computing the eigenvalues of the neutral scalar squared mass matrices evaluated in the $\Phi$-basis (under the assumption that $s_{2\beta}\neq 0$).   This computation provides a check of the results obtained in 
Sections~\ref{sec:UPibasis} and \ref{sec:GCP3basis}.  

For example, for a softly broken $\ug\otimes\Pi_2$-symmetric scalar potential, the calculation of the eigenvalues of the  neutral scalar squared-mass matrix is most easily done after rephasing $m_{12}^2$ as described below \eq{betaeq}.
In this case, one obtains $m_A^2=2m_{12}^2/s_{2\beta}$ and the
squared masses of $h$ and $H$ (with $m_H\leq m_H$) correspond~to
the eigenvalues of the  $2\times 2$ matrix,
\beq
\mathcal{M}_H^2=\begin{pmatrix} m_A^2 s_\beta^2+\lambda v^2 c_\beta^2 & \quad -s_\beta c_\beta\bigl[m_A^2-(\lambda_3+\lambda_4)v^2\bigr] \\
-s_\beta c_\beta\bigl[m_A^2-(\lambda_3+\lambda_4)v^2\bigr] & \quad m_A^2 c_\beta^2+\lambda v^2 s_\beta^2 \end{pmatrix}\,,
\eeq
with respect to the $\bigl\{\sqrt{2}\Re\Phi_1^0-vc_\beta\,,\, \sqrt{2}\Re\Phi_2^0-vs_\beta\bigr\}$ basis.
One can verify that \eqs{ineq1}{ineq2} are satisfied, as these equations are independent of the choice of scalar field basis. 

For a softly broken GCP3-symmetric scalar potential (where parameters and fields are denoted with prime superscripts), the computation of the neutral scalar squared-mass matrix in the $\Phi^\prime$-basis (where $s_{2\beta^\prime}\neq 0$) is more challenging.
After employing the GCP3 condition, $\lambda_3^\prime+\lambda_4^\prime=\lambda^\prime-\lambda_5^\prime$, the $4\times 4$ neutral scalar squared-mass matrix is given by, 
\beqa
\mathcal{M}_{N4}^2&=&\left(\begin{matrix}   \frac{s_{\beta^\prime}}{c_{\beta^\prime}}R^\prime+\lambda^\prime v^2 c_{\beta^\prime}^2 & \quad - \cos\xi^\prime\bigl[R^\prime-(\lambda^\prime-\lambda_5^\prime\sin^2\xi^\prime)v^2  s_{\beta^\prime}c_{\beta^\prime}\bigr]  \\ 
- \cos\xi^\prime\bigl[R^\prime-(\lambda^\prime-\lambda_5^\prime\sin^2\xi^\prime)v^2  s_{\beta^\prime}c_{\beta^\prime}\bigr] & \quad \frac{c_{\beta^\prime}}{s_{\beta^\prime}}R^\prime+(\lambda^\prime  s_\beta^2\cos^2\xi^\prime+\lambda_5^\prime c_\beta^2\sin^2\xi^\prime)v^2 \\
\half \lambda_5^\prime v^2 s_{\beta^\prime}^2 \sin 2\xi^\prime & \quad  \sin\xi^\prime\bigl[R^\prime+\lambda_5^\prime v^2 s_{\beta^\prime}c_{\beta^\prime}\sin^2\xi^\prime\bigr] \\
-\sin\xi^\prime\bigl[R^\prime-(L'-\lambda_5^\prime\cos^2\xi^\prime)v^2 s_{\beta^\prime}c_{\beta^\prime} \bigr] & \quad
 \half \lambda^\prime v^2 s_{\beta^\prime}^2\sin 2\xi^\prime \end{matrix}\right. \nonumber \\[15pt]
 && \left.
 \begin{matrix}  \half \lambda_5^\prime v^2 s^2_{\beta^\prime}\sin 2\xi^\prime & \quad -\sin\xi^\prime\bigl[R^\prime-(L'-\lambda_5^\prime\cos^2\xi^\prime)v^2 s_{\beta^\prime}c_{\beta^\prime}\bigr] \\  \sin\xi^\prime\bigl[R^\prime+\lambda_5^\prime v^2 s_{\beta^\prime}c_{\beta^\prime}\sin^2\xi^\prime\bigr] & \quad \half \lambda^\prime v^2 s_{\beta^\prime}^2\sin 2\xi^\prime \\ 
\frac{s_{\beta^\prime}}{c_{\beta^\prime}}R^\prime-\lambda_5^\prime v^2 s_{\beta^\prime}^2 \cos 2\xi^\prime & \quad 
-\cos\xi^\prime\bigl[ R^\prime-\lambda_5^\prime v^2 s_{\beta^\prime}c_{\beta^\prime}\cos^2\xi^\prime\bigr] \\
  -\cos\xi^\prime\bigl[R^\prime-\lambda_5^\prime v^2 s_{\beta^\prime}c_{\beta^\prime}\cos^2\xi^\prime\bigr] & \quad \frac{c_{\beta^\prime}}{s_{\beta^\prime}}R^\prime+(\lambda^\prime  s^2_{\beta^\prime} \sin^2\xi^\prime-\lambda_5^\prime c^2_{\beta^\prime}\cos^2\xi^\prime)v^2 \end{matrix}\right),
\eeqa
with respect to the {$\bigl\{\sqrt{2}\Re\Phi_1^{\prime 0}-vc_{\beta^\prime}, \sqrt{2}\Re\Phi_2^{\prime 0}-vs_{\beta^\prime}\cos\xi^\prime, \sqrt{2}\Im\Phi_1^{\prime 0}, \sqrt{2}\Im\Phi_2^{\prime 0}-vs_{\beta^\prime}\sin\xi^\prime\bigr\}$} basis, where
\beq
R^\prime\equiv  \Re(m_{12}^{\prime\,2}e^{i\xi^\prime})\,,\qquad \quad L^\prime\equiv \lambda^\prime-2\lambda_5^\prime\sin^2\xi^\prime\,. 
\eeq

The next step is to identify the neutral Goldstone boson, which resides in the Higgs basis field $\mathcal{H}_1$, and 
corresponds to the eigenvector of $\mathcal{M}_{N4}^2$ with zero eigenvalue,
\beq
\frac{1}{\sqrt{2}}\,G^0=\Im\mathcal{H}_1^0=-s_{\beta^\prime}\sin\xi^\prime\Re\Phi_2^{\prime 0}+c_{\beta^\prime}\Im\Phi_1^{\prime 0}+s_{\beta^\prime}\cos\xi^\prime\Im\Phi_2^{\prime 0}\,.
\eeq
Defining the real orthogonal matrix,
\beq
\mathcal{R}=\begin{pmatrix} 1 & \quad \phm  0 & \quad \phm 0 & \quad \phm 0 \\ 0 & \quad \phm \cos\xi^\prime& \quad \phm 0 & \quad \phm \sin\xi^\prime \\ 0 & \quad \phm c_{\beta^\prime}\sin\xi^\prime & \quad \phm  s_{\beta^\prime} & \quad -c_{\beta^\prime}\cos\xi^\prime \\
0 & \quad -s_{\beta^\prime}\sin\xi^\prime &\quad \phm c_{\beta^\prime} & \quad \phm s_{\beta^\prime} \cos\xi^\prime\end{pmatrix}\,,
\eeq
one then finds that $\mathcal{R}\mathcal{M}_{N4}^2 \mathcal{R}^{\T}$ is a matrix with respect to the rotated basis whose fourth row and column consists entirely of zeros (due to the Goldstone boson).   Removing the fourth row and fourth column yields a $3\times 3$ squared-mass matrix,
\beq \label{M2N3}
\mathcal{M}^2_{N3}=\begin{pmatrix} \frac{s_{\beta^\prime}}{c_{\beta^\prime}}R^\prime+\lambda^\prime v^2 c_{\beta^\prime}^2 & \quad - R^\prime+L^\prime v^2  s_{\beta^\prime}c_{\beta^\prime} & \quad \half \lambda_5^\prime v^2 s_{\beta^\prime}\sin 2\xi^\prime \\
- R^\prime+L^\prime v^2  s_{\beta^\prime}c_{\beta^\prime} & \quad \frac{c_{\beta^\prime}}{s_{\beta^\prime}}R^\prime+\lambda^\prime v^2 s_\beta^2
& \quad \half \lambda_5^\prime v^2 c_{\beta^\prime}\sin 2\xi^\prime \\ \half \lambda_5^\prime v^2 s_{\beta^\prime}\sin 2\xi^\prime & \quad \half \lambda_5^\prime v^2 c_{\beta^\prime}\sin 2\xi^\prime & \quad  \frac{R^\prime}{s_{\beta^\prime} c_{\beta^\prime}}-\lambda_5^\prime v^2\cos 2\xi^\prime\end{pmatrix},
\eeq
with respect to the {$\bigl\{\sqrt{2}\Re\Phi_1^{\prime 0}-vc_{\beta^\prime}, \sqrt{2}\Re(e^{-i\xi^\prime}\Phi_2^{\prime 0})-vs_{\beta^\prime}, \sqrt{2}\bigl[s_{\beta^\prime}\Im\Phi_1^{\prime 0}-c_{\beta^\prime}\Im(e^{-i\xi^\prime}\Phi_2^{\prime 0})\bigr]\bigr\}$} basis.

The normalized eigenstate corresponding to $A$ is given by,
\beq \label{Aeigen}
A=\frac{1}{\sqrt{1-s^2_{2\beta^\prime}\sin^2\xi^\prime}}\begin{pmatrix} \phm\! s_{\beta^\prime}\cos\xi^\prime \\ -c_{\beta^\prime} \cos\xi^\prime \\ -c_{2\beta^\prime}\sin\xi^\prime\end{pmatrix}\,,
\eeq
with corresponding eigenvalue,
\beq \label{mhatwo}
m_A^2=\frac{\Re(m_{12}^{\prime\,2}e^{i\xi^\prime})}{s_{\beta^\prime}c_{\beta^\prime}}+\lambda_5^\prime v^2\sin^2\xi^\prime\,,
\eeq
in agreement with \eq{masq}.  In particular,
\beqa
A&=&
(1-s^2_{2\beta^\prime}\sin^2\xi^\prime)^{-1/2}\sqrt{2}
\bigl\{s_{\beta^\prime}\bigl(\cos\xi^\prime\Re\Phi_1^{\prime 0}-c_{2\beta^\prime}\sin\xi^\prime\Im\Phi_1^{\prime\,0}\bigr)
\nonumber \\
&& \qquad\qquad \qquad\qquad\qquad +c_\beta\bigl[c_{2\beta^\prime}\sin\xi^\prime\Im(e^{-i\xi^\prime}\Phi_2^{\prime 0})-\cos\xi^\prime\Re(e^{-i\xi^\prime}\Phi_2^{\prime 0})\bigr]\bigr\} \nonumber \\
&=&(1-s^2_{2\beta^\prime}\sin^2\xi^\prime)^{-1/2}\sqrt{2}\Im\bigl[i(\cos\xi^\prime+ic_{2\beta^\prime}\sin\xi^\prime)(s_{\beta^\prime}\Phi_1^{\prime 0}-c_{\beta^\prime}e^{-i\xi^\prime}\Phi_2^{\prime 0})\bigr]\nonumber \\[3pt]
&=&\sqrt{2}\Im\bigl[ie^{i\psi}\bigl(s_{\beta^\prime}\Phi_1^{\prime 0}-c_{\beta^\prime}e^{-i\xi^\prime}\Phi_2^{\prime 0})\bigr]=
 \sqrt{2}\Im\bigl[e^{i\eta^\prime}\bigl(-s_{\beta^\prime}e^{i\xi^\prime}\Phi_1^{\prime 0}+c_{\beta^\prime}\Phi_2^{\prime 0}\bigr)\bigr]\nonumber \\[3pt]
&=& \sqrt{2}\Im\mathcal{H}_2^0\,, \label{A}
\eeqa
after making use of \eqss{invhiggs}{psidef}{etapr}.

One can then identify two other mutually orthogonal normalized vectors orthogonal to~$A$,
\beq \label{others}
\sqrt{2}\Re\mathcal{H}_1^0-v=\begin{pmatrix} c_{\beta^\prime} \\ s_{\beta^\prime} \\ 0\end{pmatrix}\quad \text{and} \quad  \sqrt{2}\Re\mathcal{H}_2^0=\frac{1}{\sqrt{1-s^2_{2\beta^\prime}\sin^2\xi^\prime}}\begin{pmatrix} -s_{\beta^\prime}c_{2\beta^\prime} \sin\xi^\prime \\ \phm c_{\beta^\prime}c_{2\beta^\prime} \sin\xi^\prime \\ -\cos\xi^\prime\end{pmatrix},
\eeq
where the identification of the vectors above with the Higgs basis fields follows the same procedure that 
yielded \eq{A}.  Hence, if we construct
 a $3\times 3$ real orthogonal  matrix $\mathcal{O}$ whose rows are given by the transposes of the column vectors exhibited in \eqs{others}{Aeigen}, respectively, then it is straightforward to 
 verify that $\mathcal{O}\mathcal{M}^2_{N3}\mathcal{O}^{\T}$ is a block diagonal 
 matrix with respect to the $\bigl\{\sqrt{2}\Re\mathcal{H}_1^0-v, \sqrt{2}\Re\mathcal{H}_2^0,\sqrt{2}\Im\mathcal{H}_2^0\bigr\}$ basis. 
 The upper $2\times 2$~block of 
$\mathcal{O}\mathcal{M}^2_{N3}\mathcal{O}^{\T}$ can be identified with the $2\times 2$ CP-even neutral scalar squared-mass matrix,
\beq \label{block}
\mathcal{M}^2_H=\begin{pmatrix} (\lambda^\prime -\lambda^\prime_5 s_{2\beta^\prime}^2\sin^2\xi^\prime)v^2 & \quad\,\,\,  -\lambda_5^\prime v^2 s_{2\beta^\prime}\sin\xi^\prime(1-s^2_{2\beta^\prime}\sin^2\xi^\prime)^{1/2}  \\ -\lambda_5^\prime v^2 s_{2\beta^\prime}\sin\xi^\prime(1-s^2_{2\beta^\prime}\sin^2\xi^\prime)^{1/2}  &  \quad\,\,\,
m_A^2-\lambda_5^\prime v^2(1-s^2_{2\beta^\prime}\sin^2\xi^\prime) \end{pmatrix},
\eeq
which reproduces the result of \eq{emhtwo}, and the squared mass of the CP-odd scalar, $m_A^2$ [which is given by \eq{mhatwo}], is the
33 element of $\mathcal{O}\mathcal{M}^2_{N3}\mathcal{O}^{\T}$.

Clearly this is not the preferred method for computing the squared masses of the neutral scalars that derive from a softly broken GCP3-symmetric scalar potential, in light of
the much simpler Higgs basis computation given in Section~\ref{sec:GCP3basis}.

\section{\texorpdfstring{The IDM in the $\Phi$-basis}{The IDM in the \uppercasePhi--basis}}
\label{app:IDM}
\renewcommand{\theequation}{E.\arabic{equation}}
\setcounter{equation}{0}

The inert doublet model (IDM) can be defined as a 2HDM in which the 2HDM Lagrangian and vacuum are invariant under a $\mathbb{Z}_2$ symmetry, $\mathcal{H}_1\to \mathcal{H}_1$, $\mathcal{H}_2\to -\mathcal{H}_2$, in the Higgs basis.  In particular, $\mathcal{H}_2$ is odd under the $\mathbb{Z}_2$ symmetry, whereas all other fields of the 2HDM (i.e.,~$\mathcal{H}_1$, the gauge bosons, and the fermions) are even under the $\mathbb{Z}_2$ symmetry.   Of course, one is free to transform the scalar field basis from the Higgs basis to an arbitrary $\Phi$-basis by employing the unitary matrix $U$ given in \eq{U}.   

Suppose one is given a 2HDM scalar potential in a  $\Phi$-basis, where the vevs of the scalar fields yield $\tan\beta=|\vev{\Phi_2^0}/\vev{\Phi_1^0}|$ and $\xi=\arg\bigl[\vev{\Phi_1^0}^*\vev{\Phi_2^0}\bigr]$.   What are the conditions on the scalar potential parameters that imply that the model under consideration is the IDM?   To answer this question, we start in the Higgs basis with $Y_3=Z_6=Z_7=0$, as mandated by the $\mathbb{Z}_2$ symmetry.  Employing eqs. (A18)--(A28) of Ref.~\cite{Boto:2020wyf},
it then follows that in the  
$\Phi$-basis, the scalar potential parameters must satisfy the following conditions,
\beqa
 \Im(m_{12}^2 e^{i\xi})&=&0\,, \label{idminert1}\\
(m_{22}^2-m_{11}^2)s_{2\beta} &=& 2\Re(m_{12}^2 e^{i\xi})c_{2\beta}\,,\label{idminert2}  \\
c_{4\beta}\Re\bigl[(\lambda_6-\lambda_7)e^{i\xi}\bigr]&=&\half s_{2\beta} c_{2\beta}\bigl[\lambda_1+\lambda_2-2\bigl(\lambda_3+\lambda_4+\Re(\lambda_5 e^{2i\xi})\bigr)\bigr]\,,\label{idminert3}  \\
c_{2\beta}\Im\bigl[(\lambda_6-\lambda_7)e^{i\xi}\bigr]&=&-s_{2\beta}\Im(\lambda_5 e^{2i\xi})\,, \label{idminert4} \\
c_{2\beta}\Re\bigl[(\lambda_6+\lambda_7)e^{i\xi}\bigr]&=&\half s_{2\beta}(\lambda_1-\lambda_2)\,, \label{idminert5} \\
\Im\bigl[(\lambda_6+\lambda_7)e^{i\xi}\bigr]&=&0\,. \label{idminert6}
\eeqa
Note that if the scalar potential parameters satisfy \eq{IDMconstraints} then \eqst{idminert1}{idminert6} yield $c_{2\beta}=\sin\xi=0$ as expected based on the corresponding scalar potential minimum conditions.

When written in a generic $\Phi$-basis, 
the form of the $\mathbb{Z}_2$ symmetry that is respected by the 
scalar potential and vacuum becomes somewhat obscure.  Nevertheless, 
it is straightforward to check that if \eqst{idminert1}{idminert6} are satisfied, then the scalar potential and vacuum are invariant with respect to the following discrete symmetry of order 2 in the $\Phi$-basis,
 \beq
 \mathbb{Z}_{2H}:\qquad \Phi_1\to c_{2\beta}\Phi_1+e^{-i\xi}s_{2\beta}\Phi_2\,,\qquad\quad \Phi_2\to e^{i\xi}s_{2\beta}\Phi_1-c_{2\beta}\Phi_2\,,
 \eeq
 which can be obtained from \eq{youoneH} by setting $\theta=\half\pi$ and applying a hypercharge U(1)$_{\rm Y}$ transformation to remove an overall factor of $-i$.  
 Note that the square of the $\mathbb{Z}_{2H}$ transformation is equal to the identity, as advertised.  As a final check, note that if we set $c_{2\beta}=\sin\xi=0$ [cf.~\eqs{IDMconstraints}{IDMbasis}], then we can identify $\mathbb{Z}_{2H}$ with $\Pi_2$. 
\end{appendices}


\begin{thebibliography}{99}

\bibitem{Aad:2012tfa}
G.~Aad \textit{et al.} [ATLAS Collaboration],
``Observation of a new particle in the search for the Standard Model Higgs boson with the ATLAS detector at the LHC,''
Phys. Lett. B \textbf{716}, 1 (2012)
[arXiv:1207.7214 [hep-ex]].

\bibitem{Chatrchyan:2012ufa}
S.~Chatrchyan \textit{et al.} [CMS Collaboration],
``Observation of a New Boson at a Mass of 125 GeV with the CMS Experiment at the LHC,''
Phys. Lett. B \textbf{716}, 30 (2012)
[arXiv:1207.7235 [hep-ex]].

\bibitem{Aad:2019mbh}
G.~Aad \textit{et al.} [ATLAS Collaboration],
``Combined measurements of Higgs boson production and decay using up to $80$ fb$^{-1}$ of proton-proton collision data at $\sqrt{s}=$ 13 TeV collected with the ATLAS experiment,''
Phys. Rev. D \textbf{101}, 012002 (2020)
[arXiv:1909.02845 [hep-ex]].

\bibitem{Sirunyan:2018koj}
A.M.~Sirunyan \textit{et al.} [CMS Collaboration],
``Combined measurements of Higgs boson couplings in proton\textendash{}proton collisions at $\sqrt{s}=13\,\text {Te}\text {V} $,''
Eur. Phys. J. C \textbf{79}, 421 (2019)
[arXiv:1809.10733 [hep-ex]].

\bibitem{CMS:2020gsy}
CMS Collaboration, ``Combined Higgs boson production and decay
  measurements with up to 137 fb$^{-1}$ of proton-proton collision data at
  $\sqrt{s}=13$ TeV,'' {CMS-PAS-HIG-19-005} (January, 2020).
  
\bibitem{ATLAS:2020qdt}
ATLAS Collaboration, ``A combination of measurements of Higgs boson production and decay using up to 139 fb$^{-1}$ of proton--proton collision data at $\sqrt{s}= 13$~TeV collected with the ATLAS experiment,'' {ATLAS-CONF-2020-027} (July, 2020).

\bibitem{Ross:1975fq}
D.A.~Ross and M.J.G.~Veltman,
``Neutral Currents in Neutrino Experiments,''
Nucl. Phys. B \textbf{95}, 135(1975).

\bibitem{Veltman:1977kh}
M.J.G.~Veltman,
``Limit on Mass Differences in the Weinberg Model,''
Nucl. Phys. B \textbf{123}, 89 (1977).

\bibitem{Chanowitz:1978uj}
M.S.~Chanowitz, M.A.~Furman and I.~Hinchliffe,
``Weak Interactions of Ultraheavy Fermions,''
Phys. Lett. B \textbf{78}, 285 (1978).

\bibitem{Toussaint:1978zm}
D.~Toussaint,
``Renormalization Effects From Superheavy Higgs Particles,''
Phys. Rev. D \textbf{18}, 1626 (1978).

\bibitem{Lee:1973iz}
T.D.~Lee,
``A Theory of Spontaneous T Violation,''
Phys. Rev. D \textbf{8}, 1226 (1973).

\bibitem{Gunion:1989we}
J.F.~Gunion, H.E.~Haber, G.~Kane and S.~Dawson,
\textit{The Higgs Hunter's Guide} (Westview Press, Boulder, CO, 2000).

\bibitem{Branco:2011iw}
G.C.~Branco, P.M.~Ferreira, L.~Lavoura, M.N.~Rebelo, M.~Sher and J.P.~Silva,
``Theory and phenomenology of two-Higgs-doublet models,''
Phys. Rept. \textbf{516}, 1 (2012)
[arXiv:1106.0034 [hep-ph]].

\bibitem{Glashow:1976nt}
S.L.~Glashow and S.~Weinberg,
``Natural Conservation Laws for Neutral Currents,''
Phys. Rev. D \textbf{15}, 1958 (1977).

\bibitem{Paschos:1976ay}
E.A.~Paschos,
``Diagonal Neutral Currents,''
Phys. Rev. D \textbf{15}, 1966 (1977).

\bibitem{WahabElKaffas:2007xd}
A.~Wahab El Kaffas, P.~Osland and O.M.~Ogreid,
``Constraining the Two-Higgs-Doublet-Model parameter space,''
Phys. Rev. D \textbf{76}, 095001 (2007)
[arXiv:0706.2997 [hep-ph]].

\bibitem{Arbey:2017gmh}
A.~Arbey, F.~Mahmoudi, O.~St\r{a}l and T.~Stefaniak,
``Status of the Charged Higgs Boson in Two Higgs Doublet Models,''
Eur. Phys. J. C \textbf{78} 182 (2018)
[arXiv:1706.07414 [hep-ph]].

\bibitem{Eberhardt:2020dat}
O.~Eberhardt, A.P.~Mart\'\i{}nez and A.~Pich,
``Global fits in the Aligned Two-Higgs-Doublet model,''
JHEP \textbf{05} (2021) 005 
[arXiv:2012.09200 [hep-ph]].

\bibitem{Fayet:1974pd}
P.~Fayet,
``Supergauge Invariant Extension of the Higgs Mechanism and a Model for the electron and Its Neutrino,''
Nucl. Phys. B \textbf{90}, 104 (1975).

\bibitem{Flores:1982pr}
R.A.~Flores and M.~Sher,
``Higgs Masses in the Standard, Multi-Higgs and Supersymmetric Models,''
Annals Phys. \textbf{148}, 95 (1983).

\bibitem{Haber:1984rc}
H.E.~Haber and G.L.~Kane,
``The Search for Supersymmetry: Probing Physics Beyond the Standard Model,''
Phys. Rept. \textbf{117}, 75 (1985).

\bibitem{Gunion:1984yn}
J.F.~Gunion and H.E.~Haber,
``Higgs Bosons in Supersymmetric Models (I),''
Nucl. Phys. B \textbf{272}, 1 (1986)
[erratum:~B \textbf{402}, 567 (1993)].

\bibitem{Susskind:1982mw}
L.~Susskind,
``The Gauge Hierarchy Problem, Technicolor, Supersymmetry, and All That,''
Phys. Rept. \textbf{104}, 181 (1984).

\bibitem{Barbieri:2006dq}
R.~Barbieri, L.J.~Hall and V.S.~Rychkov,
``Improved naturalness with a heavy Higgs: An Alternative road to LHC physics,''
Phys. Rev. D \textbf{74}, 015007 (2006)
[arXiv:hep-ph/0603188].

\bibitem{LopezHonorez:2006gr}
L.~Lopez Honorez, E.~Nezri, J.F.~Oliver and M.H.G.~Tytgat,
``The Inert Doublet Model: An Archetype for Dark Matter,''
JCAP \textbf{02}, 028 (2007)
[arXiv:hep-ph/0612275].

\bibitem{Ma:2006km}
E.~Ma,
``Verifiable radiative seesaw mechanism of neutrino mass and dark matter,''
Phys. Rev. D \textbf{73}, 077301 (2006)
[arXiv:hep-ph/0601225].

\bibitem{Arina:2009um}
C.~Arina, F.S.~Ling and M.H.G.~Tytgat,
``IDM and iDM or The Inert Doublet Model and Inelastic Dark Matter,''
JCAP \textbf{10}, 018 (2009)
[arXiv:0907.0430 [hep-ph]].

\bibitem{Goudelis:2013uca}
A.~Goudelis, B.~Herrmann and O.~St\r{a}l,
``Dark matter in the Inert Doublet Model after the discovery of a Higgs-like boson at the LHC,''
JHEP \textbf{09} (2013) 106 
[arXiv:1303.3010 [hep-ph]].

\bibitem{Davidson:2005cw}
S.~Davidson and H.E.~Haber,
``Basis-independent methods for the two-Higgs-doublet model,''
Phys. Rev. D \textbf{72}, 035004 (2005)
[erratum:~D \textbf{72}, 099902 (2005)]
[arXiv:hep-ph/0504050].

\bibitem{Ginzburg:2004vp} 
  I.F.~Ginzburg and M.~Krawczyk,
``Symmetries of two Higgs doublet model and CP violation,''
  Phys.\ Rev.\ D {\bf 72}, 115013 (2005)
    [arXiv:hep-ph/0408011].
 
\bibitem{Arhrib:2010ju} 
  A.~Arhrib, E.~Christova, H.~Eberl and E.~Ginina,
``CP violation in charged Higgs production and decays in the Complex Two Higgs Doublet Model,''
  JHEP {\bf 1104}  (2011) 089
   [arXiv:1011.6560 [hep-ph]].
  
\bibitem{Barroso:2012wz} 
  A.~Barroso, P.M.~Ferreira, R.~Santos and J.P.~Silva,
 ``Probing the scalar-pseudoscalar mixing in the 125 GeV Higgs particle with current data,''
  Phys.\ Rev.\ D {\bf 86}, 015022 (2012)
  [arXiv:1205.4247 [hep-ph]].
 
\bibitem{Inoue:2014nva} 
  S.~Inoue, M.J.~Ramsey-Musolf and Y.~Zhang,
``CP-violating phenomenology of flavor conserving two Higgs doublet models,''
  Phys.\ Rev.\ D {\bf 89}, 115023 (2014)
  [arXiv:1403.4257 [hep-ph]].
  
\bibitem{Fontes:2014xva} 
  D.~Fontes, J.C.~Rom\~{a}o and J.P.~Silva,
``$h \rightarrow Z \gamma$ in the complex two Higgs doublet model,''
  JHEP {\bf 1412} (2014) 043 
   [arXiv:1408.2534 [hep-ph]].
   
\bibitem{Grzadkowski:2014ada} 
  B.~Grzadkowski, O.M.~Ogreid and P.~Osland,
``Measuring CP violation in Two-Higgs-Doublet models in light of the LHC Higgs data,''
  JHEP {\bf 1411}  (2014) 084
   [arXiv:1409.7265 [hep-ph]].

\bibitem{Fontes:2017zfn} 
  D.~Fontes, M.~M\"uhlleitner, J.C.~Rom\~{a}o, R.~Santos, J.P.~Silva and J.~Wittbrodt,
  ``The C2HDM revisited,''
  JHEP {\bf 1802}  (2018) 073
   [arXiv:1711.09419 [hep-ph]].

\bibitem{Boto:2020wyf}
R.~Boto, T.V.~Fernandes, H.E.~Haber, J.C.~Rom\~ao and J.P.~Silva,
``Basis-independent treatment of the complex 2HDM,''
Phys. Rev. D \textbf{101}, 055023 (2020)
[arXiv:2001.01430 [hep-ph]].

\bibitem{Low:2020iua}
I.~Low, N.R.~Shah and X.P.~Wang,
``Higgs alignment and novel CP-violating observables in two-Higgs-doublet models,'
Phys. Rev. D \textbf{105}, 035009 (2022)
[arXiv:2012.00773 [hep-ph]].

\bibitem{Dev:2014yca}
P.S.~Bhupal Dev and A.~Pilaftsis,
``Maximally Symmetric Two Higgs Doublet Model with Natural Standard Model Alignment,''
JHEP \textbf{12}  (2014) 024
[erratum:~\textbf{11} (2015) 147]
[arXiv:1408.3405 [hep-ph]].

\bibitem{Dev:2017org}
P.S.~Bhupal Dev and A.~Pilaftsis,
``Natural Alignment in the Two Higgs Doublet Model,''
J. Phys. Conf. Ser. \textbf{873}, 012008 (2017)
[arXiv:1703.05730 [hep-ph]].

\bibitem{Ferreira:2009wh}
P.M.~Ferreira, H.E.~Haber and J.P.~Silva,
``Generalized CP symmetries and special regions of parameter space in the two-Higgs-doublet model,''
Phys. Rev. D \textbf{79}, 116004 (2009)
[arXiv:0902.1537 [hep-ph]].

\bibitem{Gunion:2002zf}
J.F.~Gunion and H.E.~Haber,
``The CP conserving two Higgs doublet model: The Approach to the decoupling limit,''
Phys. Rev. D \textbf{67}, 075019 (2003)
[arXiv:hep-ph/0207010 [hep-ph]].

\bibitem{Craig:2012vn}
N.~Craig and S.~Thomas,
``Exclusive Signals of an Extended Higgs Sector,''
JHEP \textbf{11}  (2012) 083
[arXiv:1207.4835 [hep-ph]].

\bibitem{Craig:2013hca}
  N.~Craig, J.~Galloway and S.~Thomas,
``Searching for Signs of the Second Higgs Doublet,''
  arXiv:1305.2424 [hep-ph].
  
\bibitem{Asner:2013psa}
  D.M.~Asner {\it et al.},
``ILC Higgs White Paper,''
  arXiv:1310.0763 [hep-ph].
  
\bibitem{Carena:2013ooa}
  M.~Carena, I.~Low, N.R.~Shah and C.E.M.~Wagner,
``Impersonating the Standard Model Higgs Boson:~Alignment without Decoupling,''
  JHEP {\bf 1404} (2014) 015
  [arXiv:1310.2248 [hep-ph]].
  
  \bibitem{Haber:2013mia}
  H.E.~Haber, in Proceedings of the of the Toyama International
  Workshop on Higgs as a Probe of New Physics 2013 (HPNP2013),  
``The Higgs data and the Decoupling Limit,''
  arXiv:1401.0152 [hep-ph].
  
 \bibitem{Sikivie:1980hm}
P.~Sikivie, L.~Susskind, M.B.~Voloshin and V.I.~Zakharov,
``Isospin Breaking in Technicolor Models,''
Nucl. Phys. B \textbf{173}, 189 (1980).

 \bibitem{Mannheim:1983ti}
P.D.~Mannheim,
``Effective Low-Energy Custodial Symmetry and Weinberg Mixing,''
Phys. Lett. B \textbf{125}, 282 (1983).

\bibitem{Pomarol:1993mu}
A.~Pomarol and R.~Vega,
``Constraints on CP violation in the Higgs sector from the $\rho$ parameter,''
Nucl. Phys. B \textbf{413}, 3 (1994)
[arXiv:hep-ph/9305272].

\bibitem{Gerard:2007kn}
J.M.~Gerard and M.~Herquet,
``A Twisted custodial symmetry in the two-Higgs-doublet model,''
Phys. Rev. Lett. \textbf{98}, 251802 (2007)
[arXiv:hep-ph/0703051].

\bibitem{Grzadkowski:2010dj}
B.~Grzadkowski, M.~Maniatis and J.~Wudka,
``The bilinear formalism and the custodial symmetry in the two-Higgs-doublet model,''\!
JHEP \textbf{11} (2011) 030
[arXiv:1011.5228 [hep-ph]].

\bibitem{Haber:2010bw}
H.E.~Haber and D.~O'Neil,
``Basis-independent methods for the two-Higgs-doublet model III:~The CP-conserving limit, custodial symmetry, and the oblique parameters $S$, $T$, $U$,''
Phys. Rev. D \textbf{83}, 055017 (2011)
[arXiv:1011.6188 [hep-ph]].

\bibitem{Aiko:2020atr}
M.~Aiko and S.~Kanemura,
``New scenario for aligned Higgs couplings originated from the twisted custodial symmetry at high energies,''
JHEP \textbf{02} (2021) 046
[arXiv:2009.04330 [hep-ph]].

\bibitem{Deshpande:1977rw}
N.G.~Deshpande and E.~Ma,
``Pattern of Symmetry Breaking with Two Higgs Doublets,''
Phys. Rev. D \textbf{18}, 2574 (1978).

\bibitem{Maniatis:2006fs}
M.~Maniatis, A.~von Manteuffel, O.~Nachtmann and F.~Nagel,
``Stability and symmetry breaking in the general two-Higgs-doublet model,''
Eur. Phys. J. C \textbf{48}, 805 (2006)
[arXiv:hep-ph/0605184 [hep-ph]].

\bibitem{Ivanov:2007de}
I.P.~Ivanov,
``Minkowski space structure of the Higgs potential in 2HDM. II. Minima, symmetries, and topology,''
Phys. Rev. D \textbf{77}, 015017 (2008)
[arXiv:0710.3490 [hep-ph]].

\bibitem{Ferreira:2010yh}
P.M.~Ferreira, H.E.~Haber, M.~Maniatis, O.~Nachtmann and J.P.~Silva,
``Geometric picture of generalized-CP and Higgs-family transformations in the two-Higgs-doublet model,''
Int. J. Mod. Phys. A \textbf{26}, 769 (2011)
[arXiv:1010.0935 [hep-ph]].

\bibitem{Battye:2011jj}
R.A.~Battye, G.D.~Brawn and A.~Pilaftsis,
``Vacuum Topology of the Two Higgs Doublet Model,''
JHEP \textbf{08} (2011) 020 
[arXiv:1106.3482 [hep-ph]].

\bibitem{Peccei:1977hh}
R.D.~Peccei and H.R.~Quinn,
``Constraints Imposed by CP Conservation in the Presence of Instantons,''
Phys. Rev. D \textbf{16}, 1791 (1977).

\bibitem{Haber:2012np}
H.E.~Haber and Z.~Surujon,
``A Group-theoretic Condition for Spontaneous CP Violation,''
Phys. Rev. D \textbf{86}, 075007 (2012)
[arXiv:1201.1730 [hep-ph]].

\bibitem{Ivanov:2005hg}
I.P.~Ivanov,
``Two-Higgs-doublet model from the group-theoretic perspective,''
Phys. Lett. B \textbf{632}, 360 (2006)
[arXiv:hep-ph/0507132].

\bibitem{AbrSte}
M.~Abramowitz and I.A.~Stegun,
\textit{Handbook of Mathematical Functions}\,
(Dover Publications Inc., New York, 1972).

\bibitem{Beyond}
R.~Irvine, \textit{Beyond the Quadratic Formula} (MAA Press, Providence, RI, 2013).

\bibitem{Donoghue:1978cj}
J.F.~Donoghue and L.F.~Li,
``Properties of Charged Higgs bosons,''
Phys. Rev. D \textbf{19}, 945 (1979).

\bibitem{Georgi:1978ri}
H.~Georgi and D.V.~Nanopoulos,
``Suppression of flavor changing effects from neutral spinless meson exchange in gauge theories,''
Phys. Lett. B \textbf{82}, 95 (1979).

\bibitem{Botella:1994cs}
F.J.~Botella and J.P.~Silva,
``Jarlskog--like invariants for theories with scalars and fermions,''
Phys. Rev. D \textbf{51}, 3870 (1995)
[arXiv:hep-ph/9411288].

\bibitem{Branco:1999fs}
G.C.~Branco, L.~Lavoura, and J.P.~Silva, \textit{CP Violation} (Oxford University Press, Oxford, UK, 1999).

\bibitem{Haber:2006ue}
H.E.~Haber and D.~O'Neil,
``Basis-independent methods for the two-Higgs-doublet model II:~The significance of $\tan\beta$,''
Phys. Rev. D \textbf{74}, 015018 (2006)
[erratum:~D \textbf{74}, 059905 (2006)]
[arXiv:hep-ph/0602242].

\bibitem{Draper:2020tyq}
P.~Draper, A.~Ekstedt and H.E.~Haber,
``A natural mechanism for approximate Higgs alignment in the 2HDM,''
JHEP \textbf{05} (2021) 235 
[arXiv:2011.13159 [hep-ph]].

\bibitem{Barroso:2007rr}
A.~Barroso, P.M.~Ferreira and R.~Santos,
``Neutral minima in two-Higgs doublet models,''
Phys. Lett. B \textbf{652}, 181 (2007)
[arXiv:hep-ph/0702098].

\bibitem{Martin:1997ns}
S.P.~Martin,
``A Supersymmetry primer,''
Adv. Ser. Direct. High Energy Phys. \textbf{18}, 1 (1998)
[arXiv:hep-ph/9709356].

\bibitem{Weldon:1984wt}
H.A.~Weldon,
``The Effects of Multiple Higgs Bosons on Tree Unitarity,''
Phys. Rev. D \textbf{30}, 1547 (1984).

\bibitem{Ginzburg:2005dt}
I.F.~Ginzburg and I.P.~Ivanov,
``Tree-level unitarity constraints in the most general 2HDM,''
Phys. Rev. D \textbf{72}, 115010 (2005)
[arXiv:hep-ph/0508020].

\bibitem{Horejsi:2005da}
J.~Horejsi and M.~Kladiva,
``Tree-unitarity bounds for THDM Higgs masses revisited,''
Eur. Phys. J. C \textbf{46}, 81 (2006)
[arXiv:hep-ph/0510154].

\bibitem{Kanemura:2015ska}
S.~Kanemura and K.~Yagyu,
``Unitarity bound in the most general two Higgs doublet model,''
Phys. Lett. B \textbf{751}, 289 (2015)
[arXiv:1509.06060 [hep-ph]].

\bibitem{Goodsell:2018fex}
M.D.~Goodsell and F.~Staub,
``Improved unitarity constraints in Two-Higgs-Doublet-Models,''
Phys. Lett. B \textbf{788}, 206 (2019)
[arXiv:1805.07310 [hep-ph]].

\bibitem{Ferreira:2010jy}
P.M.~Ferreira, H.E.~Haber and J.P.~Silva,
``Basis invariant conditions for supersymmetry in the two-Higgs-doublet model,''
Phys. Rev. D \textbf{82}, 016001 (2010)
[arXiv:1004.3292 [hep-ph]].

\bibitem{Draper:2016pys}
P.~Draper and H.~Rzehak,
``A Review of Higgs Mass Calculations in Supersymmetric Models,''
Phys. Rept. \textbf{619}, 1 (2016)
[arXiv:1601.01890 [hep-ph]].

\bibitem{Haber:1993an}
H.E.~Haber and R.~Hempfling,
``Renormalization group improved Higgs sector of the minimal supersymmetric model,''
Phys. Rev. D \textbf{48}, 4280 (1993)
[arXiv:hep-ph/9307201].

\bibitem{Haber:1989xc}
H.E.~Haber and Y.~Nir,
``Multiscalar Models With a High-Energy Scale,''
Nucl. Phys. B \textbf{335}, 363 (1990).

\bibitem{Bernon:2015qea}
J.~Bernon, J.F.~Gunion, H.E.~Haber, Y.~Jiang and S.~Kraml,
``Scrutinizing the alignment limit in two-Higgs-doublet models, Part 1: $m_h=125$~GeV,''
Phys. Rev. D \textbf{92}, 075004 (2015)
[arXiv:1507.00933 [hep-ph]].

\bibitem{Haber:2015pua}
H.E.~Haber and O.~St\r{a}l,
``New LHC benchmarks for the CP-conserving two-Higgs-doublet model,''\!
Eur.~Phys.~J.~C \textbf{75}, 491 (2015)\!
[erratum:~\!C \textbf{76}, 312 (2016)]\!\!
[arXiv:1507.04281~[hep-ph]].

\bibitem{Bernon:2015wef}
J.~Bernon, J.F.~Gunion, H.E.~Haber, Y.~Jiang and S.~Kraml,
``Scrutinizing the alignment limit in two-Higgs-doublet models, Part 2: $m_H=125 $~GeV,''
Phys. Rev. D \textbf{93}, 035027 (2016)
[arXiv:1511.03682 [hep-ph]].

\bibitem{Darvishi:2020teg}
N.~Darvishi and A.~Pilaftsis,
``Natural Alignment in Multi-Higgs Doublet Models,''
PoS \textbf{CORFU2019}, 064 (2020)
[arXiv:2004.04505 [hep-ph]].

\bibitem{tHooft:1979rat}
G.~'t Hooft,
``Naturalness, chiral symmetry, and spontaneous chiral symmetry breaking,''
 in \textit{Recent Developments in Gauge Theories}, NATO Advanced Study Institute series: Series B, Physics; volume 59, edited by G.~'t Hooft et al. (Plenum Press, New York and London, 1980) pp.~135--157.
 
\bibitem{Pilaftsis:2011ed}
A.~Pilaftsis,
``On the Classification of Accidental Symmetries of the Two Higgs Doublet Model Potential,''
Phys. Lett. B \textbf{706}, 465 (2012)
[arXiv:1109.3787 [hep-ph]].

\bibitem{Darvishi:2019dbh}
N.~Darvishi and A.~Pilaftsis,
``Classifying Accidental Symmetries in Multi-Higgs Doublet Models,''
Phys. Rev. D \textbf{101}, 095008 (2020)
[arXiv:1912.00887 [hep-ph]].

\bibitem{Ferreira:2010bm}
P.M.~Ferreira and J.P.~Silva,
``A Two-Higgs Doublet Model With Remarkable CP Properties,''
Eur. Phys. J. C \textbf{69}, 45 (2010)
[arXiv:1001.0574 [hep-ph]].

\bibitem{future}
H.E.~Haber and J.P.~Silva,
in preparation.

\bibitem{Darvishi:2019ltl}
N.~Darvishi and A.~Pilaftsis,
``Quartic Coupling Unification in the Maximally Symmetric 2HDM,''
Phys. Rev. D \textbf{99}, 115014 (2019)
[arXiv:1904.06723 [hep-ph]].

\bibitem{Draper:2016cag}
P.~Draper, H.E.~Haber and J.T.~Ruderman,
``Partially Natural Two Higgs Doublet Models,''
JHEP \textbf{06} (2016) 124 
[arXiv:1605.03237 [hep-ph]].

\bibitem{Antoniadis:2006uj}
I.~Antoniadis, K.~Benakli, A.~Delgado and M.~Quiros,
``A New gauge mediation theory,''
Adv. Stud. Theor. Phys. \textbf{2}, 645 (2008)
[arXiv:hep-ph/0610265 [hep-ph]].

\bibitem{Ellis:2016gxa}
J.~Ellis, J.~Quevillon and V.~Sanz,
``Doubling Up on Supersymmetry in the Higgs Sector,''
JHEP \textbf{10} (2016) 086
[arXiv:1607.05541 [hep-ph]].

\bibitem{Benakli:2018vqz}
K.~Benakli, M.D.~Goodsell and S.L.~Williamson,
``Higgs alignment from extended supersymmetry,''
Eur. Phys. J. C \textbf{78}, 658 (2018)
[arXiv:1801.08849 [hep-ph]].

\bibitem{Benakli:2018vjk}
K.~Benakli, Y.~Chen and G.~Lafforgue-Marmet,
``R-symmetry for Higgs alignment without decoupling,''
Eur. Phys. J. C \textbf{79}, 172 (2019)
[arXiv:1811.08435 [hep-ph]].

\bibitem{Lane:2018ycs}
K.~Lane and W.~Shepherd,
``Natural stabilization of the Higgs boson\textquoteright{}s mass and alignment,''
Phys. Rev. D \textbf{99}, 055015 (2019)
[arXiv:1808.07927 [hep-ph]].

\bibitem{Eichten:2021qbm}
E.J.~Eichten and K.~Lane,
``Higgs alignment and the top quark,''
Phys. Rev. D \textbf{103}, 115022 (2021)
[arXiv:2102.07242 [hep-ph]].

\bibitem{Branco:1980sz}
G.C.~Branco,
``Spontaneous CP Nonconservation and Natural Flavor Conservation: A Minimal Model,''
Phys. Rev. D \textbf{22}, 2901 (1980).

\bibitem{Branco:1985aq}
G.C.~Branco and M.N.~Rebelo,
``The Higgs Mass in a Model With Two Scalar Doublets and Spontaneous CP Violation,''
Phys. Lett. B \textbf{160}, 117 (1985).

\bibitem{Lin:1993pa}
C.~Lin, C.~Lee and Y.-W.~Yang,
``Spontaneous CP violation in the extended standard models,''
Chin. J. Phys. \textbf{32}, 41 (1994)
[arXiv:hep-ph/9311271].

\bibitem{Gilbert}
G.T.~Gilbert, 
``Positive Definite Matrices and Sylvester's Criterion,''
The American Mathematical Monthly, {\bf 98}, 44 (1991).

\bibitem{Gunion:2005ja}
J.F.~Gunion and H.E.~Haber,
``Conditions for CP-violation in the general two-Higgs-doublet model,''
Phys. Rev. D \textbf{72}, 095002 (2005)
[arXiv:hep-ph/0506227].

\bibitem{Kronecker}
See, e.g., S.R.~Garcia and R.A.~Horn, \textit{A Second Course in Linear Algebra} (Cambridge University Press, Cambridge, UK, 2017).

\bibitem{Ecker:1987qp}
G.~Ecker, W.~Grimus and H.~Neufeld,
``A Standard Form for Generalized CP Transformations,''
J. Phys. A \textbf{20}, L807 (1987).


\bibitem{Ecker:1981wv}
G.~Ecker, W.~Grimus and W.~Konetschny,
``Quark Mass Matrices in Left-right Symmetric Gauge Theories,''
Nucl. Phys. B \textbf{191}, 465 (1981).

\bibitem{Ecker:1983hz}
G.~Ecker, W.~Grimus and H.~Neufeld,
``Spontaneous CP Violation in Left-right Symmetric Gauge Theories,''
Nucl. Phys. B \textbf{247}, 70 (1984).

\bibitem{Neufeld:1987wa}
H.~Neufeld, W.~Grimus and G.~Ecker,
``Generalized CP Invariance, Neutral Flavor Conservation and the Structure of the Mixing Matrix,''
Int. J. Mod. Phys. A \textbf{03}, 603 (1988).

\bibitem{Branco:2005em}
G.C.~Branco, M.N.~Rebelo and J.I.~Silva-Marcos,
``CP-odd invariants in models with several Higgs doublets,''
Phys. Lett. B \textbf{614}, 187 (2005)
[arXiv:hep-ph/0502118].

\bibitem{Maniatis:2007vn}
M.~Maniatis, A.~von Manteuffel and O.~Nachtmann,
``CP violation in the general two-Higgs-doublet model:~A geometric view,''
Eur.~Phys.~J.~C \textbf{57}, 719 (2008)
[arXiv:0707.3344 [hep-ph]].

\bibitem{Maniatis:2007de}
M.~Maniatis, A.~von Manteuffel and O.~Nachtmann,
``A New type of CP symmetry, family replication and fermion mass hierarchies,''
Eur.~Phys.~J.~C \textbf{57}, 739 (2008)
[arXiv:0711.3760 [hep-ph]].

\bibitem{Maniatis:2009vp}
M.~Maniatis and O.~Nachtmann,
``On the phenomenology of a two-Higgs-doublet model with maximal CP symmetry at the LHC,''
JHEP \textbf{05} (2009) 028 
[arXiv:0901.4341 [hep-ph]].

\bibitem{Maniatis:2009by}
M.~Maniatis and O.~Nachtmann,
``On the phenomenology of a two-Higgs-doublet model with maximal CP symmetry at the LHC. II. Radiative effects,''
JHEP \textbf{04} (2010) 027
[arXiv:0912.2727 [hep-ph]].


\end{thebibliography}
\end{document}